\documentclass[a4paper,12pt,conditionalcopyright,final]{uq-thesis}
\newcommand{\Tr}{\mathop{\mathrm{Tr}}}

\newcommand{\p}[1]{\phantom{#1}}
\newcommand{\mrm}[1]{\mathrm{#1}}

\newcommand{\mbb}[1]{\mathbb{#1}}
\newcommand{\mc}[1]{\mathcal{#1}}
\newcommand{\tbf}[1]{\textbf{#1}}
\newcommand{\eref}[1]{(\ref{#1})}
\newcommand{\Eref}[1]{Eq.~(\ref{#1})}
\newcommand{\fref}[1]{Fig.~\ref{#1}}
\newcommand{\tref}[1]{Table~\ref{#1}}
\newcommand{\sref}[1]{Sec.~\ref{#1}}
\newcommand{\cref}[1]{Chapter~\ref{#1}}

\newcommand{\rmd}{\mathrm{d}}
\newcommand{\rmi}{\mathrm{i}}
\newcommand{\ra}{\rangle}
\newcommand{\la}{\langle}
\newcommand{\rbimg}[3]{\raisebox{#1}{\includegraphics[width=#2]{#3}}}

\newcommand{\be}{\begin{equation}}
\newcommand{\ee}{\end{equation}}

\newcommand{\ket} [1] {| #1 \rangle}
\newcommand{\bra} [1] {\langle #1 |}
\newcommand{\braket}[2]{\langle #1 | #2 \rangle}
\newcommand{\ketbra}[2]{|#1\rangle\langle#2|}
\newcommand{\ttr}{\mathrm{tTr}}
\newcommand{\TN}{tensor network}

\newcommand{\splitt}[3]{\Upsilon^{\mbox{\tiny \,split}}_{#1\rightarrow #2,#3}}
\newcommand{\fuse}[3]{\Upsilon^{\mbox{\tiny \,fuse}}_{#1, #2 \rightarrow #3}}
\newcommand{\fuser}{\Upsilon^{\mbox{\tiny \,fuse}}}
\newcommand{\splitter}{\Upsilon^{\mbox{\tiny \,split}}}
\newcommand{\rcite}[1]{\citet{#1}} % \onlinecite for prb
\newcommand{\prbtext}[1]{}
\newcommand{\pratext}[1]{#1}
\newcommand{\textcitecomma}[1]{\citet{#1},}
\newcommand{\citecomma}[1]{ \citep{#1},}
\newcommand{\citestop}[1]{ \citep{#1}.}
\newcommand{\Mop}{M_\alpha^{\p{\alpha}\beta}}
\newcommand{\alphabeta}{_\alpha^{\p{\alpha}\beta}}

\usepackage{amssymb}	% AMS symbols package
\usepackage{amsmath}	% AMS extended maths command set
\usepackage{amsthm}		% Creation of "theorem" environments (e.g. Example 1:... \\ Example 2:...)
\usepackage{bm}			% Bold maths symbols
\usepackage{array}		% More sophisticated table column specifications - see http://en.wikibooks.org/wiki/LaTeX/Tables
\usepackage{epstopdf}	% Automatically convert eps figures to PDF if using PDFTeX (e.g. on Mac)
\usepackage{url}		% Adds URL command, e.g. \url{http://www.google.com/}, and improves 
\usepackage[sort&compress,round]{natbib}		% Better selection of cite commands (like in RevTeX). Lots of options to set up for this...
\usepackage[nottoc]{tocbibind}
\usepackage{bigstrut}	%Allows for manual control of spacing in arrays
\usepackage{booktabs}	% Package for creating better looking tables
\usepackage{titlesec}	% Provides commands used for fancy-looking section headings etc.
\usepackage{showkeys}	% Makes \label labels visible in the compiled document (except when
\usepackage[table]{xcolor}	% Permits colouring of cells in tables using the \cellcolor command.
\definecolor{darkgreen}{rgb}{0,0.55,0}
\usepackage[breaklinks=true,colorlinks=true,citecolor=darkgreen,pagebackref=true]{hyperref} % Coloured links - for viewing online
\usepackage{multirow}	% Allows tabular entries to span multiple rows
\usepackage[margin=15pt,font={small,sf},labelfont=bf,labelsep=endash]{caption}
\usepackage{hyphenat}	% Greater control of hyphenation, e.g. \nohyphens{} command.
\usepackage{rotating}	% Useful for rotated column labels in tables: Adds "sideways" environment.

\titleformat{\paragraph}[hang]{\bfseries\small}{}{0pt}{}[]

\newtheoremstyle{henrytheoremstyle}
  {\topsep}%      Space above
  {\topsep}%      Space below
  {\itshape}%         Body font
  {}%         Indent amount (empty = no indent, \parindent = para indent)
  {\bfseries}% Thm head font
  {:}%        Punctuation after thm head
  {.5em}%     Space after thm head: " " = normal interword space;
  {}%         Thm head spec (can be left empty, meaning `normal')

\theoremstyle{henrytheoremstyle}

\title{\vspace{1.6em}\Large{{Simulation of Anyons Using Symmetric Tensor Network Algorithms\vspace{.2em}\\}}}

\author{Robert N. C. Pfeifer}
\prevdegrees{M.Sc. (Physics), 2006, The University of Queensland} %\\B.M. B.S., 2001, Nottingham\\B.Med.Sci., 1999, Nottingham}
\division{}
\school{School of Mathematics and Physics}
\thesistype{Doctor of Philosophy}

\submitdate{May 2011} % If commented out, will default to the current month and year.
\copyrightyear{2011} % Ditto for copyright year

\hypersetup{
  pdfinfo={
    Title={Simulation of Anyons Using Symmetric Tensor Network Algorithms},
    Author={Robert N. C. Pfeifer},
    Keywords={tensor networks, scale invariance, conformal field theory, symmetry groups, anyons}
  }
}

\tablespagetrue % Include tables list page
\figurespagetrue % Include figures list page
\nomenclaturepagetrue  % Include nomenclature page & change default title. You will need to run the command "makeindex thesis.nlo -s nomencl.ist -o thesis.nls".

\begin{document}

% Include parts using \include. Single chapters can then be compiled using \includeonly.
% Note a \included file always begins on a new page. Also, because the thesis class is based on
% the "book" type, new chapters (created using \chapter) always begin on the right hand side of a 
% pair of facing pages.

% --- Title page, preface text, abstract, contents, etc.
% ----------------- Begin TITLE --------------->>

\beforepreface

% ----------------- End TITLE --------------->>

% ----------------- ORIGINALITY --------------->>

\originalitystatement

\textbf{\nohyphens{Statement of Contributions to Jointly Authored Works Contained in this Thesis}}

\emph{\cref{sec:SIMERA}:}~The material presented in this Chapter is based on the paper of \citet*{pfeifer2009}. The Ansatz and algorithm for optimising the scale-invariant MERA were developed in collaboration between myself, Dr.~Glen Evenbly, and Prof.~Guifr\'{e} Vidal, as were the techniques for extracting conformal data. Implementation in \textsc{matlab} code was performed by myself and Dr.~Glen Evenbly. The text and figures of this Chapter were prepared by myself, with the exception of \fref{fig:ch2:results} which was prepared by Dr.~Glen Evenbly. Numerical results in \sref{sec:ch2:results}, including those in \fref{fig:ch2:results} and \tref{tab:ch2:results}, were computed by Dr.~Glen Evenbly.

\emph{\cref{sec:abelian}:}~Section~\ref{sec:ch3:globalsym} of this Chapter has previously been published as \citet*{singh2010a}. 
The theoretical framework presented in this paper was developed mainly by Sukhwinder Singh and Prof.~Guifre Vidal. I joined the effort when the 
project 
was at an advanced stage, and contributed ideas important to the final presentation of the framework.
The text of this paper was written by Prof.~Guifr\'{e} Vidal, with figures produced by Sukhwinder Singh with some assistance from myself. Sections~\ref{sec:ch3:U1MERA}--\ref{sec:ch3c:supplement} have previously been published as \citet*{singh2011a}. The text of this paper was predominantly written by Sukhwinder Singh and Prof.~Guifr\'{e} Vidal, with assistance from myself. Figures were produced by Sukhwinder Singh and Prof.~Guifr\'{e} Vidal, with assistance from myself. \textsc{Matlab} code to implement Abelian symmetries was written by myself, and equivalent software was also developed simultaneously by Sukhwinder Singh. I achieved the first demonstrated increase in performance from the implementation of an Abelian symmetry (the $Z_2$ symmetry implementation described in \sref{sec:ch3:comp}); the results presented in \sref{sec:ch3:U1MERA} for U(1) symmetry were computed by Sukhwinder Singh. 
Section~\ref{sec:ch3:comp} was written and illustrated by myself, and relates specifically to my own implementation of Abelian symmetries for tensor networks.

\emph{\cref{sec:anyons}:}~Sections~\ref{sec:ch4:intro}--\ref{sec:ch4:summary} of this Chapter have previously been published as \citet*{pfeifer2010}. Dr.~Miguel Aguado and Dr.~Oliver Buerschaper were involved in proving that the idea was viable and self-consistent, and helped Prof.~Guifr\'{e} Vidal and myself understand the unitary braided tensor category formalism for anyon models. I developed the anyonic tensor network notation, matrix representation of anyonic operators, the formalism for anyonic tensor networks, and the Ansatz and algorithm for the anyonic MERA, with oversight and guidance through discussions with Prof.~Guifr\'{e} Vidal. The text and diagrams of this Chapter are entirely my own work. I also implemented the formalism in \textsc{matlab} and computed and plotted the results presented in \sref{sec:ch4:results}.

\emph{\cref{sec:nonabelian}:}~The specific material presented in \cref{sec:nonabelian} is entirely my own work, though it reflects ideas developed in discussion with Prof.~Guifr\'{e} Vidal and Sukhwinder Singh.

\textbf{\nohyphens{Statement of Contributions by Others to the Thesis as a Whole}}

The directions of research pursued in this Thesis were at the suggestion and under the supervision of Prof.~Guifr\'{e} Vidal.

\textbf{\nohyphens{Statement of Parts of the Thesis Submitted to Qualify for the Award of Another Degree}}

No parts of this Thesis have been submitted to qualify for the award of another degree.

\textbf{\nohyphens{Published Works by the Author Incorporated into the Thesis}}

Chapter~\ref{sec:SIMERA}, on the Scale-Invariant MERA, presents material previously published in \citeauthor*{pfeifer2009}, \emph{Physical Review A}, \textbf{79}, 040301, 2009, \copyright~(2009) by the American Physical Society. Figure~\ref{fig:ch2:results}, \tref{tab:ch2:results}, and the results presented in \sref{sec:ch2:results} are reproduced directly from this paper. However, the majority of the text has been substantially rewritten and expanded.

Section~\ref{sec:ch3:globalsym} of \cref{sec:abelian} has previously been published as \citeauthor*{singh2010a}, \emph{Physical Review A}, \textbf{82}, 050301, 2010, \copyright~(2010) by the American Physical Society.

Sections~\ref{sec:ch3:U1MERA}--\ref{sec:ch3c:supplement} of \cref{sec:abelian} have previously been published as \citeauthor*{singh2011a}, \emph{Physical Review B}, \textbf{83}, 115125, 2011, \copyright~(2011) by the American Physical Society.

Sections~\ref{sec:ch4:intro}--\ref{sec:ch4:summary} of \cref{sec:anyons} have previously been published as \citeauthor*{pfeifer2010}, \emph{Physical Review B}, \textbf{82}, 115126, 2010, \copyright~(2010) by the American Physical Society.

%\clearpage
\textbf{\nohyphens{Additional Published Works by the Author Relevant to the Thesis but not Forming Part of it}}

None.

% ----------------- ACKNOWLEDGEMENTS --------------->>

\chapter*{Acknowledgments}

As might be expected, a great many people have contributed to the development of this Thesis, both directly and indirectly. I would like to thank them all.

In particular, I would like to thank: Prof.~Guifr\'{e} Vidal, for his tireless supervision and guidance, for always being available to discuss matters when I realised I didn't understand something as well as I thought, and for being the driving force behind the University of Queensland Quantum Simulation Group. Prof.~Andrew Doherty and Dr.~Ian McCulloch, for their specialist knowledge, backup and support. Dr.~Philippe Corboz, Dr.~Luca Tagliacozzo, Dr.~Rom\'{a}n Or\'{u}s, Dr.~Glen Evenbly, Dr.~Andrew Ferris, and Sukhwinder Singh, for many interesting discussions, and productive collaborations. Everyone else whose participation in the Quantum Simulation Group at The University of Queensland made it into the lively and stimulating research centre which it proved to be.

I took a fairly indirect route into my Ph.D., first studying medicine, before moving to physics and to Australia in 2004, first for an M.Sc., then to enrol in my Ph.D. I would like to thank everybody who made this possible, including: Prof.~Halina Rubinsztein-Dunlop, the ever-energetic and supportive Head of School. Prof.~Norman Heckenberg who, as head of admissions, approved my enrolment in an M.Sc. despite my unusual background. Everybody in the physics department of The University of Queensland who has taught courses, supervised research, presented a seminar, or otherwise helped to make it the stimulating place it has proved to be again and again during my time there.

I would also like to thank everybody at the physics department of The University of Western Australia, but especially the Head of School Prof.~Ian McArthur, and my supervisors Prof.~Sergei Kuzenko and Prof.~David Blair. During my one-year interlude in Perth I learned about topology and differential geometry, expanded my knowledge of group and representation theory, general relativity, and quantum field theory, and generally acquired an education which is often the envy of my peers. If there is one thing I would like to recommend to any theoretical Ph.D. student, it would be this: To take a year to sit down and read widely over the tools of fundamental physics as we understand it today.

On a personal note, I would like to thank all of my friends and family, particularly my parents Ilona and Peter for being as excited as I am about the new direction my life has taken, and my uncle Nic Chantler for much wise advice dispensed over a partially-dismantled Landrover engine. If it weren't for your suggestion that I study physics here in Australia, my life might have taken a very different course. Thank you also to all of my fellow students in the physics department, past and present, for good company and stimulating discussions, and for helping to make The University of Queensland physics department the stimulating place it is.

And of course, thanks to the late Prof.~Thomas Parnell for the famous pitch drop experiment. I may yet complete my entire Ph.D. between the eighth and the ninth drop.

The author acknowledges the support of the Australian Research Council given to himself and his collaborators and supervisors (APA, FF0668731, DP0878830, DP1092513). This research was supported in part by the Perimeter Institute for Theoretical Physics.

% ----------------- ABSTRACT --------------->>

\chapter*{Abstract}

The study of anyons offers one of the most exciting challenges in contemporary physics. Anyons are exotic quasiparticles with non-trivial exchange statistics, which makes them difficult to simulate. However, they are of great interest as some species offer the prospect of a highly fault-tolerant form of universal quantum computation, and it has been suggested that the simplest such species may appear in the fractional quantum Hall state with filling fraction $\nu$ = 12/5. Despite the current strong interest in the development of practical quantum computing, our ability to study the collective behaviour of systems of anyons remains limited.

Meanwhile, tensor network algorithms are a relatively recent development in the field of condensed matter physics. They consist of Ans\"atze for the low-energy states of a lattice system, whose number of free parameters scale at most polynomially in the system size, and algorithms for their optimisation, manipulation, and analysis. However, many condensed matter systems possess a high degree of symmetry, which may be exploited to yield an even more efficient description of the low-energy subspace, and when I began work on my Thesis these algorithms (with the exception of DMRG) did not in general take advantage of these symmetries.

In this Thesis I develop a formalism, based on the frameworks of spin networks and category theory, whereby a tensor network acting on any lattice model exhibiting a mathematical structure corresponding to a Unitary Braided Tensor Category (UBTC) may be represented in a particularly compact and efficient manner corresponding to the exploitation of this structure. This permits the exploitation of global Abelian and non-Abelian internal group symmetries, both to facilitate the study of particular symmetry sectors of the model, and for computational gain. Furthermore, the formalism also naturally admits the study of models possessing non-trivial exchange statistics (e.g. fermions, Abelian anyons) and models possessing a UBTC structure which is not associated with a group (some Abelian and non-Abelian anyons), all for a computational cost polynomial in the system size.

In addition I also describe the development of a tensor network algorithm to exploit the spatial symmetry of scale invariance present in quantum critical lattice systems. The resulting Ansatz provides a remarkably efficient description of the low-energy subspace of an infinite quantum critical lattice model, naturally yielding the polynomial correlators typical of such a system, and providing easy access to the majority of the conformal data which describe its behaviour in the continuum limit. Combining this Ansatz with the UBTC formalism for tensor networks provides a demonstration of the flexibility of these techniques, computing the conformal data associated with the continuum limit of two non-Abelian anyonic quantum critical lattice models.

In summary, this Thesis provides a new Ansatz for the study of quantum critical lattice models, and a formalism permitting the exploitation of Abelian and non-Abelian symmetries of lattice models, allowing the analysis of many fermionic and anyonic systems in polynomial time.

No Ph.D. students were harmed during the making of this Thesis.

\textbf{Keywords:} tensor networks, scale invariance, conformal field theory, symmetry groups, anyons

\textbf{Australian and New Zealand Standard Research Classifications (ANZSRC):}
\\\p{0}020401 Condensed Matter Characterisation Technique Development (50\%)
\\\p{0}020603 Quantum Information, Computation and Communication (50\%).
%%%%%%%%%%%%%%%%%%%%%%%%%%%%%%%%%%%%%%%%%%%%%%%%%%%%%%%%%%%%%%%%%%%%%%%%%%%%%%%%%%%%%%%%%%%%%%
% Fill in text as appropriate
%%%%%%%%%%%%%%%%%%%%%%%%%%%%%%%%%%%%%%%%%%%%%%%%%%%%%%%%%%%%%%%%%%%%%%%%%%%%%%%%%%%%%%%%%%%%%%

%\chapter*{List of Abbreviations and Symbols}
%% Only if not generating this automatically using the makenomenclature package
%
%
%%%%%%%%%%%%%%%%%%%%%%%%%%%%%%%%%%%%%%%%%%%%%%%%%%%%%%%%%%%%%%%%%%%%%%%%%%%%%%%%%%%%%%%%%%%%%%%
%% Fill in text as appropriate
%%%%%%%%%%%%%%%%%%%%%%%%%%%%%%%%%%%%%%%%%%%%%%%%%%%%%%%%%%%%%%%%%%%%%%%%%%%%%%%%%%%%%%%%%%%%%%%

% ----------------- TABLES OF CONTENTS ETC. --------------->>

%\clearpage
%~
%\cleardoublepage

\afterpreface

% --- Main body of thesis:
\chapter{Introduction}

\section{How to Read This Thesis}

This Thesis presents a number of recent developments in the formalism of tensor network Ans\"atze and algorithms, in which the author played a leading role. Tensor network Ans\"atze (or simply ``tensor networks'') are mathematical tools which may be used to efficiently represent a portion of the Hilbert space of a quantum mechanical system, frequently either the ground state or the low-energy subspace with respect to a specified Hamiltonian, and tensor network algorithms are algorithms for the efficient construction and manipulation of these tensor networks. %
They are frequently used to calculate properties such as the low-energy spectra of lattice Hamiltonians, and the evolution of states as a function of time %
(see Secs.~\ref{sec:ch1:MERA} and \ref{sec:ch1:bgreading} for introductory citations).
However, for a given Hamiltonian the ability of a tensor network algorithm to accurately represent the low energy subspace depends upon a number of factors. Existing tensor network algorithms tend to favour Hamiltonians which may be written as local operators over a small number of adjacent sites on a 1D or 2D lattice, with other factors which determine whether a particular system may be efficiently analysed including the statistics of the system (e.g. bosonic, fermionic), and
whether the structure of the tensor network reflects the structure of entanglement in the states being studied \citep{evenblyvidalinprep}.

The developments presented in the following chapters greatly extend the range of systems to which tensor network techniques may be applied, including formalisms for the study of infinite systems (\cref{sec:SIMERA}), fermions (\cref{sec:abelian}), and anyons (\cref{sec:anyons}). They also expand the capabilities of existing tensor network Ans\"atze when analysing symmetric Hamiltonians (Chapters~\ref{sec:abelian} and \ref{sec:nonabelian}), and yield
substantial improvements in computational performance (Chapters~\ref{sec:SIMERA}---\ref{sec:nonabelian}).

Although continuous tensor network algorithms do exist \citep{verstraete2010}, in this Thesis I will consider only tensor networks for lattice models. With the exception of \cref{sec:SIMERA}, however, most of the results presented in this Thesis are completely general and may in principle be applied to any tensor network Ansatz or algorithm. However, when providing examples and demonstrations I will favour tensor networks of the Multi-scale Entanglement Renormalisation Ansatz (MERA) type. \nomenclature{\tbf{MERA}}{Multi-scale Entanglement Renormalisation Ansatz/Ans\"atze: A class of tensor network Ans\"atze designed for the representation of states close to the fixed points of a renormalisation group flow, including both lightly-entangled states (close to the ground state) and heavily-entangled states (close to a critical point). MERA in 1D and 2D may be encountered in the present literature.} I therefore begin with a brief review of tensor network Ans\"atze in general and the 1D MERA in particular.

As described in the Statements of Contributions, much of the work presented in this Thesis has previously been published in international peer-reviewed journals. Rather than re-invent the wheel, this material has largely been reproduced verbatim in the Thesis. 
As a result
the Thesis has a modular structure, where individual Chapters of the Thesis are essentially self-contained. Inevitably this approach comes at the expense of some repetition of background material, with individual Chapters, and sometimes Sections within those Chapters, often having their own introduction drawing the reader's attention to the relevant parts of this material. These are supplemented by material in the present Chapter, which provides context for the Thesis as a whole. Supplementary text at the beginnings or ends of the Chapters serves to bring out connections between the different research topics, and to place the individual research areas into a larger context. %

\textbf{A Note on the Use of Personal Pronouns in this Thesis}

In this Thesis, use has been made of both the singular personal pronoun (``I''), and the plural (``we''), with the latter being used in different contexts to indicate either the author and the reader, or the author and his collaborators. Choice of personal pronoun should not therefore be treated as an indicator of whether work was performed independently or in collaboration. This information may be found in the Statements of Contributions in the preface to this text.

\section{A General Introduction to Tensor Networks}

The idea that the state of a lattice model may be represented by a network of tensors may be motivated as follows: First, consider an $n$-site lattice $\mc{L}_0$. If each site of this lattice is described by a $d$-dimensional Hilbert space $\mc{H}_\mrm{site}$, then the Hilbert space of the $n$-site lattice is $\mc{H}_\mrm{lattice}=(\mc{H}_\mrm{site})^{\otimes n}$. We may write a general state $|\psi\ra$ on lattice $\mc{L}_0$ as
\begin{equation}
|\psi\ra = \sum_{i_1,\ldots,i_n=1}^d c_{i_1\ldots i_n} |i_1,\ldots,i_n\ra\label{eq:ch1:state}
\end{equation}
where $|1\ra\ldots |d\ra$ constitute an orthonormal basis of $\mc{H}_\mrm{site}$, and $|i_1,\ldots,i_n\ra=|i_1\ra\otimes|i_2\ra\otimes\ldots\otimes|i_n\ra$. For a fixed basis of $\mc{H}_\mrm{site}$, a state of the lattice may be entirely specified by giving the tensor $c_{i_1\ldots i_n}$. It is convenient to introduce a graphical notation whereby a tensor with $x$ indices may be represented by a blob with $x$ legs, so that $c_{i_1\ldots i_n}$ is represented graphically by
\begin{equation}
\rbimg{-35pt}{246.0pt}{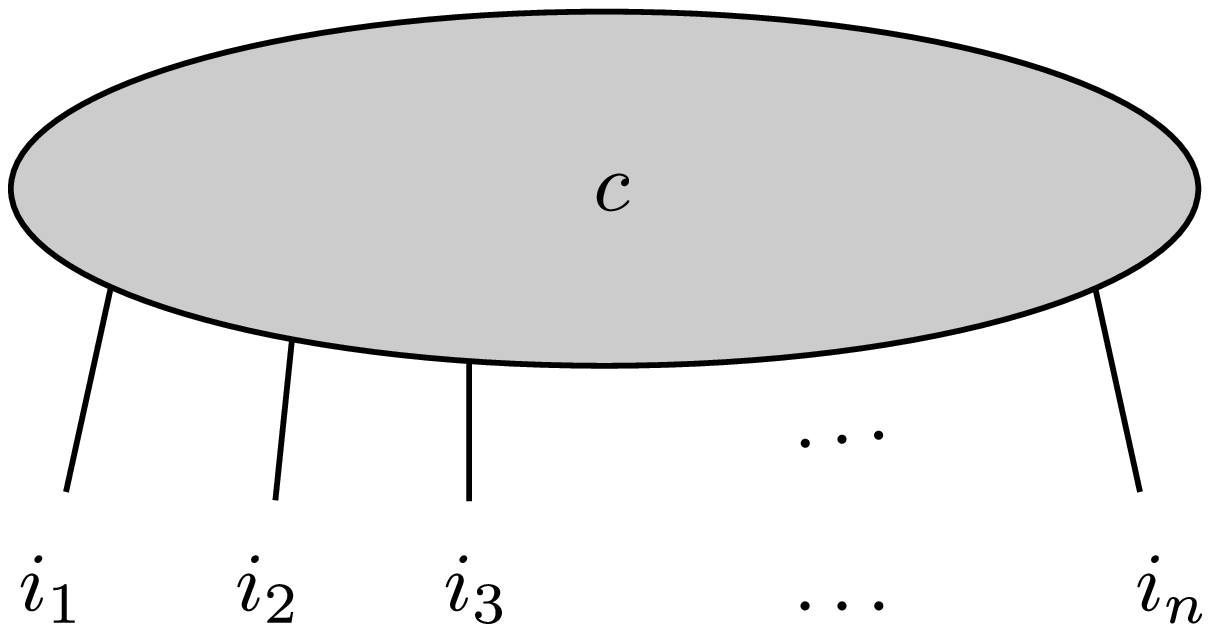}.
\end{equation}
Optionally we may choose to give each leg a vertical orientation and identify upgoing legs with upper indices and downgoing legs with lower indices, but this practice is seldom followed for tensor networks which do not exploit internal symmetries as the metric, and hence conversion between upgoing and downgoing legs, is trivial. This notation may be understood as a simplified form of the Penrose graphical calculus \citep{penrose1971}.
The distinction between upgoing and downgoing legs will 
become more important %
in the scheme presented in Chapters~\ref{sec:anyons}--\ref{sec:nonabelian} for systems with non-Abelian symmetries and for anyons, and a different but related graphical notation will be introduced in \cref{sec:anyons} for the description of tensor networks with non-Abelian symmetries.

The product of multiple tensors may similarly be represented by multiple blobs, with open legs corresponding to free indices and shared legs corresponding to summed indices. For example, matrix multiplication may be represented as
\begin{equation}
A^\alpha_\beta B^\beta_\gamma \equiv \rbimg{-19pt}{246.0pt}{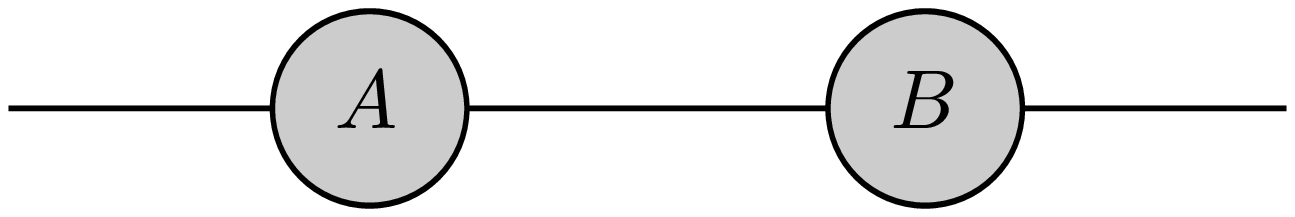}\raisebox{-8pt}{.}
\end{equation}
Unless otherwise specified, in this Thesis I will employ the Einstein summation convention, so $\beta$ is summed over in the left-hand side of the above expression. If we now evaluate this sum over $\beta$, we can write $A^\alpha_\beta B^\beta_\gamma = C^\alpha_\gamma$,
\begin{equation}
\rbimg{-30pt}{369.0pt}{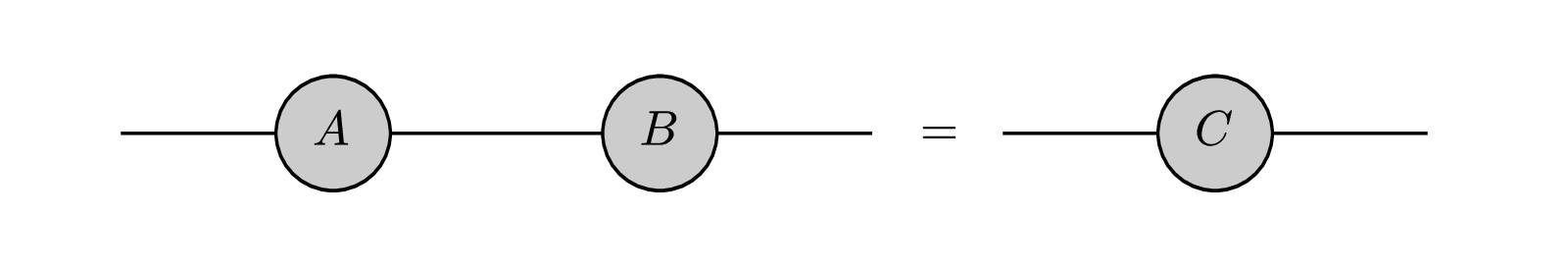}\raisebox{-10pt}{.}\label{eq:ch1:ABC}
\end{equation}
The idea behind tensor network Ans\"atze is that for any tensor $c_{i_1\ldots i_n}$ describing a state on a lattice, we may write down an equivalent collection of tensors linked by summed indices, such that when all these sums are evaluated, we recover the original tensor $c_{i_1\ldots i_n}$. Indeed, \Eref{eq:ch1:ABC} may be thought of as a simple example where $A$ and $B$ form a tensor network which evaluates to give $C$. In general a tensor network diagram will contain multiple separate tensors, and the process of evaluating this diagram to obtain a single tensor (or, where there are no free indices, a number) is known as \emph{tensor network contraction}. Similarly, taking two tensors within such a network and replacing them by a single, equivalent tensor (such that on contracting the entire network, the same resulting tensor is obtained) is termed \emph{contracting} these two tensors. For example, $A$ and $B$ in \Eref{eq:ch1:ABC} might in fact constitute part of a larger tensor network, such as the simple one shown below, and \Eref{eq:ch1:ABC} then describes the contraction of $A$ with $B$ such that
\begin{equation}
\rbimg{-50pt}{432.96pt}{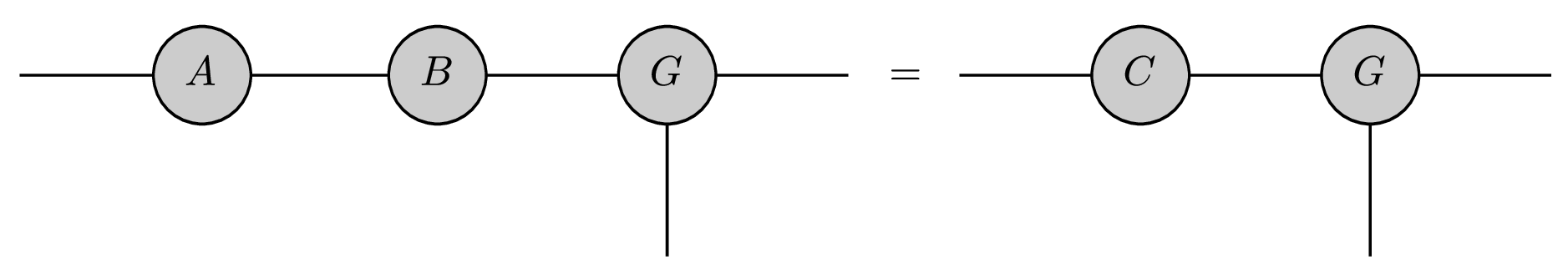}\raisebox{-40pt}{.}
\end{equation}
In general, any tensor network may be completely contracted by a series of pairwise contractions of its component tensors.

But why use tensor networks at all? The answer is simple: Efficiency. If a tensor network is to be capable of representing any state in the Hilbert space $\mc{H}_\mrm{lattice}$, then it must contain at least as many free parameters as $c_{i_1\ldots i_n}$, i.e. $d^n$, and in general will be even less convenient for computation than the form of \Eref{eq:ch1:state}. However, there exist particular choices of tensor network having less than $d^n$ parameters which are nevertheless capable of providing an accurate description of an interesting subregion of this Hilbert space, for example the low energy subspace with respect to a particular Hamiltonian. Using these tensor networks, we may therefore numerically study the properties of a physical system governed by this Hamiltonian, typically at a fraction of the computational cost we would have incurred if we had chosen to retain the full description afforded by $c_{i_1\ldots i_n}$ in \Eref{eq:ch1:state}. This Thesis assumes a basic familiarity with the use of tensor network states %
and their associated algorithms, although for the reader desiring further material, a brief recapitulation of the MERA is provided in \sref{sec:ch1:MERA}, and a number of references for further reading are listed in Secs.~\ref{sec:ch1:MERA} and \ref{sec:ch1:bgreading}.

\section{Multi-scale Entanglement Renormalisation Ans\"atze\label{sec:ch1:MERA}}

Introduced in \citet{vidal2007} and \citet{vidal2008a}, the family of tensor network Ans\"atze known as MERA are motivated by the idea of implementing a real-space renormalisation group transformation on the lattice. They represent a state $|\psi\ra$ using a layered structure, where each layer $i$ may be considered as map between an initial $n_{i-1}$-site lattice $\mc{L}_{i-1}$ and  a coarse-grained $n_i$-site lattice $\mc{L}_i$, for $n_{i-1}>n_i$. In addition to the tensors which perform this coarse-graining, each layer %
of a MERA also incorporates a number of unitary tensors which act on the lattice %
to remove entanglement before coarse-graining takes place.

\begin{figure}
\begin{center}
\includegraphics[width=400.0pt]{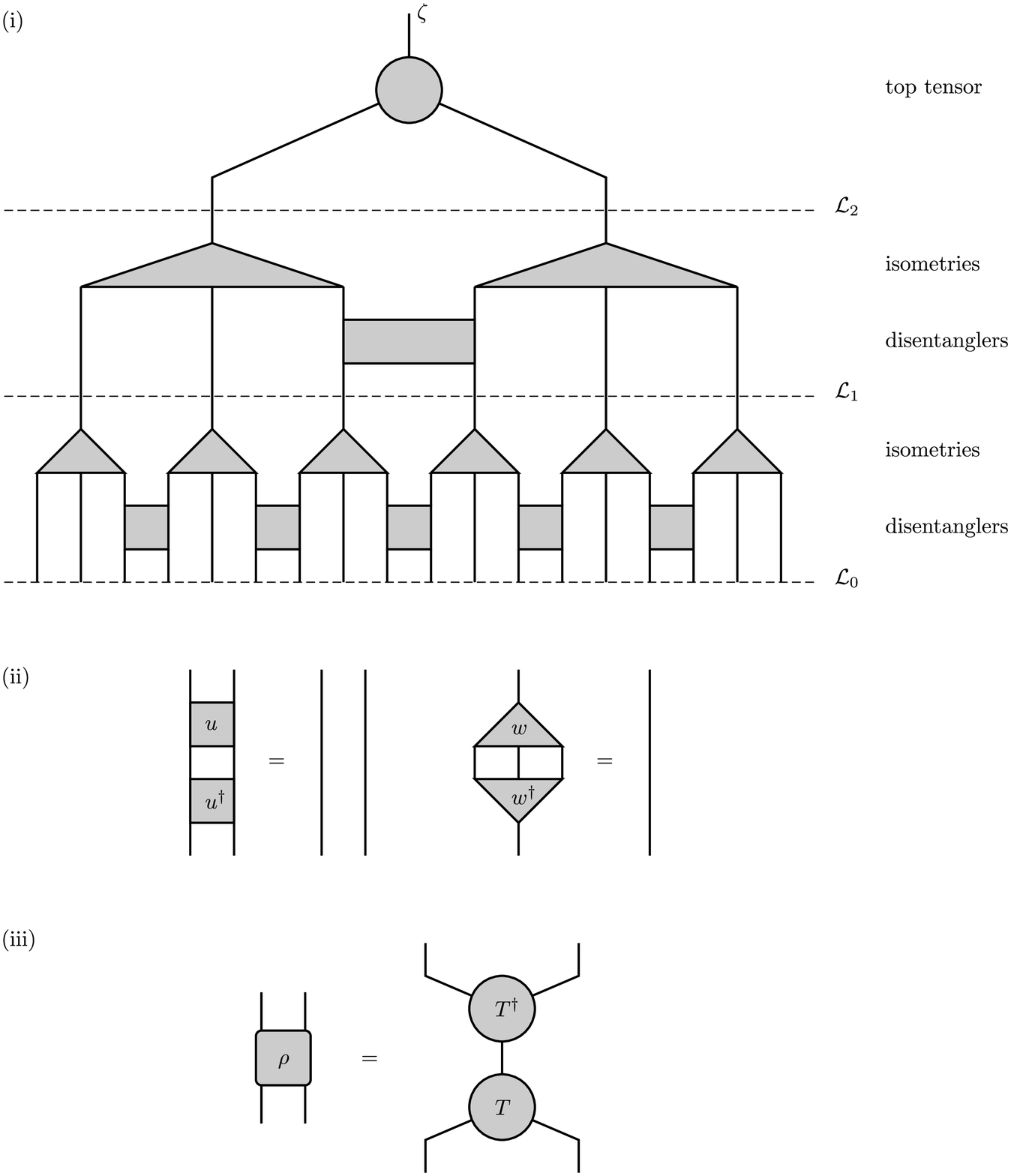}
\caption[(i)~Tensor network for a 3:1 1D MERA representing a subspace of dimension $\chi_\mrm{top}$ on a lattice of 18 sites. (ii)~Constraints on the disentanglers and isometry tensors of the 3:1 1D MERA. (iii)~A reduced density matrix $\rho$ may be constructed from the top tensor $T$ and its conjugate $T^\dagger$ as shown.]{(i)~Tensor network for a 3:1 1D MERA representing a subspace of dimension $\chi_\mrm{top}$ on a lattice of 18 sites. The top index $\zeta$ of the tensor network ranges from 1 to $\chi_\mrm{top}$. (ii)~Constraints on the disentanglers and isometry tensors of the 3:1 1D MERA, denoted $u$ and $w$ respectively. (iii)~A reduced density matrix $\rho$ may be constructed from the top tensor $T$ and its conjugate $T^\dagger$ as shown.\label{fig:ch1:31MERA}}
\end{center}
\end{figure}%
As an example, consider the 3:1 1D MERA (\fref{fig:ch1:31MERA}). %
The open bonds at the bottom of \fref{fig:ch1:31MERA}(i) correspond to indices $i_1,\ldots,i_{18}$, the physical sites of the lattice. The open index at the top of the diagram enables the MERA to represent multiple sites within the Hilbert space, where the value of this index enumerates the represented states. This index ranges from 1 to $\chi_\mrm{top}$. For a fixed value of this index $\zeta$, the network represents a single state $|\psi^\zeta\ra$ and may be contracted to the tensor $c_{i_1\ldots i_{18}}$ which specifies the coefficients of this state as per \Eref{eq:ch1:state}. A MERA having $\chi_\mrm{top}=1$ consequently represents only a single state $|\psi^1\ra$.
Finally, we limit the dimension of each of the summed indices to at most $\chi$, where $\chi$ is a tunable parameter determining the number of free parameters in the Ansatz.

The diagrammatic counterpart of Hermitian conjugation is implemented by vertically reflecting a tensor and complex conjugating all of its entries. Thus if we denote a disentangler by $u^{\alpha\beta}_{\gamma\delta}$ and its Hermitian conjugate by $u^{\dagger\gamma\delta}_{\alpha\beta} = (u^{\alpha\beta}_{\gamma\delta})^*$, then their diagrammatic representations are
\begin{equation}
\rbimg{-35pt}{45pt}{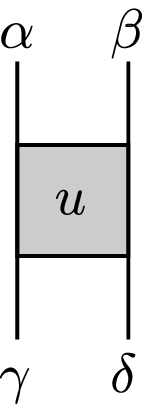}~~\textrm{and}~~\rbimg{-35pt}{45pt}{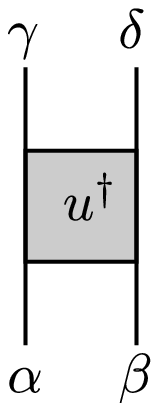}
\end{equation} 
respectively. Similarly, the MERA for a bra, $\la\psi^\zeta|$, is constructed by vertically reflecting the ket, $|\psi^\zeta\ra$, of \fref{fig:ch1:31MERA}(i)%
, and complex conjugating the coefficients of all the tensors in the network.

For discussion of MERA algorithms to approximate the ground state of a system, %
see \citet{dawson2008}, \citet{rizzi2008}, and \citet{evenbly2009}. %
A pedagogical introduction to the MERA, predominantly in one dimension, may be found in \citet{vidal2010}, presenting interpretations of the MERA formalism both in terms of the real-space renormalisation group transformation (coarse-graining) described in this section, and also as a quantum circuit. Applications of the MERA to 2D systems may be found in e.g.  \citet{evenbly2009}, \citet{evenbly2009b}, \citet{evenbly2010}, \citet{evenbly2010c}, \citet{evenbly2010d}, \citet{cincio2008}, \cite{aguado2008}, and \citet{konig2009}.
Note that \citet{vidal2010} also includes material on the scale-invariant MERA, which is the subject of \cref{sec:SIMERA} of this Thesis.

\section{Other Tensor Networks\label{sec:ch1:bgreading}}

There also exist a number of other tensor network Ans\"atze and algorithms. This section lists a selection of introductory references and example papers for a few of the more popular.

\subsection{Matrix Product States}
\nomenclature{\tbf{MPS}}{Matrix Product State: A 1D tensor network Ansatz capable of efficiently representing lightly entangled states. Correlators under this Ansatz decay exponentially as a function of $r$.}
One of the most common tensor network Ans\"atze in use today is the Matrix Product State (MPS), which is the Ansatz underlying the Density Matrix Renormalisation Group (DMRG) technique developed by \citet{white1992} for computation of ground states.\nomenclature{\tbf{DMRG}}{Density Matrix Renormalisation Group: A real-space renormalisation group technique developed by \protect{\citet{white1992}}, which may be interpreted in terms of the Matrix Product State tensor network Ansatz.}
Time evolution may be simulated using the Time Evolving Block Decimation (TEBD) algorithm of \citet{vidal2004}. \nomenclature{\tbf{TEBD}}{Time Evolving Block Decimation algorithm: An algorithm for simulating the time evolution of Matrix Product States.}
For further reading on DMRG and MPS, see \citet{white1992a,white1992,noack1993,white1993,white2004,schollwock2005,perez-garcia2007,schollwock2011}, %
and for TEBD of infinite chains, see \citet{vidal2007}.
 
\subsection{Tree Tensor Networks\label{sec:ch1:TTN}}
\begin{figure}
\begin{center}
\includegraphics[width=400.0pt]{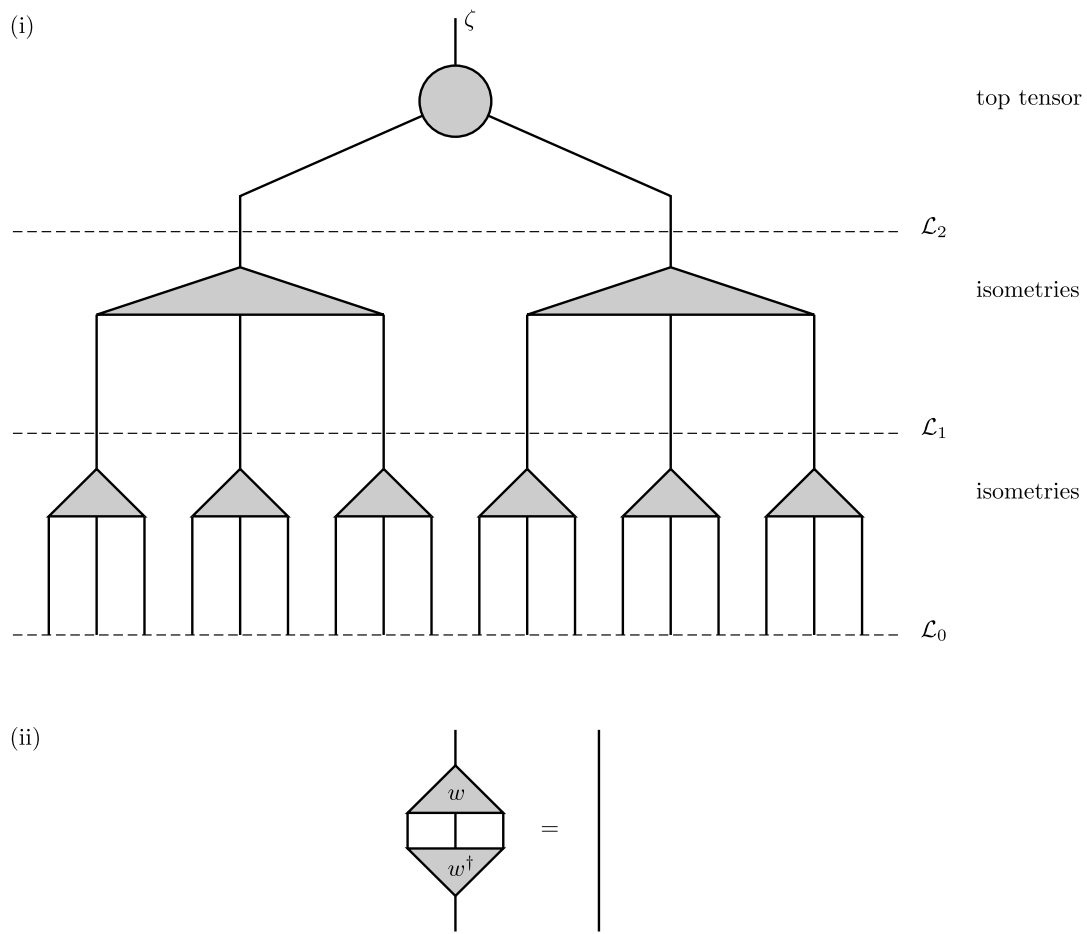}
\caption[(i)~Tensor network for a 3:1 1D TTN representing a subspace of dimension $\chi_\mrm{top}$ on a lattice of 18 sites. (ii)~Constraint on the isometries of the TTN.]{(i)~Tensor network for a 3:1 1D TTN representing a subspace of dimension $\chi_\mrm{top}$ on a lattice of 18 sites. The top index $\zeta$ of the tensor network ranges from 1 to $\chi_\mrm{top}$. Compare with the 1D MERA of \protect{\fref{fig:ch1:31MERA}}. (ii)~Constraint on the isometries of the TTN (denoted $w$).\label{fig:ch1:TTN}}
\end{center}
\end{figure}

\nomenclature{\tbf{TTN}}{Tree Tensor Network: A heirarchical tensor network Ansatz which is capable of efficiently representing lightly entangled states, typically in one or two dimensions.}
The Tree Tensor Network (TTN) may be used to represent states on lattices of arbitrary dimension. It has a heirarchical structure, but is not well suited to the representation of large critical systems due to the need for large bond dimensions (indices with large ranges) towards the top of the tree. Structurally, a TTN may be thought of as a MERA without disentanglers (\fref{fig:ch1:TTN}). 
For further reading, see \citet{shi2006}, and also \citet{fannes1992,otsuka1996,niggemann1997,friedman1997,lepetit2000,martin-delgado2002,nagaj2008,tagliacozzo2009}. %

\subsection{Projected Entangled Pair States}
\nomenclature{\tbf{PEPS}}{Projected Entangled Pair State: The 2D generalisation of the Matrix Product State Ansatz.}
The Projected Entangled Pair State (PEPS) Ansatz is a generalisation of the Matrix Product State Ansatz to two dimensions \citep{verstraete2004}. See also \citet{sierra1998,maeshima2001,nishio2004,murg2007,jordan2008,murg2009}. %

\chapter{Scale-Invariant MERA\label{sec:SIMERA}}

\section{Introduction} %

In this Chapter of the Thesis, we see how MERA-type tensor network Ans\"atze may be used to study the properties of infinite scale-invariant systems. Right from the start \citep{vidal2007}, the MERA has been constructed to implement a real-space renormalisation group (RSRG) \nomenclature{\tbf{RSRG}}{Real-Space Renormalisation Group: A real-space renormalisation group transformation on the lattice is a mapping from an infinite lattice $\mc{L}_0$ to an infinite lattice $\mc{L}_1$ where $n_0$ sites of local dimension $d_0$ on lattice $\mc{L}_0$ are mapped into $n_1$ sites of local dimension $d_1$ on $\mc{L}_1$, for $n_1$ and $d_1$ such that $n_1\,d_1<n_0\,d_0$. Otherwise known as a coarse-graining transformation.}
transformation. It is by exploiting this property of the MERA that an Ansatz for scale-invariant systems may be constructed, as described by \citet{giovannetti2008} and \citet{pfeifer2009}. Further publications studying and applying the scale-invariant MERA include those by \citet{evenbly2009b}, \citet{montangero2009}, and \citet{giovannetti2009}. \citet{evenbly2010a} address the application of the scale-invariant MERA to half-infinite and bounded 1D chains, and most recently, \citet{pfeifer2010} applies the scale-invariant MERA to a quantum critical system of anyons (see also \cref{sec:anyons} of this Thesis).

The material presented in this Chapter is based upon research first published as \citet*{pfeifer2009}. The numerical results of \sref{sec:ch2:results}, including Table~\ref{tab:ch2:results} and Fig.~\ref{fig:ch2:results}, are reproduced or adapted from this reference and are \copyright~(2009) by the American Physical Society.

\subsection{Real-Space Renormalisation Group Transformations}

Real-space renormalisation group transformations have a long history in condensed matter physics, dating back to Kadanoff's spin-blocking technique \citep{kadanoff1966} and Wilson's solution of the Kondo problem \citep{wilson1975}. However, such techniques really came of age with the development of the DMRG algorithm by \citet{white1992,white1993}. The defining feature of such techniques is that there exists some procedure whereby a theory on an initial lattice $\mc{L}_0$ may be subject to some numerical coarse-graining process to yield an effective description on a new lattice $\mc{L}_1$, where each lattice site $i$ on $\mc{L}_1$ corresponds to some region on $\mc{L}_0$, say sites $j_1,\ldots,j_n$, and 
\begin{equation}
\textrm{dim}(i)\leq\prod_{a=1}^n\textrm{dim}(j_a). \label{eq:truncHspace}
\end{equation}

This concept was originally proposed by \citet{kadanoff1966} in a classical context, with the idea of replacing a group of spins with a single effective spin chosen to be %
representative of the group. The first successful quantum mechanical application of an RSRG approach was the treatment of the Kondo problem by \citet{wilson1975}, in which a coarse-graining procedure was chosen so that 
the retained portion of the Hilbert space corresponded locally to the low-energy eigenstates of \emph{individual terms} of the Hamiltonian, e.g. $\hat h_{i,i+1}$ in a Hamiltonian of the form
\begin{equation}
\hat H=\sum_i \hat h_{{i,i+1}}
\end{equation}
where $i$ and $i$+1 are sites of a 1D lattice. However, %
the approach did not appear to generalise well to other problems.

The development of DMRG by \citeauthor{white1992} in \citeyear{white1992} provided the next crucial insight---that the ground state wavefunction which minimises $\la\hat H\ra$ does not necessarily also minimise the expectation value of each local term $\la\hat h_{{i,i+1}}\ra$, and the retained portion of the Hilbert space must therefore be chosen in a way which takes into account the total Hamiltonian%
. DMRG is one algorithm which satisfies this requirement.

More generally, any tensor network which admits a description as a procedure mapping between a series of increasingly coarse-grained infinite lattices may be understood as defining an RSRG transformation, and one can define a \emph{cost function} [e.g. $\Tr(\hat\rho\hat H)$] whose minimum corresponds to the desired state or subspace, and attempt to numerically optimise the defined RSRG transformation so as to extremise this cost function.
For example, applying this philosophy to construct a quantum mechanical version of the spin-blocking technique, one obtains the Tree Tensor Network formalism of \citet{shi2006} (see \sref{sec:ch1:TTN}, above). 

When an RSRG transformation is applied to a Hamiltonian $\hat H_0$ on a lattice $\mc{L}_0$, it yields a Hamiltonian $\hat H_1$ on the coarse-grained lattice $\mc{L}_1$. Repeated application of the RSRG transformation therefore causes the set of Hamiltonians $\hat H_i$ to describe a trajectory in the space of Hamiltonians, termed a Renormalisation Group (RG) flow. \nomenclature{\tbf{RG}}{Abbreviation for ``Renormalisation Group''.}
If the Hamiltonians $\hat H_0$ and $\hat H_1$ satisfy $\hat H_1=\hat H_0$, then the Hamiltonian remains unchanged under repeated application of the RSRG transformation, and we term $\hat H_0$ a \emph{fixed point} of the RG flow.

\subsection{Lattice Models Exhibiting Scale Invariance\label{sec:ch2:sc-inv}}

In order to exhibit scale invariance, a lattice model must be free of any characteristic length scales. Consequently, when a Hamiltonian is a fixed point of the RG flow defined by an RSRG transformation $\mc{R}$, then the correlation length $\xi$ for all operators in a system must be either zero or infinite.

Recall now that our tensor network Ansatz must not only describe the RSRG transformation $\mc{R}$, but also provide an accurate description of the low-energy subspace of the Hilbert space of the system. Let us consider the different possible types of scale-invariant lattice model we may encounter:
\begin{enumerate}
\item $\xi=0$, ground state is a product state: Unentangled. Product states may trivially be described by any tensor network.
\item $\xi=0$, ground state is a topologically ordered state: Entangled. Experience indicates that such states may be efficiently described by a MERA with finite bond dimension \citep{aguado2008,konig2009}.
\item $\xi=\infty$, quantum critical system: Highly entangled. In the ground state, entanglement of a contiguous region $A$ of length $L$ with the rest of the lattice (as measured by the von Neumann entropy $S$) scales as $S\sim\log(L)$ in 1D, and $S\sim L$ or $S\sim L\log(L)$ in 2D, depending on the model under consideration.
\end{enumerate}
Obviously, we may simulate a product state using any tensor network we like, including the MERA. Prior experience shows that for topologically ordered systems, the MERA may once again be a good choice. Finally, what about quantum critical systems? We require a tensor network capable of encoding a bipartite entanglement entropy which in 1D should scale as $\log(L)$ when evaluated for a contiguous region $A^{(1)}$ having length $L$, and in 2D should scale at least as $L$, and preferably as $L\log(L)$, for a region $A^{(2)}$ having dimension $L\times L$.
To understand why the MERA is once again the natural choice, it is necessary to briefly examine how entanglement entropy scales for any tensor network.

Consider a state $|\psi\ra$ represented by a single tensor $c_{i_1\ldots i_n}$ as in \Eref{eq:ch1:state}. Suppose we wish to investigate the entanglement between two regions of the lattice, sites $i_1\ldots i_a$ and $i_{a+1}\ldots i_n$. If we perform a Schmidt decomposition of state $|\psi\ra$, then we write
\begin{equation}
c_{i_1\ldots i_n}=\sum_j c'_{i_1\ldots i_aj}\lambda_j c''_{ji_{a+1}\ldots i_n}\label{eq:ch2:Schmidt}
\end{equation}
where $\lambda_j$ is a list of strictly positive coefficients. In terms of linear algebra, this is equivalent to performing a singular value decomposition on a matrix $c_{k_1k_2}$, where indices $k_1$ and $k_2$ enumerate states on indices $i_1,\ldots, i_a$ and $i_{a+1},\ldots,i_n$ respectively. The dimension of index $j$ may range from 1 (for a product state) to the lesser of $\dim(k_1)=\dim(i_1)\times\ldots\times\dim(i_a)$ and $\dim(k_2)=\dim(i_{a+1})\ldots\times\dim(i_n)$ (for a highly entangled state). Assuming that the coefficients $\lambda_j$ are sorted in decreasing order of magnitude, then the more entangled the state, the higher the value of $j$ before $\lambda_j\ll\lambda_1$.

\begin{figure}
\begin{center}
\includegraphics[width=400.0pt]{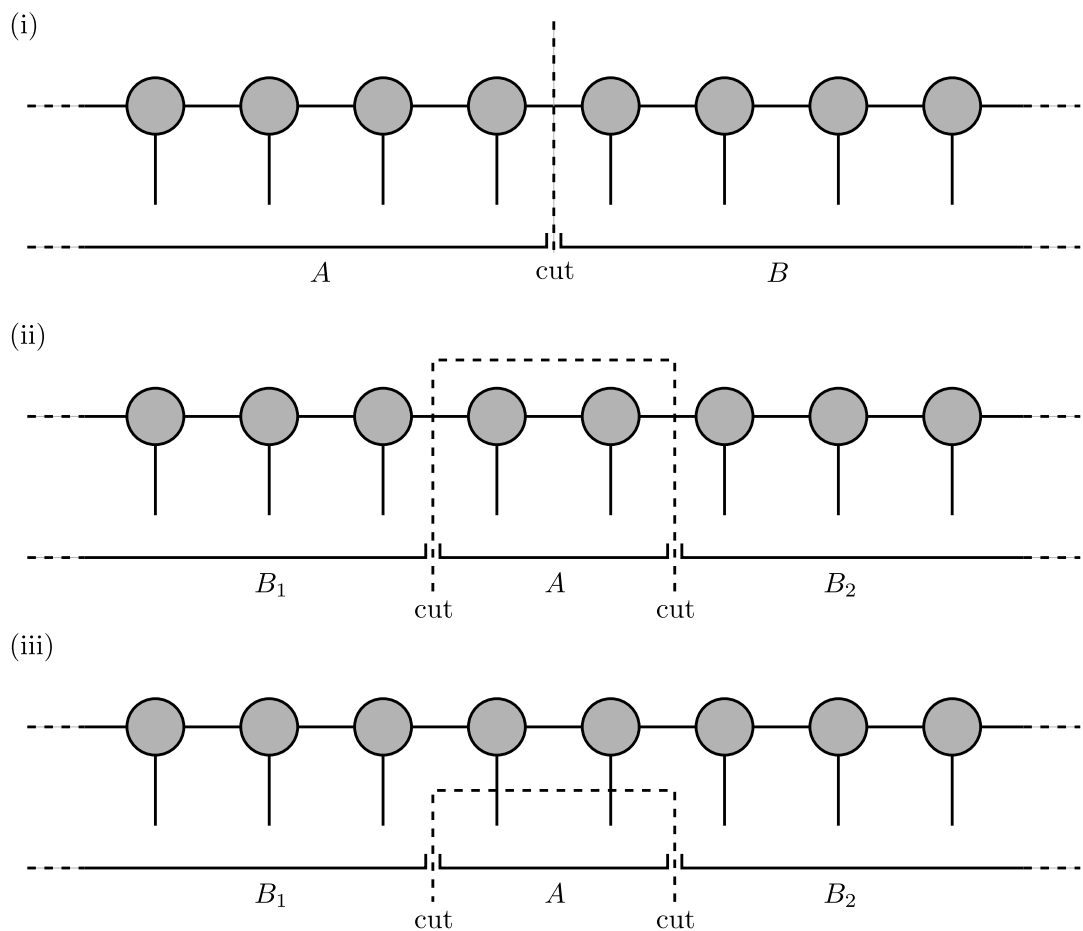}
\caption[Bipartitions of an MPS into two regions $A$ and $B$. The dimensions of the bonds traversing the partitioning of the system limit the amount of entanglement between the two regions.]{(i)~Bipartition of an MPS into two regions $A$ and $B$. The dimension of the bond traversing the partitioning of the system limits the amount of entanglement between the two regions. (ii)~Entanglement of a region $A$ with the rest of the system ($B_1$ and $B_2$) is limited by the product of the dimensions of the bonds which must be cut to separate $A$ from the rest of the system. Multiple such cuts exist for a given region $A$, for example the alternative shown in~(iii), each placing an upper bound on the amount of entanglement. Only the most stringent such bound is therefore of interest.\label{fig:ch2:cutMPS}}
\end{center}
\end{figure}

Now, consider as an example a 1D state represented by an MPS, where the range of the indices in the network has been limited to $\chi$. If we perform a bipartition of such a state, as shown in the \fref{fig:ch2:cutMPS}(i), then the range of $j$ in \Eref{eq:ch2:Schmidt} will be limited to at most $\chi$, and this provides an upper bound on the amount of entanglement which such a state may represent.

In general, a crude quantification of the maximum entanglement a tensor network may encode between a region $A$ and the rest of the lattice is therefore given by taking the product of the dimensions of the bonds which one must cut to separate the tensor network into two regions, one contacting the physical lattice only within region $A$, and the other contacting the lattice only outside of $A$ [e.g. \fref{fig:ch2:cutMPS}(ii)]. (Taking the logarithm of one over this value yields an upper bound on the von Neumann or entanglement entropy, $S$.) 
In general, multiple such cuts exist, and the maximum amount of entanglement which may be encoded is determined by the cut giving the smallest value.

By studying the scaling of $S$ with the size of region $A$, we observe that for an MPS, the maximum amount of entanglement which may be encoded is independent of the size of region $A$, $S\sim const.$ For a TTN, the amount of entanglement which may be encoded exhibits a more complicated dependence on the size and position of region $A$. However, for an $n$-into-1 tree it is possible for any $k\in\mbb{Z}^+$ to choose a region $A$ of linear size $L=kn$, having the same entropy as an appropriately chosen region of length $L=n$. The overall performance of the Ansatz is limited by this worst-case scenario, and consequently a TTN of constant $\chi$ also exhibits an entropy scaling $S\sim const.$ In contrast, the 1D MERA exhibits a scaling $S\sim \log(L)$, making it well-suited to the study of quantum critical systems. In 2D the situation is less ideal, with the 2D MERA exhibiting a scaling $S\sim L$, and thus only being suited to the study of critical systems which do not display a logarithmic correction to the entanglement entropy. However, recent development of a \emph{branching} MERA algorithm \citep{evenblyvidalinprep} with entropy scaling as $S\sim L\log(L)$ or better suggests that once again, some form of the MERA may prove to be a good choice of Ansatz for all critical systems in 2D.

This is not to say that the MPS and TTN cannot be used to calculate properties of quantum critical systems. They can, and in many situations may yield excellent numerical approximations to ground state energies and short-range correlators. However, due to their limited capacity for encoding entanglement within the structure of the tensor network, attempts to construct a RSRG transformation will fail at sufficiently large length scales (and we shall see in \sref{sec:ch2:algorithmComputeProperties} that many interesting properties of quantum critical systems may be computed in the large-length-scale, or infra-red, limit). Consider as an example the application of an $n$-into-1 TTN to a highly entangled lattice model on a lattice $\mc{L}_0$. A single layer of the TTN coarse-grains $\mc{L}_0$ into an effective lattice $\mc{L}_1$, and each site in $\mc{L}_1$ now corresponds directly to $n$ sites in $\mc{L}_0$.
The entanglement between a single site of $\mc{L}_1$ and the rest of the lattice will thus be the same as between those $n$ sites on $\mc{L}_0$ and the rest of the lattice. If the entanglement entropy of the model exhibits any dependence on $L$, then repeated coarse-graining of the lattice will cause this entanglement to continually increase. If the dimension of the indices of the TTN is bounded by some value $\chi$, then after some number of coarse-graining steps, the approximation made in imposing this limit on index dimension will lead to a failure of the TTN to accurately reproduce the properties of the ground state over large length scales. Alternatively, the index dimension would have to increase with each layer of the coarse-graining procedure, eventually becoming infinite. First, this is computationally unfeasible, and second, we anyway desire that the structure of our tensor network should reflect the scale-invariant nature of the ground state.

In contrast, the MERA may be thought of as a TTN supplemented by additional tensors known as \emph{disentanglers} (for illustration of this in 1D, compare Figs.~\ref{fig:ch1:31MERA} and \ref{fig:ch1:TTN}). When the MERA is interpreted as an RSRG transformation, then the disentanglers in each level act to remove short-range entanglement from the ground state. We anticipated that in 1D, for a well-optimised MERA representation of the low-energy subspace of a quantum critical system, they would do so to a sufficient extent that the entanglement entropy of a region $A'$ of length $L$ on lattice $\mc{L}_x$ would be the same as the entanglement entropy of a region $A$ of length $L$ on lattice $\mc{L}_0$, for any coarse-grained lattice $\mc{L}_x$, even when the entanglement entropy on an individual lattice scales as $S\sim \log(L)$. This proposition was based on the observation that the 1D MERA is constructed to be %
capable of encoding an entanglement entropy which scales as $S\sim \log(L)$, and
its predictions have been borne out by subsequent experience.

\subsection{Interesting Properties of 1D Quantum Critical Systems\label{sec:ch2:qcritproperties}}

In studying quantum critical systems, we are particularly interested in their behaviours in the infra-red limit. When we take the continuum limit of a 1D quantum critical system, we obtain a Conformal Field Theory (CFT) in 1+1D \citep{cardy1996,difrancesco1997} which describes the infra-red behaviour of the system in the vicinity of the associated phase transition, and a useful question to ask is whether we can extract from our Ansatz sufficient data to identify and fully characterise the CFT.\nomenclature{\tbf{CFT}}{Conformal Field Theory: A field theory invariant under translation, rotation, rescaling, and special conformal boosts. Taking the continuum limit of a 1D quantum critical system yields an associated 1+1D CFT which describes the infra-red behaviour of the associated quantum phase transition.}
 
In the operator formalism, a 1+1D CFT may be described in terms of an infinite number of operator-valued fields $\hat\phi_\alpha(x,t)$. It is conventional to 
first define the theory
on the cylinder, with $t\in[-\infty,+\infty]$ and $x\in[0,L)$, before mapping to the complex plane via the reparameterisation
\begin{equation}
z=\mrm{e}^{2\pi(t+\rmi x)/L}, \quad \bar{z}=\mrm{e}^{2\pi(t-\rmi x)/L},
\end{equation}
with the fields of the theory now being denoted $\hat\phi_\alpha(z,\bar{z})$.
Under the action of a conformal mapping $z\rightarrow z'=f(z)$,
the correlators of these fields %
transform as
\begin{equation}
\la\hat\phi_1(z_1,\bar z_1)\hat\phi_2(z_2,\bar z_2)\ldots\ra = \prod_\alpha f'(z_\alpha)^{h_\alpha}\bar f'(z_\alpha)^{\bar h_\alpha}\la\hat\phi_1(z'_1,\bar z'_1)\hat\phi_2(z'_2,\bar z'_2)\ldots\ra,
\end{equation}
where each field $\hat\phi_\alpha$ is associated with a holomorphic and an antiholomorphic \emph{conformal dimension}, $h_\alpha$ and $\bar h_\alpha$ respectively. These in turn may be combined to give the \emph{scaling dimension} of the field, $\Delta_\alpha=h_\alpha + \bar h_\alpha$, and the \emph{conformal spin}, $s_\alpha=h_\alpha-\bar h_\alpha$.

For any 1+1D CFT, these fields $\hat\phi_\alpha$, which we will term \emph{scaling fields}, 
may be organised into \emph{conformal families}, each consisting of an infinite number of fields.
Within each conformal family, the field with the smallest scaling dimension is termed the \emph{primary field}, with all others being termed \emph{descendants}. We may associate with each operator field $\hat\phi_\alpha$ a state $|\phi_\alpha\ra$ generated by acting with $\hat\phi_\alpha$ on the origin of the vacuum state (which corresponds to $t=-\infty$),
\begin{equation}
|\phi_\alpha\ra=\hat\phi_\alpha(0,0)|0\ra.
\end{equation}
If we define the operators
\begin{equation}
\hat L_n=\frac{1}{2\pi}\oint z^{n+1}\hat T(z)\rmd z,   \qquad\hat{\bar{L}}_n=\frac{1}{2\pi}\oint \bar z^{n+1}\hat T(\bar z)\rmd \bar z,
\end{equation}
where $\hat T$ is the energy-momentum tensor, and the contour integration is performed over any contour which encircles the origin%
, then these operators $\hat L_n$ and $\hat{\bar{L}}_n$ form representations of the Virasoro algebra. They obey the commutation relations
\begin{align}
[\hat L_m,\hat L_n]&=(m-n)\hat L_{m+n} + \frac{c}{12}(m^3-m)\delta_{m+n,0},\label{eq:ch2:virasoro1}\\
[\hat{\bar L}_m,\hat{\bar L}_n]&=(m-n)\hat{\bar L}_{m+n} + \frac{c}{12}(m^3-m)\delta_{m+n,0},\label{eq:ch2:virasoro2}\\
[\hat L_m,\hat{\bar L}_n]&=0,\label{eq:ch2:virasoro3}
\end{align}
where parameter $c$ is a constant known as the \emph{central charge} of the CFT,
and repeated application of $\hat L_n,~n<0$ and $\hat{\bar L}_m,~m<0$ to the state $|\phi_\alpha\ra$ associated with the primary field of any conformal family will generate all other states associated with members of that family. 
There may be a finite or an infinite number of conformal families, but of greatest interest to us will be the CFTs known as \emph{minimal models}, for which the number of conformal families is finite.

The identity operator is always the primary field for one of the conformal families, and
we may always choose %
our operator fields to satisfy %
\begin{align}
\la\hat\phi_\alpha(z,\bar z)\ra &= \delta_{\alpha\mbb{I}},\label{eq:ch2:onepoint}\\
\la\hat\phi_\alpha(z_\alpha,\bar z_\alpha)\hat\phi_\beta(z_\beta,\bar z_\beta)\ra &= \frac{C_{\alpha\beta}}{(z_\alpha-z_\beta)^{2h_\alpha}(\bar z_\alpha-\bar z_\beta)^{2\bar h_\alpha}},\qquad
C_{\alpha\beta}=\delta_{\alpha\beta},
\end{align}
where $C_{\alpha\beta}=\delta_{\alpha\beta}$ corresponds to a particular choice of normalisation, and $\delta_{\alpha\mbb{I}}=1$ if $\hat\phi_\alpha=\hat{\mbb{I}}$ and 0 otherwise.
We must also
specify the coefficients $C_{\alpha\beta\gamma}$ of the three-point function,
\begin{align}
\begin{split}
\la\hat\phi_\alpha(z_\alpha,\bar z_\alpha)\hat\phi_\beta(z_\beta,\bar z_\beta)\hat\phi_\gamma(z_\gamma,\bar z_\gamma)\ra &= \frac{C_{\alpha\beta\gamma}}{
z_{\alpha\beta}^{\p{\alpha\beta}h_\alpha+h_\beta-h_\gamma}z_{\beta\gamma}^{\p{\beta\gamma}h_\beta+h_\gamma-h_\alpha}z_{\gamma\alpha}^{\p{\gamma\alpha}h_\gamma+h_\alpha-h_\beta}}\\
&\times
\frac{1}{\bar z_{\alpha\beta}^{\p{\alpha\beta}\bar{h}_\alpha+\bar{h}_\beta-\bar{h}_\gamma}\bar z_{\beta\gamma}^{\p{\beta\gamma}\bar{h}_\beta+\bar{h}_\gamma-\bar{h}_\alpha}\bar z_{\gamma\alpha}^{\p{\gamma\alpha}\bar{h}_\gamma+\bar{h}_\alpha-\bar{h}_\beta} }
\end{split}\label{eq:ch2:threepoint}\\
z_{\alpha\beta}=|z_\alpha-z_\beta|\qquad&\qquad
\bar z_{\alpha\beta}=|\bar z_\alpha-\bar z_\beta|.
\end{align}
Whereas $C_{\alpha\beta}$ was a normalisation factor which we were free to choose as we liked%
, the values of $C_{\alpha\beta\gamma}$ 
form part of the description of
the CFT under consideration.

Expression~\eref{eq:ch2:threepoint} implies an algebra known as the Operator Product Expansion (OPE),
\begin{equation}
\hat\phi_\alpha(z_\alpha,\bar z_\alpha)\hat\phi_\beta(z_\beta,\bar z_\beta)=\sum_\gamma C_{\alpha\beta\gamma}(z_\alpha-z_\beta)^{-h_\alpha-h_\beta+h_\gamma}(\bar z_\alpha-\bar z_\beta)^{-\bar h_\alpha-\bar h_\beta+\bar h_\gamma}\hat\phi_\gamma(z_\alpha,\bar z_\alpha)+\ldots,
\end{equation}
which may be inserted into higher-order correlation functions with higher-order terms vanishing in the limit that $|z_\alpha-z_\beta|$ is much smaller than any other separation in the correlator.
\nomenclature{\tbf{OPE}}{Operator Product Expansion: An operator identity in conformal field theory, valid within (and frequently used in the evaluation of) higher-order correlators. See \protect{\sref{sec:ch2:qcritproperties} of this Thesis, \citet{cardy1996}, \citet{difrancesco1997}, and \citet{cardy2006}.}}

To fully describe a 1+1D CFT in the operator formalism, it %
suffices to specify
\begin{enumerate}
\item The primary fields, $\phi_\alpha$.
\item Their scaling dimensions $\Delta_\alpha$ and conformal spins $s_\alpha$.
\item The central charge $c$ of the Virasoro algebra.
\item The coefficients $C_{\alpha\beta\gamma}$ of the operator algebra for the primary fields.
\end{enumerate}
We will see that it is possible to extract all of these data from the Scale-Invariant MERA, with the exception of the conformal spin. However, the data which can be obtained are nevertheless frequently sufficient to uniquely identify the CFT describing the infra-red limit of a particular quantum critical system.

\section{Scale-Invariant MERA Algorithm\label{sec:ch2:SIMERAalgorithm}}

In \sref{sec:ch2:algorithm}, I describe the algorithm for constructing an infinite, scale-invariant MERA. This approach may be applied to either the 1D or the 2D MERA, but in this Thesis I will primarily address the study of 1D quantum critical systems, whose infra-red limits correspond to the interesting and highly-studied 1+1D CFTs. Material on the extraction of conformal data in \sref{sec:ch2:algorithmComputeProperties} is addressed primarily to these systems, and to computation of the parameters described in \sref{sec:ch2:qcritproperties}.

An example application of the Scale-Invariant MERA algorithm to infinite 2D lattice models may be found in \citet{evenbly2009b}.

\subsection{Construction of MERA for the Low Energy Subspace\label{sec:ch2:algorithm}}

\subsubsection{Overview}

For a finite system, a MERA normally consists of a finite number of layers of tensors, each layer consisting of a row of disentanglers and a row of isometries (\fref{fig:ch1:31MERA}). Each layer performs a coarse-graining procedure, mapping from a lattice $\mc{L}_{i-1}$ to a coarser lattice $\mc{L}_{i}$. This process incorporates a truncation of the Hilbert space, such that after all layers of the MERA have been applied, the dimension of the Hilbert space on the maximally coarse-grained lattice is sufficiently small to exactly diagonalise. Numerical optimisation of the MERA \citep{dawson2008,rizzi2008,evenbly2009} is performed to ensure that the Hilbert space of the final coarse-grained lattice exhibits maximal overlap with the interesting region of the Hilbert space of the original lattice, typically the low-energy subspace of a system.

For an infinite system, this procedure obviously requires some modification. No matter how many times we apply a coarse-graining transformation to an infinite lattice $\mc{L}_0$, the result is always an infinite lattice, and the system never becomes small enough to exactly diagonalise. However, the scale-invariant property of quantum critical systems comes to our rescue. To see how this works, let us assume that we have a Hamiltonian $\hat H_{0,\mrm{fp}}$ which is constructed on lattice $\mc{L}_0$ and lies exactly at the fixed point of an RG flow.

We know that if we were able to construct a MERA with an infinite number of layers which represented the low-energy subspace of this Hamiltonian, then because $\hat H_{0,\mrm{fp}}$ %
is a fixed point of the RG flow, application of a layer of the MERA to perform a coarse-graining from $\mc{L}_0$ to $\mc{L}_1$ would map $\hat H_{0,\mrm{fp}}$ into an identical operator $\hat H_{1,\mrm{fp}}$ on the coarse-grained lattice. 
An object which maps operators into operators is termed a \emph{superoperator}, and we may therefore define the \emph{scaling superoperator} $\mc{S}$ as the superoperator implemented by this layer of the MERA, which maps operators from lattice $\mc{L}_0$ to $\mc{L}_1$ for our scale-invariant system. Of course, because $\hat H_{0,\mrm{fp}}$ and $\hat H_{1,\mrm{fp}}$ are identical, the layer of MERA constructed on $\mc{L}_1$ will be identical to that constructed on $\mc{L}_0$, and we may equally well define $\mc{S}$ %
with reference to any layer of this infinite MERA.
Because the Hamiltonians are similarly identical, we will drop the lattice index, and simply write $\hat H_{\mrm{fp}}$ for the fixed-point Hamiltonian on any lattice $\mc{L}_i$. 

Because the Hamiltonian is identical on all lattices $\mc{L}_i$, the reduced density matrix $\hat\rho_\mrm{fp}$ which minimises the energy $\mrm{Tr}(\hat \rho_\mrm{fp}\hat H_\mrm{fp})$ is similarly also identical on every layer of coarse-graining. However, in a MERA we may always calculate the reduced density matrix on a lattice $\mc{L}_{i-1}$ from the reduced density matrix on lattice $\mc{L}_{i}$ \citep{evenbly2009}. Let us 
denote by $\mc{S}^*$ the superoperator which is the
dual of $\mc{S}$, and maps operators on $\mc{L}_i$ into operators on $\mc{L}_{i-1}$. Because 
$\hat \rho_\mrm{fp}$
is identical on every layer, it must be an eigenoperator of $\mc{S}^*$, and 
because $\Tr(\rho_\mrm{fp})=1$ on every layer, 
it must have eigenvalue 1. Provided there exists only one eigenoperator of $\mc{S}^*$ which has eigenvalue 1, this then suffices to uniquely define the fixed-point reduced density matrix $\hat \rho_\mrm{fp}$. When $\mc{S}^*$ has only one eigenoperator with eigenvalue 1, knowledge of the scaling superoperator $\mc{S}$ and its dual are sufficient to compute the reduced density matrix, and these superoperators in turn may be constructed from any layer of this infinite, scale-invariant MERA.

Finally, because all layers of this MERA are identical, we need only describe the disentanglers and isometries of one layer in order to describe the state of the entire system. Assuming also translation invariance, we need only one disentangler and one isometry in order to describe the entirety of this infinite MERA, or compute the reduced density matrix on any lattice $\mc{L}_i$. What will be presented in this Section is therefore an algorithm for determining exactly these tensors: The disentangler and isometry of the scale-invariant MERA.

\subsubsection{A Less Idealised Situation}

In the above discussion, it was assumed that the Hamiltonian of the system was precisely the fixed point Hamiltonian of the RSRG transform; that is, $\hat H_1=\hat H_0=\hat H_\mrm{fp}$. For this to be true, $\hat H_\mrm{fp}$ must correspond to a scaling field, or sum of scaling fields, of the associated CFT, all with identical scaling dimension $\Delta_\alpha$.

In practice, the Hamiltonian of the quantum critical system may not be exactly the fixed point Hamiltonian, but may also include additional scaling fields, provided these fields have scaling dimension $\Delta_\beta > \Delta_\alpha$. On repeated coarse-graining these fields are suppressed relative to the Hamiltonian. While these fields will in theory never vanish completely, and on repeated coarse-graining $\hat H_i$ will only approach the fixed point of the RG flow asymptotically,
\begin{equation}
\lim_{i\rightarrow\infty} \hat H_i \stackrel{\mrm{RG~flow}}{\longrightarrow} \hat H_\mrm{fp},
\end{equation}
we will assume that they decay sufficiently rapidly that their existence may be neglected after some finite number $\tau$ of applications of the coarse-graining process. We therefore construct our Ansatz to consist of $\tau$ layers of ordinary MERA, acting on lattices $\mc{L}_0$ to $\mc{L}_{\tau-1}$, after which the difference between $\hat H_\tau$ on lattice $\mc{L}_\tau$ and $\hat H_\mrm{fp}$ is negligible, and all subsequent layers of the MERA will be essentially identical. We therefore surmount the layers $1\ldots \tau$ of the MERA by one further layer $\tau+1$, which acts on lattice $\mc{L}_\tau$, and is assumed to be repeated an infinite number of times (as layers $\tau+2$ and above, acting on lattices $\mc{L}_{\tau+1}$ to $\mc{L}_\infty$). It is this layer $\tau+1$ of the MERA which is then used in the construction of the scaling superoperator.

We will call an operator a \emph{scaling operator} if it is an eigenoperator of $\mc{S}$, and in Secs.~\ref{sec:ch2:algorithmComputeProperties}--\ref{sec:ch2:results} we will endeavour to identify these operators with the scaling fields $\hat\phi_\alpha$ of the CFT associated with our quantum critical lattice model.
As a note of terminology, scaling fields which decay more rapidly than the fixed point Hamiltonian under the action of the RSRG transformation are termed \emph{irrelevant}, as are the associated fields of CFT. Those which decay at the same rate are termed \emph{marginal}, and those which decay less rapidly are termed \emph{relevant}. We will adopt the same terminology for scaling operators. The Hamiltonian of a quantum critical system will only ever contain marginal and possibly irrelevant terms.

\subsubsection{The Algorithm Itself}

I now present explicitly a practical algorithm for optimisation of the Scale-Invariant MERA for a local quantum critical Hamiltonian on a lattice, which may contain irrelevant terms. This algorithm will be described in general language applicable to both 1D and 2D systems, though accompanying illustrations will refer specifically to the 3:1 MERA in 1D.

It is assumed that the Hamiltonian under consideration is nearest-neighbour, next-to-nearest neighbour, etc., as appropriate to the MERA being employed (for example, the 3:1 MERA on the 1D lattice is constructed for the study of nearest-neighbour Hamiltonians). This may always be achieved by means of some initial coarse-graining onto an effective lattice of higher site dimension if required. As an example, we will subsequently consider the critical Ising model on a 1D lattice, which is a nearest-neighbour Hamiltonian and thus is directly suitable for analysis using the 3:1 MERA.

\begin{figure}
\begin{center}
\includegraphics[width=345.0pt]{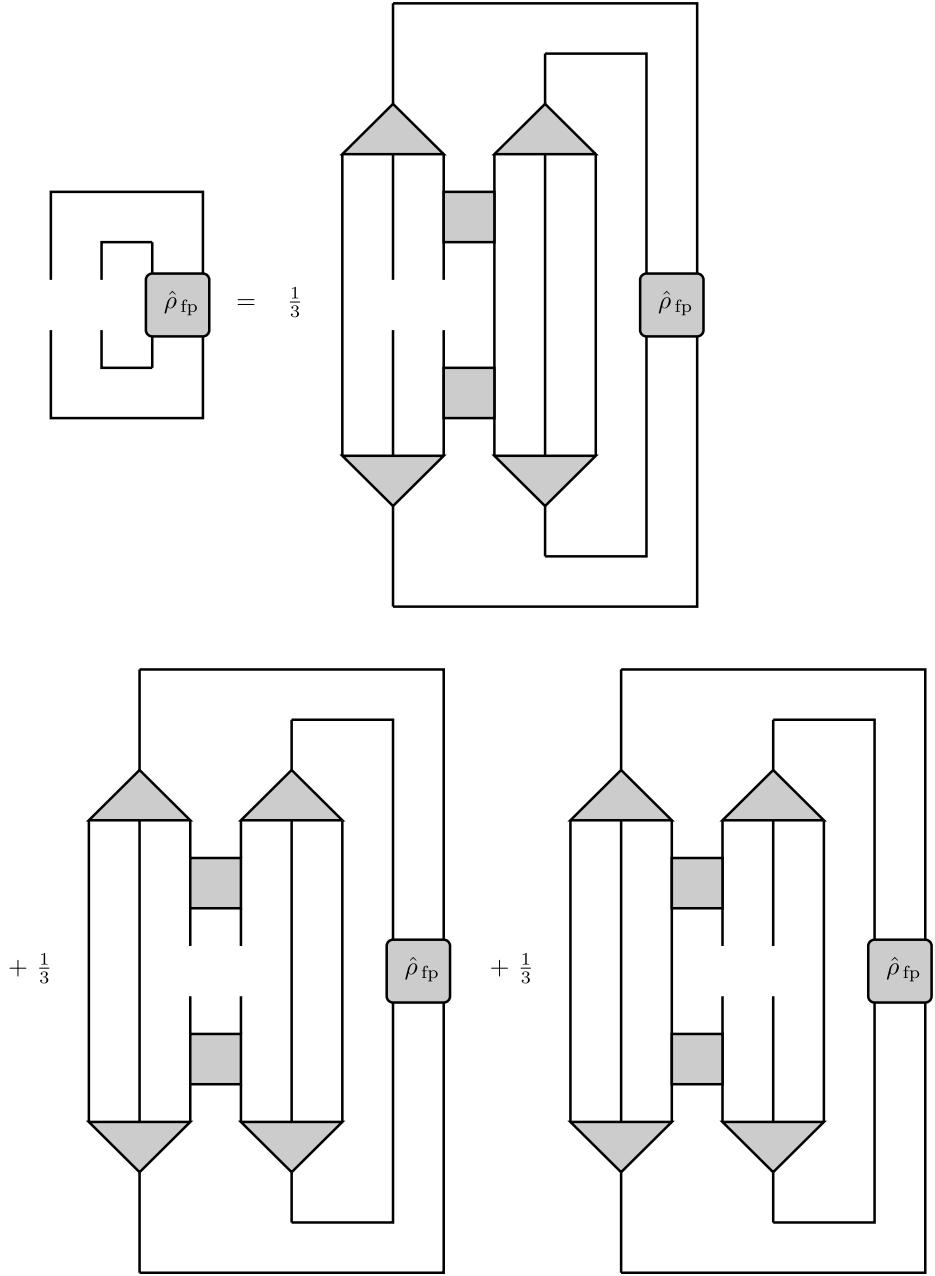}
\caption[Construction of the two-site fixed-point reduced density matrix $\hat\rho_{\,\mrm{fp}}$ of a 1D 3:1 MERA by diagonalising the dual of the scaling superoperator $\mc{S}^*$.]{Construction of the two-site fixed-point reduced density matrix $\hat\rho_{\,\mrm{fp}}$ of a 1D 3:1 MERA by diagonalising the dual of the scaling superoperator $\mc{S}^*$. The reduced density matrix is found by solving the above graphical equation, where the disentanglers and isometries come from the scale-invariant layer $\tau+1$ of the MERA. We assume that the solution is unique (i.e. that $\mc{S}^*$ has precisely one eigenoperator with eigenvalue 1), an assumption which proves valid for the systems studied in this Thesis. A reduced density matrix which satisfies this equation remains unchanged upon being ``descended'' from layer to layer within the scale-invariant region of the MERA.\label{fig:ch2:diagSstar}}
\end{center}
\end{figure}

The MERA is initially constructed to consist of some small number of free layers $1\ldots \tau$, where $\tau$ is typically around 1 or 2, over which the local dimension of the lattice increases from $d$ on $\mc{L}_0$, to $\chi\geq d$ on $\mc{L}_{\tau-1}$. These are surmounted by the scale-invariant layer, which maps from lattice $\mc{L}_{\tau-1}$ to lattice $\mc{L}_\tau$, both of dimension $\chi$. Initial choices of tensor for the disentanglers and isometries are comparatively unimportant, and may be chosen randomly within the constraints of \fref{fig:ch1:31MERA}(ii), or assigned systematically to some known initial configuration. Optimisation then proceeds as follows:
\begin{enumerate}
\item Construct the fixed-point reduced density matrix, $\hat\rho_{\,\mrm{fp}}\equiv\hat\rho_{\tau+1}$ on lattice $\mc{L}_{\tau+1}$, by diagonalising the dual of the scaling superoperator $\mc{S}^*$ (see \fref{fig:ch2:diagSstar} for an example).
\item ``Descend'' $\hat\rho_{\tau+1}$ to obtain $\hat\rho_\tau\ldots\hat\rho_0$ in the usual manner \citep{evenbly2009}.
\item Proceeding row by row from layer 1 to layer $\tau$ of the MERA, for each layer $i$:
\begin{enumerate}
\item Update the disentanglers in the usual manner \citep{evenbly2009}.
\item Update the isometries in the usual manner.
\item ``Lift'' the Hamiltonian from $\mc{L}_{i-1}$ to $\mc{L}_i$ in the usual manner.
\end{enumerate}
\item The Hamiltonian $\hat H_{\tau+1}$ now closely resembles the fixed point Hamiltonian $\hat H_\mrm{fp}$. Optionally, we may now ``lift'' $\hat H_{\tau+1}$ a few more times, using the tensors of the scale-invariant layer $\tau+1$ of the MERA, to obtain a tensor $\hat H^*_{\tau+1}$ which is even more close to $\hat H_\mrm{fp}$. Using $\hat H^*_{\tau+1}$ in lieu of $\hat H_{\tau+1}$ may yield more accurate computation of critical exponents, but at the cost of slower convergence of the Scale-Invariant MERA.
\item Using $\hat H_{\tau+1}$ (or $\hat H^*_{\tau+1}$, if preferred) and the reduced density matrix $\hat\rho_{\tau+2}$ from one layer further up the infinite MERA (which is taken to be the same as $\hat\rho_{\tau+1}$), update the disentanglers of layer $\tau+1$ of the MERA (the scale-invariant layer).
\item Compute numerical properties (e.g. ground state energy, scaling dimensions; see \sref{sec:ch2:algorithmComputeProperties}).
\item Repeat all steps until the chosen cost function is satisfactorily converged.
\end{enumerate}
Following initial convergence of the MERA, the quality of the numerical results (e.g. ground state energy, scaling dimensions, etc.) may be increased by adding more free layers below the scale-invariant layer. To do so, copy the tensors of the scale-invariant layer (denoted $\tau+1$) to obtain a layer $\tau+2$. Layer $\tau+2$ is now the scale-invariant layer, and the above optimisation procedure is now repeated with layers $1\ldots \tau+1$ optimised in the usual manner for a standard MERA, and layer $\tau+2$ being used to construct the fixed-point reduced density matrix and the scaling superoperator. This process may be repeated until insertion of additional layers no longer causes a significant change in the computed properties of the MERA.

The above algorithm serves as an illustrative example as to how a scale-invariant MERA may be converged. In practice, a significant time saving may be made by modifying the computation of $\hat\rho_{\tau+1}$. Rather than computing the dominant eigenoperator of $\mc{S}^*$ exactly on every iteration, we instead assume that after updating the scale-invariant MERA layer, $\hat\rho_{\tau+1}$ from the previous iteration has a non-trivial overlap with the dominant eigenoperator of the new $\mc{S}^*$. We therefore take $\hat\rho_{\tau+1}$ from the previous iteration, and apply the dual of the scaling superoperator once (i.e. we descend this operator using the \emph{scale-invariant layer} $\tau+1$). We then take the resulting operator to be the new $\hat\rho_{\tau+1}\ldots\hat\rho_\infty$. In the limit that the MERA converges (assuming, as always, that this limit exists---an assumption borne out well in practice), $\mc{S}^*$ remains constant from iteration to iteration, and thus is repeatedly applied to $\hat\rho_{\tau+1}$, which will thus gradually converge to the dominant eigenoperator of $\mc{S}^*$ as required. In practice, this process leads to a co-ordinated convergence of $\mc{S}^*$ and $\hat\rho_{\tau+1}$, and requires less time than exactly computing $\hat\rho_{\tau+1}$ on every iteration.

\subsection{Extraction of Conformal Data\label{sec:ch2:algorithmComputeProperties}}

As described in \sref{sec:ch2:qcritproperties}, we may associate the infra-red (large-scale) behaviour of a 1D quantum critical theory with a 1+1D CFT. To extract the conformal data describing this CFT, we must first identify the objects in the quantum critical theory which correspond to the scaling fields of the CFT. These are objects which remain invariant under the action of an RSRG transformation, and consequently may be identified with operators which are eigenoperators of the scaling superoperator. Note that in this Section, we are interested in calculating properties in the large-scale regime of the quantum system, and consequently all disentanglers, isometries, reduced density matrices, etc. are drawn from the scale-invariant layer of the MERA, which is assumed to be repeated an infinite number of times and therefore describes the behaviour of the quantum critical system on all larger length scales.

\begin{figure}
\begin{center}
\includegraphics[width=400.0pt]{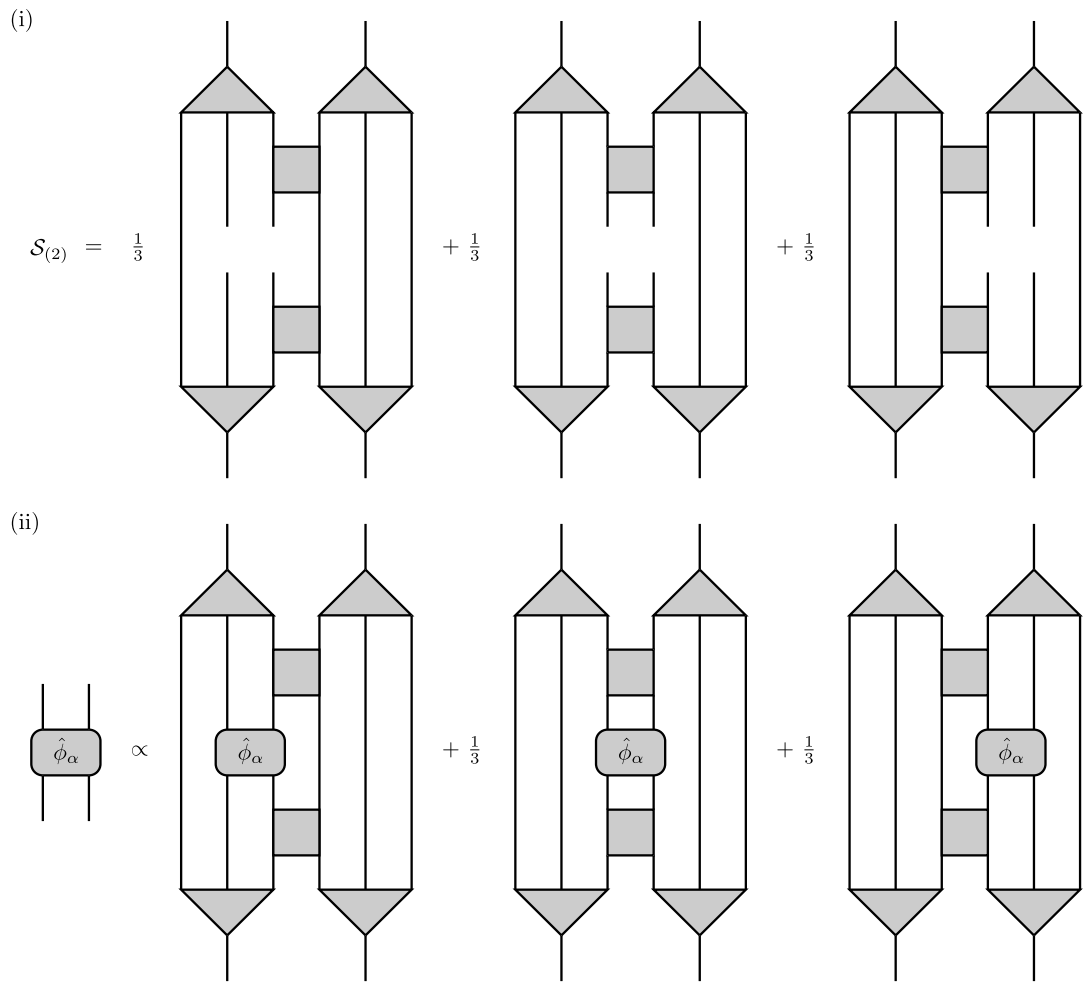}
\caption[(i)~The two-site scaling superoperator $\mc{S}_{(2)}$ of the 3:1 MERA. (ii)~Two-site local scaling operators are eigenoperators of $\mc{S}_{(2)}$.]{(i)~The two-site scaling superoperator $\mc{S}_{(2)}$ of the 3:1 MERA. It is constructed from the isometries and disentanglers of the scale-invariant layer of the MERA. 
(ii)~Two-site local scaling operators are eigenoperators of $\mc{S}_{(2)}$. Note that the scaling superoperator is the average of three diagrams. This is because a two-site operator on the coarse-grained lattice receives contributions from three distinct pairs of sites on the fine-grained lattice, and this must be taken into account in the construction of the scaling superoperator. 
\label{fig:ch2:2sitesuperop}}
\end{center}
\end{figure}

\begin{figure}
\begin{center}
\includegraphics[width=400.0pt]{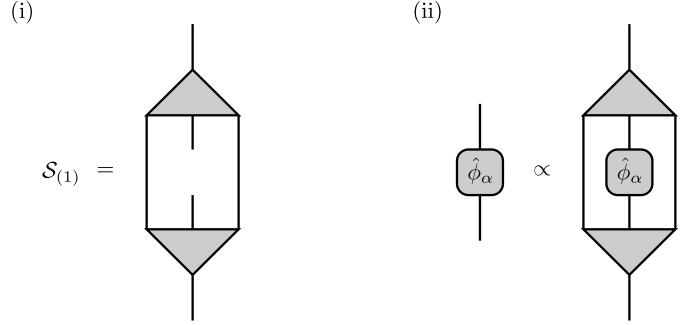}
\caption{(i)~The one-site scaling superoperator, $\mc{S}_{(1)}$, constructed from the isometries of the scale-invariant region of the MERA. (ii)~One-site scaling operators are eigenoperators of $\mc{S}_{(1)}$.\label{fig:ch2:1sitesuperop}}
\end{center}
\end{figure}

For the 3:1 MERA, the causal cone has a width of two sites, and consequently we may construct a two-site scaling superoperator $\mc{S}_{(2)}$ %
[\fref{fig:ch2:2sitesuperop}(i)] whose eigenoperators are two-site scaling operators [\fref{fig:ch2:2sitesuperop}(ii)]. Note that the scaling superoperator is the average of three diagrams. This is because a two-site operator on the coarse-grained lattice receives contributions from operators on three distinct pairs of sites on the fine-grained lattice, and this must be taken into account in the construction of the scaling superoperator. The two-site reduced density matrix is similarly an eigenoperator of $\mc{S}^*_{(2)}$, the dual of $\mc{S}_{(2)}$, with eigenvalue 1, as shown in \fref{fig:ch2:diagSstar}.
However, we also note that on privileged sites of the 3:1 MERA, it is also possible to consider one-site scaling operators which remain invariant under the action of the RSRG transformation. These operators are eigenoperators of the one-site scaling superoperator, $\mc{S}_{(1)}$, as shown in \fref{fig:ch2:1sitesuperop}.

Note that due to the contraints on the disentanglers and isometries [\fref{fig:ch1:31MERA}(ii)], both scaling superoperators are necessarily \emph{unital}, $\mc{S}(\mbb{I})=\mbb{I}$, so that the identity operator is always a scaling operator with eigenvalue $\lambda_\mbb{I}=1$, and \emph{contractive}, $|\lambda_\alpha|\leq 1~\forall~\lambda_\alpha$ \citep{bratteli1979}.

\begin{figure}
\begin{center}
\includegraphics[width=400.0pt]{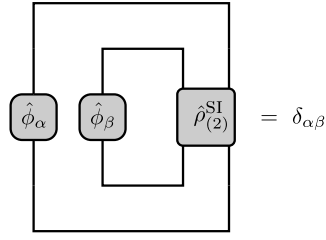}
\caption{Diagrammatic expression for the normalisation of $\la\hat\phi_\alpha(0)\,\hat\phi_\beta(1)\ra$.\label{fig:ch2:normScOps}}
\end{center}
\end{figure}

As the sites of the MERA are spacelike-separated, the one- and two-site scaling operators will satisfy isotemporal versions of the correlators \eref{eq:ch2:onepoint}--\eref{eq:ch2:threepoint},
\begin{align}
\la\hat\phi_\alpha(x_\alpha)\ra &= \delta_{\alpha\mbb{I}},\\
\la\hat\phi_\alpha(x_\alpha)\hat\phi_\beta(x_\beta)\ra &= \frac{C_{\alpha\beta}}{r_{\alpha\beta}^{\p{\alpha\beta}(\Delta_\alpha+\Delta_\beta)}}\label{eq:ch2:twopointspatial}\\
\la\hat\phi_\alpha(x_\alpha)\hat\phi_\beta(x_\beta)\hat\phi_\gamma(x_\gamma)\ra &= \frac{C_{\alpha\beta\gamma}}{
r_{\alpha\beta}^{\p{\alpha\beta}\Delta_\alpha+\Delta_\beta-\Delta_\gamma}r_{\beta\gamma}^{\p{\beta\gamma}\Delta_\beta+\Delta_\gamma-\Delta_\alpha}r_{\gamma\alpha}^{\p{\gamma\alpha}\Delta_\gamma+\Delta_\alpha-\Delta_\beta}}
\end{align}
where $x_\alpha$ is a purely spatial co-ordinate, and
\begin{equation}
r_{\alpha\beta} = |x_\alpha-x_\beta|.
\end{equation}
For convenience, we shall now choose to work specifically with the one-site scaling operators.
We may normalise these scaling operators by imposing \Eref{eq:ch2:twopointspatial} with $C_{\alpha\beta}=\delta_{\alpha\beta}$. When two scaling operators are placed on consecutive lattice sites, the correlator \eref{eq:ch2:twopointspatial} reduces to 
\begin{equation}
\la\hat\phi_\alpha(0)\hat\phi_\beta(1)\ra=\delta_{\alpha\beta},
\end{equation}
where the expectation value is computed with respect to the two-site reduced density matrix on the scale-invariant portion of the MERA, which we will denote $\hat\rho^\textsc{si}_{(2)}$. For the 3:1 MERA, this is the same as the fixed-point reduced density matrix $\hat\rho_{\,\mrm{fp}}$ calculated during optimisation.
The local scaling operators $\hat\phi_\alpha$ must therefore be (ortho)normalised to satisfy
\begin{equation}
\Tr{[(\hat\phi_\alpha\otimes\hat\phi_\beta)\hat\rho^\textsc{si}_{(2)}]}=\delta_{\alpha\beta},
\end{equation}
as represented graphically in \fref{fig:ch2:normScOps}.

\begin{figure}
\begin{center}
\includegraphics[width=400.0pt]{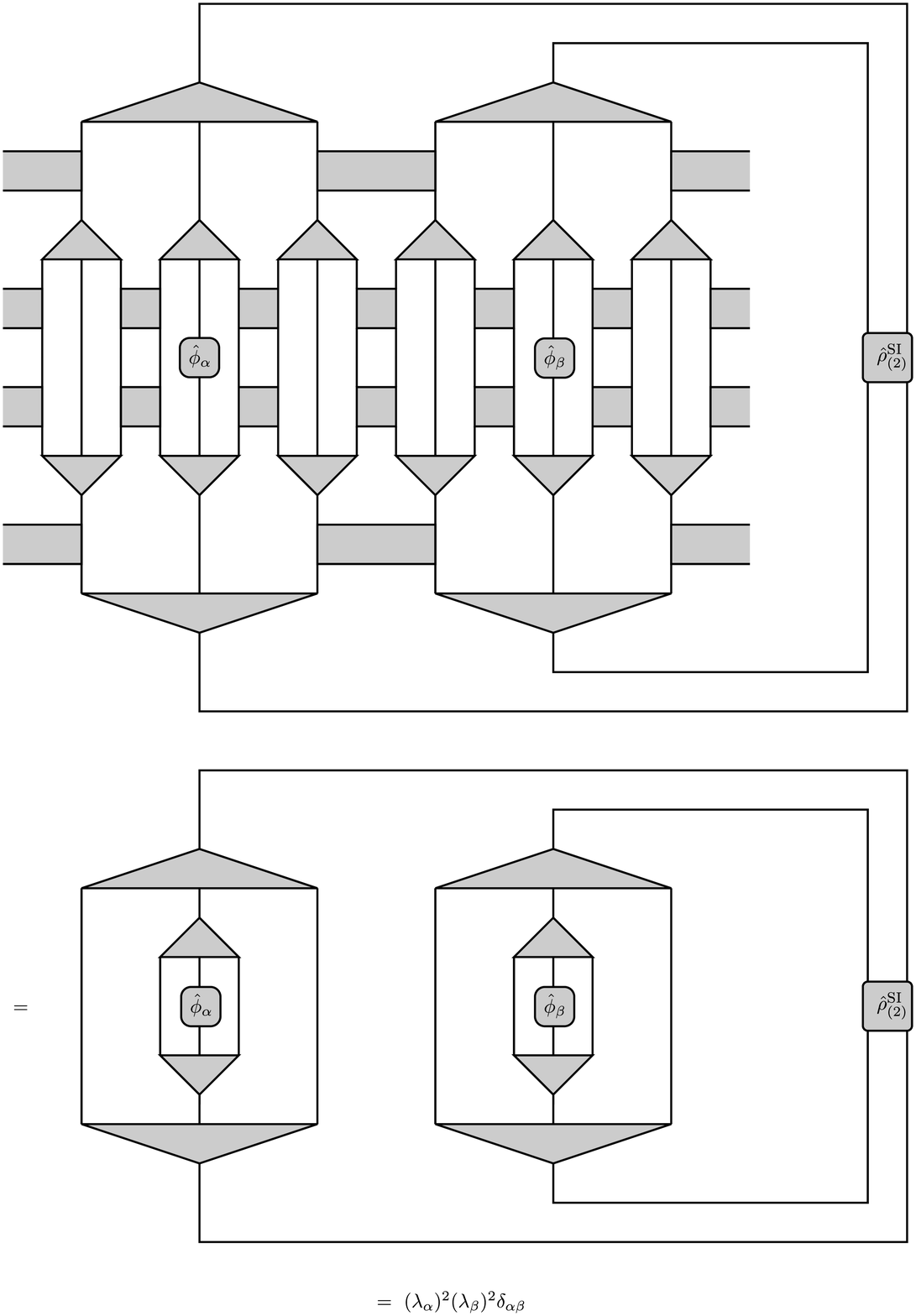}
\caption[Example of a longer-ranged correlator: Computation of $\la\hat\phi_\alpha(x_0)\,\hat\phi_\beta(x_0+9)\ra$ for one-site scaling operators on privileged sites with causal cones of width 1.]{Example of a longer-ranged correlator: Computation of $\la\hat\phi_\alpha(x_0)\,\hat\phi_\beta(x_0+9)\ra$ for one-site scaling operators on privileged sites with causal cones of width 1. Note that all tensors, on both layers, correspond to those of the scale-invariant region of the MERA.\label{fig:ch2:longercorrelator}}
\end{center}
\end{figure}

We may now compute correlators for pairs of one-site local scaling operators, provided these operators are located at a separation such that each application of the MERA maps a one-site local scaling operator into a one-site local scaling operator (e.g. \fref{fig:ch2:longercorrelator}). We find that under these conditions, the correlators scale as
\begin{equation}
\la\hat\phi_\alpha(x_\alpha)\hat\phi_\beta(x_\beta)\ra = \delta_{\alpha\beta}(\lambda_\alpha)^{2\log_3(r_{\alpha\beta})}.
\end{equation}
Using the identity
\begin{equation}
x^{\log{y}} = y^{\log{x}}
\end{equation}
(try taking the log of both sides), we see that the scaling dimensions of the primary fields may be computed according to
\begin{equation}
\Delta_\alpha=-\log_3 \lambda_\alpha.
\end{equation}
(More generally, for an $n$-into-1 MERA, the same argument yields $\Delta_\alpha=-\log_n \lambda_\alpha$.)

\begin{figure}
\begin{center}
\includegraphics[width=400.0pt]{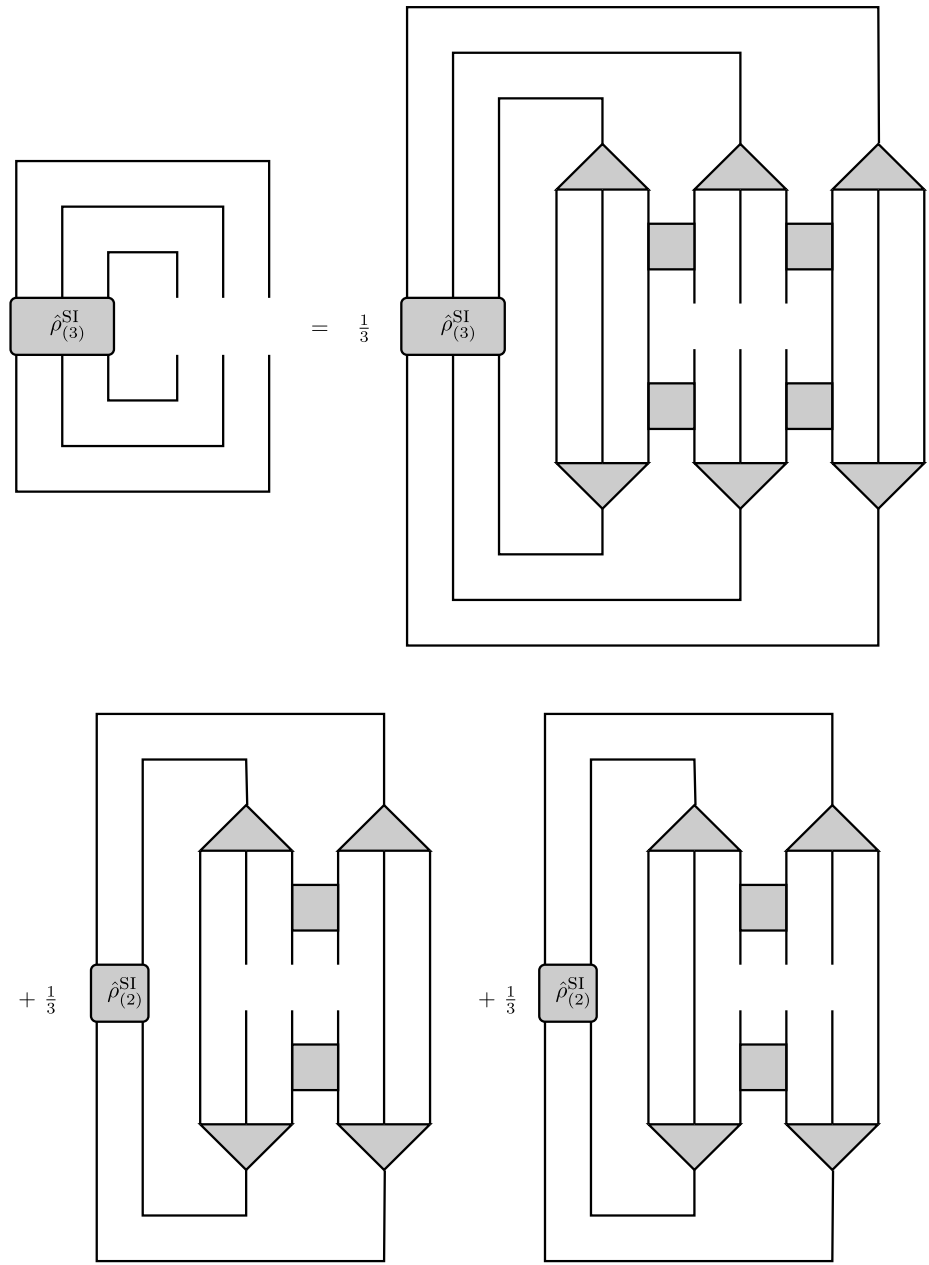}
\caption[Expression to be solved for $\hat\rho^\textsc{si}_{(3)}$.]{Expression to be solved for $\hat\rho^\textsc{si}_{(3)}$. All disentanglers, isometries, and reduced density matrices are those from the scale-invariant region of the MERA.\label{fig:ch2:rho3}}
\end{center}
\end{figure}

\begin{figure}
\begin{center}
\includegraphics[width=400.0pt]{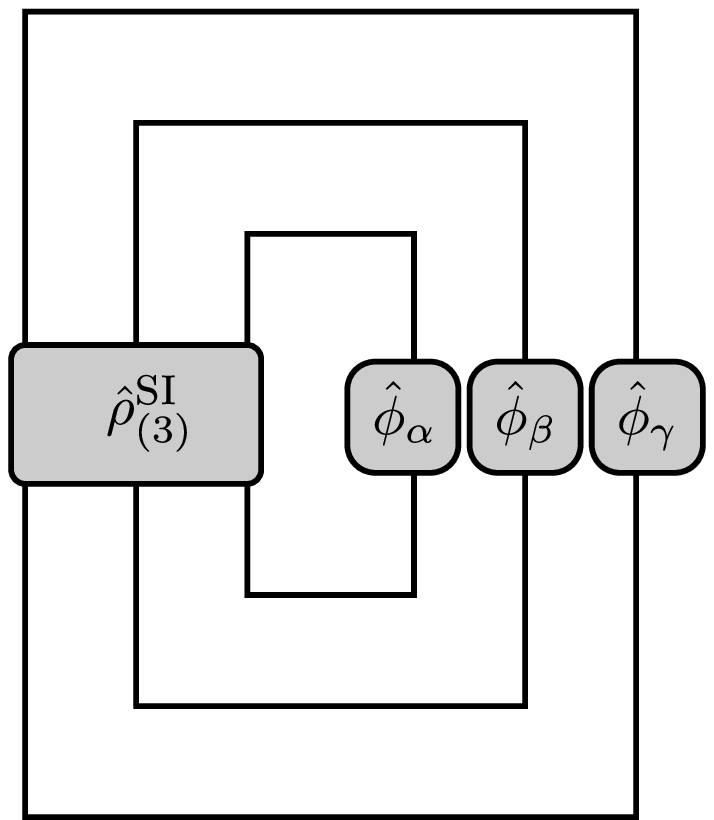}
\caption{Correlator $\la\hat\phi_\alpha(0)\hat\phi_\alpha(1)\hat\phi_\alpha(2)\ra$ evaluated using $\hat\rho^\textsc{si}_{(3)}$.\label{fig:ch2:expec3a}}
\end{center}
\end{figure}

We may also use correlators to calculate the OPE coefficients $C_{\alpha\beta\gamma}$. Although we do not have direct access to a translation-invariant scale-invariant reduced density matrix $\hat\rho^\textsc{si}_{(3)}$ in direct analogy to the two-site reduced density matrix $\hat\rho^\textsc{si}_{(2)}$, we may nevertheless easily compute three-point correlators $\la\hat\phi_\alpha(x_0-1)\,\hat\phi_\beta(x_0)\,\hat\phi_\gamma(x_0+1)\ra$ for certain privileged locations on the lattice. Assuming translation invariance of $\la\hat\phi_\alpha(x_0-1)\,\hat\phi_\beta(x_0)\,\hat\phi_\gamma(x_0+1)\ra$ and symmetrising across the two diagrams given in \fref{fig:ch2:expec3} suffices to give us the values of $C_{\alpha\beta\gamma}$:
\begin{figure}
\begin{center}
\includegraphics[width=400.0pt]{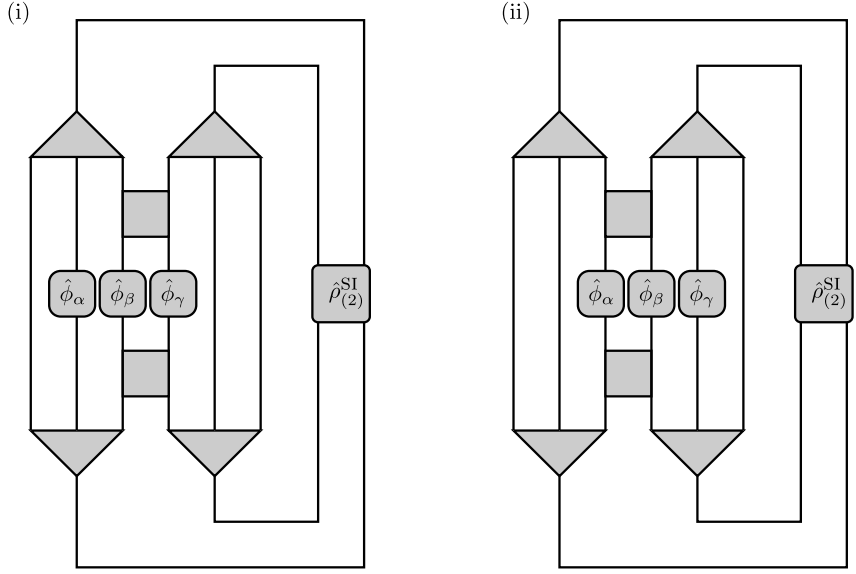}
\caption[Graphical expressions for the three-point correlator $\la\hat\phi_\alpha(x_0-1)\,\hat\phi_\beta(x_0)\,\hat\phi_\gamma(x_0+1)\ra$ at two privileged locations on the 1D lattice.]{Graphical expressions for the three-point correlator $\la\hat\phi_\alpha(x_0-1)\,\hat\phi_\beta(x_0)\,\hat\phi_\gamma(x_0+1)\ra$ at two privileged locations on the 1D lattice. All disentanglers, isometries, and reduced density matrices are those from the scale-invariant region of the MERA.
\label{fig:ch2:expec3}}
\end{center}
\end{figure}%
\begin{align}
C_{\alpha\beta\gamma}&=2^{\Delta_\gamma+\Delta_\alpha-\Delta_\beta}\la\hat\phi_\alpha(0)\hat\phi_\alpha(1)\hat\phi_\alpha(2)\ra.%
\end{align}
If sufficient computational resources are available, a more rigorous evaluation of $C_{\alpha\beta\gamma}$ may be achieved by computing $\hat\rho^\textsc{SI}_{(3)}$ according to \fref{fig:ch2:rho3}, and then determining $\la\hat\phi_\alpha(0)\hat\phi_\alpha(1)\hat\phi_\alpha(2)\ra$ using $\hat\rho^\textsc{SI}_{(3)}$ according to \fref{fig:ch2:expec3a}.

\begin{figure}
\begin{center}
\includegraphics[width=400.0pt]{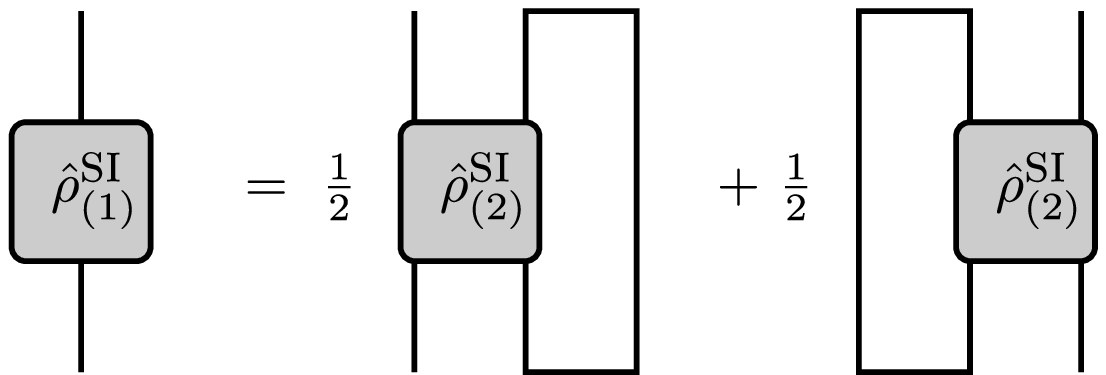}
\caption{The one-site scale-invariant reduced density matrix $\hat\rho^\textsc{si}_{(1)}$ may be obtained by symmetrising over the two different ways of tracing out one site of the two-site scale-invariant reduced density matrix $\hat\rho^\textsc{si}_{(2)}$.\label{fig:ch2:getrho1}}
\end{center}
\end{figure}

We may also compute the central charge, which is obtained from the von Neumann entropies associated with the one- and two-site reduced density matrices of the scale-invariant layer of the MERA:
\begin{align}
S(\hat\rho)&=-\Tr{\left(\hat\rho\log_2\hat\rho\right)}\\
c&=3\left[S\left(\hat\rho^\textsc{si}_{(2)}\right)-S\left(\hat\rho^\textsc{si}_{(1)}\right)\right].
\end{align}
(The one-site scale-invariant reduced density matrix $\hat\rho^\textsc{si}_{(1)}$ may be obtained by symmetrising over the two ways of tracing out one site of $\hat\rho^\textsc{si}_{(2)}$---see \fref{fig:ch2:getrho1}.)

How well may these calculations be expected to work in practice? It is important to recognise that for any CFT, there are always an infinite number of local scaling operators. However, for any MERA with finite bond dimension $\chi$, there are only ever a finite number of eigenoperators of the scaling superoperator. We anticipated, and this is borne out in practice, that by constructing a MERA which represents well the low-energy subspace of the quantum critical theory, we would obtain to a high level of accuracy the conformal parameters associated with the scaling fields of lowest scaling dimension, but that this accuracy would decrease on going to larger scaling dimensions, with an inevitable truncation at some finite scaling dimension $\Delta_\mrm{max}$. The space of states $|\hat\phi_\alpha\ra=\hat\phi_\alpha(0)|0\ra$ associated with the local scaling operators of the MERA is therefore a finite-dimensional vector space, on which exists at best only an approximate, truncated representation of the Virasoro algebra describing the associated CFT (see Eqs.~\ref{eq:ch2:virasoro1}--\ref{eq:ch2:virasoro3} of \sref{sec:ch2:qcritproperties}), becoming exact in the limit $\chi\rightarrow\infty$. Nevertheless, we find that even for relatively modest $\chi$, it is frequently possible to construct a MERA which yields reasonable accuracies for the conformal data.

\section{Results\label{sec:ch2:results}}

\begin{figure}
\begin{center}
\includegraphics[width=317.0pt]{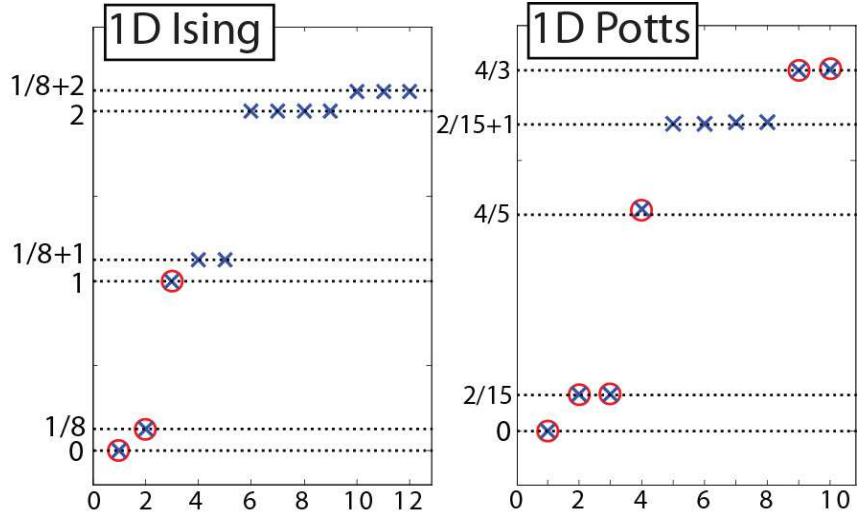}
\caption[Scaling dimensions $\Delta_\alpha$ obtained from the spectra of the scaling superoperators $\mc{S}$. Left: Ising model. Right: Three-state Potts model.]{Scaling dimensions $\Delta_\alpha$ obtained from the spectra of the scaling superoperators $\mc{S}$. Circles indicate primary fields. Left: For the Ising model we can identify the scaling dimensions of the three primary fields, the so-called identity $(\mbb{I})$, spin $(\sigma)$, and energy $(\epsilon)$ fields, together with several of their descendants. Right: The spectrum of $\mc{S}$ for the three-state Potts model shows the primary fields $\mbb{I}$ $(\Delta_\mbb{I}=0)$, $\sigma_1$ and $\sigma_2$ $(\Delta_{\sigma_1}=\Delta_{\sigma_2}=2/15)$, $\epsilon$ $(\Delta_\epsilon=4/5)$, and $Z_1$ and $Z_2$ $(\Delta_{Z_1}=\Delta_{Z_2}=4/3)$, along with the first descendants of $\sigma_1$ and $\sigma_2$. Figure reproduced from \citeauthor*{pfeifer2009}, Physical Review A, \textbf{79}, 040301, \citeyear{pfeifer2009}, \copyright~(2009) by the American Physical Society.\label{fig:ch2:results}}
\end{center}
\end{figure}

This Section presents results demonstrating the capabilities of the Scale-Invariant MERA. Two systems were studied: The Ising model, and the three-state Potts model, which are known to be associated with CFT minimal models $\mc{M}(4,3)$ and $\mc{M}(6,5)$ respectively. Using a $\chi=22$ MERA, scaling dimensions of the primary fields were obtained to within a relative error of 0.01\% for the Ising model, and 2.5\% for the three-state Potts model respectively (\tref{tab:ch2:results}), with appropriate multiplicities for all primary fields and also the lower-scaling-dimension secondary fields (\fref{fig:ch2:results}).
\begin{table}%
\begin{center}
\begin{tabular}{!~cccc}
\toprule
Field\rule{0pt}{2.6ex} & $\Delta^\mrm{CFT}$ & $\Delta^\mrm{MERA}$ & Relative error (\%) \\
[0.5ex]\midrule
\multicolumn{4}{!~l}{Ising model}\\
$\mbb{I}$ & 0 & 0 & -- \\
$\sigma$ & $1/8=0.125$ & 0.124997 & 0.002 \\
$\epsilon$ & 1 & 1.0001 & 0.01 \\
[0.5ex]\midrule
\multicolumn{4}{!~l}{Potts model}\\
$\mbb{I}$ & 0 & 0 & -- \\
$\sigma_1$ & $2/15=0.1\hat{3}$ & 0.1339 & 0.4 \\
$\sigma_2$ & $2/15=0.1\hat{3}$ & 0.1339 & 0.4 \\
$\epsilon$ & $4/5=0.8$ & 0.8204 & 2.5 \\
$Z_1$ & $4/3=1.\hat{3}$ & 1.3346 & 0.1 \\
$Z_2$ & $4/3=1.\hat{3}$ & 1.3351 & 0.1 \\
[0.5ex]\bottomrule
\end{tabular}
\caption[Scaling dimensions, exact $(\Delta^\mrm{CFT})$ and computed $(\Delta^\mrm{MERA})$, for the primary fields of the 1D Ising and three-state Potts models.]{Scaling dimensions, exact $(\Delta^\mrm{CFT})$ and computed $(\Delta^\mrm{MERA})$, for the primary fields of the 1D Ising and three-state Potts models. Numerical results were obtained using a 3:1 Scale-Invariant MERA with $\chi=22$. Table adapted from \citeauthor*{pfeifer2009}, Physical Review A, \textbf{79}, 040301, \citeyear{pfeifer2009}, \copyright~(2009) by the American Physical Society.\label{tab:ch2:results}}
\end{center}
\end{table}%

The computed central charges of $c_\mrm{Ising}=0.5007$ and $c_\mrm{Potts}=0.806$ closely reflected the exact values of $c^\mrm{CFT}_\mrm{Ising}=0.5$ and $c^\mrm{CFT}_\mrm{Potts}=0.8$ respectively. OPEs were also computed for the Ising model and compared with the exact figures of
\begin{equation}
C^\mrm{CFT}_{\alpha\beta\mbb{I}}=\delta_{\alpha\beta}\qquad C^\mrm{CFT}_{\sigma\sigma\epsilon}=\frac{1}{2}\qquad C^\mrm{CFT}_{\sigma\sigma\sigma}=C^\mrm{CFT}_{\epsilon\epsilon\epsilon}=C^\mrm{CFT}_{\epsilon\epsilon\sigma}=0,
\end{equation}
and permutations of the indices thereon, with errors in all values being bounded by $3\times 10^{-4}$ \citep{pfeifer2009}.

These data are easily sufficient to identify the CFTs associated with these lattice models to a high degree of accuracy, and confirm the hypothesis that the Scale-Invariant MERA is an effective Ansatz for the description of quantum critical systems in one dimension. An example of the application of the scale-invariant MERA algorithm to infinite 2D lattice models may be found in \citet{evenbly2009b}.

\section{Adding a Boundary}

The author of this Thesis was also briefly involved in the development of the scale-invariant MERA with a boundary described in \citet{evenbly2010a}. The concept of a boundary scale-invariant MERA was proposed by G.~Vidal, with attempts at implementation by R.~N.~C.~Pfeifer and G.~Evenbly, and theoretical support from V.~Pic\'o, S.~Iblisdir, L.~Tagliacozzo, and I.~McCulloch. The author's implementation attempted to optimise both the bulk and the boundary of the MERA simultaneously, and was not overly successful. It was subsequently laid aside in favour of the implementation by G.~Evenbly described in the above reference. The interested reader is directed to this paper for further information.

%\section{Supplementary material: Computing Scaling of the Entanglement Entropy of Tensor Networks\label{sec:ch2:maxentangle}}
\chapter{Abelian Symmetries of Spin Systems\label{sec:abelian}}

In \cref{sec:SIMERA}, I explained how it is possible to exploit the scale invariance of a quantum critical system on a lattice to construct an efficient Ansatz for the description of the low energy subspace, and how this could be used to extract the conformal data describing the behaviour of this system in the infra-red (large length-scale) limit. Although other numerical techniques and different Ans\"atze have previously been employed to study such systems, the approach developed in \cref{sec:SIMERA} was unique in the way in which it reflects the underlying scaling symmetry of the system. This resulted in an Ansatz which naturally reproduced the polynomially decaying correlators of a quantum critical system, and provided easy access to the data of the associated conformal field theory, as well as providing a numerical description of an infinite physical system which is remarkably compact.

But scale invariance is not the only symmetry exhibited by quantum lattice models. Frequently a Hamiltonian will exhibit additional internal global symmetries which may be described by a group. For example, the Hamiltonian of the Ising model,
\begin{equation}
\hat H_\mrm{Ising}=-\sum_s \sigma_x^{(s)}\sigma_x^{(s+1)}-h\sigma_z^{(s)}\label{eq:ch3c:HIsing}
\end{equation}
is invariant under a rotation of $\pi$ radians about the $z$ axis,
\begin{align}
\sigma_x^{(s)}&\rightarrow -\sigma_x^{(s)}\\
\sigma_y^{(s)}&\rightarrow -\sigma_y^{(s)}\\
\sigma_z^{(s)}&\rightarrow \p{-}\sigma_z^{(s)}\\
\begin{split}
\hat H_\mrm{Ising}&\rightarrow -\sum_s (-\sigma_x^{(s)})(-\sigma_x^{(s+1)})-h\sigma_z^{(s)}\\
&=-\sum_s \sigma_x^{(s)}\sigma_x^{(s+1)}-h\sigma_z^{(s)},
\end{split}
\end{align}
giving it a Z$_2$ symmetry. The $XX$ model, 
\begin{equation}
\hat{H}_{XX} = -J\sum_{s=1}^{L} 
\left(\hat{\sigma}_{x}^{(s)} \hat{\sigma}_{x}^{(s+1)} + \hat{\sigma}_{y}^{(s)} \hat{\sigma}_{y}^{(s+1)}\right)-h\sum_{s=1}^{L} \sigma_z^{(s)},
\end{equation}
similarly exhibits a U(1) symmetry corresponding to invariance under rotation by any angle about the $z$ axis, and the Heisenberg model,
\begin{equation}
\hat H_\mrm{Heisenberg} = -\sum_s \left(J_x\sigma_x^{(s)}\sigma_x^{(s+1)}+J_y\sigma_y^{(s)}\sigma_y^{(s+1)}+J_z\sigma_z^{(s)}\sigma_z^{(s+1)}+h\sigma_z^{(s)}\right),
\end{equation}
has a U(1) symmetry about the $z$ axis for $J_x=J_y$ (known as the Heisenberg $XXZ$ model), and an SU(2) symmetry if $J_x=J_y=J_z$ and $h=0$ (the Heisenberg $XXX$ model).
The natural question to ask was whether global symmetries such as these can also be exploited, either to facilitate the study of a particular symmetry sector of the model, or for computational advantage. (As an example of the former, in the Heisenberg $XXX$ model, a U(1) subgroup of the SU(2) symmetry may be identified with particle number, and it may be desireable to study a lattice with a particular number of particles present. This may be achieved approximately by using a chemical potential, but the resulting model is still subject to fluctuations in particle number. It would be useful to be able to fix exactly either the particle number density or the total particle number on the lattice, and we shall see that this can indeed be done in \sref{sec:ch3:U1MERA}.)

In this Chapter I will therefore discuss the exploitation of global symmetries of the Hamiltonian of a quantum lattice model, with particular attention to Abelian symmetries. This Chapter is divided into three parts:
\begin{enumerate}
\item A concise summary of how these internal symmetries manifest in tensor network Ans\"atze (\sref{sec:ch3:globalsym}).
\item An example, being a self-contained presentation of the exploitation of U(1) symmetry in the MERA (\sref{sec:ch3:U1MERA}).
\item A discussion of the practicalities of how these symmetries may be efficiently implemented for an Abelian symmetry group, and how this is modified in the presence of fermionic exchange statistics (\sref{sec:ch3:comp}).
\end{enumerate}
The focus of \sref{sec:ch3:globalsym} will be predominantly on development of the formalisms and techniques respectively whereby the global internal symmetries of a Hamiltonian may be exploited. Section~\ref{sec:ch3:U1MERA} puts this material into practice, presenting in considerable detail how the formalisms of the preceding Sections are applied in the construction of a U(1)-invariant MERA, culminating in demonstrations of both the ability to select out any symmetry sectors of a system which may be desired, and a substantial (approximately eight- to tenfold) decrease in computational cost when compared with the standard MERA, confirming that it is indeed both feasible and useful to exploit these symmetries in %
the MERA. Section~\ref{sec:ch3:comp} provides an additional level of implementation detail not present in \sref{sec:ch3:U1MERA}, and discusses the extension of the approach presented here to systems of fermions, which have non-trivial exchange statistics in addition to a $Z_2$ parity symmetry.

\begin{center}
\rule{0.75\linewidth}{0.3mm}
\end{center}

Section~\ref{sec:ch3:globalsym} of this Chapter has previously been published as \citeauthor*{singh2010a}, \emph{Physical Review A}, \textbf{82}, 050301, 2010, \copyright~(2010) by the American Physical Society.

Sections~\ref{sec:ch3:U1MERA}--\ref{sec:ch3c:supplement} of this Chapter have previously been published as \citeauthor*{singh2011a}, \emph{Physical Review B}, \textbf{83}, 115125, 2011, \copyright~(2011) by the American Physical Society.

\clearpage

\section{Tensor Network Decompositions in the Presence of a Global Symmetry\label{sec:ch3:globalsym}}

Tensor network decompositions offer an efficient description of certain many-body states of a lattice system and are the basis of a wealth of numerical simulation algorithms. In this Section I discuss how to incorporate a global symmetry, given by a compact, completely reducible group $\mathcal{G}$, into tensor network decompositions and algorithms. This is achieved by considering tensors that are invariant under the action of the group $\mathcal{G}$. Each symmetric tensor decomposes into two types of tensors: \emph{degeneracy tensors}, containing all the degrees of freedom, and \emph{structural tensors}, which depend only on the symmetry group. In numerical calculations, the use of symmetric tensors ensures the preservation of the symmetry, allows selection of a specific symmetry sector, and significantly reduces computational costs. On the other hand, the resulting tensor network may also be interpreted as a superposition of exponentially many spin networks. Spin networks are used extensively in loop quantum gravity, where they represent states of quantum geometry. This work highlights their importance also in the context of tensor network algorithms, thus setting the stage for cross-fertilization between these two areas of research.

\subsection{Introduction}

Locality and symmetry are pivotal concepts in the formulation of physical theories. In a quantum many-body system, locality implies that the dynamics are governed by a Hamiltonian $\hat H$ that decomposes as the sum of terms involving only a small number of particles, and whose strength decays with the distance between the particles. In turn, a symmetry of the Hamiltonian $\hat H$ allows us to organize the kinematic space of the theory according to the irreducible representations of the symmetry group.

Both symmetry and locality can be exploited to obtain a more compact description of many-body states and to reduce computational costs in numerical simulations. In the case of symmetries, this has long been understood. Space symmetries, such as invariance under translations or rotations, as well as internal symmetries, such as particle number conservation or spin isotropy, divide the Hilbert space of the theory into sectors labeled by quantum numbers or charges. The Hamiltonian $\hat H$ is by definition block-diagonal in these sectors. If, for instance, the ground state is known to have zero momentum, it can be obtained by just diagonalizing the (comparatively small) zero-momentum block of $\hat H$. 

In recent times, the far-reaching implications of locality for our ability to describe many-body systems have also started to unfold.
The local character of the Hamiltonian $\hat H$ limits the amount of entanglement that low-energy states may have, and in a lattice system, restrictions on entanglement can be exploited to succinctly describe these states with a tensor network %
decomposition. 
Examples of \TN{} decompositions include Matrix Product States \citep{ostlund1995,fannes1992}, Projected Entangled-Pair States \citep{verstraete2004,sierra1998}, and the MERA \citep{vidal2007,vidal2008a}. Importantly, in a lattice made of $N$ sites, where the Hilbert space dimension grows exponentially with $N$, \TN{} decompositions often offer an efficient description (with costs that scale roughly as $N$). This allows for scalable simulations of quantum lattice systems, even in cases that are beyond the reach of standard Monte Carlo sampling techniques. As an example, the MERA has been recently used to investigate ground states of frustrated antiferromagnets \citep{evenbly2010}.

In this Section we investigate how to incorporate a global symmetry into a \TN{}, so as to be able to simultaneously exploit both the locality and the symmetries of physical Hamiltonians to describe many-body states. Specifically, in order to represent a symmetric state that has a limited amount of entanglement, we use a \TN{} made of symmetric tensors. This leads to an \emph{approximate}, efficient decomposition that preserves the symmetry \emph{exactly}. Moreover, a more compressed description is obtained by breaking each symmetric tensor into several degeneracy tensors (containing all the degrees of freedom of the original tensor) and structural tensors (completely fixed by the symmetry). This decomposition leads to a substantial reduction in computational costs and reveals a connection between \TN{} algorithms and the formalism of spin networks \citep{penrose1971a} used in loop quantum gravity \citep{rovelli1995,rovelli2008}.

In the case of an MPS, global symmetries have already been studied by many authors in the context of both one-dimensional quantum systems and two-dimensional classical systems (see e.g.~\citealt{ostlund1995,fannes1992,white1992,sierra1997,mcculloch2002,mcculloch2007,singh2010,perez-garcia2008,sanz2009}). An MPS is a trivalent \TN{} (i.e., each tensor has at most three indices) and symmetries are comparatively easy to characterize. The present analysis applies to the more challenging case of a generic \TN{} decomposition (where tensors typically have more than three indices).

\subsection{Symmetric Decomposition of a Tensor Network\label{sec:ch3a:symdc}}

\begin{figure}
\begin{center}
\includegraphics[width=300pt]{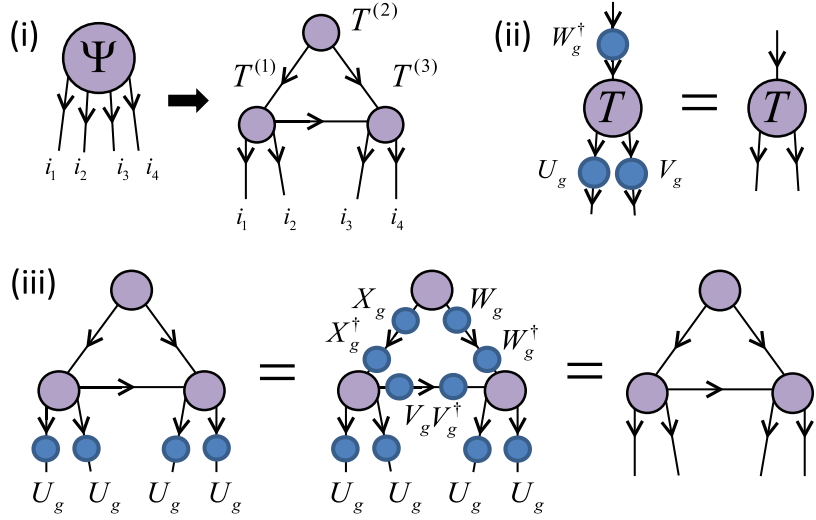}
\caption[(i)~Four-site state $\Psi$ expressed in terms of a tensor network made of three tensors connected according to a directed graph. (ii)~Invariance of tensor $T$ in \Eref{eq:ch3a:symmT}. (iii)~Invariance of a tensor network of symmetric tensors.]{(i)~Four-site state $\Psi$ expressed in terms of a tensor network made of three tensors connected according to a directed graph. (ii)~Invariance of tensor $T$ in \Eref{eq:ch3a:symmT}. (iii)~Invariance of a tensor network of symmetric tensors, \Eref{eq:ch3a:symm}.\label{fig:ch3a:symmTN}}
\end{center}
\end{figure}

We consider a lattice $\mathcal{L}$ made of $N$ sites, where each site is described by a complex vector space $\mathbb{V}$ of finite dimension $d$. A pure state $\ket{\Psi}\in\mathbb{V}^{\otimes N}$ of the lattice can be expanded as
\begin{equation}
	\ket{\Psi} = \sum_{i_1,i_2, \ldots, i_N=1}^d (\Psi)_{i_1i_2\ldots i_N} \ket{i_1 i_2 \ldots i_N}, 
\end{equation}
where $\ket{i_s}$ denotes a basis of $\mathbb{V}$ for site $s \in \mathcal{L}$.
For our purposes, a \TN{} decomposition for $\ket{\Psi}$ consists of a set of tensors $T^{(v)}$ and a network pattern or graph characterized by a set of vertices and a set of directed edges. Each tensor $T^{(v)}$ sits at a vertex $v$ of the graph, and is connected with neighboring tensors by \emph{bond} indices according to the edges of the graph. The graph also contains $N$ open edges, corresponding to the $N$ \emph{physical} indices $i_1,i_2, \ldots, i_N$. The $d^N$ coefficients $(\Psi)_{i_1i_2\ldots i_N}$ are expressed as [\fref{fig:ch3a:symmTN}(i)]
\begin{equation}
	(\Psi)_{i_1 i_2 \ldots i_N} = \ttr \left(\bigotimes_{v} T^{(v)}\right),
	\label{eq:ch3a:TN}
\end{equation}
namely as the tensor product of the tensors $T^{(v)}$ on all the vertices $v$, where the \emph{tensor trace} $\ttr$ contracts all bond indices, so that only the physical indices $i_1, i_2, \ldots i_N$ remain on the r.h.s. of \Eref{eq:ch3a:TN}.

We also introduce a compact, completely reducible group $\mathcal{G}$.  This includes finite groups as well as Lie groups such as O($n$), SO($n$), U($n$), and SU($n$). Let $U:\mathcal{G}\rightarrow L(\mathbb{V})$ be a unitary matrix representation of $\mathcal{G}$ on the space $\mathbb{V}$ of one site, so that for each $g\in \mathcal{G}$, $U_g:\mathbb{V} \rightarrow \mathbb{V}$ denotes a unitary matrix and $U_{g_1g_2} = U_{g_1}U_{g_2}$. Here we are interested in states $\ket{\Psi}$ that are invariant under transformations of the form $U_g^{\otimes N}$,\footnote{A set of states $\ket{\Psi_{t}}$ that transform \textit{covariantly}, $(U_g)^{\otimes N} \ket{\Psi_t} = \sum_{t'} (W_g)_{tt'}\ket{\Psi_{t'}}$, where $W$ is a unitary representation of $\mathcal{G}$, can be represented by an invariant pure state $\ket{\Phi} \propto \sum_t \ket{\Psi_t}\ket{t}$ of lattice $\mathcal{L}$ and one additional site on which the group acts with $W_g^{\dagger}$. The same is true for an invariant mixed state $\rho \propto \sum_t \ketbra{\Psi_t}{\Psi_t}$, with $(U_g)^{\otimes N} \rho (U_g^{\dagger})^{\otimes N}$.\label{sec:ch3a:fn1}}
\begin{equation}
	(U_g)^{\otimes N} \ket{\Psi} = \ket{\Psi},~~~~~\forall~g\in \mathcal{G}.
	\label{eq:ch3a:symm}
\end{equation}
The space $\mathbb{V}$ of one site decomposes as the direct sum of irreducible representations (irreps) of $\mathcal{G}$,
\begin{equation}
	 \mathbb{V} \cong \bigoplus_a d_a \mathbb{V}^{a} \cong  \bigoplus_a \left( \mathbb{D}^{a} \otimes \mathbb{V}^{a}\right), 
	 \label{eq:ch3a:irreps}
\end{equation}
where $\mathbb{V}^{a}$ denotes the irrep labeled with charge $a$ and $d_a$ is the number of times $\mathbb{V}^{a}$ appears in $\mathbb{V}$. We denote by $a=0$ the charge corresponding to the trivial irrep, so that $\mathbb{V}^{0} \cong \mathbb{C}$ and $U^{0}_g=1$. In \Eref{eq:ch3a:irreps} we have also rewritten the same decomposition in terms of a $d_a$-dimensional degeneracy space $\mathbb{D}^{a}$. We choose a local basis $\ket{i} = \ket{a,\alpha_a,m_a}$ in $\mathbb{V}$, where $\alpha_a$ labels states within the degeneracy space $\mathbb{D}^{a}$ (i.e.~$\alpha_a=1,\ldots,d_a$) and $m_{a}$ labels states within irrep $\mathbb{V}^{a}$. In this basis, $U_g$ reads
\begin{equation}
	U_g = \bigoplus_a \left( \mathbb{I}^{a} \otimes U_g^{a}\right).
\end{equation}
Recall that an operator $M:\mathbb{V}\rightarrow \mathbb{V}$ that commutes with the group, $[M,U_g]=0$ for all $g\in\mathcal{G}$, decomposes as \citep{cornwell1997}
\begin{equation}
	M = \bigoplus_a \left( M^{a} \otimes \tilde{\mathbb{I}}^{a}\right)%
	\label{eq:ch3a:Schur}
\end{equation}
(Schur's lemma).

\begin{figure}
\begin{center}
\includegraphics[width=300pt]{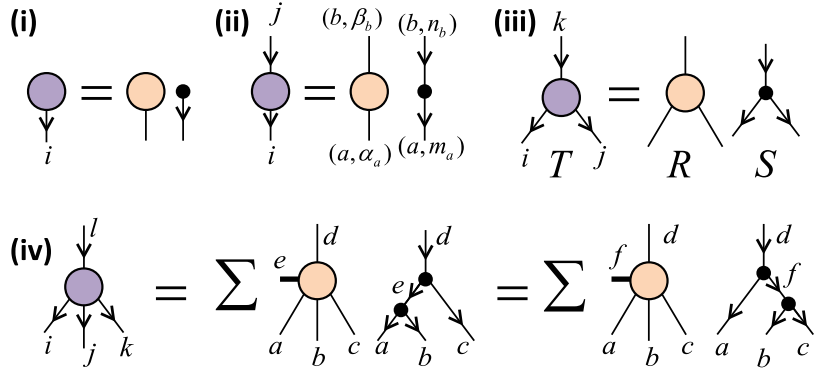}
\caption[Decomposition of tensors with one to four indices.]{Decomposition of tensors with one to four indices. The sums in (iv) run over the intermediate indices $(e, \epsilon_e, q_e)$ and $(f,\zeta_f,r_f)$ in Eqs.~\eref{eq:ch3a:T4}--\eref{eq:ch3a:T4b}.\label{fig:ch3a:T1234}}
\end{center}
\end{figure}

Our goal is to characterize a \TN{} made of symmetric tensors, namely, tensors that are invariant under the simultaneous action of $\mathcal{G}$ on all their indices. A symmetric tensor $T$ with for example, two outgoing indices $i$ and $j$ and one incoming index $k$, fulfills [\fref{fig:ch3a:symmTN}(ii)]
\begin{equation}
	\sum_{ijk} (U_g)_{i'i} (V_g)_{j'j}  (T)_{ijk} (W^{\dagger}_g)_{kk'}= (T)_{i'j'k'},  ~~ \forall\,g\in \mathcal{G},
	\label{eq:ch3a:symmT}
\end{equation}
where $U$, $V$, and $W$ denote unitary matrix representations of $\mathcal{G}$. Clearly, this choice guarantees that \Eref{eq:ch3a:symm} is satisfied [\fref{fig:ch3a:symmTN}(iii)]. Standard group representation theory results \citep{cornwell1997} imply that each symmetric tensor can be further decomposed in such a way that the degrees of freedom that are not fixed by the symmetry can be isolated (\fref{fig:ch3a:T1234}). Next we discuss the cases of 
tensors with a 
small number of indices. Recall that an index $i$ of a tensor is associated with a vector space that decomposes as in \Eref{eq:ch3a:irreps}; therefore we can write $i = (a,\alpha_a,m_a)$, $j = (b, \beta_b, n_b)$, $k=(c, \gamma_c, o_c)$, and so on.

\emph{One leg.---} A tensor $T$ with only one index $i$ is invariant only if $\mathcal{G}$ acts on it trivially, so %
the only relevant irrep is $a=0$, and index $i=\alpha_{0}$ labels states within the degeneracy space $\mathbb{V}^{0}$. 

\emph{Two legs.---} Schur's lemma \citep{cornwell1997} establishes that a symmetric tensor $T$ with one outgoing index $i$ and one incoming index $j$ decomposes as [cf. \Eref{eq:ch3a:Schur}]
\begin{equation}
	(T)_{ij} = (P^{ab})_{\alpha_a \beta_b} (Q^{ab})_{m_an_b}, ~~~~Q^{ab}= \delta_{ab} \delta_{m_a n_b}.
\label{eq:ch3a:T2}
\end{equation}
Thus, for fixed values of the charges $a$ and $b$, $(T)_{ij}$ breaks into a \emph{degeneracy tensor} $P^{ab}$ (where only $a=b$ is relevant) and another tensor $Q^{ab}$. $P^{ab}$ contains all the degrees of freedom of $T$ that are not fixed by the symmetry, whereas $Q^{ab}$ is completely determined by $\mathcal{G}$. 
Another combination of outgoing and incoming indices, for example two incoming indices, leads to a different form for tensor $Q^{ab}$.

\emph{Three legs.---} The tensor product of two irreps with charges $a$ and $b$ can be decomposed as the direct sum of irreps,
\begin{equation}
	\mathbb{V}^{a}\otimes \mathbb{V}^{b} \cong \bigoplus_{c} N_{ab}^c \mathbb{V}^{c} ,
\end{equation}
where $N_{ab}^c$ denotes the number of copies of $\mathbb{V}^{c}$ that appear in the tensor product. For notational simplicity, from now on we assume that $\mathcal{G}$ is multiplicity-free,\footnote{In non-multiplicity-free groups, such as SU(3), where $N_{ab}^c$ might be larger than 1, the coupled basis $\ket{c,o_c,\mu}$ and tensor $S^{abc}_{\mu}$ must include an extra index $\mu=1,\ldots,N_{ab}^c$. See, for example, \protect{\cref{sec:anyons}} of this Thesis, and \protect{\citet{pfeifer2010}}.} that is, $N_{ab}^c\leq 1$, and denote by $(Q^{abc})_{m_a n_b o_c}$ the change of basis between the product basis $\ket{a,m_{a}}\otimes\ket{b,n_b}$ and the coupled basis $\ket{c,o_c}$. The Wigner-Eckart theorem states that a symmetric tensor $T$ with, for example, two outgoing indices $i,j$ and one incoming index $k$, then decomposes as 
\begin{equation}
	(T)_{ijk} = (P^{abc})_{\alpha_a \beta_b \gamma_c} (Q^{abc})_{m_an_bo_c}.
	\label{eq:ch3a:T3}
\end{equation}
As before, for fixed values of the charges $a$, $b$, and $c$, $(T)_{ijk}$ factorizes into degeneracy tensors $P^{abc}$ with all the degrees of freedom and structural tensors $Q^{abc}$ (the Clebsch-Gordan coefficients) completely determined by the group $\mathcal{G}$. An analogous decomposition with different $Q^{abc}$ holds for other combinations of incoming and outgoing indices. 

\emph{Four legs.---}The tensor product of three irreps $\mathbb{V}^{a}\otimes \mathbb{V}^b \otimes \mathbb{V}^c $ may contain several copies of an irrep $\mathbb{V}^d$. Let $e$ be the charge that results from fusing $a$ and $b$, $\mathbb{V}^{a}\otimes \mathbb{V}^b = \bigoplus_e N_{ab}^{e} \mathbb{V}^e$. We can use the values of $e$ for which $N_{ab}^{e}N_{ec}^d \neq 0$ (i.e., such that $a$ and $b$ fuse to $e$, and $e$ and $c$ fuse to $d$) to label the different copies of $\mathbb{V}^d$ that appear in $\mathbb{V}^{a}\otimes \mathbb{V}^b \otimes \mathbb{V}^c$. Let $(Q^{abcd}_e)_{m_an_bo_cp_dq_e}$ denote the change of basis between the product basis $\ket{a m_a} \otimes \ket{b n_b} \otimes \ket{c o_c}$ and the coupled basis $\ket{d p_d; e}$ obtained by fusing to the intermediate basis $\ket{eq_e} \in \mathbb{V}^e$. Then a symmetric tensor $T$ with three outgoing indices $i$, $j$, and $k$ and one incoming index $l = (d,\delta_d,p_d)$ decomposes as\footnote{Note that \protect{\Eref{eq:ch3a:T4}} differs from Eq.~(11) of \protect{\citet{singh2010a}}. This is due to an error in the published paper, where this equation was mistakenly given as
\begin{equation}
	(T)_{ijkl} = \sum_{e,\epsilon_e,q_e}(P^{abcd}_e)_{\alpha_a \beta_b \gamma_c \delta_d \epsilon_e} (Q^{abcd}_e)_{m_an_bo_cp_dq_e},\tag{11}
\end{equation}
with explicit degeneracy indices $\epsilon_e$ and $q_e$ associated with the intermediate charge index $e$.
For a given value of $\epsilon_e$, $(P^{abcd}_e)_{\alpha_a \beta_b \gamma_c \delta_d \epsilon_e}$ is non-zero for precisely one set of values $\{\alpha_a,\beta_b,\gamma_c,\delta_d\}$, and similarly for $q_e$, $(Q^{abcd}_e)_{m_an_bo_cp_dq_e}$, and $\{m_a,n_b,o_c,p_d\}$. Consequently, to obtain the most efficient representation of $(T)_{ijkl}$ we would always evaluate the sum over $\epsilon_e$ and $q_e$ in the above expression, reducing it to \Eref{eq:ch3a:T4}. Similar corrections have been made to Eqs.~\eref{eq:ch3a:T4b} and \eref{eq:ch3a:decoT}.} %
\begin{equation}
	(T)_{ijkl} = \sum_{e}(P^{abcd}_e)_{\alpha_a \beta_b \gamma_c \delta_d} (Q^{abcd}_e)_{m_an_bo_cp_d},
\label{eq:ch3a:T4}
\end{equation}
where the sum is over all relevant values of the intermediate charge $e$. %
Alternatively, $T$ can be decomposed as
\begin{equation}
	(T)_{ijkl} = \sum_{f}(\tilde{P}^{abcd}_f)_{\alpha_a \beta_b \gamma_c \delta_d} (\tilde{Q}^{abcd}_f)_{m_an_bo_cp_d},
\label{eq:ch3a:T4b}
\end{equation}
where $(\tilde{Q}^{abcd}_f)_{m_an_bo_cp_d}$ denotes the change of basis to another coupled basis $\ket{d p_d; f}$ of $\mathbb{V}^d$ obtained by fusing first $b$ and $c$ into $f$, and then $a$ and $f$ into $d$, involving a different intermediate charge index $f$. The two coupled bases are related by a unitary transformation given by the 6-index tensor $F$ [related to the 6-$j$ symbols for e.g.~$\mathcal{G} =\mathrm{SU(2)}$; see \Eref{eq:ch5:SU2F} of \cref{sec:nonabelian}] such that
\begin{equation}
	\tilde{Q}^{abcd}_f = \sum_{e} (F^{abc}_d)^{e}_{f} Q^{abcd}_{e}.
	\label{eq:ch3a:Fmove}
\end{equation}
Since Eqs.~\eref{eq:ch3a:T4} and \eref{eq:ch3a:T4b} represent the same tensor $T$, the degeneracy tensors $P$ and $\tilde{P}$ are related by
\begin{equation}
	\tilde{P}^{abcd}_f = \sum_{e} ({F^{abc}_d}^{*})^{e}_{f} P^{abcd}_{e}.
	\label{eq:ch3a:Fmove2}
\end{equation}

More generally, a symmetric tensor $T$ with $t$ indices $i_{s}=(a_s,\alpha_{a_s},m_{a_s})$, where $s=1,\ldots, t$, decomposes as
\begin{equation}
	(T)_{i_1 i_2 \ldots i_t} = \sum (P^{a_1 \ldots a_t}_{e_1 \ldots e_{t'}})_{\alpha_{a_1} \ldots \alpha_{a_t}} (Q^{a_1\ldots a_t}_{e_1 \ldots e_{t'}})_{m_{a_1} \ldots m_{a_t}},
	\label{eq:ch3a:decoT}
\end{equation}
where the sum is over the intermediate charges $e_k$,~$k=1,\ldots, t'$. The degeneracy tensors $P^{a_1 \ldots a_t}_{e_1 \ldots e_{t'}}$ contain all the degrees of freedom of $T$, whereas the structural tensors $Q^{a_1 \ldots a_t}_{e_1 \ldots e_{t'}}$ are completely determined by the symmetry. Here $e_{1}, e_{2},\ldots, e_{t'}$ are intermediate charges that decorate the inner branches of a trivalent tree used to label a basis in the space of intertwining operators between the tensor products of incoming and outgoing irreps. A different choice of tree will produce different sets of tensors $\tilde{P}$ and $\tilde{Q}$, related to $P$ and $Q$ by F-moves.\footnote{When $\mathcal{G}$ is an Abelian group, such as U(1), the tensor product $\mathbb{V}^{a}\otimes \mathbb{V}^{b}$ of two irreps only gives rise to one irrep $\mathbb{V}^{c}$, so that no intermediate charges $e_1, e_2, \ldots, e_{t'}$ need to be specified in \protect{\Eref{eq:ch3a:decoT}}, simplifying significantly the decomposition of symmetric tensors.}

\begin{figure}
\begin{center}
\includegraphics[width=300pt]{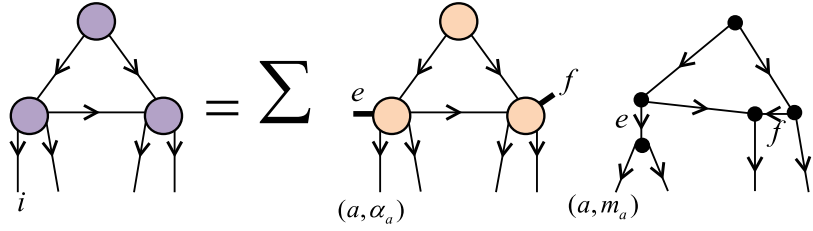}
\caption[A \TN{} for a symmetric state $\ket{\Psi}\in \mathbb{V}^{\otimes N}$ of lattice $\mathcal{L}$ (\fref{fig:ch3a:symmTN}) is expressed as a linear superposition of spin networks.]{A \TN{} for a symmetric state $\ket{\Psi}\in \mathbb{V}^{\otimes N}$ of lattice $\mathcal{L}$ (\fref{fig:ch3a:symmTN}) is expressed as a linear superposition of spin networks. The sum runs over the intermediate indices that carry charges $e$ and $f$ (shown explicitly) as well as all indices shared by two tensors.\label{fig:ch3a:spinNetwork}}
\end{center}
\end{figure}%

We can now investigate how the \TN{} decomposes if we write each of its tensors $T$ in the $(P,Q)$ form of \Eref{eq:ch3a:decoT} (see \fref{fig:ch3a:spinNetwork}). For any fixed value of all the charges, the whole \TN{} factorizes into two terms. The first one is a network of degeneracy tensors. The second one is a directed graph with edges labeled by irreps of $\mathcal{G}$ and vertices labeled by intertwining operators. This is nothing other than a \emph{spin network} \citep{penrose1971a}, a well-known object in mathematical physics and, especially, in loop quantum gravity \citep{rovelli1995,rovelli2008}, where it is used to describe states of quantum geometry. Accordingly, a symmetric \TN{} for the state $\ket{\Psi}\in \mathbb{V}^{\otimes N}$ of a lattice $\mathcal{L}$ of $N$ sites can be regarded as a linear superposition of spin networks with $N$ open edges. The number of spin networks in the linear superposition grows exponentially with the size of the \TN{}. The expansion coefficients are given by the degeneracy tensors.

\subsection{Applications of Symmetric Tensor Networks}

Computationally, the present characterization of a symmetric \TN{} is of interest for several reasons. First%
, it allows us to describe a state $\ket{\Psi}^{\otimes N}$ with specific quantum numbers, which are preserved exactly during approximate numerical simulations. Let us consider as an example the group U(1), with charge $n$ corresponding to particle number ($n=0,\pm 1,\pm 2, \ldots$), and the group SU(2), with charge $j$ corresponding to the spin ($j=0,1/2,1,3/2, \ldots$). The symmetric \TN{} can be used to describe a state with, for example, zero particles ($n=0$) and zero spin ($j=0$), respectively---or, more generally, covariant states with any value of $n$ and $j$.\footnote{See footnote~\protect{\ref{sec:ch3a:fn1}}, above.}

Second, the ($P,Q$)-decomposition (\ref{eq:ch3a:decoT}) concentrates all the degrees of freedom of a symmetric tensor $T$ in the degeneracy tensors $P$, producing a more compact description. For instance, for the U(1) and SU(2) groups, an approximation 
of the ground state of the antiferromagnetic Heisenberg spin-$\frac{1}{2}$ chain with a MERA of bond dimension $\chi=21$ requires five and thirty-five times less parameters than with nonsymmetric tensors, respectively \citep{singh2011a,singhvidalinprep}. 

\begin{figure}
\begin{center}
\includegraphics[width=300pt]{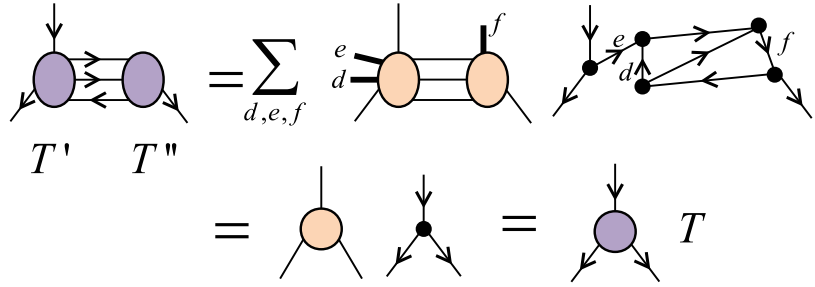}
\caption[Product of two symmetric tensors. Only the intermediate charges $d$, $e$, and $f$ are explicitly shown. The computation involves evaluating spin networks.]{Product of two symmetric tensors. Only the intermediate charges $d$, $e$, and $f$ are explicitly shown. Additional sums apply to all indices shared by two tensors. The computation involves evaluating spin networks.\label{fig:ch3a:T1T2}}
\end{center}
\end{figure}%

In addition, the ($P,Q$)-decomposition (\ref{eq:ch3a:decoT}) lowers the cost of simulations significantly. Consider the multiplication of two tensors (\fref{fig:ch3a:T1T2}) which is central to most \TN{} algorithms. Cost reductions come from two fronts:
\begin{enumerate}
\item \emph{Block-sparse matrices.---} The most costly step in multiplying two tensors $T'$ and $T''$ consists of multiplying two matrices $M'$ and $M''$ obtained from $T'$ and $T''$. These matrices are of the form of \Eref{eq:ch3a:Schur}, and therefore their multiplication can be done blockwise: 
\begin{equation}
M = M'M''= \bigoplus_{a} \left[ (M'^a M''^a) \otimes \tilde{\mathbb{I}}^a\right].
\end{equation} 
\item \emph{Pre-computation.---} Given a $(P,Q)$-decomposition of tensor $T$, another ($\tilde{P},\tilde{Q}$)-decomposition (as required, e.g., to obtain the matrices $M'$ and $M''$ above) involves a linear map $\Gamma$: 
\begin{equation}
\tilde{P} = \Gamma(P).
\label{eq:ch3a:Gamma}	
\end{equation}
This map $\Gamma$, of which \Eref{eq:ch3a:Fmove2} is an example, is completely determined by the symmetry. In those \TN{} algorithms that proceed by repeating a sequence of manipulations, map $\Gamma$ can be %
computed once and stored in memory for repeated usage.
\end{enumerate}

More detailed explanations of algorithmic details, as well as practical examples of the gains obtained using invariant tensors, are presented in \citet{singh2011a} (also \sref{sec:ch3:U1MERA} of this Thesis) and \citet{singhvidalinprep} for the groups U(1) and SU(2), respectively. \citet{evenbly2010} exploited the U(1) symmetry in a 2D MERA calculation that involved tensors with up to twelve indices.

Finally, the connection between symmetric \TN{}s and spin networks allows us to import into the context of \TN{} algorithms techniques developed to evaluate spin networks in loop quantum gravity. Such techniques can be used, for example, to compute the linear map $\Gamma$ of \Eref{eq:ch3a:Gamma}. Conversely, \TN{} algorithms may also prove useful in loop quantum gravity, since they allow (for example) %
the efficient manipulation of
superpositions of an exponentially large number of spin networks.

\section{Example: U(1)-Symmetric MERA\label{sec:ch3:U1MERA}}

Tensor network decompositions offer an efficient description of certain many-body states of a lattice system and are the basis of a wealth of numerical simulation algorithms. In \sref{sec:ch3:globalsym} %
I discussed how to incorporate a global internal symmetry, given by a compact, completely reducible group $\mathcal{G}$, into tensor network decompositions and algorithms. Here I specialize to the case of Abelian groups and, for concreteness, to a U(1) symmetry, associated, e.g., with particle number conservation. I will consider tensor networks made of tensors that are invariant (or covariant) under the symmetry, and explain how to decompose and manipulate such tensors in order to exploit their symmetry. In numerical calculations, the use of U(1)-symmetric tensors allows selection of a specific number of particles, ensures the exact preservation of particle number, and significantly reduces computational costs. I illustrate all these points in the context of the multi-scale entanglement renormalization Ansatz.

\subsection{Introduction\label{sec:ch3c:intro}}
  
Tensor networks are becoming increasingly popular as a tool to represent wave-functions of quantum many-body systems. Their success is based on the ability to %
{efficiently} describe the ground state of a broad class of local Hamiltonians on the lattice. Tensor network states are used both as a variational Ansatz to numerically approximate ground states and as a theoretical framework to characterize and classify quantum phases of matter.
  
Examples of tensor network states for one dimensional systems include the Matrix Product State or MPS\footnote{\protect{\citet{fannes1992a,ostlund1995,perez-garcia2007}}}, which results naturally from both Wilson's numerical renormalization group \citep{wilson1975} and White's Density Matrix Renormalization Group (DMRG),\footnote{\protect{\citet{white1992, white1993, schollwock2005,schollwock2011, mcculloch2008}}} and is also used as a basis for simulation of time evolution, e.g.~with the time evolving block decimation (TEBD) algorithm \citep{vidal2003,vidal2004,vidal2007b} and variations thereof, often collectively referred to as time-dependent DMRG;\footnote{\protect{\citet{vidal2003,vidal2004,daley2004,white2004,schollwock2005a,vidal2007b}}} the Tree Tensor Network (TTN) \citep{shi2006}, which follows from coarse-graining schemes where the spins are blocked hierarchically; and the Multi-scale Entanglement Renormalization Ansatz (MERA),\footnote{\protect{\citet{vidal2007, vidal2008a, evenbly2009, giovannetti2008, pfeifer2009, vidal2010}}} which results from a renormalization group procedure known as entanglement renormalization \citep{vidal2007,vidal2010}. For two dimensional lattices there are generalizations of these three tensor network states, namely projected entangled pair  states (PEPS),\footnote{\protect{\citet{verstraete2004, sierra1998, nishino1998, nishio2004, murg2007, jordan2008, gu2008, jiang2008a, xie2009, murg2009}}} 2D TTN \citep{tagliacozzo2009, murg2010}, and 2D MERA,\footnote{\protect{\citet{evenbly2010c, evenbly2010d, aguado2008, cincio2008, evenbly2009b, konig2009, evenbly2010}}} respectively. As variational Ans\"atze, PEPS and 2D MERA are particularly interesting since they can be used to address large two-dimensional lattices, including systems of frustrated spins \citep{murg2009, evenbly2010} and interacting fermions,\footnote{\protect{\citet{corboz2010, kraus2010, pineda2010, corboz2009a, barthel2009, shi2009, li2010, corboz2010a, pizorn2010, gu2010}}} where Monte Carlo techniques fail due to the sign problem. 
 
A many-body Hamiltonian $\hat H$ may be invariant under transformations that form a group of symmetries \citep{cornwell1997}. The symmetry group divides the Hilbert space of the theory into symmetry sectors labeled by quantum numbers or conserved charges. On a lattice one can distinguish between \textit{space} symmetries, which correspond to some permutation of the sites of the lattice, and \textit{internal} symmetries, which act on the vector space of each site. An example of space symmetry is invariance under translations by some unit cell, which leads to conservation of momentum. An example of internal symmetry is SU(2) invariance, e.g.~spin isotropy in a quantum spin model. An internal symmetry can in turn be \textit{global}, if it transforms the space of each of the lattice sites according to the same transformation (e.g.~a spin-independent rotation); or \textit{local}, if each lattice site is transformed according to a different transformation (e.g.~a spin-dependent rotation), as it is for gauge symmetric models. A global internal SU(2) symmetry gives rise to conservation of total spin. By targetting a specific symmetry sector during a calculation, computational costs can often be significantly reduced while explicitly preserving the symmetry. It is therefore not surprising that symmetries play an important role in numerical approaches.

In \sref{sec:ch3:globalsym} %
I described a formalism for incorporating global internal symmetries into a generic tensor network algorithm. Both Abelian and non-Abelian symmetries were considered. The purpose of \sref{sec:ch3:U1MERA} is to address, at a pedagogical level, the implementation of Abelian symmetries into tensor networks. We will also discuss several more practical aspects of the exploitation of Abelian symmetries not covered in \sref{sec:ch3:globalsym}. For concreteness this Section concentrates on U(1) symmetry, but extending these results to any Abelian group is straightforward. A similar analysis of non-Abelian groups will be considered in \citet{singhvidalinprep}, as well as being discussed briefly in \cref{sec:nonabelian} of this Thesis. 

In tensor network approaches, the exploitation of global internal symmetries has a long history, especially in the context of MPS. %
Both Abelian and non-Abelian symmetries have been thoroughly incorporated into DMRG code and have been exploited to obtain computational gains.\footnote{\protect{\citet{ostlund1995,white1992,schollwock2005a,ramasesha1996,sierra1997,tatsuaki2000,mcculloch2002,bergkvist2006,pittel2006,mcculloch2007,perez-garcia2008,sanz2009}}} 
Symmetries have also been used in more recent proposals to simulate time evolution with MPS.\footnote{\protect{\citet{vidal2004,daley2004,white2004,schollwock2005a,vidal2007b,daley2005,danshita2007,muth2010,mishmash2009a,singh2010,cai2010}}}

When considering symmetries, it is important to notice that an MPS is a trivalent tensor network. That is, in an MPS each tensor has at most three indices. The Clebsch--Gordan coefficients, or coupling coefficients, of a symmetry group are also trivalent \citep{cornwell1997}, and this makes incorporating the symmetry into an MPS by considering symmetric tensors particularly simple. In contrast, tensor network states with a more elaborate network of tensors, such as MERA or PEPS, consist of tensors having a larger number of indices. In this case a more general formalism is required in order to exploit the symmetry. As explained in \sref{sec:ch3:globalsym}%
, a generic symmetric tensor can be decomposed into a \textit{degeneracy} part, which contains all degrees of freedom not determined by symmetry, and a \textit{structural} part, which is completely determined by symmetry and can be further decomposed as a trivalent network of Clebsch--Gordan coefficients.

The use of symmetric tensors in more complex tensor networks has also been discussed in \citet{perez-garcia2010} and \citet{zhao2010}. In particular, \citet{perez-garcia2010} has shown that under convenient conditions (injectivity), a PEPS that represents a symmetric state can be represented with symmetric tensors, generalizing similar results for MPS obtained in \citet{perez-garcia2008}. Notice that these studies are not concerned with how to decompose symmetric tensors so as to computationally protect or exploit the symmetry. On the other hand, exploitation of U(1) symmetry for computational gain in the context of PEPS was reported in \citet{zhao2010}, although no implementation details were provided. Finally, several aspects of \textit{local} internal symmetries in tensor network algorithms have been addressed in \citet{schuch2010}, \citet{swingle2010}, \citet{chen2010b}, and \citet{tagliacozzo2010}.

The discussion of the U(1)-symmetric MERA is organized into Sections as follows:

Section \ref{sec:ch3c:tensor} contains a review of the tensor network formalism and introduces the nomenclature and diagrammatical representation of tensors used in the rest of the Chapter. It also describes a set $\mathcal{P}$ of primitives for manipulating tensor networks, consisting of manipulations that involve a single tensor (permutation, fusion and splitting of the indices of a tensor) and matrix operations (multiplication and factorization). 

Section~\ref{sec:ch3c:symmetry} reviews basic notions of representation theory of the Abelian group U(1). The action of the group is analysed first on a single vector space, where U(1)-symmetric states and U(1)-invariant operators are decomposed in a compact, canonical manner. %
This canonical form allows us to identify the degrees of freedom which are not constrained by the symmetry. The action of the group is then also analysed on the tensor product of two vector spaces and, finally, on the tensor product of a finite number of vector spaces.
 
Section~\ref{sec:ch3c:symTN} explains how to incorporate the U(1) symmetry into a generic tensor network algorithm, by considering U(1)-invariant tensors in a canonical form, and by adapting the set $\mathcal{P}$ of primitives for manipulating tensor networks. These include the multiplication of two U(1)-invariant matrices in their canonical form, which is at the core of the computational savings obtained by exploiting the symmetry in tensor network algorithms.

Section~\ref{sec:ch3c:MERA} illustrates the practical exploitation of the U(1) symmetry in a tensor network algorithm by presenting %
MERA calculations of the ground state and low energy states of two quantum spin chain models.

The canonical form offers a more compact description of U(1)-invariant tensors, and leads to faster matrix multiplications and factorizations. However, there is also an additional cost associated with maintaining an invariant tensor in its canonical form while reshaping (fusing and/or splitting) its indices. In some situations, this cost may offset the benefits of using the canonical form. In \sref{sec:ch3c:supplement} we discuss a scheme to lower this additional cost in tensor network algorithms that are based on iterating a repeated sequence of transformations. This is achieved by identifying, in the manipulation of a tensor, operations which only depend on the symmetry. Such operations can be \textit{precomputed} once at the beginning of a simulation. Their result, stored in memory, can be re-used at each iteration of the simulation. Section~\ref{sec:ch3c:supplement} describes two such specific precomputation schemes.

\subsection{Review: Tensor Network Formalism\label{sec:ch3c:tensor}}

In this Section we review background material concerning the formalism of tensor networks, without reference to symmetry. We introduce basic definitions and concepts, as well as the nomenclature and graphical representation for tensors, tensor networks, and their manipulations, that will be used in \sref{sec:ch3:U1MERA}.

\subsubsection{Tensors\label{sec:ch3c:tensor:tensor}}

A tensor $\hat{T}$ is a multidimensional array of complex numbers $\hat{T}_{i_{1}i_{2}\cdots i_{k}} \in \mathbb{C}$. The \textit{rank} of tensor $\hat{T}$ is the number $k$ of indices. For instance, a rank-0 tensor ($k=0$) is a complex number. Similarly, rank-1 ($k=1$) and rank-2 ($k=2$) tensors %
represent vectors and matrices, respectively. The \textit{size} of an index $i$, denoted $|i|$, is the number of values that the index takes, $i \in \left\{1, 2, \ldots, |i| \right\}$. The size of a tensor $\hat{T}$, denoted $|\hat{T}|$, is the number of complex numbers it contains, namely $|\hat{T}| = |i_1|\times |i_2| \times \ldots \times |i_k|$.
In this discussion of the U(1)-symmetric MERA, we will use the hat ($\hat{~}$) to indicate that an object is a tensor. Vectors are included in this convention, writing their components as, e.g., $\hat \Psi_i$, although for simplicity we will omit the hat when a vector is written in bra or ket form, e.g.~$|\Psi\rangle$.

\begin{figure}
\begin{center}
\includegraphics[width=300pt]{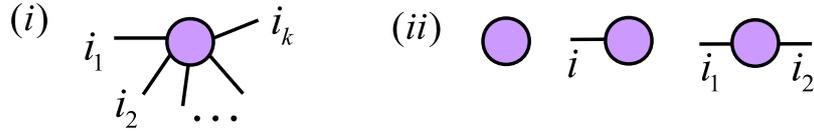}
\caption[(i)~Graphical representation of a tensor $\hat{T}$ of rank $k$ and components $\hat{T}_{i_1 i_2 \ldots i_k}$. (ii)~Graphical representation of tensors with rank $k=0,1$ and $2$, corresponding to a complex number $c \in \mathbb{C}$, a vector $\ket{v} \in \mathbb{C}^{|i|}$ and a matrix $\hat{M} \in \mathbb{C}^{|i_1|\times |i_2|}$, respectively.]{(i)~Graphical representation of a tensor $\hat{T}$ of rank $k$ and components $\hat{T}_{i_1 i_2 \ldots i_k}$. The tensor is represented by a shape (circle) with $k$ emerging lines corresponding to the $k$ indices $i_1, i_2, \ldots, i_k$. Notice that the indices emerge in counterclockwise order. (ii)~Graphical representation of tensors with rank $k=0$, 1, and 2, corresponding to a complex number $c \in \mathbb{C}$, a vector $\ket{v} \in \mathbb{C}^{|i|}$, and a matrix $\hat{M} \in \mathbb{C}^{|i_1|\times |i_2|}$, respectively.\label{fig:ch3c:tensor}} 
\end{center}
\end{figure}

It is convenient to use a graphical representation of tensors, as introduced in Fig.~\ref{fig:ch3c:tensor}, where a tensor $\hat{T}$ is depicted as a circle (more generally some shape, e.g.~a square) and each of its indices is represented by a line emerging from it.
 In order to specify which index corresponds to which emerging line, we follow the prescription that the lines corresponding to indices $\{i_1, i_2, \ldots, i_k\}$ emerge in counterclockwise order. Unless stated otherwise, the first index will correspond to the line emerging at nine o'clock (or the first line encountered while proceeding counterclockwise from nine o'clock).

\begin{figure}
\begin{center}
\includegraphics[width=300pt]{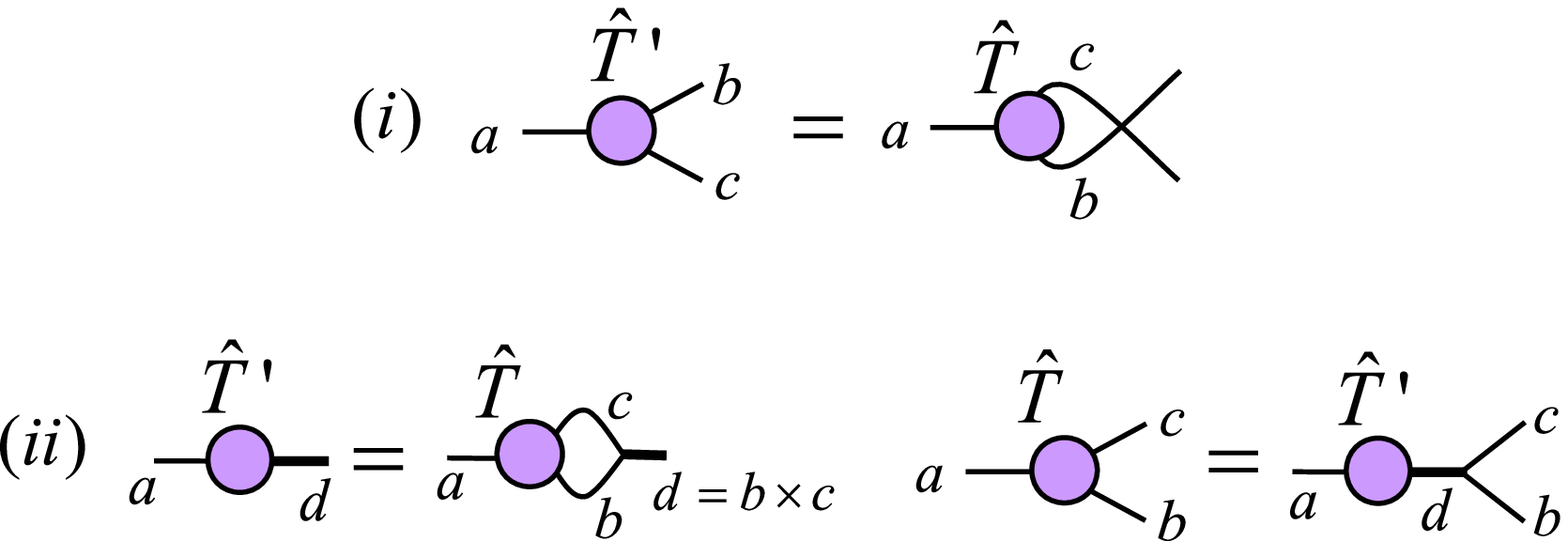}
\caption{%
Transformations of a tensor: (i)~Permutation of indices $b$ and $c$. (ii)~Fusion of indices $b$ and $c$ into $d = b \times c$; splitting of index $d=b \times c$ into $b$ and $c$.\label{fig:ch3c:tensorman}} 
\end{center}
\end{figure}

Two elementary ways in which a tensor $\hat{T}$ can be transformed are by \textit{permuting} and \textit{reshaping} its indices. A \textit{permutation} of indices corresponds to creating a new tensor $\hat{T}'$ from $\hat{T}$ by simply changing the order in which the indices appear, e.g.
\begin{equation}
	(\hat{T}')_{acb} = \hat{T}_{abc}.
	\label{eq:ch3c:permute}
\end{equation}
On the other hand, a tensor $\hat{T}$ can be \textit{reshaped} into a new tensor $\hat{T}'$ by ``fusing'' and/or ``splitting'' some of its indices. For instance, in 
\begin{eqnarray}
	(\hat{T}')_{ad} = \hat{T}_{abc},~~~~~~~d = b\times c,
	\label{eq:ch3c:fuse}
\end{eqnarray}
tensor $\hat{T}'$ is obtained from tensor $\hat{T}$ by fusing indices $b \in \left\{1, \ldots, |b|\right\}$ and $c \in \left\{1, \ldots, |c|\right\}$ together into a single index $d$ of size $|d| = |b| \cdot |c|$ that runs over all pairs of values of $b$ and $c$, i.e.~$d \in \left\{ (1,1), (1,2), \ldots, (|b|, |c|-1), (|b|,|c|) \right\}$, whereas in
\begin{eqnarray}
	\hat{T}_{abc} = (\hat{T}')_{ad},~~~~~~~d = b\times c,
	\label{eq:ch3c:split}
\end{eqnarray}
tensor $\hat{T}$ is recovered from $\hat{T}'$ by splitting index $d$ of $\hat{T}'$ back into indices $b$ and $c$. The permutation and reshaping of the indices of a tensor have a straightforward graphical representation; see Fig.~\ref{fig:ch3c:tensorman}. 

\subsubsection{Multiplication of Two Tensors\label{sec:ch3c:tensor:multiply}}

\begin{figure}
\begin{center}
\includegraphics[width=225pt]{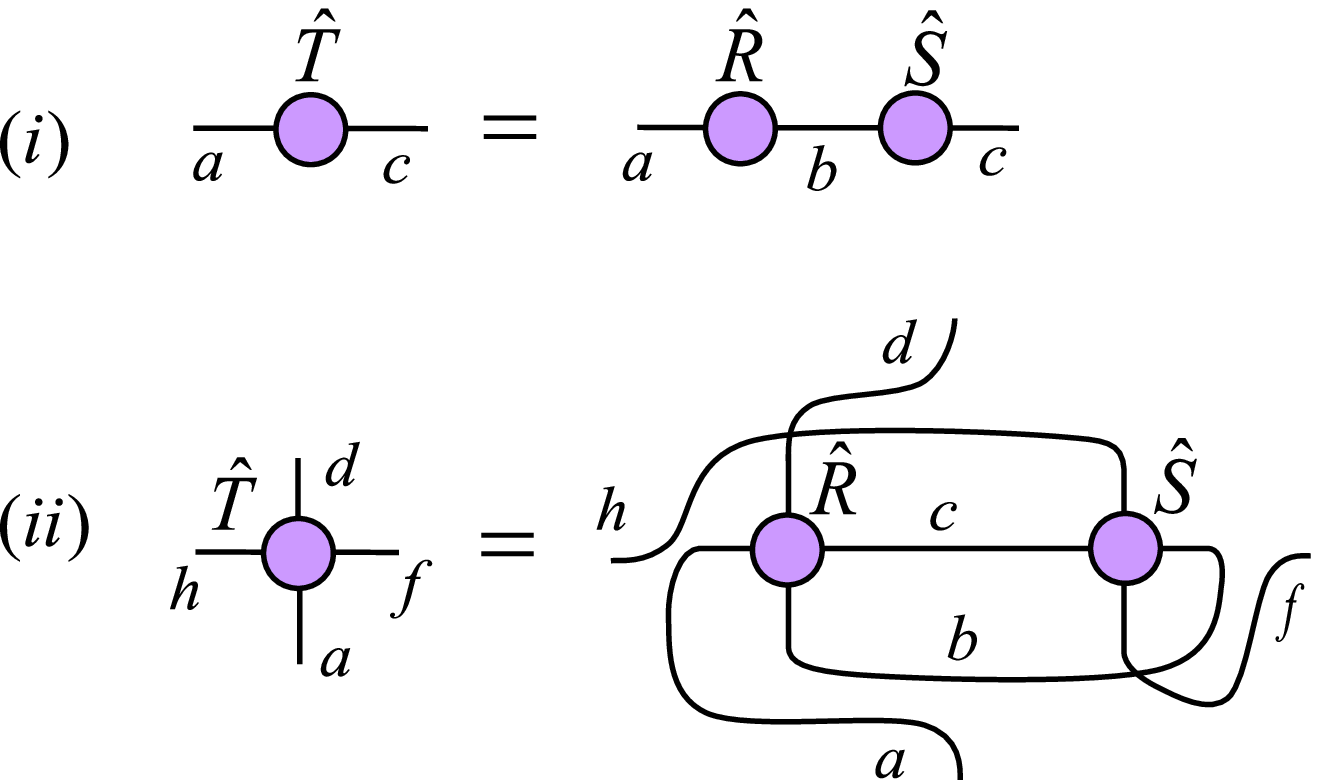}
\caption[(i)~Graphical representation of the multiplication of two matrices $\hat{R}$ and $\hat{S}$ to give a new matrix $\hat{T}$ (ii)~Graphical representation of an example contraction of two tensors $\hat{R}$ and $\hat{S}$ into a new tensor $\hat{T}$.]{%
(i)~Graphical representation of the multiplication of two matrices $\hat{R}$ and $\hat{S}$ to give a new matrix $\hat{T}$ \protect{\eref{eq:ch3c:Mmultiply}} (ii)~Graphical representation of an example contraction of two tensors $\hat{R}$ and $\hat{S}$ into a new tensor $\hat{T}$ \protect{\eref{eq:ch3c:multiply}}. \label{fig:ch3c:multiply1}} 
\end{center}
\end{figure}

Given two matrices $\hat{R}$ and $\hat{S}$ with components $\hat{R}_{ab}$ and $\hat{S}_{bc}$, we can multiply them together to obtain a new matrix $\hat{T}$, $\hat{T} = \hat{R}\cdot \hat{S}$, with components
\begin{equation}
	\hat{T}_{ac} = \sum_{b} \hat{R}_{ab}\hat{S}_{bc},
	\label{eq:ch3c:Mmultiply}
\end{equation}
by summing over or \textit{contracting} index $b$. The multiplication of matrices $\hat{R}$ and $\hat{S}$ is represented graphically by connecting together the emerging lines of $\hat{R}$ and $\hat{S}$ corresponding to the contracted index, as shown in Fig.~\ref{fig:ch3c:multiply1}(i).

Matrix multiplication can be generalized to tensors. For instance, given tensors $\hat{R}$ and $\hat{S}$ with components $\hat{R}_{abcd}$ and $\hat{S}_{cfbh}$, we can define a tensor $\hat{T}$ with components $\hat{T}_{hafd}$ given by
\begin{equation}
	\hat{T}_{hafd} = \sum_{bc} \hat{R}_{abcd}\hat{S}_{cfbh}.
\label{eq:ch3c:multiply}
\end{equation}
Again the multiplication of two tensors can be graphically represented by connecting together the lines corresponding to indices that are being contracted [indices $b$ and $c$ in \Eref{eq:ch3c:multiply}]; see Fig.~\ref{fig:ch3c:multiply1}(ii).

\begin{figure}
\begin{center}
\includegraphics[width=300pt]{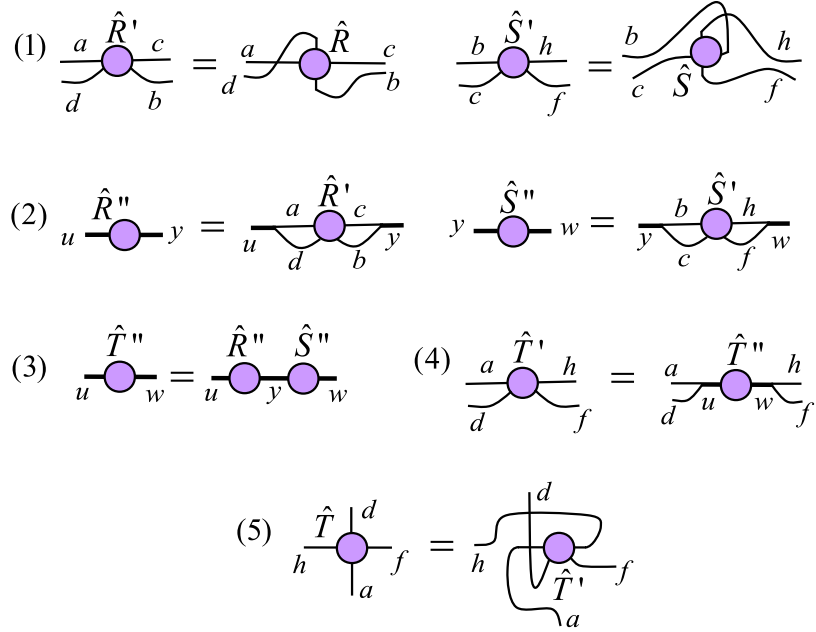}
\caption[Graphical representations of the five elementary steps (1)--(5) into which one can decompose the contraction of the tensors of \protect{\fref{fig:ch3c:multiply1}}(ii).]{%
Graphical representations of the five elementary steps 1-5 into which one can decompose the contraction of the tensors of \protect{\Eref{eq:ch3c:multiply}}.\label{fig:ch3c:multiply2}} 
\end{center}
\end{figure}

The multiplication of two tensors can be broken down into a sequence of elementary steps by transforming the tensors into matrices, multiplying the matrices together, and then transforming the resulting matrix back into a tensor. These steps are now described for the contraction given in \Eref{eq:ch3c:multiply}. They are illustrated in Fig.~\ref{fig:ch3c:multiply2}.

\begin{enumerate}
	\item \textit{Permute} the indices of tensor $\hat{R}$ in such a way that the indices to be contracted, $b$ and $c$, appear in the last positions and in a given order, e.g.~$bc$; similarly, permute the indices of $\hat{S}$ so that the indices to be contracted, again $b$ and $c$, appear in the first positions and in the same order $bc$: 
	\begin{align}
	(\hat{R}')_{ad ~bc} &= \hat{R}_{abc d}   \nonumber \\
	(\hat{S}')_{bc ~fh} &= \hat{S}_{c f b h} 
	\end{align}
	
	\item \textit{Reshape} tensor $\hat{R}'$ into a matrix $\hat{R}''$ by fusing into a single index $u$ all the indices that are not going to be contracted, $u = a\times d$, and into a single index $y$ all indices to be contracted, $y = b \times c$. Similarly, reshape tensor $\hat{S}'$ into a matrix $\hat{S}''$ with indices $y = b\times c$ and $w = f\times h$,
		\begin{align}\label{eg1}
		(\hat{R}'')_{uy} &= (\hat{R}')_{adbc} \nonumber \\
		(\hat{S}'')_{y w} &= (\hat{S}')_{b c fh}.
	\end{align}
	
	\item \textit{Multiply} matrices $\hat{R}''$ and $\hat{S}''$ to obtain a matrix $\hat{T}''$, with components
	\begin{equation}
	(\hat{T}'')_{uw} = \sum_{y} (\hat{R}'')_{uy} ~~(\hat{S}'')_{y w}
	\end{equation}
	
	\item \textit{Reshape} matrix $\hat{T}''$ into a tensor $\hat{T}'$ by splitting indices $u = a\times d$ and $w = f\times h$,
		\begin{equation}
	(\hat{T}')_{adfh} = (\hat{T}'')_{uw}
	\end{equation}

	\item \textit{Permute} the indices of $\hat{T}'$ into the order in which they appear in $\hat{T}$,
	\begin{equation}
	\hat{T}_{hafd} = (\hat{T}')_{adfh}.\label{eq:ch3c:endmultiply}
	\end{equation}
\end{enumerate}

Note that breaking down a multiplication of two tensors into elementary steps is not necessary---one can simply implement the contraction of \Eref{eq:ch3c:multiply} as a single process. However, it is often more convenient to compose the above elementary steps since, for instance, in this way one can use existing linear algebra libraries for matrix multiplication. In addition, it can be seen that the leading computational cost in multiplying two large tensors is not changed when decomposing the contraction in the above steps. In Sec.~\ref{sec:ch3c:symTN:discussion} this subject will be discussed in more detail for U(1)-invariant tensors. 

\subsubsection{Factorization of a Tensor\label{sec:ch3c:tensor:factorize}}

\begin{figure}
\begin{center}
  \includegraphics[width=225pt]{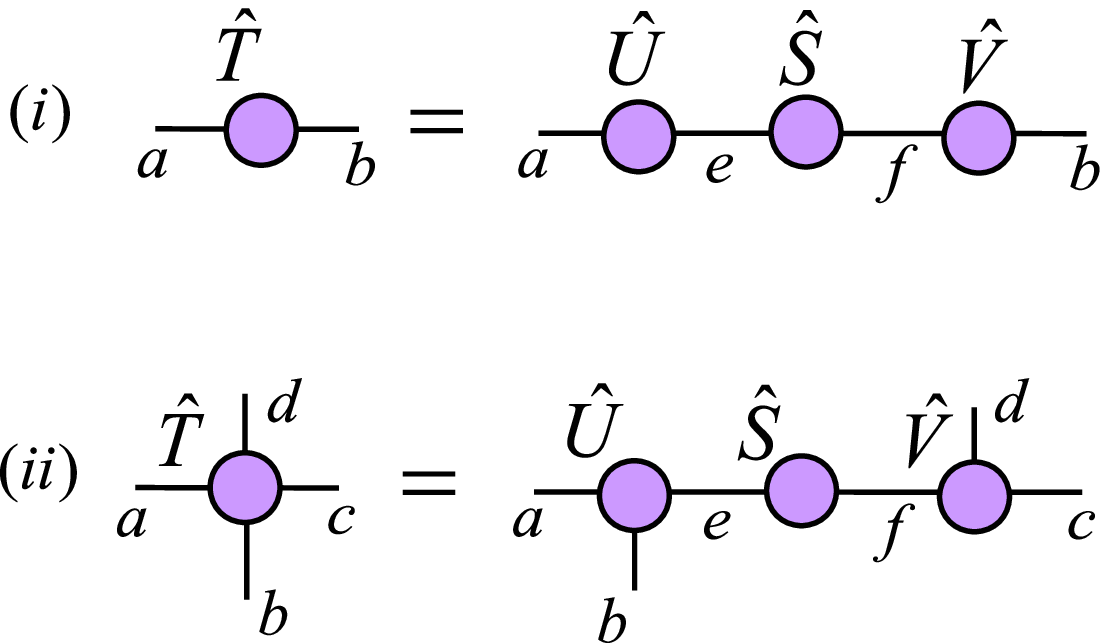}
\caption[(i)~Factorization of a matrix $\hat{T}$ according to a singular value decomposition. (ii)~Factorization of a rank-4 tensor $\hat{T}$ according to one of several possible singular value decompositions.]{%
(i)~Factorization of a matrix $\hat{T}$ according to a singular value decomposition \protect{\eref{eq:ch3c:singular}}. (ii)~Factorization of a rank-4 tensor $\hat{T}$ according to one of several possible singular value decompositions. \label{fig:ch3c:decompose}} 
\end{center}
\end{figure}

A matrix $\hat{T}$ can be factorized into the product of two (or more) matrices in one of several canonical forms. For instance, the \textit{singular value decomposition}
\begin{equation}
	\hat{T}_{ab} = \sum_{c,d} \hat{U}_{ac}\hat{S}_{cd}\hat{V}_{db} 
	= \sum_{c} \hat{U}_{ac}s_{c}\hat{V}_{cb}
	\label{eq:ch3c:singular}
\end{equation}
factorizes $\hat{T}$ into the product of two unitary matrices $\hat{U}$ and $\hat{V}$, and a diagonal matrix $\hat{S}$ with non-negative diagonal elements $s_c = \hat{S}_{cc}$ known as the \textit{singular values} of $\hat{T}$; see Fig.~\ref{fig:ch3c:decompose}(i).
 On the other hand, the \textit{eigenvalue} or \textit{spectral decomposition} of a square matrix $\hat{T}$ is of the form
\begin{equation}
	\hat{T}_{ab} = \sum_{c,d} \hat{M}_{ac}D_{cd}(\hat{M}^{-1})_{db} 
	= \sum_{c} \hat{M}_{ac}\lambda_{c}(\hat{M}^{-1})_{cb},
	\label{eq:ch3c:spectral}
\end{equation}
where $\hat{M}$ is an invertible matrix whose columns encode the eigenvectors $\ket{\lambda_c}$ of $\hat{T}$, 
\begin{equation}
	\hat{T} \ket{\lambda_{c}} = \lambda_c \ket{\lambda_c},
\end{equation}
$\hat{M}^{-1}$ is the inverse of $\hat{M}$, and $\hat{D}$ is a diagonal matrix, with the eigenvalues $\lambda_c=\hat{D}_{cc}$ on its diagonal. Other useful factorizations include the LU decomposition, the QR decomposition, etc. We refer to any such decomposition generically as a \textit{matrix factorization}.

A tensor $\hat{T}$ with more than two indices can be converted into a matrix in several ways, by specifying how to join its indices into two subsets. After specifying how tensor $\hat{T}$ is to be regarded as a matrix, we can factorize $\hat{T}$ according to any of the above matrix factorizations, as illustrated in Fig.~\ref{fig:ch3c:decompose}(ii) for a singular value decomposition. This requires first permuting and reshaping the indices of $\hat{T}$ to form a matrix, then decomposing the latter, and finally restoring the open indices of the resulting matrices into their original form by undoing the reshapes and permutations.

\subsubsection{Tensor Networks and Their Manipulation\label{sec:ch3c:tensor:TN}}

\begin{figure}
\begin{center}
  \includegraphics[width=262.5pt]{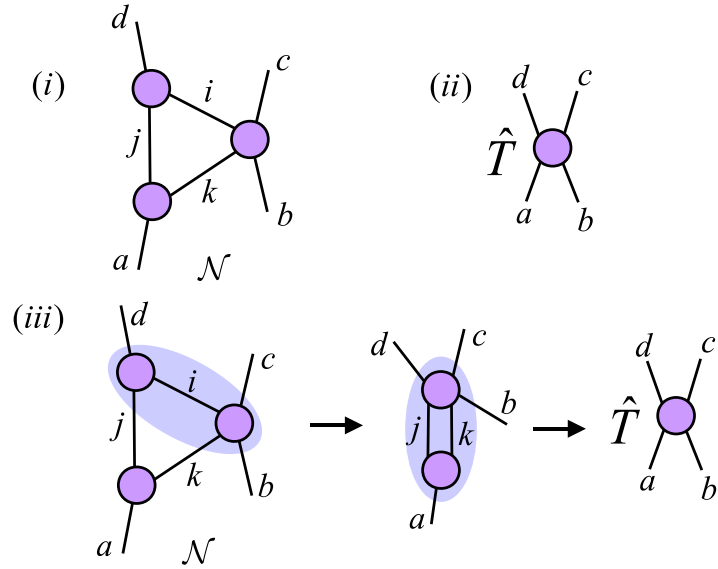}
\caption[(i)~Example of a tensor network $\mathcal{N}$. (ii)~Tensor $\hat{T}$ of which the tensor network $\mathcal{N}$ could be a representation. (iii)~Tensor $\hat{T}$ can be obtained from $\mathcal{N}$ through a sequence of contractions of pairs of tensors.]{%
(i)~Example of a tensor network $\mathcal{N}$. (ii)~Tensor $\hat{T}$ of which the tensor network $\mathcal{N}$ could be a representation. (iii)~Tensor $\hat{T}$ can be obtained from $\mathcal{N}$ through a sequence of contractions of pairs of tensors. Shading indicates the two tensors to be multiplied together at each step.\label{fig:ch3c:TN}} 
\end{center}
\end{figure}

A \textit{tensor network} $\mathcal{N}$ is a set of tensors whose indices are connected according to a network pattern, e.g.~Fig.~\ref{fig:ch3c:TN}. 

Given a tensor network $\mathcal{N}$, a single tensor $\hat{T}$ can be obtained by contracting all the indices that connect the tensors in $\mathcal{N}$ [\fref{fig:ch3c:TN}(ii)]. Here, the indices of tensor $\hat{T}$ correspond to the open indices of the tensor network $\mathcal{N}$. We then say that the 
network $\mathcal{N}$ is a tensor network decomposition of $\hat{T}$. One way to obtain $\hat{T}$ from $\mathcal{N}$ is through a sequence of contractions involving two tensors at a time [Fig.~\ref{fig:ch3c:TN}(iii)].

From a tensor network decomposition $\mathcal{N}$ for a tensor $\hat{T}$, another tensor network decomposition for the same tensor $\hat{T}$ can be obtained in many ways. One possibility is to replace two tensors in $\mathcal{N}$ with the tensor resulting from contracting them together, as is done in each step of Fig.~\ref{fig:ch3c:TN}(iii). Another way is to replace a tensor in $\mathcal{N}$ with a decomposition of that tensor (e.g.~with a singular value decomposition). In this Chapter, we will be concerned with manipulations of a tensor network that, as in the case of multiplying two tensors or decomposing a tensor, can be broken down into a sequence of operations from the following list:
\begin{enumerate}
	\item Permutation of the indices of a tensor, \Eref{eq:ch3c:permute}.
	\item Reshape of the indices of a tensor, Eqs.~\eref{eq:ch3c:fuse}--\eref{eq:ch3c:split}.
	\item Multiplication of two matrices, \Eref{eq:ch3c:Mmultiply}.
	\item Decomposition of a matrix [e.g.~singular value decomposition \protect{\eref{eq:ch3c:singular}} or spectral decomposition \protect{\eref{eq:ch3c:spectral}}].
\end{enumerate}

These operations constitute a set $\mathcal{P}$ of \textit{primitive} operations for tensor network manipulations (or, at least, for the type of manipulations we will be concerned with). 

In \sref{sec:ch3c:symTN} we will discuss how this set $\mathcal{P}$ of primitive operations can be generalized to tensors that are symmetric under the action of the group U(1).

\subsubsection{Tensor Network States for Quantum Many-Body Systems\label{sec:ch3c:tensor:TNstates}}

As mentioned in \sref{sec:ch3c:intro}, tensor networks are used as a means to represent the wave-function of certain quantum many-body systems on a lattice. Let us consider a lattice $\mathcal{L}$ made of $L$ sites, each described by a complex vector space $\mathbb{V}$ of dimension $d$. A generic pure state $\ket{\Psi} \in \mathbb{V}^{\otimes L}$ of $\mathcal{L}$ can always be expanded as
\begin{equation}
\label{eq:ch3c:purePsi}
\ket{\Psi} = \sum_{i_{1}, i_{2},\ldots, i_{L}} \hat{\Psi}_{i_{1} i_{2}\ldots i_{L}} \ket{i_{1}}\ket{ i_{2}} \ldots \ket{i_{L}},
\end{equation}
where $i_{s} = 1, \ldots, d$ labels a basis $\ket{i_s}$ of $\mathbb{V}$ for site $s \in \mathcal{L}$. Tensor $\hat{\Psi}$, with components $\hat\Psi_{i_{1} i_{2}\ldots i_{L}}$, contains $d^L$ complex coefficients. This is a number that grows exponentially with the size $L$ of the lattice. Thus, the representation of a \textit{generic} pure state $\ket{\Psi} \in \mathbb{V}^{\otimes L}$ is \textit{inefficient}. However, it turns out that an \textit{efficient} representation of \textit{certain} pure states can be obtained by expressing tensor $\hat{\Psi}$ in terms of a tensor network.

\begin{figure}
\begin{center}
  \includegraphics[width=300pt]{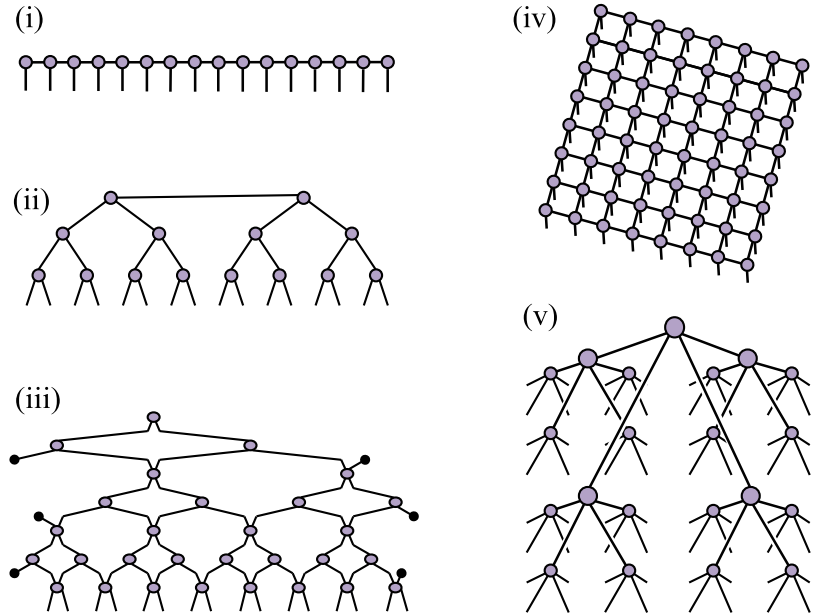}
\caption[Examples of tensor network states for 1D systems: (i)~matrix product state (MPS), (ii)~tree tensor network (TTN), (iii)~multi-scale entanglement renormalization Ansatz (MERA). Examples of tensor network states for 2D systems: (iv)~projected entangled-pair state PEPS, (v)~2D TTN.]{%
Examples of tensor network states for 1D systems: (i)~matrix product state (MPS), (ii)~tree tensor network (TTN), (iii)~multi-scale entanglement renormalization Ansatz (MERA). Examples of tensor network states for 2D systems: (iv)~projected entangled-pair state PEPS, (v)~2D TTN. (2D MERA not depicted).\label{fig:ch3c:TNs}}
\end{center}
\end{figure}

Figure~\ref{fig:ch3c:TNs} shows several popular tensor network decompositions used to approximately describe the ground states of local Hamiltonians $H$ of lattice models in one or two spatial dimensions. 
The open indices of each of these tensor networks correspond to the indices $i_1, i_2, \ldots, i_L$ of tensor $\hat{\Psi}$. Notice that all the tensor networks of Fig.~\ref{fig:ch3c:TNs} contain $O(L)$ tensors. If $p$ is the rank of the tensors in one of these tensor networks, and $\chi$ is the size of their indices, then the tensor network depends on $O(L\chi^p)$ complex coefficients. For a fixed value of $\chi$ this number grows linearly in $L$, and not exponentially. It therefore does indeed offer an efficient description of the pure state $\ket{\Psi} \in \mathbb{V}^{\otimes L}$ that it represents. Of course only a subset of pure states can be decomposed in this way. Such states, often referred to as tensor network states, are used as variational Ans\"atze, with the $O(L\chi^p)$ complex coefficients as the variational parameters.

Given a tensor network state, a variety of algorithms\footnote{see e.g.~\protect{\citet{wilson1975,white1992,white1993,schollwock2005,schollwock2011,mcculloch2008,vidal2003,vidal2004,vidal2007b,daley2004,white2004,schollwock2005a,shi2006,vidal2007,vidal2008a,evenbly2009,giovannetti2008,pfeifer2009,vidal2010,verstraete2004,sierra1998,nishino1998,nishio2004,murg2007,jordan2008,gu2008,jiang2008a,xie2009,murg2009,tagliacozzo2009,murg2010,evenbly2010c,evenbly2010d,aguado2008,cincio2008,evenbly2009b,konig2009,evenbly2010,corboz2010,kraus2010,pineda2010,corboz2009a,barthel2009,shi2009,li2010,corboz2010a,gu2010}}.\label{footnote:ch3c:fn1}} are used for tasks such as: (i)~computation of the expectation value $\bra{\Psi}\hat o\ket{\Psi}$ of a local observable $\hat o$, (ii)~optimization of the variational parameters so as to minimize the expectation value of the energy $\bra{\Psi}\hat{H}\ket{\Psi}$, or (iii)~simulation of time evolution, e.g.~$e^{-\rmi\hat H t}\ket{\Psi}$. These tasks are accomplished by manipulating tensor networks. 

On most occasions, all required manipulations can be reduced to a sequence of primitive operations in the set $\mathcal{P}$ introduced in Sec.~\ref{sec:ch3c:tensor:TN}. Thus, in order to adapt tensor network algorithms such as those listed in footnote~\ref{footnote:ch3c:fn1}, above, to the presence of a symmetry, we only need to modify the set $\mathcal{P}$ of primitive tensor network operations. This will be done in Sec.~\ref{sec:ch3c:symTN}.

\subsubsection{Tensors as Linear Maps\label{sec:ch3c:tensor:linear}}

\begin{figure}
\begin{center}
  \includegraphics[width=262.5pt]{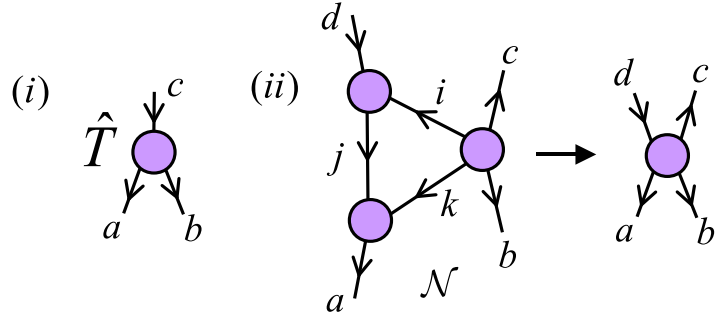}
\caption[(i)~Tensor $\hat{T}$ with one incoming index and two outgoing indices, denoted by incoming and outgoing arrows respectively. (ii)~A tensor network $\mathcal{N}$ with directed links can be interpreted as a linear map between incoming and outgoing spaces (of the incoming and outgoing indices) obtained by composing the linear maps associated with each of the tensors in $\mathcal{N}$.]{%
(i)~Tensor $\hat{T}$ with one incoming index and two outgoing indices, denoted by incoming and outgoing arrows respectively \protect{\eref{eq:ch3c:Tabc}}. (ii)~A tensor network $\mathcal{N}$ with directed links can be interpreted as a linear map between incoming and outgoing spaces (of the incoming and outgoing indices) obtained by composing the linear maps associated with each of the tensors in $\mathcal{N}$.\label{fig:ch3c:arrow}} 
\end{center}
\end{figure}

A tensor can be used to define a linear map between vector spaces in the following way. 
First, notice that an index $i$ can be used to label a basis $\{\ket{i}\}$ of a complex vector space $\mathbb{V}^{[i]} \cong \mathbb{C}^{|i|}$ of dimension $|i|$. On the other hand, given a tensor $\hat{T}$ of rank $k$, we can attach a direction ``in'' or ``out'' to each index $i_{1}, i_{2}, \ldots, i_k$. This direction divides the indices of $\hat{T}$ into the subset $I$ of \textit{incoming} indices and the subset $O$ of \textit{outgoing} indices. We can then build input and output vector spaces given by the tensor product of the spaces of incoming and outgoing indices,
\begin{equation}
	\mathbb{V}^{[\text{in}]} = \bigotimes_{i_l \in I} \mathbb{V}^{[i_l]},~~~~~~~
	\mathbb{V}^{[\text{out}]} = \bigotimes_{i_l \in O} \mathbb{V}^{[i_l]},
	\label{eq:ch3c:inout}
\end{equation}
and use tensor $\hat{T}$ to define a linear map between $\mathbb{V}^{[\text{in}]}$ and $\mathbb{V}^{[\text{out}]}$. For instance, if a rank-3 tensor $\hat{T}_{abc}$ has one incoming index $c \in I$ and two outgoing indices $a,b \in O$, then it defines a linear map $\hat{T} : \mathbb{V}^{[c]} \rightarrow \mathbb{V}^{[a]}\otimes \mathbb{V}^{[b]}$ given by
\begin{equation}
	\hat{T} = \sum_{a,b,c} \hat{T}_{abc}  \ket{a}\ket{b}  \bra{c} 
	\label{eq:ch3c:Tabc}
\end{equation}
Graphically, we denote the direction of an index by means of an arrow; see Fig.~\ref{fig:ch3c:arrow}(i).

By decorating the lines of a tensor network $\mathcal{N}$ with arrows [Fig.~\ref{fig:ch3c:arrow}(ii)], this can be regarded as a composition of linear maps---namely, one linear map for each tensor in $\mathcal{N}$. While arrows might be of limited relevance in the absence of a symmetry, they will play an important role when we consider symmetric tensors since they specify how the group acts on each index of a given tensor.

\subsection {Review: Representation Theory of the Group U(1)\label{sec:ch3c:symmetry}}

In this Section we review basic background material concerning the representation theory of the group U(1). We first consider the action of U(1) on a vector space $\mathbb{V}$, which decomposes into the direct sum of (possibly degenerate) irreducible representations. We then consider vectors of $\mathbb{V}$ that are symmetric (invariant or covariant) under the action of U(1), as well as linear operators that are U(1)-invariant. Then we consider the action of U(1) on the tensor product of two vector spaces, and its generalization to the tensor product of an arbitrary number of vector spaces.

\subsubsection{Decomposition Into Direct Sum of Irreducible Representations\label{sec:ch3c:symmetry:irreps}}

Let $\mathbb{V}$ be a finite-dimensional space and let $\varphi \in [0,2\pi)$ label a set of linear transformations $\hat{W}_\varphi$, 
\begin{equation}
\hat{W}_{\varphi}:\mathbb{V}\rightarrow \mathbb{V}, 
\end{equation}
that are a unitary representation of the group U(1). That is,
\begin{align}
&\hat{W}_{\varphi}^{~\dagger} \hat{W}_{\varphi} = \hat{W}_{\varphi} \hat{W}_{ \varphi}^{~\dagger} = \mathbb{I}, &&\forall\ \varphi \in [0,2\pi),\\
&\hat{W}_{\varphi_1}\hat{W}_{\varphi_2} = \hat{W}_{\varphi_2}\hat{W}_{\varphi_1} = \hat{W}_{\varphi_1+\varphi_2|_{2\pi}} &&\forall\ \varphi_1,\varphi_2 \in [0,2\pi).
\end{align}
Then $\mathbb{V}$ decomposes as the direct sum of (possibly degenerate) one-dimensional irreducible representations (or \textit{irreps}) of U(1),
\begin{equation}
\mathbb{V} \cong \bigoplus_n \mathbb{V}_{n},
\label{eq:ch3c:decoV}
\end{equation}
where $\mathbb{V}_{n}$ is a subspace of dimension $d_{n}$, made of $d_n$ copies of an irrep of U(1) with charge $n\in \mathbb{Z}$. We say that irrep $n$ is $d_n$-fold degenerate and that $\mathbb{V}_n$ is the degeneracy space. For concreteness, in this Section we identify the integer charge $n$ as labelling the number of particles (another frequent identification is with the $z$ component of the spin, in which case semi-integer numbers may be considered). 
The representation of group U(1) is generated by the particle number operator $\hat{n}$,
\begin{equation}
\hat{n} \equiv \sum_n n \hat{P}_n,~~~~~\hat{P}_n \equiv \sum_{t_{n}=1}^{d_{n}} \ketbra{nt_{n}}{nt_{n}},
\end{equation}
where $\hat{P}_n$ is a projector onto the subspace $\mathbb{V}_n$ of particle number $n$, and the vectors $\ket{n t_n}$, 
\begin{equation}
	\hat{n}\ket{nt_{n}} = n\ket{nt_{n}},~~~~~~t_n=1,\ldots, d_n,
\label{eq:ch3c:eigen}
\end{equation}
are an orthonormal basis of $\mathbb{V}_n$. 
In terms of $\hat{n}$, the transformations $\hat{W}_{\varphi}$ read
\begin{equation}
	\hat{W}_{\varphi} = e^{-\rmi\hat{n}\varphi}.
\end{equation}
It then follows from \Eref{eq:ch3c:eigen} that
\begin{equation}
	\hat{W}_{\varphi}\ket{nt_{n}} = e^{-\rmi n\varphi}\ket{nt_{n}}~~~~~~\forall\ \varphi \in [0,2\pi).
\end{equation}
The dual basis $\left\{\bra{nt_{n}}\right\}$ is transformed by the \textit{dual representation} of U(1), with elements $\hat{W}_{\varphi}^{~\dagger}$, as
\begin{equation}	
	\bra{nt_{n}} \hat{W}_{\varphi}^{~\dagger} = e^{\rmi n\varphi} \bra{nt_{n}}~~~~~~\forall\ \varphi \in [0,2\pi).
\end{equation}

\textbf{Example 1:} Consider a two-dimensional space $\mathbb{V}$ that decomposes as $\mathbb{V} \cong \mathbb{V}_{0} \oplus \mathbb{V}_{1}$, where the irreps $n = 0$ and $n= 1$ are non-degenerate (i.e.~$d_0=d_1=1$). Then the orthogonal vectors $\left\{\ket{n=0, t_0 = 1}, \ket{n=1, t_1 = 1}\right\}$ form a basis of $\mathbb{V}$. In column vector notation,
\begin{equation}
\begin{pmatrix} 1 \\ 0 \end{pmatrix}  \equiv \;  \ket{n=0, t_0 = 1},~~~~
\begin{pmatrix} 0 \\ 1 \end{pmatrix}  \equiv \; \ket{n=1, t_1 = 1},
\end{equation}
the particle number operator $\hat{n}$ and transformation $\hat{W}_{\varphi}$ read
\begin{equation}
\hat{n} \equiv \; \begin{pmatrix} 0 & 0 \\ 0 & 1 \end{pmatrix},~~~~~~
\hat{W}_{\varphi} \equiv \; \begin{pmatrix} 1 & 0 \\ 0 & e^{-\rmi\varphi} \end{pmatrix}.
\end{equation}

\textbf{Example 2:} Consider a four-dimensional space $\mathbb{V}$ that decomposes as $\mathbb{V} \cong \mathbb{V}_{0} \oplus \mathbb{V}_{1} \oplus \mathbb{V}_{2}$, where $d_0=d_2=1$ and $d_1=2$, so that now irrep $n = 1$ is twofold degenerate. Let $\left\{\ket{n=1, t_1 = 1}, \ket{n=1, t_1 = 2}\right\}$ form a basis of $\mathbb{V}_1$. In column vector notation,
\begin{equation}
\begin{pmatrix} 1 \\ 0 \\ 0 \\ 0 \end{pmatrix}  \equiv \;  \ket{n=0, t_0 = 1},\qquad
\begin{pmatrix} 0 \\ 1 \\ 0 \\ 0 \end{pmatrix}  \equiv \; \ket{n=1, t_1 = 1},
\end{equation}
\begin{equation}
\begin{pmatrix} 0 \\ 0 \\ 1 \\ 0 \end{pmatrix}  \equiv \;  \ket{n=1, t_1 = 2},\qquad
\begin{pmatrix} 0 \\ 0 \\ 0 \\ 1 \end{pmatrix}  \equiv \; \ket{n=2, t_2 = 1},
\end{equation}
the particle number operator $\hat{n}$ and transformation $\hat{W}_{\varphi}$ read
\begin{equation}
\hat{n} \equiv \; \begin{pmatrix} 0 & 0 & 0 & 0 \\ 0 & 1 & 0 & 0 \\ 0 & 0 & 1 & 0 \\ 0 & 0 & 0 & 2  \end{pmatrix},~~~
\hat{W} \equiv \; \begin{pmatrix} 1 & 0 & 0 & 0 \\ 0 & e^{-\rmi\varphi} & 0 & 0 \\ 0 & 0 & e^{-\rmi\varphi} & 0 \\ 0 & 0 & 0 & e^{-\rmi2\varphi}  \end{pmatrix}. 
\end{equation}

\subsubsection{Symmetric States and Operators\label{sec:ch3c:symmetry:states}}

In this work we are interested in states and operators that have a simple transformation rule under the action of U(1). A pure state $\ket{\Psi} \in \mathbb{V}$ is \textit{symmetric} if it transforms as
\begin{equation}
	\hat{W}_{\varphi}\ket{\Psi} = e^{-\rmi n \varphi} \ket{\Psi}~~~~~~\forall\ \varphi \in [0,2\pi).
	\label{eq:ch3c:nPsi1}
\end{equation}
The case $n=0$ corresponds to an \textit{invariant} state, $\hat{W}_{\varphi}\ket{\Psi} = \ket{\Psi}$, which transforms trivially under U(1), whereas for $n\neq 0$ the state is \textit{covariant}, with $\ket{\Psi}$ being multiplied by a non-trivial phase $e^{-\rmi n\varphi}$. Notice that a symmetric state $\ket{\Psi}$ is an eigenstate of $\hat{n}$: that is, it has a well-defined particle number $n$. $\ket{\Psi}$ can thus be expanded in terms of a basis of the relevant subspace $\mathbb{V}_n$,
\begin{equation}
	\hat{n}\ket{\Psi} = n \ket{\Psi},~~~~~~~\ket{\Psi} = \sum_{t_n=1}^{d_n} (\hat\Psi_n)_{t_n} \ket{n t_n},
	\label{eq:ch3c:nPsi2}
\end{equation}
where we have introduced a charge label $_n$ on the state coefficients of $\ket{\Psi}$ so that we can explicitly associate each coefficient $(\hat\Psi_n)_{t_n}$ 
with its corresponding basis vector $\ket{nt_n}$.
 
A linear operator $\hat{T}: \mathbb{V} \rightarrow \mathbb{V}$ is invariant if it commutes with the generator $\hat{n}$ ,
\begin{equation}
	[\hat{T}, \hat{n}] = 0,
	\label{eq:ch3c:commutator}
\end{equation}
or equivalently if it commutes with the action of the group,
\begin{equation}
	\hat{W}_{\varphi} \hat{T} \hat{W}_{\varphi}^{~\dagger} = \hat{T}~~~~~~~~~\forall\ \varphi \in [0,2\pi).
	\label{eq:ch3c:commutator2}
\end{equation}
It follows that $\hat{T}$ decomposes as (Schur's lemma)
\begin{equation}
\hat{T} = \bigoplus_{n} \hat{T}_{n}
\label{eq:ch3c:Schur}
\end{equation}
where $\hat{T}_{n}$ is a $d_n\times d_n$ matrix that acts on the subspace $\mathbb{V}_n$ in \Eref{eq:ch3c:decoV}.

Notice that the operator $\hat{T}$ in \Eref{eq:ch3c:Schur} transforms vectors with a well-defined particle number $n$ into vectors with the same particle number. That is, U(1)-invariant operators \textit{conserve particle number}. 

\textbf{Example 1 revisited:} In Example 1 above, symmetric vectors must be proportional to either $\ket{n=0, t_0=1}$ or $\ket{n=1,t_1=1}$. An invariant operator $\hat{T} = \hat{T}_0 \oplus \hat{T}_1$ is of the form
\begin{equation}
\hat{T} = \; \begin{pmatrix} \alpha_0 & 0 \\ 0 & \alpha_1 \end{pmatrix},~~~ \alpha_0,\alpha_1 \in \mathbb{C}.
\label{eq:ch3c:ex1rev}
\end{equation}

\textbf{Example 2 revisited:} In Example 2 above, a symmetric vector $\ket{\Psi}$ must be of the form
\begin{equation}
	\ket{\Psi} = \begin{pmatrix} \alpha_0 \\ 0 \\ 0 \\ 0 \end{pmatrix}, ~~~~~
\ket{\Psi} = \begin{pmatrix} 0 \\ \alpha_1 \\ \beta_1 \\ 0 \end{pmatrix}, ~~~\mbox{or}~~~
\ket{\Psi} = \begin{pmatrix} 0 \\ 0 \\ 0 \\ \alpha_2 \end{pmatrix}, 
	\label{eq:ch3c:ex2rev}
\end{equation}
where $\alpha_0, \alpha_1, \beta_1, \alpha_2 \in \mathbb{C}$. An invariant operator $\hat{T} = \hat{T}_0 \oplus \hat{T}_1 \oplus \hat{T}_2$ is of the form
\begin{equation}
	\hat{T} = \; \begin{pmatrix} \alpha_0 & 0 & 0 & 0 \\ 0 & \alpha_1 & \beta_1 & 0 \\ 0 & \gamma_1 & \delta_1 & 0 \\ 0 & 0 & 0 & \alpha_2 \end{pmatrix}
	\label{eq:ch3c:ex2rev2}
\end{equation} 
where $\hat{T}_1$ corresponds to the $2\times 2$ central block 
and $\alpha_0, \alpha_1, \beta_1, \gamma_1, \delta_1, \alpha_2 \in \mathbb{C}$. 

The above examples illustrate that the symmetry imposes constraints on vectors and operators. By using an eigenbasis $\{\ket{n t_n}\}$ of the particle number operator $\hat{n}$, these constraints imply the presence of the zeros in Eqs.~\eref{eq:ch3c:ex1rev}--\eref{eq:ch3c:ex2rev2}. Thus, a reduced number of complex coefficients is required in order to describe U(1)-symmetric vectors and operators. As we will discuss in Sec.~\ref{sec:ch3c:symTN}, performing manipulations on symmetric tensors can also result in a significant reduction in computational costs.

\subsubsection{Tensor Product of Two Representations\label{sec:ch3c:symmetry:tp}}

Let $\mathbb{V}^{(A)}$ and $\mathbb{V}^{(B)}$ be two spaces that carry representations of U(1), as generated by particle number operators $\hat{n}^{(A)}$ and $\hat{n}^{(B)}$, and let 
\begin{equation}
\mathbb{V}^{(A)} \cong \bigoplus_{n_{A}} \mathbb{V}^{(A)}_{n_A},~~~~~~\mathbb{V}^{(B)} \cong \bigoplus_{n_{B}} \mathbb{V}^{(B)}_{n_B}
\label{eq:ch3c:AandB}
\end{equation}
be their decompositions as a direct sum of (possibly degenerate) irreps. Let us also consider the action of U(1) on the tensor product $\mathbb{V}^{(AB)} \cong \mathbb{V}^{(A)}\otimes \mathbb{V}^{(B)}$ as generated by the \textit{total particle number operator}  
\begin{equation}
	\hat{n}^{(AB)} \equiv \hat{n}^{(A)}\otimes \mathbb{I} + \mathbb{I} \otimes \hat{n}^{(B)},
\end{equation}
that is, implemented by unitary transformations 
\begin{equation}
	\hat{W}_{\varphi}^{(AB)} \equiv e^{-\rmi\hat{n}^{(AB)}\varphi}.
\end{equation}

The space $\mathbb{V}^{(AB)}$ also decomposes as the direct sum of (possibly degenerate) irreps,
\begin{equation}
\mathbb{V}^{(AB)} \cong \bigoplus_{n_{AB}} \mathbb{V}^{(AB)}_{n_{AB}}. 
\label{eq:ch3c:decoVAB}
\end{equation}
Here the subspace $\mathbb{V}^{(AB)}_{n_{AB}}$, with total particle number $n_{AB}$, corresponds to the direct sum of all products of subspaces $\mathbb{V}^{(A)}_{n_A}$ and $\mathbb{V}^{(B)}_{n_B}$ such that $n_A + n_B = n_{AB}$,
\begin{equation}
	\mathbb{V}^{(AB)}_{n_{AB}} \cong \bigoplus_{n_A,n_B |_{n_A+n_B = n_{AB}}} \mathbb{V}^{(A)}_{n_A} \otimes \mathbb{V}^{(B)}_{n_B}.
\end{equation}

For each subspace $\mathbb{V}^{(AB)}_{n_{AB}}$ in \Eref{eq:ch3c:decoVAB} we introduce a \textit{coupled} basis $\{\ket{n_{AB} t_{n_{AB}}}\}$,  
\begin{equation}
	\hat{n}^{(AB)}\ket{n_{AB} t_{n_{AB}}} = n_{AB} \ket{n_{AB} t_{n_{AB}}},
\end{equation}
where each vector $\ket{n_{AB} t_{n_{AB}}}$ corresponds to the tensor product $\ket{n_{A}t_{n_A};n_B t_{n_B}} \equiv \ket{n_{A}t_{n_A}} \otimes \ket{n_B t_{n_B}}$ of a unique pair of vectors $\ket{n_{A}t_{n_A}}$ and $\ket{n_B t_{n_B}}$, with $n_A+n_B = n_{AB}$. Let table $\fuser$, with components
\begin{equation}
\fuse{n_{A}t_{n_A}}{ n_{B}t_{n_B}}{n_{AB}t_{n_{AB}}} \equiv \braket{n_{AB}t_{n_{AB}}}{n_{A}t_{n_A};n_{B}t_{n_B}},
\label{eq:ch3c:u1fuse}
\end{equation}
encode this one-to-one correspondence. Notice that each component of $\fuser$ is either a 0 or a 1. Then
\begin{equation}
	\ket{n_{AB}t_{n_{AB}}} = \!\!\sum_{n_A t_{n_A} n_B t_{n_B}}\!\!\!\fuse{n_{A}t_{n_A}}{ n_{B}t_{n_B}}{n_{AB}t_{n_{AB}}}\ \ket{n_{A}t_{n_A};n_{B}t_{n_B}}.\label{eq:ch3c:u1fuse2}
\end{equation}

For later reference (\sref{sec:ch3c:supplement}), we notice that $\fuser$ can be decomposed into two pieces. The first piece expresses a basis $\{\ket{n_{A}t_{n_{A}}; n_{B}t_{n_{B}}}\}$ of $\mathbb{V}^{(AB)}$ in terms of the basis $\{\ket{n_{A}t_{n_A}}\}$ of $\mathbb{V}^{(A)}$ and the basis $\{\ket{n_{B}t_{n_B}}\}$ of $\mathbb{V}^{(B)}$. This assignment occurs as in the absence of the symmetry, where one creates a composed index $d = b \times c$ by running, for example, fast over index $c$ and slowly over index $b$ as in \Eref{eq:ch3c:fuse}. Note that this procedure does not always lead to the set $\{\ket{n_{A}t_{n_{A}}; n_{B}t_{n_{B}}}\}$ being ordered such that states corresponding to the same total particle number $n_{AB}=n_A+n_B$ are adjacent to each other within the set. This ordering is achieved by the second piece: a permutation of basis elements that reorganizes them according to their total particle number $n_{AB}$, so that they are identified in an one-to-one correspondence with the coupled states $\{\ket{n_{AB}t_{n_{AB}}}\}$.

Finally, the product basis can be expressed in terms of the coupled basis
\begin{equation}
	\ket{n_{A}t_{n_A};n_{B}t_{n_B}} = \!\!\sum_{n_{AB} t_{n_{AB}}}\!\!\!\splitt{n_{AB} t_{n_{AB}}}{n_{A}t_{n_{A}}}{n_{B}t_{n_{B}}}\ \ket{n_{AB}t_{n_{AB}}}, 
\label{eq:ch3c:u1split2}
\end{equation}
with
\begin{equation}
\splitt{n_{AB} t_{n_{AB}}}{n_{A}t_{n_{A}}}{n_{B}t_{n_{B}}} = \fuse{n_{A}t_{n_A}}{ n_{B}t_{n_B}}{n_{AB}t_{n_{AB}}}.
\label{eq:ch3c:u1split}
\end{equation}

\textbf{Example 3: } Consider the case where both $\mathbb{V}^{(A)}$ and $\mathbb{V}^{(B)}$ correspond to the space of Example 1, that is
 $\mathbb{V}^{(A)} \cong \mathbb{V}^{(A)}_{0} \oplus \mathbb{V}^{(A)}_{1}$ and $\mathbb{V}^{(B)} \cong \mathbb{V}^{(B)}_{0} \oplus \mathbb{V}^{(B)}_{1}$, where $\mathbb{V}^{(A)}_{0}$, $\mathbb{V}^{(A)}_{1}$, $\mathbb{V}^{(B)}_{0}$, and $\mathbb{V}^{(B)}_{1}$ all have dimension 1. Then $\mathbb{V}^{(AB)}$ corresponds to the space in Example 2, namely
\begin{align}
\mathbb{V}^{(AB)} &\cong \mathbb{V}^{(A)} \otimes \mathbb{V}^{(B)} \nonumber \\
 &\cong \left(\mathbb{V}^{(A)}_{0} \oplus \mathbb{V}^{(A)}_{1}\right) \otimes \left(\mathbb{V}^{(B)}_{0} \oplus \mathbb{V}^{(B)}_{1}\right) \nonumber \\
 &\cong \mathbb{V}^{(AB)}_{0} \oplus \mathbb{V}^{(AB)}_{1} \oplus \mathbb{V}^{(AB)}_{2},
\end{align}
where 
\begin{align}
\mathbb{V}^{(AB)}_{0} &\cong \mathbb{V}^{(A)}_{0} \otimes \mathbb{V}^{(B)}_{0}\\
\mathbb{V}^{(AB)}_{1} &\cong 
\left( \mathbb{V}^{(A)}_{0} \otimes \mathbb{V}^{(B)}_{1} \right) 
\oplus \left( \mathbb{V}^{(A)}_{1} \otimes \mathbb{V}^{(B)}_{0} \right)\\
\mathbb{V}^{(AB)}_{2} &\cong \mathbb{V}^{(A)}_{1} \otimes \mathbb{V}^{(B)}_{1}.
\end{align}
The coupled basis $\left\{\ket{n_{AB}t_{n_{AB}}}\right\}$ reads
\begin{align}
&\ket{n_{AB} = 0, t_0 = 1} ~=~ \ket{n_A=0, t_0 = 1} \otimes \ket{n_B=0, t_0 = 1} \nonumber\\
&\ket{n_{AB} = 1, t_1 = 1} ~=~ \ket{n_A=0, t_0 = 1} \otimes \ket{n_B=1, t_1 = 1} \nonumber\\
&\ket{n_{AB} = 1, t_1 = 2} ~=~ \ket{n_A=1, t_1 = 1} \otimes \ket{n_B=0, t_0 = 1} \nonumber\\
&\ket{n_{AB} = 2, t_2 = 1} ~=~ \ket{n_A=1, t_1 = 1} \otimes \ket{n_B=1, t_1 = 1}, \label{eg03} 
\end{align}
where we emphasize that the degeneracy index $t_{n_{AB}}$ takes two possible values for $n_{AB} = 1$, i.e.~$t_1\in \{1,2\}$, since there are two states $\ket{n_{A}t_{n_A}} \otimes \ket{n_B t_{n_B}}$ with $n_A + n_B = 1$. The components $\fuse{n_{A}t_{A}}{ n_{B}t_{B}}{n_{AB}t_{AB}}$ of the tensor $\fuser$ that encodes this change of basis are all zero except for
\begin{align}
	\fuse{01}{01}{01} = \fuse{01}{11}{11} = \fuse{11}{01}{12} = \fuse{11}{11}{21} \;\; &= 1 \nonumber.
	\label{eq:ch3c:Ex3fuser}
\end{align}

\subsubsection{Lattice Models With U(1) Symmetry\label{sec:ch3c:symmetry:lattice}}

The action of U(1) on the threefold tensor product 
\begin{equation}
	\mathbb{V}^{(ABC)} \cong \mathbb{V}^{(A)}\otimes \mathbb{V}^{(B)} \otimes \mathbb{V}^{(C)},
\end{equation}
as generated by the total particle number operator
\begin{equation}
	\hat{n}^{(ABC)} = \hat{n}^{(A)} \otimes \mathbb{I} \otimes \mathbb{I} + \mathbb{I} \otimes \hat{n}^{(B)}\otimes \mathbb{I} + \mathbb{I} \otimes \mathbb{I} \otimes \hat{n}^{(C)},
\end{equation}
induces a decomposition
\begin{equation}
\mathbb{V}^{(ABC)} \cong \bigoplus_{n_{ABC}} \mathbb{V}^{(ABC)}_{n_{ABC}} 
\label{eq:ch3c:decoVABC}
\end{equation}
in terms of irreps $\mathbb{V}^{(ABC)}_{n_{ABC}}$ which we can now relate to $\mathbb{V}^{(A)}_{n_{A}}$, $\mathbb{V}^{(B)}_{n_{B}}$ and $\mathbb{V}^{(C)}_{n_{C}}$. For example, we can consider first the product $\mathbb{V}^{(AB)}_{n_{AB}} \cong \mathbb{V}^{(A)}_{n_{A}}\otimes \mathbb{V}^{(B)}_{n_{B}}$ and then the product $\mathbb{V}^{(ABC)}_{n_{ABC}} \cong \mathbb{V}^{(AB)}_{n_{AB}}\otimes \mathbb{V}^{(C)}_{n_{C}}$, using a different table $\fuser$ at each step to relate the coupled basis to the product basis as discussed in the previous Section. Similarly we could consider the action of U(1) on four tensor products, and so on.

In particular we will be interested in a lattice $\mathcal{L}$ made of $L$ sites with vector space $\mathbb{V}^{\otimes L}$, where for simplicity we will assume that each site $s\in \mathcal{L}$ is described by the same finite-dimensional vector space $\mathbb{V}$ (see Sec.~\ref{sec:ch3c:tensor:TNstates}). Given a particle number operator $\hat{n}$ defined on each site, we can consider the action of U(1) generated by the total particle number operator
\begin{equation}
	\hat{N} \equiv \sum_{s=1}^{L} \hat{n}^{(s)},
\label{eq:ch3c:hatN}
\end{equation}
which corresponds to unitary transformations
\begin{equation}
	W^{[L]}_{\varphi} \equiv e^{-\rmi\hat{N}\varphi} = (e^{-\rmi\hat{n}\varphi})^{\otimes L} = \left( \hat{W}_{\varphi} \right)^{\otimes L}.
\end{equation}
The tensor product space $\mathbb{V}^{\otimes L}$ decomposes as
\begin{equation}
\mathbb{V}^{\otimes L} \cong \bigoplus_{N} \mathbb{V}_{N}
\end{equation}
and we denote by $\left\{\ket{Nt_N}\right\}$ the particle number basis in $\mathbb{V}^{\otimes L}$.

We say that a lattice model is \emph{U(1)-symmetric} if its Hamiltonian $\hat{H}: \mathbb{V} \rightarrow \mathbb{V}$ commutes with the action of the group. That is,
\begin{equation}
	[\hat{H}, \hat{N}] = 0,
\label{eq:ch3c:ham0}
\end{equation}
or equivalently 
\begin{equation}
	\left(\hat{W}_{\varphi}\right)^{\otimes L} \hat{H} \left(\hat{W}_{\varphi}^{~\dagger}\right)^{\otimes L} = \hat{H} ~~~~\forall\ \varphi \in [0, 2\pi).
\label{eq:ch3c:ham}
\end{equation}

One example of a U(1)-symmetric model is the hard core Bose--Hubbard model, with Hamiltonian
\begin{equation}\label{hcbh}
\hat{H}_\mrm{HCBH} \equiv \sum_{s=1}^{L}\left(\hat{a}_{s}^{\dagger}\hat{a}_{s+1} + \hat{a}_{s}\hat{a}_{s+1}^{\dagger} + \gamma \hat{n}_{s}\hat{n}_{s+1}\right) -\mu \sum_{s=1}^L \hat{n}_s,
\end{equation}
where we consider periodic boundary conditions (by identifying sites $L+1$ and $1$), and $\hat{a}_{s}^{\dagger}$ and $\hat{a}_{s}$ are hard-core bosonic creation and annihilation operators, respectively. In terms of the basis introduced in Example 1 these operators are defined as
\begin{equation}
\hat{a} \equiv \begin{pmatrix} 0 &1 \\ 0 &0 \end{pmatrix}, ~~~~~~~ \hat{n} \equiv \hat{a}^{\dagger}\hat{a} = \begin{pmatrix} 0 &0 \\ 0 &1 \end{pmatrix}. \nonumber
\end{equation}
To see that $\hat{H}_\mrm{HCBH}$ commutes with the action of the group, we first observe that for two sites 
\begin{equation}
	\left[\hat{a}_{1}^{\dagger} \hat{a}_{2} + \hat{a}_{2}^{\dagger} \hat{a}_{1} \;,\; \hat{n}_{1} + \hat{n}_{2} \right] = 0,
\end{equation}
from which it readily follows that $\left[\hat{H}_\mrm{HCBH}, \hat{N}\right ] = 0$. 

Notice that the chemical potential term $-\mu\sum_s \hat{n}_s = -\mu \hat{N}$ also commutes with the rest of the Hamiltonian. The ground state $\ket{\Psi_N^{\tiny\mbox{GS}}}$ of $\hat{H}_\mrm{HCBH}$ in a particular subspace $\mathbb{V}_{N}$ or particle number sector can be turned into the absolute ground state by tuning the chemical potential $\mu$. This fact can be used to find the ground state $\ket{\Psi_N^{\tiny\mbox{GS}}}$ of any particle number sector through an algorithm which can only minimize the expectation value of $\hat{H}_\mrm{HCBH}$. However, we will later see that the use of symmetric tensors in the context of tensor network states will allow us to directly minimize the expectation value of $\hat{H}_\mrm{HCBH}$ in a given particle number sector by restricting the search to states
\begin{equation}
	\ket{\Psi_N} = \sum_{t_N=1}^{d_N} (\hat\Psi_N)_{t_N} \ket{N t_N}
\end{equation}
with the desired particle number $N$.

Finally, by making the identifications 
\begin{equation}
\hat{n} = \frac{\mathbb{I} - \hat{\sigma}_{z}}{2},~~~~~~~ \hat{a} = \frac{\hat{\sigma}_{x} + i\hat{\sigma}_{y}}{2} \nonumber
\end{equation}
where $\hat{\sigma}_{x}, \hat{\sigma}_{y}, \hat{\sigma}_{z}$ are the Pauli matrices 
\begin{equation}
\hat{\sigma}_x \equiv \begin{pmatrix} 0 & 1 \\ 1 & 0 \end{pmatrix}, ~~~~~
\hat{\sigma}_y \equiv \begin{pmatrix} 0 & -i \\ i &0 \end{pmatrix},~~~~~
\hat{\sigma}_z \equiv \begin{pmatrix} 1 & 0 \\ 0 & -1 \end{pmatrix},
\end{equation}
one can map $\hat{H}_\mrm{HCBH}$ to the spin-$\frac{1}{2}$ $XXZ$ quantum spin chain
\begin{equation}\label{eq:ch3c:XXZ}
\hat{H}_{XXZ} \equiv  \sum_{s=1}^{L} 
\left( \hat{\sigma}_{x}^{(s)}\hat{\sigma}_{x}^{(s+1)} 
+ \hat{\sigma}_{y}^{(s)} \hat{\sigma}_{y}^{(s+1)} 
+ \Delta \hat{\sigma}_{z}^{(s)} \hat{\sigma}_{z}^{(s+1)}\right),
\end{equation}
where we have ignored terms proportional to $\hat{N}$ and set $\Delta \equiv \gamma/4$. In particular, for $\Delta = 0$ we obtain the quantum $XX$ spin chain 
\begin{equation}\label{eq:ch3c:XX}
\hat{H}_{XX} \equiv  \sum_{s=1}^{L} 
\left(\hat{\sigma}_{x}^{(s)} \hat{\sigma}_{x}^{(s+1)} 
+ \hat{\sigma}_{y}^{(s)} \hat{\sigma}_{y}^{(s+1)}\right),
\end{equation}
and for $\Delta = 1$, the quantum Heisenberg spin chain
\begin{equation}\label{eq:ch3c:XXX}
\hat{H}_{XXX} \equiv  \sum_{s=1}^{L} 
\left( \hat{\sigma}_{x}^{(s)}\hat{\sigma}_{x}^{(s+1)} 
+ \hat{\sigma}_{y}^{(s)} \hat{\sigma}_{y}^{(s+1)} 
+ \hat{\sigma}_{z}^{(s)} \hat{\sigma}_{z}^{(s+1)}\right).
\end{equation}
In Sec.~\ref{sec:ch3c:MERA}, the quantum spin models \eref{eq:ch3c:XX} and \eref{eq:ch3c:XXX} will be used to benchmark the performance increase resulting from the use of symmetries in tensor networks algorithms.

\subsection{Tensor Networks With U(1) Symmetry\label{sec:ch3c:symTN}}

In this Section we will consider U(1)-symmetric tensors and tensor networks. I will explain how to decompose U(1)-symmetric tensors in a compact, canonical form that exploits their symmetry, and then discuss how to adapt the set $\mathcal{P}$ of primitives for tensor network manipulations in order to work in this form. We will also analyse how working in the canonical form affects computational costs.

\subsubsection{U(1)-Symmetric Tensors\label{sec:ch3c:symTN:tensor}}

Let $\hat{T}$ be a rank-$k$ tensor with components $\hat{T}_{i_1 i_2 \ldots i_k}$. As in Sec.~\ref{sec:ch3c:tensor:linear}, we regard tensor $\hat{T}$ as a linear map between the vector spaces $\mathbb{V}^{[\text{in}]}$ and $\mathbb{V}^{[\text{out}]}$ \eref{eq:ch3c:inout}. This implies that each index is either an incoming or outgoing index. On each space $\mathbb{V}^{[i_l]}$, associated with index $i_l$, we introduce a particle number operator $\hat{n}^{(l)}$ that generates a unitary representation of U(1) given by matrices $\hat{W}_{\varphi}^{(l)} \equiv e^{-\rmi\hat{n}^{(l)}\varphi}$, $\varphi \in [0,2\pi)$. In the following, we use $\hat{W}_{\varphi}^{(l)~*}$ to denote the complex conjugate of $\hat{W}_{\varphi}^{(l)}$. 

\begin{figure}
\begin{center}
  \includegraphics[width=300pt]{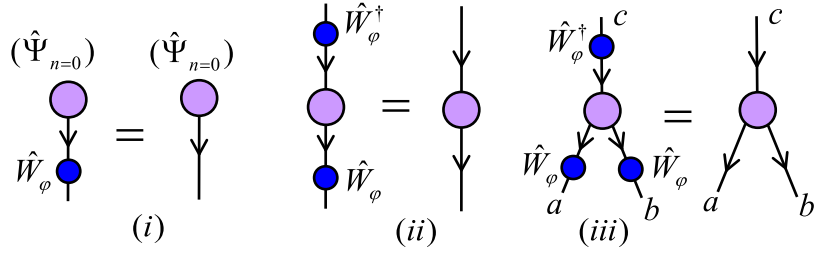}
\caption[(i)~Constraint fulfilled by a U(1)-invariant vector. The only allowed particle number on the single index is $n=0$. (ii)~Constraint fulfilled by a U(1)-invariant matrix. The matrix is block-diagonal in particle number. (iii)~Constraint fulfilled by a rank-three tensor with one incoming index and two outgoing indices.]{%
(i)~Constraint fulfilled by a U(1)-invariant vector. The only allowed particle number on the single index is $n=0$. (ii)~Constraint fulfilled by a U(1)-invariant matrix. It follows from Schur's lemma that the matrix is block-diagonal in particle number. (iii)~Constraint fulfilled by a rank-three tensor with one incoming index and two outgoing indices.\label{fig:ch3c:invariant}}
\end{center}
\end{figure}

Let us consider the action of U(1) on the space
\begin{equation}
	\mathbb{V}^{[i_1]} \otimes \mathbb{V}^{[i_2]} \otimes \ldots \otimes  \mathbb{V}^{[i_k]}
	\label{eq:ch3c:prodspace}
\end{equation}
given by
\begin{equation}
	\hat{X}^{(1)}_{\varphi}\otimes \hat{X}^{(2)}_{\varphi}\otimes \ldots \otimes \hat{X}^{(k)}_{\varphi},
\label{eq:ch3c:Xtrans}
\end{equation}
where 
\begin{equation}
\hat{X}^{(l)}_{\varphi} = \left\{ 
	\begin{array}{cc} \hat{W}^{(l)~*}_{\varphi}& ~~~\mbox{ if } i_l \in I,\\ 
	 									\hat{W}^{(l)}_{\varphi}& ~~~~\mbox{ if } i_l \in O,
	\end{array} \right.
\end{equation}
That is, $\hat{X}^{(l)}_{\varphi}$ acts differently depending on whether index $i_l$ of tensor $\hat{T}$ is an incoming or outgoing index.
We then say that tensor $\hat{T}$, with components $T_{i_1 i_2 \ldots i_k}$, is \textit{U(1)-invariant} if it is invariant under the transformation of \Eref{eq:ch3c:Xtrans}, 
\begin{equation}
	\sum_{i_1, i_2, \ldots, i_k} 	\left(\hat{X}^{(1)}_{\varphi}\right)_{i_1'i_1} \left(\hat{X}^{(2)}_{\varphi} \right)_{i_2' i_2} \ldots \left( \hat{X}^{(k)}_{\varphi} \right)_{i_k' i_k} \hat{T}_{i_1 i_2 \ldots i_k} = \hat{T}_{i_1' i_2' \ldots i_k'},
\end{equation}
for all $\varphi \in [0,2\pi)$. This is depicted in Fig.~\ref{fig:ch3c:invariant}.
 
\textbf{Example 4:} A U(1)-invariant vector $\ket{\Psi}$---that is, a vector with $\hat{n}\ket{\Psi} = 0$ and components $(\hat\Psi_{n=0})_{t_0}$ in the subspace $\mathbb{V}_{n=0}$ which corresponds to vanishing particle number $n=0$ [cf. \Eref{eq:ch3c:nPsi2}]---fulfills
\begin{equation}
	(\hat\Psi_{n=0})_{{t_0}'} = \sum_{t_0} \left(\hat{W}_{\varphi}\right)_{{t_0}'t_0} (\hat\Psi_{n=0})_{t_0} ~~~~~~\forall\ \varphi \in [0,2\pi),
\end{equation}
in accordance with \Eref{eq:ch3c:nPsi1}, as shown in Fig.~\ref{fig:ch3c:invariant}(i).

\textbf{Example 5:} A U(1)-invariant matrix $\hat{T}$  \eref{eq:ch3c:Schur} fulfills
\begin{eqnarray}
	\hat{T}_{a'b'} &=& \sum_{a,b} \left(\hat{W}_{\varphi}\right)_{a'a} \left( \hat{W}^{~*}_{\varphi} \right)_{b'b} \hat{T}_{ab} \\
	&=& \sum_{a,b} \left(\hat{W}_{\varphi}\right)_{a'a} \hat{T}_{ab} \left(\hat{W}^{~\dagger}_{\varphi}\right)_{bb'} ~~~\forall\ \varphi \in [0,2\pi),~~~
\end{eqnarray}
in accordance with \Eref{eq:ch3c:commutator2} [see Fig.~\ref{fig:ch3c:invariant}(ii)].
 
\textbf{Example 6:} Tensor $\hat{T}$ in \Eref{eq:ch3c:Tabc}, with components $\hat{T}_{abc}$ where $a$ and $b$ are outgoing indices and $c$ is an incoming index, is U(1)-invariant iff
\begin{eqnarray}
	\hat{T}_{a' b' c'} &=& \sum_{a,b,c} \left(\hat{W}^{(1)}_{\varphi}\right)_{a'a}\left(\hat{W}^{(2)}_{\varphi} \right)_{b'b} \left( \hat{W}^{(3)~*}_{\varphi} \right)_{c'c} \hat{T}_{abc}~~~~\\
&=& \sum_{a,b,c} \left(\hat{W}^{(1)}_{\varphi}\right)_{a'a}\left(\hat{W}^{(2)}_{\varphi} \right)_{b'b} \hat{T}_{abc} \left(\hat{W}^{(3)~\dagger}_{\varphi}\right)_{cc'} ~~~~
\end{eqnarray}
for all $\varphi \in [0,2\pi)$ [see Fig.~\ref{fig:ch3c:invariant}(iii)].

Further, we say that a tensor $\hat{Q}$ with components $\hat{Q}_{i_1 i_2 \ldots i_k}$ is \textit{U(1)-covariant} if under the transformation of \Eref{eq:ch3c:Xtrans} it acquires a non-trivial phase $e^{-\rmi n\varphi}$,
\begin{equation}
	\sum_{i_1, i_2, \ldots, i_k} 	\left(\hat{X}^{(1)}_{\varphi}\right)_{i_1'i_1} \left(\hat{X}^{(2)}_{\varphi} \right)_{i_2 i_2'} \ldots \left( \hat{X}^{(k)}_{\varphi} \right)_{i_k' i_k} \hat{Q}_{i_1 i_2 \ldots i_k} = e^{-\rmi n\varphi} \hat{Q}_{i_1' i_2' \ldots i_k'},
\label{eq:ch3c:Tcov}
\end{equation}
for all $\varphi \in [0,2\pi)$. 

\begin{figure}
\begin{center}
  \includegraphics[width=300pt]{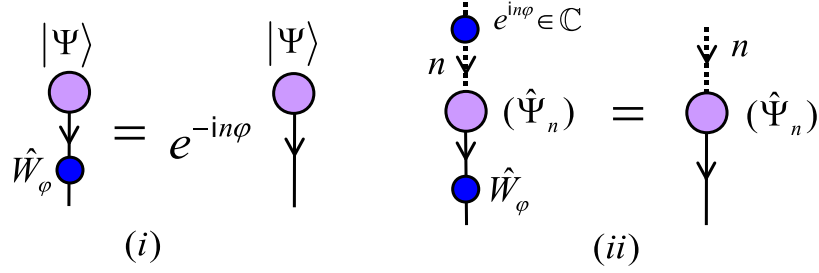}
\caption[(i)~Under the action of U(1), a U(1)-covariant vector $\ket{\Psi}$ acquires a phase $e^{-\rmi n\varphi}$. (ii)~The U(1)-covariant vector $\ket{\Psi}$, with components $(\Psi_n)_{t_n}$, can be represented by a U(1)-invariant matrix $\hat{T}$ with components $\hat{T}_{i_1 i} = (\Psi_n)_{i_1}$, where $i$ is a trivial index with charge $n$ and is decorated by the opposite arrow to $i_1$.]{%
(i)~U(1)-covariant vector $\ket{\Psi}$, with some non-vanishing particle number $n\neq 0$. Under the action of U(1) on its index, the covariant vector $\ket{\Psi}$ acquires a phase $e^{-\rmi n\varphi}$ \protect{\eref{eq:ch3c:coPsi}}. (ii)~The U(1)-covariant vector $\ket{\Psi}$, with components $(\Psi_n)_{t_n}$, can be represented by a U(1)-invariant matrix $\hat{T}$ with components $\hat{T}_{i_1 i} = (\Psi_n)_{i_1}$, where $i$ is a trivial index ($|i|=1$) with charge $n$ and is decorated by the opposite arrow to $i_1$. \label{fig:ch3c:covariant}}
\end{center}
\end{figure}

\textbf{Example 7:} A U(1)-covariant vector $\ket{\Psi}$---that is, one which satisfies $\hat{n}\ket{\Psi} = n\ket{\Psi}$ for some $n\neq 0$, and has nonzero components $(\hat\Psi_{n})_{t_n}$ only in the relevant subspace $\mathbb{V}_{n}$ [cf. \Eref{eq:ch3c:nPsi2}]---fulfills
\begin{equation}
	 \sum_{t_n} \left(\hat{W}_{\varphi}\right)_{{t'_n}t_n} (\hat\Psi_{n})_{t_n} = e^{-\rmi n\varphi}(\hat\Psi_{n})_{{t'_n}} ~~~\forall\ \varphi \in [0,2\pi),
	 \label{eq:ch3c:coPsi}
\end{equation}
in accordance with \Eref{eq:ch3c:nPsi1}. (See also Fig.~\ref{fig:ch3c:covariant}.)

Notice that we can describe the rank-$k$ covariant tensor $\hat{Q}$ above by a rank-$(k+1)$ invariant tensor $\hat{T}$ with components
\begin{equation}
	\hat{T}_{i_1i_2\ldots i_k i} \equiv \hat{Q}_{i_1i_2\ldots i_k}\qquad|i|=1.
	\label{eq:ch3c:TQ}
\end{equation}
This is built from $\hat{Q}$ by adding an extra incoming index $i$, where index $i$ has fixed particle number $n$ and no degeneracy (i.e., $i$ is associated to a trivial space $\mathbb{V}^{[i]} \cong \mathbb{C}$). We refer to both \textit{invariant} and \textit{covariant} tensors as \textit{symmetric} tensors. By using the above construction, in this work we will represent all U(1)-symmetric tensors by means of U(1)-invariant tensors. In particular, we represent the non-trivial components $(\hat\Psi_{n})_{t_n}$ of the covariant vector $\ket{\Psi_n}$ in Eqs.~\eref{eq:ch3c:nPsi1}--\eref{eq:ch3c:nPsi2} as an  invariant matrix $\hat{T}$ of size $|t_n|\times 1$ with components $\hat{T}_{t_n 1} = (\hat\Psi_n)_{t_n}$.
Consequently, from now on we will mostly consider only invariant tensors. 

\subsubsection{Canonical Form For U(1)-Invariant Tensors\label{sec:ch3c:symTN:canonical}}

Let us now write a tensor $\hat{T}$ in a particle number basis on each factor space in \Eref{eq:ch3c:prodspace}. That is, each index $i_1$, $i_2$, $\ldots$, $i_k$ is decomposed into a particle number index $n$ and a degeneracy index $t_n$, $i_1 = (n_1, t_{n_1})$,  $i_2 = (n_2, t_{n_2})$, $\ldots$, $i_k = (n_k, t_{n_k})$, and
\begin{equation}
	\hat{T}_{i_1 i_2 \ldots i_k} \equiv \left(\hat{T}_{n_1 n_2 \ldots n_k}\right)_{t_{n_1} t_{n_2} \ldots t_{n_k}}. \label{eq:ch3c:tdec}
\end{equation}
Here, for each set of particle numbers $n_1, n_2, \ldots, n_k,$ we regard $\hat{T}_{n_1n_2\ldots n_k}$ as a tensor with components  $\left(\hat{T}_{n_1 n_2 \ldots n_k}\right)_{t_{n_1} t_{n_2} \ldots t_{n_k}}$. Let $N_{\text{in}}$ and $N_{\text{out}}$ denote the sum of particle numbers corresponding to incoming and outgoing indices,
\begin{equation}
	N_{\text{in}} \equiv \sum_{n_l\in I} n_l,~~~~~~~~N_{\text{out}}\equiv \sum_{n_l\in O} n_l.
\end{equation}
The condition for a non-vanishing tensor of the form $\hat{T}_{n_1n_2\ldots n_k}$ to be invariant under U(1), \Eref{eq:ch3c:Xtrans}, is simply that the sum of incoming particle numbers equals the sum of outgoing particle numbers. Therefore, a U(1)-invariant tensor $\hat{T}$ satisfies
\begin{equation}
	\hat{T} = \bigoplus_{n_1, n_2, \ldots, n_k} \hat{T}_{n_1 n_2 \ldots n_k} \delta_{N_{\text{in}},N_{\text{out}}}.
	\label{eq:ch3c:Tcanon}
\end{equation}
[We use the direct sum symbol $\bigoplus$ to denote that the different tensors $\hat{T}_{n_1n_2 \ldots n_k}$ are supported on orthonormal subspaces of the tensor product space of \Eref{eq:ch3c:prodspace}.]
In components, the above expression reads
\begin{equation}
	\hat{T}_{i_1 i_2 \ldots i_k} \equiv \left(\hat{T}_{n_1 n_2 \ldots n_k}\right)_{t_{n_1} t_{n_2} \ldots t_{n_k}}\delta_{N_{\text{in}},N_{\text{out}}}.
	\label{eq:ch3c:Tcanon2}
\end{equation}
Here, $\delta_{N_{\text{in}},N_{\text{out}}}$ implements particle number conservation: if $N_{\text{in}} \neq N_{\text{out}}$, then all components of $\hat{T}_{n_1n_2\ldots n_{k}}$ must vanish. This generalizes the block structure of U(1)-invariant matrices in \Eref{eq:ch3c:Schur} (where $\hat{T}_{nn}$ is denoted $\hat{T}_n$) to tensors of arbitrary rank $k$. The canonical decomposition in \Eref{eq:ch3c:Tcanon} is important, in that it allows us to identify the degrees of freedom of tensor $\hat{T}$ that are not determined by the symmetry. Expressing tensor $\hat{T}$ in terms of the tensors $\hat{T}_{n_1 n_2 \ldots n_k}$ with $N_{\text{in}} = N_{\text{out}}$ ensures that we store $\hat{T}$ in the most compact way possible.

Notice that the canonical form of \Eref{eq:ch3c:Tcanon} is a particular case of the canonical form presented in \Eref{eq:ch3a:T4} of \sref{sec:ch3a:symdc} for more general (possibly non-Abelian) symmetry groups. There, a symmetric tensor was decomposed into \textit{degeneracy} tensors [analogous to tensors $\hat{T}_{n_1n_2 \ldots n_k}$ in \Eref{eq:ch3c:Tcanon}] and structural tensors [generalizing the term $\delta_{N_{\text{in}},N_{\text{out}}}$ in \Eref{eq:ch3c:Tcanon}] which can in general be expanded as a trivalent network of Clebsch--Gordan (or coupling) coefficients of the symmetry group. In the case of non-Abelian groups, where some irreps have dimension larger than 1, the structural tensors are highly non-trivial. However, for the group U(1) discussed in this Section (as for any other Abelian group) all irreps are one dimensional and the structural tensors are always reduced to a simple expression such as $\delta_{N_{\text{in}},N_{\text{out}}}$ in \Eref{eq:ch3c:Tcanon}. (Nevertheless, in \sref{sec:ch3c:supplement} we will resort to a more elaborate decomposition of the structural tensors in order to better exploit the presence of symmetry 
in those tensor network algorithms based on iterating a fixed sequence of manipulations.)

\subsubsection{U(1)-Symmetric Tensor Networks\label{sec:ch3c:symTN:TN}}

\begin{figure}
\begin{center}
  \includegraphics[width=300pt]{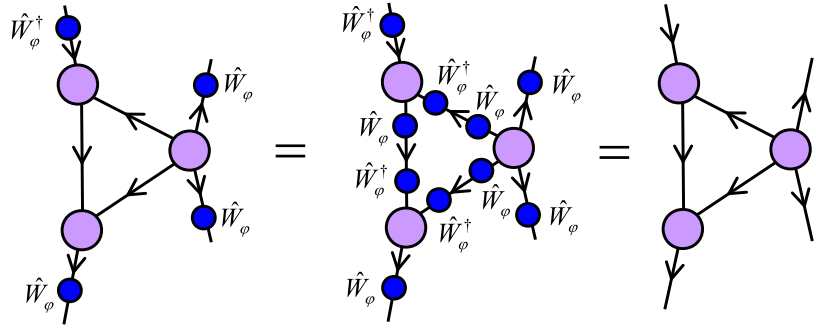}
\caption[A tensor network $\mathcal{N}$ made of U(1)-invariant tensors is shown to represent a U(1)-invariant tensor $\hat{T}$.]{%
A tensor network $\mathcal{N}$ made of U(1)-invariant tensors represents a U(1)-invariant tensor $\hat{T}$. This is seen by means of two equalities. The first equality is obtained by inserting resolutions of the identity $\mathbb{I} = \hat W_{\varphi} \hat W^{\dagger}_{\varphi}$ on each index connecting two tensors in $\mathcal{N}$. The second equality follows from the fact that each tensor in $\mathcal{N}$ is U(1)-invariant. \label{fig:ch3c:symTN}}
\end{center}
\end{figure}

In Sec.~\ref{sec:ch3c:tensor:linear} we saw that a tensor network $\mathcal{N}$ where each line has a direction (represented with an arrow) can be interpreted as a collection of linear maps composed into a single linear map $\hat{T}$ of which $\mathcal{N}$ is a tensor network decomposition. By introducing a particle number operator on the vector space associated to each line of $\mathcal{N}$, we can define a unitary representation of U(1) on each index of each tensor in $\mathcal{N}$. Then we say that $\mathcal{N}$ is a U(1)-invariant tensor network if all its tensors are U(1)-invariant. Notice that, by construction, if $\mathcal{N}$ is a U(1)-invariant tensor network, then the resulting linear map $\hat{T}$ is also U(1)-invariant. This is illustrated in Fig.~\ref{fig:ch3c:symTN}. 

More generally, we can consider a U(1)-symmetric tensor network, made of tensors that are U(1)-symmetric (that is, either invariant or covariant). Recall, however, that any covariant tensor can be represented as an invariant tensor by adding an extra index \eref{eq:ch3c:TQ}. Therefore without loss of generality we can restrict our attention to invariant tensor networks.

\subsubsection{Tensor Network States and Algorithms With U(1) Symmetry\label{sec:ch3c:symTN:TNstate}}

As discussed in Sec.~\ref{sec:ch3c:tensor:TNstates}, a tensor network $\mathcal{N}$ can be used to describe certain pure states $\ket{\Psi}\in \mathbb{V}^{\otimes L}$ of a lattice $\mathcal{L}$. If $\mathcal{N}$ is a U(1)-symmetric tensor network then it will describe a pure state $\ket{\Psi}$ that has a well-defined total particle number $N$. That is, a U(1)-symmetric pure state
\begin{equation}
 \hat{N}\ket{\Psi} = N \ket{\Psi}, ~~~~~~~~~e^{-\rmi \hat{N}\varphi}\ket{\Psi} = e^{-\rmi N\varphi} \ket{\Psi}.	
\end{equation}
In this way we can obtain a more refined version of popular tensor network states such as MPS, TTN, MERA, PEPS, etc. %
As a variational Ansatz, a symmetric tensor network state is more constrained than a regular tensor network state, and consequently it can represent less states $\ket{\Psi} \in \mathbb{V}^{\otimes L}$. However, it also depends on fewer parameters. This implies a more economical description, as well as the possibility of reducing computational costs during its manipulation.

The rest of this Section is devoted to explaining how one can achieve a reduction in computational costs. This is based on storing and manipulating U(1)-invariant tensors expressed in the canonical form of Eqs.~\eref{eq:ch3c:Tcanon}--\eref{eq:ch3c:Tcanon2}. We next explain how to adapt the set $\mathcal{P}$ of four primitive operations for the tensor network manipulations discussed in Sect \ref{sec:ch3c:tensor:TN}, namely permutation and reshaping of indices, matrix multiplication, and factorization.

\subsubsection{Permutation of Indices\label{sec:ch3c:symTN:permutation}}

Given a U(1)-invariant tensor $\hat{T}$ expressed in the canonical form of Eqs.~\eref{eq:ch3c:Tcanon}--\eref{eq:ch3c:Tcanon2}, permuting two of its indices is straightforward. It is achieved by %
swapping the position of the two particle numbers of $\hat{T}_{n_1n_2 \ldots n_k}$ involved, and also the corresponding degeneracy indices. For instance, if the rank-$3$ tensor $\hat{T}$ of \Eref{eq:ch3c:Tabc} is U(1)-invariant and has components
\begin{equation}
	\hat{T}_{abc} = \left(\hat{T}_{n_A n_B n_C}\right)_{t_{n_A}t_{n_B}t_{n_C}} \delta_{n_A+n_B, n_C}
	\label{eq:ch3c:TabcSym}
\end{equation}
when expressed in the particles number basis $a = (n_{A}, t_{n_A})$, $b = (n_B, t_{n_B})$, $c = (n_C, t_{n_C})$, then tensor $\hat{T}'$ of \Eref{eq:ch3c:permute}, obtained from $\hat{T}$ by permuting the last two indices, has components
\begin{equation}
	(\hat{T}')_{acb} = \left(\hat{T}_{n_A n_C n_B}'\right)_{t_{n_A}t_{n_C}t_{n_B}} \delta_{n_A+n_B, n_C}
\end{equation}
where
\begin{equation}
\left(\hat{T}_{n_A n_C n_B}'\right)_{t_{n_A}t_{n_C}t_{n_B}} = \left(\hat{T}_{n_A n_B n_C}\right)_{t_{n_A}t_{n_B}t_{n_C}}.
\end{equation}

\begin{figure}
\begin{center}
  \includegraphics[width=240pt]{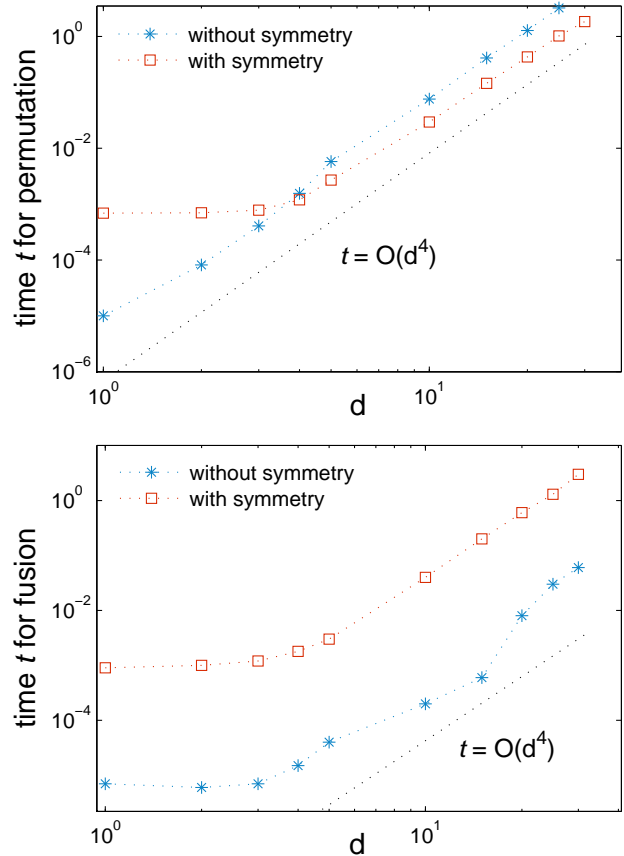}
\caption[Computation times (in seconds) required to permute and fuse two indices of a rank-4 tensor $\hat{T}$, as a function of the size of the indices.]{%
Computation times (in seconds) required to permute and fuse two indices of a rank-four tensor $\hat{T}$, as a function of the size of the indices. All four indices of $\hat{T}$ have the same size, $5d$, and therefore the tensor contains $|\hat{T}|=5^4d^4$ coefficients. The figures compare the time required to perform these operations using a regular tensor and a U(1)-invariant tensor, where in the second case each index contains five different values of the particle number $n$ (each with degeneracy $d$) and the canonical form of Eqs.~\eref{eq:ch3c:Tcanon}--\eref{eq:ch3c:Tcanon2} is used. The upper figure shows the time required to permute two indices: For large $d$, exploiting the symmetry of a U(1)-invariant tensor by using the canonical form results in shorter computation times. The lower figure shows the time required to fuse two adjacent indices. In this case, maintaining the canonical form requires more computation time. Notice that in both figures the asymptotic cost scales as $O(d^4)$, or the size of $\hat{T}$, since this is the number of coefficients which need to be rearranged.
We note that the fixed-cost overheads associated with symmetric manipulations could potentially vary substantially with choice of programming language, compiler, and machine architecture. The results given here show the performance of a \textsc{matlab} implementation of U(1) symmetry.
\label{fig:ch3c:permFuse}}
\end{center}
\end{figure}

Notice that since we only need to permute the components of those $\hat{T}_{n_A n_B n_C}$ such that $n_A+n_B = n_C$, implementing the permutation of indices requires less computation time than a regular index permutation. This is shown in Fig.~\ref{fig:ch3c:permFuse}, corresponding to a permutation of indices using \textsc{matlab}.

\subsubsection{Reshaping of Indices\label{sec:ch3c:symTN:reshape}}

The indices of a U(1)-invariant tensor can be reshaped (fused or split) in a similar manner to those of a regular tensor. However, maintaining the convenient canonical form of Eqs.~\eref{eq:ch3c:Tcanon}--\eref{eq:ch3c:Tcanon2} requires additional steps. Two adjacent indices can be fused together using the table $\fuser$ of \Eref{eq:ch3c:u1fuse}, which is a sparse tensor made of ones and zeros. Similarly an index can be split into two adjacent indices by using its inverse, the sparse tensor $\splitter$ of \Eref{eq:ch3c:u1split}.

\textbf{Example 8:} Let us consider again the rank-$3$ tensor $\hat{T}$ of \Eref{eq:ch3c:Tabc} with components given by \Eref{eq:ch3c:TabcSym}, where $a$ and $b$ are outgoing indices and $c$ is an incoming index. We can fuse outgoing index $b$ and incoming index $c$ into an (e.g.~incoming) index $d$, obtaining a new tensor $\hat{T}'$ with components
\begin{equation}
	(\hat{T}')_{ad} = \left(\hat{T}'_{n_An_D}\right)_{t_{n_A}t_{t_{n_D}}} \delta_{n_A,n_D},
\end{equation}
where $n_{D}= - n_B + n_C$. (The sign in front of $n_B$ comes from the fact that $d$ is an incoming index and $b$ is an outgoing index.) The components of $\hat{T}'$ are in one-to-one correspondence with those of $\hat{T}$ and follow from the transformation
\begin{equation}
	\left(\hat{T}'_{n_{A} n_D}\right)_{t_{n_A} t_{n_D}} = \sum_{n_B,t_{n_B},n_C,t_{n_C}} \left(\hat{T}_{n_A n_B n_C}\right)_{t_{n_A} t_{n_B}t_{n_C}}\fuse{n_B t_{n_B}}{n_C t_{n_C}}{n_D t_{n_D}},
\label{eq:ch3c:TabcFuse}
\end{equation}
where only the case $n_A=n_D$ needs to be considered.
To complete the example, let us assume that the index $a$ is described by the vector space $\mathbb{V}^{(A)}\cong \mathbb{V}_0 \oplus \mathbb{V}_1 \oplus \mathbb{V}_2$ with degeneracies $d_0 = 1$, $d_{1}=2$, and $d_{2}=1$; index $b$ is described by a vector space $\mathbb{V}^{(B)} \cong \mathbb{V}_{-1} \oplus \mathbb{V}_0$ without degeneracies, i.e.~$d_{-1}=d_{0}=1$; and index $c$ is described by a vector space $\mathbb{V}^{(C)} \cong \mathbb{V}_{0} \oplus \mathbb{V}_1$ also without degeneracies, $d_{-1}=d_{0}=1$. Then $\mathbb{V}^{(D)} \cong \mathbb{V}^{(B)}\otimes\mathbb{V}^{(C)}$ (and in this example, also $\mathbb{V}^{(D)}\cong\mathbb{V}^{(A)}$) and \Eref{eq:ch3c:TabcFuse} amounts to 
\begin{eqnarray}
\left(\hat{T}'_{00}\right)_{11} &=& \left(\hat{T}_{000}\right)_{111}, \nonumber\\
\left(\hat{T}'_{11}\right)_{11} &=& \left(\hat{T}_{101}\right)_{111}, \nonumber\\
\left(\hat{T}'_{11}\right)_{12} &=& \left(\hat{T}_{101}\right)_{211}, \nonumber\\
\left(\hat{T}'_{11}\right)_{21} &=& \left(\hat{T}_{1(-1)0}\right)_{111}, \nonumber\\
\left(\hat{T}'_{11}\right)_{22} &=& \left(\hat{T}_{1(-1)0}\right)_{211}, \nonumber\\
\left(\hat{T}'_{22}\right)_{11} &=& \left(\hat{T}_{2(-1)1}\right)_{111}, \nonumber
\end{eqnarray}
where we notice that tensor $\hat{T'}$ is a matrix as in \Eref{eq:ch3c:ex2rev2}. Similarly, we can split incoming index $d$ of tensor $\hat{T}'$ back into outgoing index $b$ and incoming index $c$ of tensor $\hat{T}$ according to
\begin{equation}
	\left(\hat{T}_{n_A n_B n_C}\right)_{t_{n_A} t_{n_B} t_{n_C}} = \sum_{n_D,t_{n_D}} \left(\hat{T}_{n_A n_D} ' \right)_{t_{n_A} t_{n_D}} \splitt{n_D t_{n_D}}{n_B t_{n_B}}{n_C t_{n_C}}
\label{eq:ch3c:TadSplit}
\end{equation}
which, again, is non-trivial only for $-n_{B}+n_{C}=n_{D}$ and $n_A+n_B=n_C$.

This example illustrates that fusing and splitting indices while maintaining the canonical form of Eqs.~\eref{eq:ch3c:Tcanon}--\eref{eq:ch3c:Tcanon2} requires more work than reshaping regular indices. Indeed, after taking indices $b$ and $c$ into $d=b\times c$ by listing all pairs of values $b\times c$, we still need to reorganize the resulting basis elements according to their particle number $n_D$. Although this can be done by following the simple table given by $\fuser$, it may add significantly to the overall computational cost associated with reshaping a tensor. For instance, Fig.~\ref{fig:ch3c:permFuse} shows that fusing indices of invariant tensors can be more expensive than fusing indices of regular tensors.

\subsubsection{Multiplication of Two Matrices\label{sec:ch3c:symTN:multiply}}

By permuting and reshaping the indices of a U(1)-invariant tensor, we can convert it into a U(1)-invariant matrix $\hat{T}= \bigoplus_{n n'} \hat{T}_{nn'} \delta_{n,n'}$, or simply
\begin{equation}
	\hat{T} = \bigoplus_{n} \hat{T}_{n},
\label{eq:ch3c:Mcanon}
\end{equation}
where $\hat{T}_{n} \equiv \hat{T}_{nn}$. In components, matrix $\hat{T}$ reads
\begin{equation}
	(\hat{T})_{ab} = \left(\hat{T}_{n}\right)_{t_n t_n'}, 
	\label{eq:ch3c:Mcanon2}
\end{equation}
where $a=(n,t_n)$ and $b = (n, t_n')$. In particular, similar to the discussion in Sec.~\ref{sec:ch3c:tensor:multiply} for regular tensors, the multiplication of two tensors invariant under the action of U(1) can be reduced to the multiplication of two U(1)-invariant matrices.

Let $\hat{R}$ and $\hat{S}$ be two U(1)-invariant matrices, with canonical forms 
\begin{equation}
	\hat{R} = \bigoplus_n \hat{R}_n, ~~~~ \hat{S} = \bigoplus_n \hat{S}_n.
\end{equation}
Their product $\hat{T} = \hat{R}\cdot \hat{S}$, \eref{eq:ch3c:Mmultiply}, is then another matrix $\hat{T}$ which is also block diagonal,
\begin{equation}
	\hat{T} = \bigoplus_n \hat{T}_n,
\end{equation}
such that each block $\hat{T}_n$ is obtained by multiplying the corresponding blocks $\hat{R}_n$ and $\hat{S}_n$,
\begin{equation}
	\hat{T}_n = \hat{R}_n\cdot \hat{S}_n.
\label{eq:ch3c:TRSblock}
\end{equation}

Equations~\eref{eq:ch3c:Mcanon} and \eref{eq:ch3c:TRSblock} make evident the potential reduction of computational costs that can be achieved by manipulating U(1)-invariant matrices in their canonical form. First, a reduction in memory space follows from only having to store the diagonal blocks in \Eref{eq:ch3c:Mcanon}. Second, a reduction in computational time is implied by only having to multiply these blocks in \Eref{eq:ch3c:TRSblock}. This is illustrated in the following example.

\textbf{Example 9:} Consider a U(1)-invariant matrix $\hat{T}$ which is a linear map in a space $\mathbb{V}$ that decomposes into $q$ irreps $\mathbb{V}_n$, each of which has the same degeneracy $d_n=d$.
That is, $\hat{T}$ is a square matrix of dimensions $dq\times dq$, with the block-diagonal form of \Eref{eq:ch3c:Mcanon}. Since there are $q$ blocks $\hat{T}_n$ and each block has size $d\times d$, the U(1)-invariant matrix $\hat{T}$ contains $qd^2$ coefficients. For comparison, a regular matrix of the same size contains $q^2d^2$ coefficients, a number greater by a factor of $q$.

Let us now consider multiplying two such matrices. We use an algorithm that requires $O(l^3)$ computational time to multiply two matrices of size $l\times l$. The cost of performing $q$ multiplications of $d\times d$ blocks in \Eref{eq:ch3c:TRSblock} scales as $O(qd^3)$. In contrast the cost of multiplying two regular matrices of the same size scales as $O(q^3d^3)$, requiring $q^2$ times more computation time. 

\begin{figure}
\begin{center}
  \includegraphics[width=265pt]{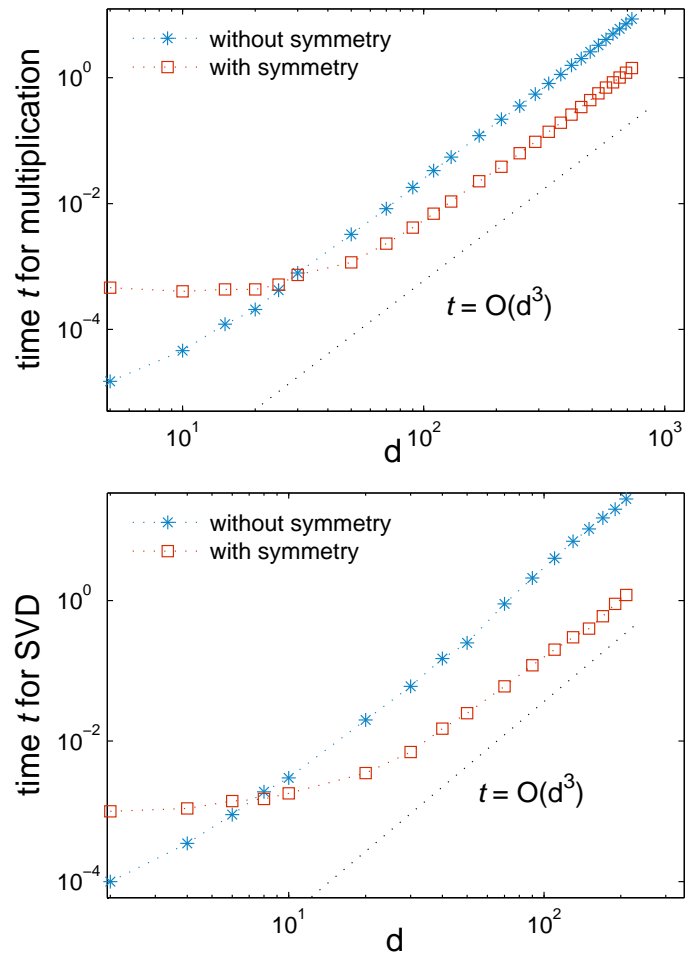}
\caption[Computation times (in seconds) required to multiply two matrices (upper panel) and to perform a singular value decomposition (lower panel), as a function of the size of the indices.]{%
Computation times (in seconds) required to multiply two matrices (upper panel) and to perform a singular value decomposition (lower panel), as a function of the size of the indices. Matrices of size $5d \times 5d$ are considered.
The figures compare the time required to perform these operations using regular matrices and U(1)-invariant matrices, where for the U(1)-invariant matrices each index contains five different values of the particle number $n$, each with degeneracy $d$, and the canonical form of Eqs.~\eref{eq:ch3c:Mcanon}--\eref{eq:ch3c:Mcanon2} is used. That is, each matrix decomposes into five blocks of size $d\times d$. 
For large $d$, exploiting the block diagonal form of U(1)-invariant matrices results in shorter computation time both for multiplication and for singular value decomposition. The asymptotic cost scales with $d$ as $O(d^3)$, while the size of the matrices grows as $O(d^2)$.
We note that the fixed-cost overheads associated with symmetric manipulations could potentially vary substantially with choice of programming language, compiler, and machine architecture. The results given here show the performance of a \textsc{matlab} implementation of U(1) symmetry.
\label{fig:ch3c:multSvd}}
\end{center}
\end{figure}

Figure~\ref{fig:ch3c:multSvd} shows a comparison of computation times when multiplying two matrices for both U(1)-symmetric and regular matrices.

\subsubsection{Factorization of a Matrix\label{sec:ch3c:symTN:factorize}}

The factorization of a U(1)-invariant matrix $\hat{T}$ \eref{eq:ch3c:Mcanon} can also benefit from the block-diagonal structure. Consider, for instance, the singular value decomposition $\hat{T} = \hat{U}\hat{S}\hat{V}$ of \Eref{eq:ch3c:singular}. In this case we can obtain the matrices
\begin{equation}
	\hat{U} = \bigoplus_n \hat{U}_n 
	~~~~ \hat{S} = \bigoplus_n \hat{S}_n 
	~~~~ \hat{V} = \bigoplus_n \hat{V}_n
\end{equation}
by performing the singular value decomposition of each block $\hat{T}_n$ independently,
\begin{equation}
	\hat{T}_n = \hat{U}_n \hat{S}_n \hat{V}_n.
\end{equation}

The computational savings are analogous to those described in Example 9 above for the multiplication of matrices. Figure~\ref{fig:ch3c:multSvd} also shows a comparison of computation times required to perform a singular value decomposition on U(1)-invariant and regular matrices using \textsc{matlab}.

\subsubsection{Discussion\label{sec:ch3c:symTN:discussion}}

In this Section we have seen that U(1)-invariant tensors can be written in the canonical form of Eqs.~\eref{eq:ch3c:Tcanon}--\eref{eq:ch3c:Tcanon2}, and that this canonical form is of interest because it offers a compact description in terms of only those coefficients which are not constrained by the symmetry. We have also seen that maintaining the canonical form during tensor manipulations adds some computational overhead when reshaping (fusing or splitting) indices, but reduces computation time when permuting indices (for sufficiently large tensors) and when multiplying or factorizing matrices (for sufficiently large matrix sizes).

The cost of reshaping and permuting indices is proportional to the size $|\hat{T}|$ of the tensors, whereas the cost of multiplying and factorizing matrices is a larger power of the matrix size, for example $|\hat{T}|^{3/2}$. The use of the canonical form when manipulating large tensors therefore frequently results in an overall reduction in computation time, making it a very attractive option in the context of tensor network algorithms. This is exemplified in the next Section, where we apply the MERA to study the ground state of quantum spin models with a U(1) symmetry.

On the other hand, however, the cost of maintaining invariant tensors in the canonical form becomes more relevant when dealing with smaller tensors. In the next Section we will also see that in some situations, this additional cost may significantly reduce, or even offset, the benefits of using the canonical form. In this event, and in the specific context of algorithms where the same tensor manipulations are iterated many times, it is possible to significantly decrease the additional cost by \textit{precomputing} the parts of the tensor manipulations that are repeated on each iteration. Precomputation schemes are described in more detail in \sref{sec:ch3c:supplement}. Their performance is illustrated in the next Section.

\subsection[Tensor Network Algorithms With U(1) Symmetry: A Practical \nohyphens{Example}]{Tensor Network Algorithms With U(1) Symmetry:\\A Practical Example\label{sec:ch3c:MERA}}

In previous Sections we have described a strategy to incorporate a U(1) symmetry into tensors, tensor networks, and their manipulations. To further illustrate how the strategy works in practice, in this Section we demonstrate its use in the context of the multi-scale entanglement renormalization Ansatz, or MERA, and present numerical results from our reference implementation of the U(1) symmetry in \textsc{matlab}.

\subsubsection{Multi-Scale Entanglement Renormalization Ansatz\label{sec:ch3c:MERA:ansatz}}

\begin{figure}
\begin{center}
  \includegraphics[width=300pt]{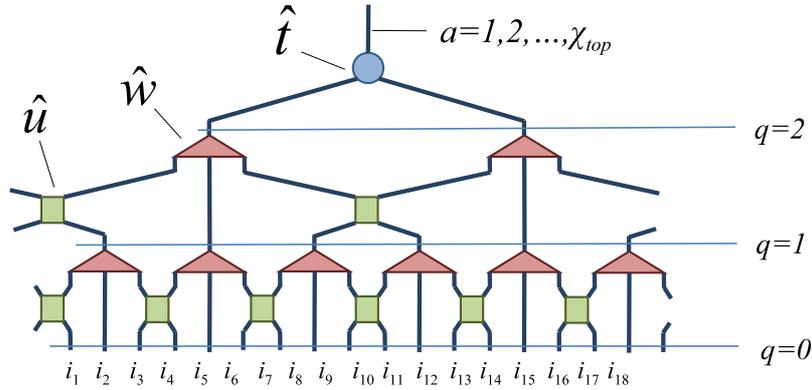}
\caption{%
MERA for a system of $L=2\times 3^{2}= 18$ sites, made of two layers of disentanglers $\hat{u}$ and isometries $\hat{w}$, and a top tensor $\hat{t}$.\label{fig:ch3c:MERA}}
\end{center}
\end{figure}

Figure~\ref{fig:ch3c:MERA} shows a MERA that represent states $\ket{\Psi}\in \mathbb{V}^{\otimes L}$ of a lattice $\mathcal{L}$ made of $L=18$ sites (see Sec.~\ref{sec:ch3c:tensor:TNstates}).
 Recall that the MERA is made of layers of isometric tensors, known as disentanglers $\hat{u}$ and isometries $\hat{w}$, that implement a coarse-graining transformation. In this particular scheme, isometries map three sites into one and the coarse-graining transformation reduces the $L=18$ sites of $\mathcal{L}$ into two sites using two layers of tensors. A collection of states on these two sites is then encoded in a top tensor $\hat{t}$, whose upper index $a=1,2,\ldots, \chi_{\tiny \mbox{top}}$ is used to label $\chi_{\tiny \mbox{top}}$ states $\ket{\Psi_a} \in \mathbb{V}^{\otimes L}$. This particular arrangement of tensors corresponds to the 3:1 MERA described in \citet{evenbly2009}.

In this Section we will consider a MERA analogous to that of Fig.~\ref{fig:ch3c:MERA} but with $Q$ layers of disentanglers and isometries, which we will use to describe states on a lattice $\mathcal{L}$ made of $2\times 3^{Q}$ sites. We will use this variational Ansatz to obtain an approximation to the ground state and first excited states of two quantum spin chains that have a global internal U(1) symmetry, namely the spin-$1/2$ quantum $XX$ chain of \Eref{eq:ch3c:XX} and the spin-$1/2$ antiferromagnetic quantum Heisenberg chain of \Eref{eq:ch3c:XXX}. Each spin-1/2 degree of freedom of the chain is described by a vector space spanned by two orthonormal states $\{\ket{\downarrow}, \ket{\uparrow}\}$. Here we will represent them by the states $\{\ket{0}, \ket{1}\}$ corresponding to zero and one particles, as in Example 1 of Sec.~\ref{sec:ch3c:symmetry:irreps}. For computational convenience, we will consider a lattice $\mathcal{L}$ where each site contains two spins, or states, $\{\ket{\downarrow \downarrow}, \ket{\downarrow \uparrow}, \ket{\uparrow \downarrow},  \ket{\uparrow \uparrow}\}$. Therefore each site of $\mathcal{L}$ is described by a space $\mathbb{V} \cong \mathbb{V}_0 \oplus \mathbb{V}_1 \oplus \mathbb{V}_2$, where $d_0=d_2=1$ and $d_1=2$, as in Example 2 of Sec.~\ref{sec:ch3c:symmetry:irreps}. Thus, a lattice $\mathcal{L}$ made of $L$ sites corresponds to a chain of $2L$ spins. In such a system, the total particle number $N$ ranges from $0$ to $2L$. (Equivalently, the $z$ component of the total spin $S_z$ ranges from $-L$ to $L$, with $S_z = N-L$.)
 
\subsubsection{MERA With U(1) Symmetry}

A U(1)-invariant version of the MERA, or U(1) MERA for short, is obtained by simply considering U(1)-invariant versions of all of the isometric tensors, namely the disentanglers $\hat{u}$, isometries $\hat{w}$, and the top tensor $\hat{t}$. This requires assigning a particle number operator to each index of the MERA. Each open index of the first layer of disentanglers corresponds to one site of $\mathcal{L}$. The particle number operator on any such index is therefore given by the quantum spin model under consideration. We can characterize the particle number operator by two vectors, $\vec{n}$ and $\vec{d}$: a list of the different values the particle number takes and the degeneracy associated with each such particle number, respectively. In the case of the vector space $\mathbb{V}$ for each site of $\mathcal{L}$ described above, $\vec{n} = \{0, 1, 2\}$ and $\vec{d}=\{1, 2, 1\}$. For the open index of the tensor $\hat{t}$ at the very top of the MERA, the assignment of charges is also straightforward. For instance, to find an approximation to the ground state and first seven excited states of the quantum spin model with particle number $N$, we choose $\vec{n} = \{N\}$ and $\vec{d} = \{8\}$. (In particular, a vanishing $S_z$ corresponds to $N=L$.)

For each of the remaining indices of the MERA, the assignment of the pair $(\vec{n},\vec{d})$ needs careful consideration and a final choice may only be possible after numerically testing several options and selecting the one which produces the lowest expectation value of the energy. Table \ref{tab:ch3c:degdist} shows the assignment of particle numbers and degeneracies made to represent the ground state and several excited states in a system of $L=2\times 3^3 = 54$ sites (that is, $108$ spins) with total particle number $N=L=54$ (or $S_z=0$).
\begin{table}
\centering %
\begin{tabular}{c c c} %
\toprule
Level $q$ & Particle numbers $\vec{n}$ & Degeneracies $\vec{d}$ \\ [0.5ex] %
\midrule
top & $\{N=54\}$ & $\{\chi_{\tiny \mbox{top}}\}$ \\ 
3 & $\left\{25, 26, 27, 28, 29\right\}$ & $\left\{1, 3, 5, 3, 1\right\}$  \\
2 & $\left\{7, 8, 9, 10, 11\right\}$ & $\left\{1, 3, 5, 3, 1\right\}$  \\
1 & $\left\{1, 2, 3, 4, 5\right\}$ & $\left\{1, 3, 5, 3, 1\right\}$  \\
0 & $\left\{0, 1, 2\right\}$ & $\left\{1, 2, 1\right\}$  \\ %
[0.5ex]%
\bottomrule
\end{tabular}
\caption[Example of particle number assignment in a U(1) MERA for $L = 54$ sites (or $108$ spins).]{
Example of particle number assignment in a U(1) MERA for $L = 54$ sites (or $108$ spins). The total bond dimension is $\chi = 1+3+5+3+1 = 13$. The value of $\chi_{\tiny \mbox{top}}$ is set as described in the text.\label{tab:ch3c:degdist} %
}
\end{table}
 Notice that at level $q$ of the MERA ($q=1,2,3$), each index effectively corresponds to a block of $n_q \equiv 3^q$ sites of $\mathcal{L}$. Therefore having exactly $n_q$ particles in a block of $n_q$ sites corresponds to a density of $1$ particle per site of $\mathcal{L}$. The assigned particle numbers of Table \ref{tab:ch3c:degdist}, namely $[n_q-2,n_q-1,n_q,n_q+1,n_q+2]$ for level $q$, then correspond to allowing for fluctuations of up to two particles with respect to the average density. The sum of corresponding degeneracies $\vec{d}=\{d_{n_q-2},d_{n_q-1},d_{n_q},d_{n_q+1},d_{n_q+2}\}$ gives the bond dimension $\chi$, which in the example is $\chi=13$.

In order to find an approximation to the ground state of either $\hat{H}_{XX}$ or $\hat{H}_{XXX}$ in Eqs.~\eref{eq:ch3c:XX}--\eref{eq:ch3c:XXX}, we set $\chi_{\tiny\mbox{top}}=1$ and optimize the tensors in the MERA so as to minimize the expectation value 
\begin{equation}
	\bra{\Psi} \hat{H} \ket{\Psi},
\end{equation}
where $\ket{\Psi}\in \mathbb{V}^{\otimes L}$ is the pure state represented by the MERA and $\hat{H}$ is the relevant Hamiltonian. In order to find an approximation to the $\chi_{\tiny\mbox{top}}>1$ eigenstates of $\hat{H}$ with lowest energies, we optimize the tensors in the MERA so as to minimize the expectation value
\begin{equation}
	\sum_{a=1}^{\chi_{\tiny\mbox{top}}}\bra{\Psi_a} \hat{H} \ket{\Psi_a},~~~~\braket{\Psi_a}{\Psi_{a'}} = \delta_{aa'}.
\end{equation}
The optimization is carried out using the MERA algorithm described in \citet{evenbly2009}, which requires contracting tensor networks (by sequentially multiplying pairs of tensors) and performing singular value decompositions. In the present example, all of these operations will be performed exploiting the U(1) symmetry.

\begin{figure}
\begin{center}
  \includegraphics[width=300pt]{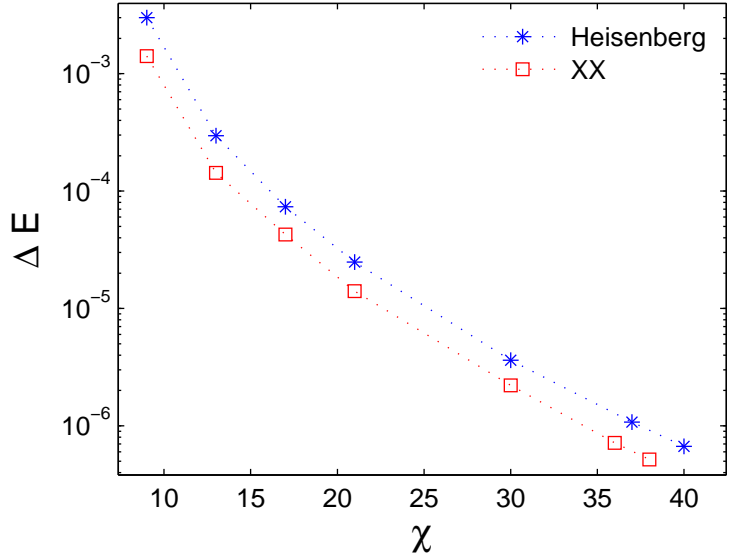}
\caption[Error in ground state energy $\Delta E$ as a function of $\chi$ for the $XX$ and Heisenberg models with $2L=108$ spins and periodic boundary conditions, in the particle number sector $N=L$ (or $S_z=0$).]{Error in ground state energy $\Delta E$ as a function of $\chi$ for the $XX$ and Heisenberg models with $2L=108$ spins and periodic boundary conditions, in the particle number sector $N=L$ (or $S_z=0$). The error is calculated with respect to the exact solutions, and is seen to decay exponentially with $\chi$.\label{fig:ch3c:gserror}}
\end{center}
\end{figure}

Figure~\ref{fig:ch3c:gserror} shows the error in the ground state energy as a function of the bond dimension $\chi$, for assignments of degeneracies similar to those in Table \ref{tab:ch3c:speedup}.
 The error is seen to decay exponentially with increasing $\chi$, indicating increasingly accurate approximations to the ground state.
\begin{table}%
\centering %
\begin{tabular}{cc r r c} %
\toprule
$\chi$ & Degeneracy           & no. of~~~ & no. of~~~~ & ratio \\ 
       &            $\vec{d}$ & coefficients    & coefficients     & \\
       &                      & (regular)~    & (symmetric)     & \\
[0.5ex] %
\midrule
~~4~~ & $\left[0, 1, 2, 1, 0\right]$  & %
1552~~ & 426~~ & ~~3.6~:~1~   \\ %
~~8~~ & $\left[0, 2, 4, 2, 0\right]$  & 17216~~ & 4714~~  & ~~3.7~:~1~   \\ %
~~13~~ & $\left[1, 3, 5, 3, 1\right]$  & 115501~~ & 21969~~ & ~~5.3~:~1~  \\
~~17~~ & $\left[1, 4, 7, 4, 1\right]$  & 335717~~ & 68469~~ & ~~5.0~:~1~ \\
~~21~~ & $\left[1, 5, 9, 5, 1\right]$  & 779965~~ & 166901~~ & ~~4.7~:~1~   \\
~~30~~ & $\left[2, 7, 12, 7, 2\right]$ & 3243076~~ & 639794~~ & ~~5.1~:~1~   \\ 
[0.5ex]%
\bottomrule
\end{tabular}
\caption[Number of coefficients required to specify the tensors of a MERA for $L=54$ as a function of the bond dimension $\chi$, decomposed according to a degeneracy vector $\vec{d}$: comparison between regular tensors and U(1)-invariant tensors.]{
Number of coefficients required to specify the tensors of a MERA for $L=54$ as a function of the bond dimension $\chi$, decomposed according to a degeneracy vector $\vec{d}$. A comparison is made between regular tensors and U(1)-invariant tensors.\label{tab:ch3c:speedup} %
}
\end{table}

\subsubsection{Exploiting the Symmetry}

We now discuss some of the advantages of using the U(1) MERA.

\paragraph{Selection of Particle Number Sector}

An important advantage of the U(1) MERA is that it exactly preserves the U(1) symmetry. In other words, the states resulting from a numerical optimization are exact eigenvectors of the total particle number operator $\hat{N}$ \eref{eq:ch3c:hatN}. In addition, the total particle number $N$ can be pre-selected at the onset of optimization by specifying it in the open index of the top tensor $\hat{t}$. 

\begin{figure}
\begin{center}
  \includegraphics[width=300pt]{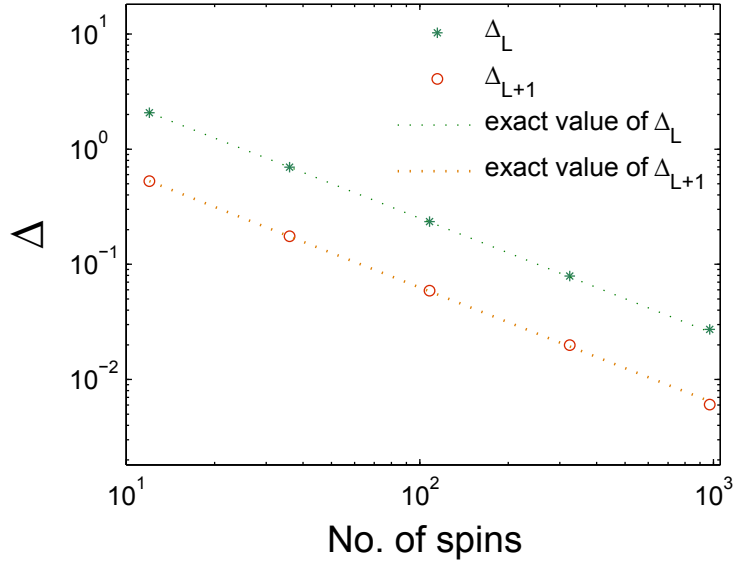}
\caption[Decay of energy gaps $\Delta$ with system size $L$ in the $XX$ model.]{Decay of energy gaps $\Delta$ with system size $L$ in the $XX$ model. The upper line corresponds to the energy gap $\Delta_L$ between the ground state and the first excited state in the $N=L$ particle number (or $S_z=0$) sector. The lower line corresponds to the energy gap $\Delta_{L+1}$ between the ground states of the $N=L$ and $N=L+1$ particle number sectors.
\label{fig:ch3c:XXgaps1}}
\end{center}
\end{figure}

Figure~\ref{fig:ch3c:XXgaps1} shows the energy gap between the ground state and two excited states of an $XX$ chain with $2L$ spins (or $L$ sites), for $N=L$ particles ($S_z=0$).
 One is the first excited state which also has $N=L$ particles. The other is the ground state in the sector with $N=L+1$ particles. The two energy gaps are seen to decay with the system size as $L^{-1}$. The ability to pre-select a given particle number $N$ means that only two optimizations were required: one MERA optimization for $N=L$ with $\chi_{\tiny \mbox{top}}=2$ in order to obtain an approximation to the ground state and first excited state of $\hat{H}_{XX}$ in that particle number sector; and one MERA optimization for $N=L+1$ with $\chi_{\tiny \mbox{top}}=1$ in order to obtain an approximation to the ground state of $\hat{H}_{XX}$ in the particle number sector $N=L+1$. 

Similar results can be obtained with the regular MERA. For instance, one can obtain an approximation to the ground state of a given particle number sector by adding a chemical potential term  $-\mu\sum_{s}\hat{n}^{(s)}$ to the Hamiltonian and carefully tuning the chemical potential term $\mu$ until the expectation value of the particle number $\hat{N}$ is the desired one. However, the regular MERA cannot guarantee that the states obtained in this way are exact eigenvectors of $\hat{N}$. Instead the resulting states are likely to have particle number fluctuations.

\begin{figure}
\begin{center}
  \includegraphics[width=318.75pt]{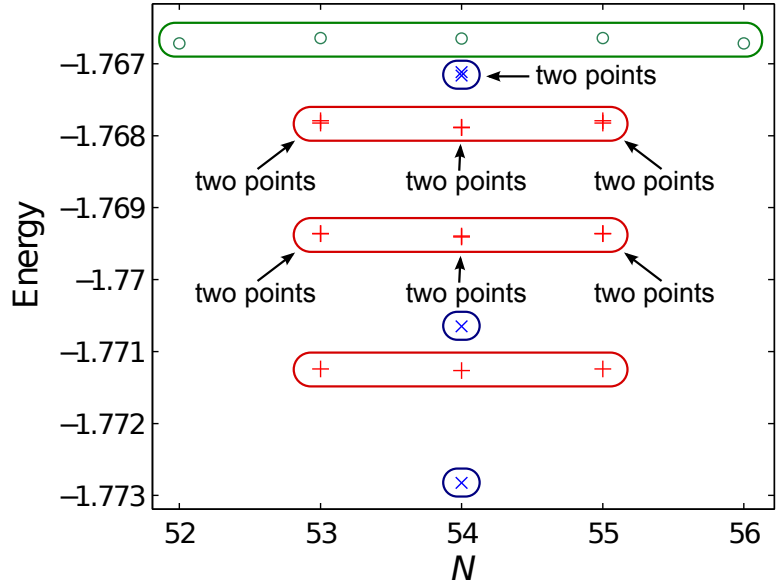}
\caption[Low energy spectrum of $\hat{H}_{XXX}$ with $L=54$ sites (=108 spins). States are labelled according to total spin and number of particles.]{%
Low energy spectrum of $\hat{H}_{XXX}$ with $L=54$ sites (=108 spins). Depicted states have spins of zero ($\times$, blue loops), one (+, red loops), or two ($\circ$, green loop), and total number of particles ($N$) between 52 and 56. Note that the second and third spin-1 triplets are twofold degenerate.\label{fig:ch3c:h3spectra}}
\end{center}
\end{figure}

Figure~\ref{fig:ch3c:h3spectra} shows the low energy spectrum of the Heisenberg model $\hat{H}_{XXX}$ for a periodic system of $L=54$ sites (or $108$ spins), including the ground state and several excited states both in the particle sector $N=54$ (or $S_z=0$), and in neighboring particle sectors.
 Recall that $\hat{H}_{XXX}$ is actually invariant under a global internal SU(2) symmetry, of which particle number is a U(1) subgroup. Correspondingly the spectrum is organized according to irreps of SU(2), namely singlets (total spin $0$), triplets (total spin $1$), quintuplets (total spin $2$), etc. Again, using the U(1) MERA, the five particle number sectors $N=52,53,54,55$, and $56$ can be addressed with independent computations. This implies, for instance, that in order to find the gap between the first and fourth singlets, we can simply set $N=54$ and $\chi_{\tiny \mbox{top}} = 9$ on the open index of the top tensor $\hat{t}$, to accommodate the first four spin-0 states and five spin-1 states in the $N=54$ sector, as seen in \fref{fig:ch3c:h3spectra}. In order to capture the fourth singlet using the regular MERA, we would need to consider at least $\chi_{\tiny\mbox{top}} = 19$ (at a larger computational cost and possibly lower accuracy), since this state has only the $19$\textsuperscript{th} lowest energy overall.

\paragraph{Reduction of Computational Costs}

The use of U(1)-invariant tensors in the MERA also results in a reduction of computational costs. 

First, U(1)-invariant tensors, when written in the canonical form of Eqs.~\eref{eq:ch3c:Tcanon}--\eref{eq:ch3c:Tcanon2}, are block-diagonal and therefore require less storage space. Table \ref{tab:ch3c:speedup} compares the number of MERA coefficients that need to be stored in the regular and symmetric case, for different choices of particle number assignments relevant to the present examples. 

\begin{figure}
\begin{center}
  \includegraphics[width=356.25pt]{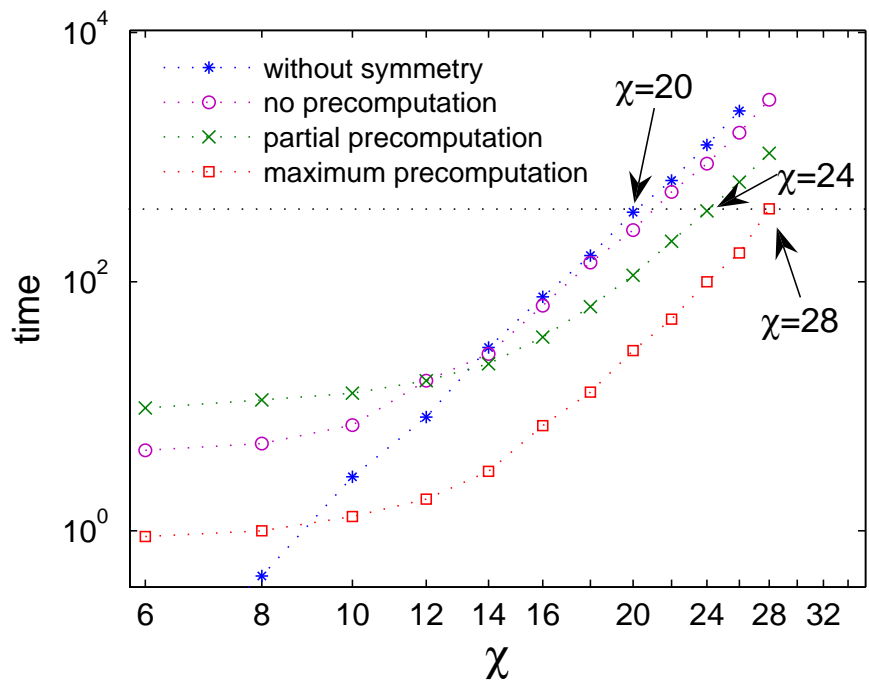}
\caption[Computation time (in seconds) for one iteration of the MERA energy minimization algorithm, as a function of the bond dimension $\chi$, with and without exploitation of U(1) symmetry.]{%
Computation time (in seconds) for one iteration of the MERA energy minimization algorithm, as a function of the bond dimension $\chi$. For sufficiently large $\chi$, exploiting the U(1) symmetry leads to reductions in computation time.\label{fig:ch3c:all1}
The horizontal line on this graph shows that this reduction in computation time equates to the ability to evaluate MERAs with a higher bond dimension $\chi$: For the same cost per iteration incurred when optimising a standard MERA in \textsc{matlab} with bond dimension $\chi=20$, one may choose instead to optimise a U(1)-symmetric MERA with partial precomputation and $\chi=24$, or with full precomputation and $\chi=28$.}
\end{center}
\end{figure}

Second, the computation time required to manipulate tensors is also reduced when using U(1)-invariant tensors in the canonical form. Figure~\ref{fig:ch3c:all1} shows the computation time required for one iteration of the energy minimization algorithm of \citet{evenbly2009} (during which all tensors in the MERA are updated once), as a function of the total bond dimension $\chi$. 
The plot compares the time required using regular tensors and U(1)-invariant tensors. For U(1)-invariant tensors, we display the time per iteration for three different levels of precomputation, as described in \sref{sec:ch3c:supplement}.
The figure shows that for sufficiently large $\chi$, using U(1)-invariant tensors %
leads to a shorter time per iteration of the optimization algorithm. 

In our reference implementation (written in \textsc{matlab}), using the symmetry without precomputation is seen to only reduce the computation time by about a factor of 2 for the largest $\chi$ under consideration. This is because maintaining the canonical form for U(1)-invariant tensors still imposes a significant overhead for the values of $\chi$ considered. 
In contrast, when using precomputation we obtained times shorter by a factor of 10 or more.

The magnitude of the overhead imposed by maintaining the canonical form will depend on factors such as programming language and machine architecture, but in general more significant gains can be obtained by making full use of precomputation. This option, however, requires a significant amount of additional memory (see \sref{sec:ch3c:supplement}), and a more convenient middle ground can be obtained by using a partial precomputation scheme.

\section{Conclusions\label{sec:ch3c:conclusions}}

In this Chapter we have provided a detailed explanation of how a global internal Abelian symmetry may be incorporated into any tensor network algorithm.
We considered tensor networks constructed from tensors which were invariant under the action of the internal symmetry, and showed how each tensor may be decomposed according to a canonical form %
into \textit{degeneracy} tensors (which contain all the degrees of freedom that are not affected by the symmetry) and \textit{structural} tensors (which are completely determined by the symmetry). We then introduced a set of primitive operations $\mathcal{P}$ which may be used to carry out tensor network algorithms using Ans\"atze such as MPS, PEPS, and MERA, and showed how each of these operations 
can be implemented in such a way that the canonical form is both preserved and exploited for computational gain. 

We then demonstrated the implementation of this decomposition for tensors with an internal U(1) symmetry, and computed multiple benchmarks demonstrating the computational costs and speed-ups inherent in this approach. We found that although maintaining the canonical form imposed additional costs when combining or splitting tensor indices, for simulations of a sufficiently large scale these costs can be offset by the gains made when performing permutations, matrix multiplications, and matrix decompositions. 

Finally, we implemented the 
MERA on a quantum spin chain with U(1) symmetry. We showed that exploitation of this symmetry can lead to a decrease in the computational cost by a factor of 10 or more.
These gains may be used either to reduce overall computation time or to permit substantial increases in the MERA bond dimension $\chi$, and consequently in the accuracy of the results obtained. 

Although in this Chapter we have focused 
on an example which is a continuous Abelian group, the formalism presented here may equally well be applied 
to a finite Abelian group. In particular let us consider a cyclic group $Z_q$, $q\in\mbb{Z}^+$.\footnote{The fundamental theorem of Abelian groups states that every finite Abelian group may be expressed as a direct sum of cyclic subgroups of prime-power order.}
As in the case of U(1), the Hilbert space decomposes under the action of the group
into a direct sum of one dimensional irreps which are each characterized by an integer charge $a$,
and consequently most of the analysis presented in this Chapter remains 
unchanged. 
In particular, matrices which are invariant under the action of the group will be block diagonal in the basis labeled by charge according to \Eref{eq:ch3c:Schur}, and symmetric tensors enjoy the canonical decomposition stated in Eqs.~\eref{eq:ch3c:Tcanon}--\eref{eq:ch3c:Tcanon2}. The only objects which need modification are the fusion and splitting maps, which need to be altered so that they encode the fusion rules for $Z_q$ instead for U(1). For a cyclic group $Z_q$, the fusion of two charges $a$ and $a'$ gives rise to a charge $a''$ according to $a'' = (a+a')|_q$ where $|_q$ indicates that the addition is performed modulo $q$. 
For example, $Z_3$ has
charges $a=0,1,2$, and the fusion rules for $Z_3$
take the form $a\times a'\rightarrow a''$ where the value of $a''$ is given in the following table:

\begin{center}
\begin{tabular}{cc||ccc}
&&&~~~$a$~~~&\\
&&~~~0~~~&~~~1~~~&~~~2~~~\\
\hline
\hline
    &0&~~~0~~~&~~~1~~~&~~~2~~~\\
$a'$&1&~~~1~~~&~~~2~~~&~~~0~~~\\
    &2&~~~2~~~&~~~0~~~&~~~1~~
\end{tabular}
\end{center}

More generally, a generic Abelian group 
will be characterised by a set of charges $(a_1, a_2, a_3,\ldots)$. 
When fusing two such sets of charges $(a_1,a_2,a_3,\ldots)$ and $(a'_1,a'_2,a'_3,\ldots)$, each charge $a_i$ 
is combined with its counterpart $a'_i$ 
according to the fusion rule of the relevant subgroup. Once again, 
this behaviour may be encoded in a single fusion map $\fuser$ and its inverse $\splitter$. The formalism presented in this Chapter 
is therefore directly applicable to any Abelian group.

\section{\nohyphens{Supplement: Use of Precomputation in Iterative Algorithms}\label{sec:ch3c:supplement}}

We have seen that the use of the canonical form given in Eqs.~\eref{eq:ch3c:Tcanon}--\eref{eq:ch3c:Tcanon2} to represent U(1)-invariant tensors can potentially lead to substantial reductions in memory requirements and in calculation time. We also pointed out, however, that there is an additional cost in maintaining an invariant tensor in its canonical form, and that this is associated with the reshaping (fusing and/or splitting) of its indices. In some situations this additional cost may significantly reduce, or even offset, the benefits of using the canonical form. 

In this Section we investigate 
techniques for reducing this additional cost in the context of iterative tensor network algorithms. Many of the algorithms discussed in \sref{sec:ch3c:tensor:TNstates} 
are iterative algorithms, repeating the same sequence of tensor network manipulations many times over. Examples include algorithms which compute tensor network approximations to the ground state by 
minimizing the expectation value of the energy 
or by
simulating evolution in imaginary time,
with each iteration yielding an
increasingly accurate 
approximation to the ground state of the system. 

The goal of this Section is to identify calculations which depend only on the symmetry group, and are independent of the variational coefficients of such algorithms.
Where these calculations are repeated in each iteration of the algorithm, we can
effectively eliminate the associated computational cost by performing them 
only once, either during or prior to the first iteration of the algorithm, and then 
storing and reusing these \emph{precomputed} results 
in subsequent iterations.
We will illustrate this procedure by considering the precomputation of a series of operations applied to a single tensor $\hat T$. 

To do this, we begin by revisiting the fusion and splitting tables of \sref{sec:ch3c:symmetry:tp} and introducing a graphical representation of these objects. We then introduce a convenient decomposition of a symmetric tensor into a matrix accompanied by multiple fusion and/or splitting tensors, and linear maps $\Gamma$ that map one such decomposition into another. These linear maps are independent of the coefficients of the tensor being reorganized, and consequently they are precisely the objects which can be precomputed in order to quicken an iterative algorithm at the expense of additional memory cost. Finally we describe two specific precomputation schemes, differing in what is precomputed and in how the precomputed data are utilized during the execution of the algorithm, in order to illustrate the trade-off between the amount of memory needed to store the precomputation data and the associated computational speedup which may be obtained. In practice, the nature of the specific implementation employed will depend on available computational resources.

\subsection{Diagrammatic Notation of Fusing and Splitting Tensors\label{apdx:ch3c:diag:fusesplit}}

\begin{figure}
\begin{center}
  \includegraphics[width=300pt]{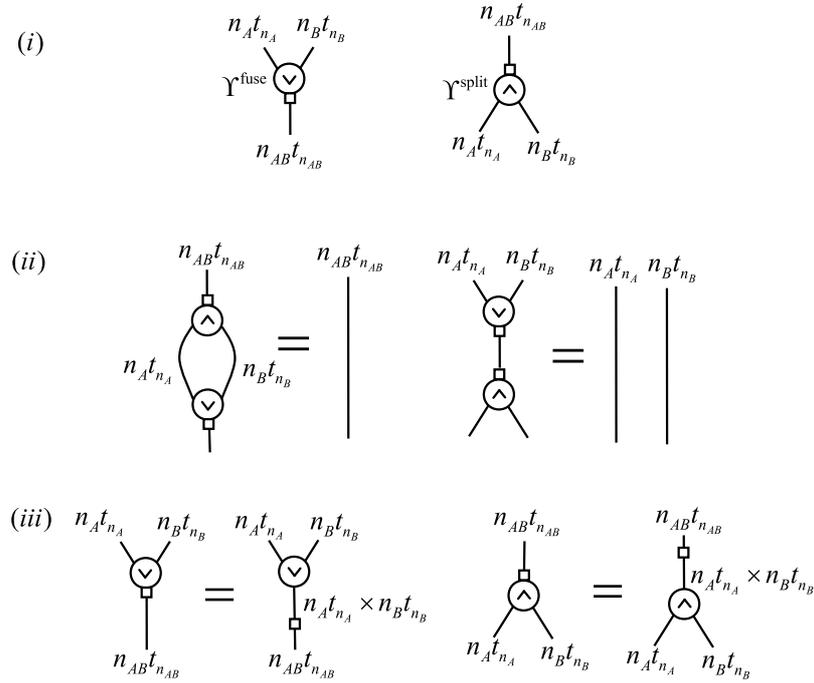}
\caption[(i)~Graphical representation of the fusion tensor $\fuser$ and the splitting tensor $\splitter$. 
(ii)~The tensors $\fuser$ and $\splitter$ are unitary, and thus yield the identity when contracted pairwise as shown.
(iii)~A fusion tensor decomposed into two parts. The first part performs the tensor product of input irreps, and the second part is a permutation that associates each pair of vectors in the input bases of $\mathbb{V}^{(A)}$ and $\mathbb{V}^{(B)}$ with a unique vector in the coupled basis of $\mathbb{V}^{(AB)}$.]
{%
(i)~Graphical representation of the fusion tensor $\fuser$ and the splitting tensor $\splitter$. 
(ii)~The tensors $\fuser$ and $\splitter$ are unitary, and thus yield the identity when contracted pairwise as shown.
(iii)~A fusion tensor decomposed into two parts. The first part (indicated by a circle with an arrow) performs the tensor product of input irreps, $n_A t_A \times n_B t_B$. The result is an index that labels pairs $(n_A t_A, n_B t_B)$. The second part (indicated by a rectangle) is a permutation that associates each pair $(n_A t_A, n_B t_B)$ with a unique $(n_{AB} t_{n_{AB}})$, corresponding to a vector in the coupled basis of $\mathbb{V}^{(AB)}$.\label{fig:ch3c:u1fuse}}
\end{center}
\end{figure}

In describing how we can precompute repeated manipulations of this tensor $\hat T$, we will find it useful to employ diagrammatic representations of the fusion and splitting tables $\fuser$ and $\splitter$ introduced in \sref{sec:ch3c:symmetry:tp}. These tables implement a linear map between a pair of indices and their fusion product, and thus can be understood as trivalent tensors having two input legs and one output leg (or vice versa) in accordance with \sref{sec:ch3c:tensor:linear}. We choose to represent them graphically as shown in \fref{fig:ch3c:u1fuse}(i), where the arrow within the circle always points toward the coupled index.
The linear maps $\fuser$ and $\splitter$ are unitary, and consequently we impose that the tensors of \fref{fig:ch3c:u1fuse}(i) must satisfy the identities given in \fref{fig:ch3c:u1fuse}(ii), corresponding to unitarity under the action of the conjugation operation employed in diagrammatic tensor network notation (vertical reflection of a tensor and the complex conjugation of its components, typically denoted $^\dagger$).
Our notation also reflects the property, first noted in \sref{sec:ch3c:symmetry:tp}, that $\fuser$ and $\splitter$ may be decomposed into two pieces [Fig. \ref{fig:ch3c:u1fuse}(iii)]. For the fusion tensor, we identify the first piece (represented by a circle containing an arrow) with the creation of a composed index using the manner we would employ in the absence of symmetry (\ref{eq:ch3c:fuse}). The second piece, represented by the small square, permutes the basis elements of the composed index, reorganizing them according to total particle number. The two components of the splitting tensor are then uniquely defined by consistency with the process of conjugation for the diagrammatic representation of tensors, 
and with the unitarity condition of \fref{fig:ch3c:u1fuse}(ii).

\begin{figure}
\begin{center}
  \includegraphics[width=195pt]{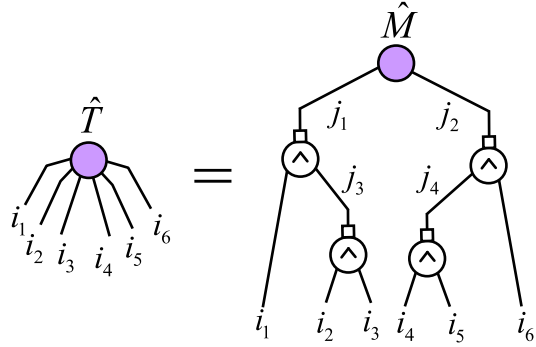}
\caption[Binary tree decomposition of a symmetric tensor $\hat{T}$ having components $\hat{T}_{i_1 i_2 i_3 i_4 i_5 i_6}$.]{%
Binary tree decomposition of a symmetric tensor $\hat{T}$ having components $\hat{T}_{i_1 i_2 i_3 i_4 i_5 i_6}$. The tree $\mathcal{T}$ is comprised of a matrix $\hat{M}$ as the root node, %
four splitting tensors as internal nodes, and $i_1, i_2, ..., i_6$ as its leaf indices. No incoming or outgoing arrows are indicated on the indices in the figure, as the decomposition is valid for any such assignment of directional arrows.\label{fig:ch3c:treeDeco}}
\end{center}
\end{figure}

These requirements have an important consequence. Suppose the first part of $\fuser$ implements $b\times c\rightarrow d$ by iterating rapidly over the values of $b$ and more slowly over the values of $c$, and $b$ lies clockwise of $c$ on the graphical representation of $\fuser$. This then means that on the graphical representation of $\splitter$ which implements $d\rightarrow b\times c$, index $b$ must lie \emph{counterclockwise} of $c$. It is therefore vitally important to distinguish between the splitting tensor and a rotated depiction of the fusing tensor. To this end we require that when using this diagrammatic notation,
all tensors (with the exception of the fusion and splitting tensors) must be drawn with only downward-going legs, as seen for example in \fref{fig:ch3c:treeDeco}, though the legs are still free to carry either incoming or outgoing arrows as before.

\subsection{Tree Decomposition}

We find it convenient to decompose a rank-$k$, U(1)-invariant tensor $\hat{T}$, having components $\hat{T}_{i_1i_2\ldots i_k}$, as a binary tree tensor network $\mathcal{T}$ consisting of a matrix $\hat{M}$ which we will call the \textit{root node}, and of $k-2$ splitting tensors $\splitter$ as branching \textit{internal nodes}, with the \textit{leaf} indices of tree $\mathcal{T}$ corresponding to the indices $\{i_1, i_2, \ldots, i_k\}$  of tensor $\hat{T}$. We refer to decomposition $\mathcal{T}$ as a tree decomposition of $\hat{T}$. Figure~\ref{fig:ch3c:treeDeco} shows an example of tree decomposition for a rank-6 tensor. It is of the form
\begin{equation}
\hat{T}_{i_1 i_2 i_3 i_4 i_5 i_6} = \sum_{j_1, j_2, j_3, j_4} \hat{M}_{j_1 j_2} \splitt{j_1}{i_1}{j_3} \splitt{j_2}{j_4}{i_6} \splitt{j_3}{i_2}{i_3} \splitt{j_4}{i_4}{i_5},\label{decomposeob}
\end{equation}
where $\{j_1,j_2, j_3, j_4\}$ are the internal indices of the tree.

\begin{figure}
\begin{center}
  \includegraphics[width=277.125pt]{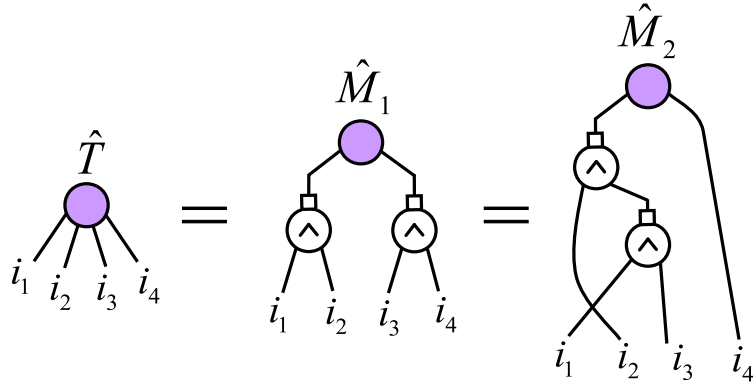}
\caption[Two possible tree decompositions of a rank-4 tensor $\hat{T}$. Different choices lead to different matrix representations of the same tensor.]{%
Two possible tree decompositions of a rank-4 tensor $\hat{T}$. Different choices $\mathcal{T}_1, \mathcal{T}_2, \ldots$ of tree decomposition for tensor $\hat{T}$ lead to different matrices $\hat{M}_1, \hat{M}_2, \ldots$ for the same tensor.\label{fig:ch3c:treeDeco2}}
\end{center}
\end{figure}

\begin{figure}
\begin{center}
  \includegraphics[width=262.5pt]{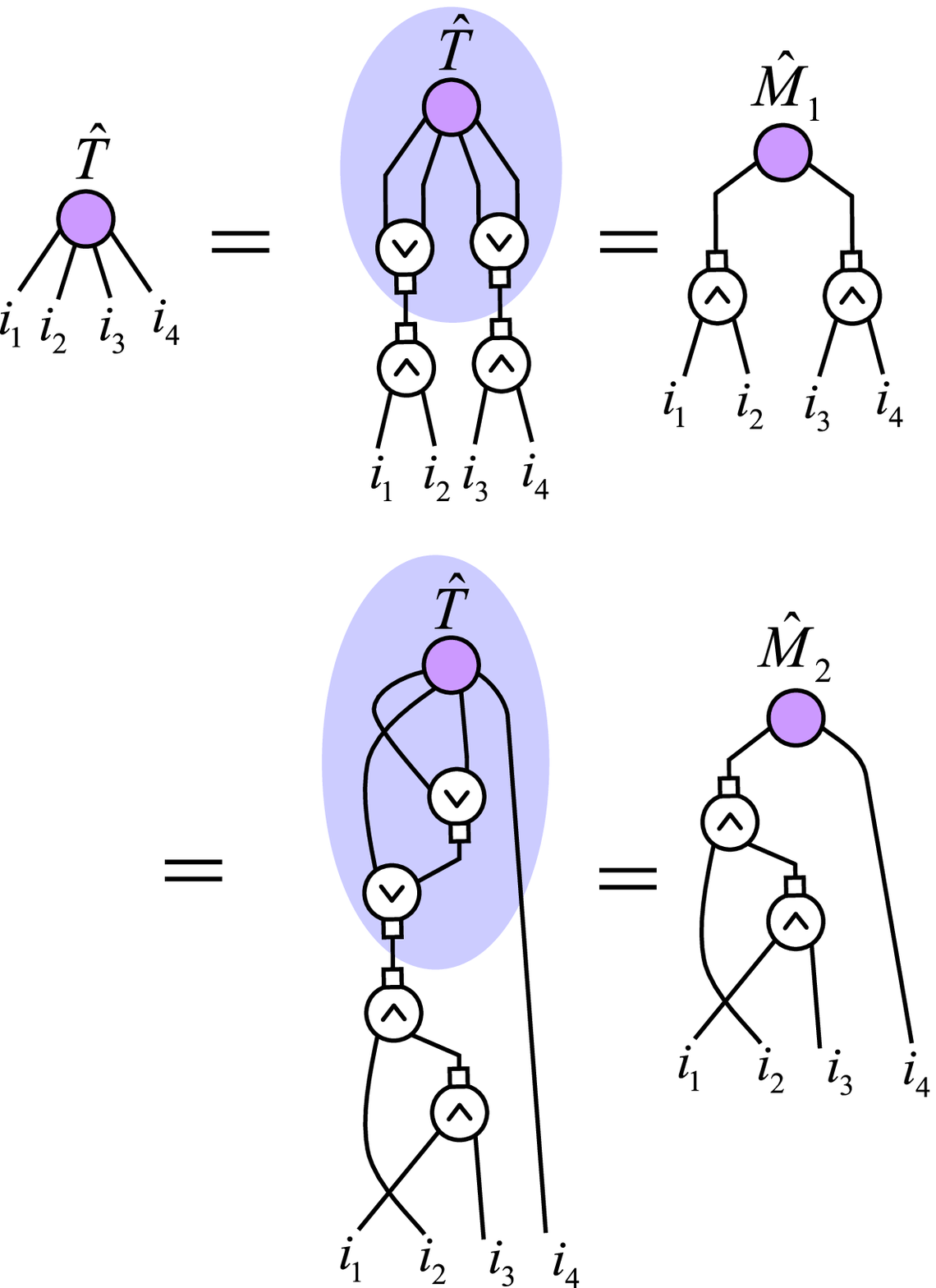}
\caption[Tree decompositions of tensor $\hat{T}$ are obtained by contracting the tensor with an appropriate resolution of the identity on its indices, selected according to the desired choice of the fusion tree $\mathcal{T}$.]{Tree decompositions of tensor $\hat{T}$ are obtained by contracting the tensor with an appropriate resolution of the identity on its indices, selected according to the desired choice of the fusion tree $\mathcal{T}$. In each instance, evaluation of the contents of the shaded region yields the appropriate matrix $\hat M$.\label{fig:ch3c:treeDeco3}}
\end{center}
\end{figure}

The same tensor $\hat{T}$ may be decomposed as a tree in many different ways, corresponding to different choices of the fusion tree. As an example we show two different but equivalent decompositions of a rank-4 tensor in Fig. \ref{fig:ch3c:treeDeco2}.
Different choices $\mathcal{T}_1, \mathcal{T}_2, \ldots$ of tree decomposition for tensor $\hat{T}$ will lead to different matrix representations $\hat{M}_1, \hat{M}_2, \ldots$ of the same tensor. Finally, Fig.~\ref{fig:ch3c:treeDeco3} shows how to obtain the tree decompositions from $\hat{T}_{i_1i_2i_3i_4}$ by introducing an appropriate resolution of the identity, constructed from pairs of fusion operators $\fuser$ and splitting operators $\splitter$ in accordance with Fig.~\ref{fig:ch3c:u1fuse}(ii). 
 
The representation of a tensor $\hat T$ by means of a tree decomposition is particularly useful because many tensor network algorithms may be understood as a sequence of operations carried out on tensors reduced to matrix form. 
For example, consider tensor network algorithms such as MPS, MERA, and PEPS. When tensors are updated in these algorithms, the new tensor is typically created as a matrix, to which operations from the primitive set $\mc{P}$ of \sref{sec:ch3c:tensor:TN} are then applied. When they are decomposed or contracted with other tensors, this may once again take place with the tensor in matrix form. Any such matrix form may always be understood as the matrix component of an appropriate tree decomposition $\mathcal{T}$ of tensor $\hat T$, where the sequence of operations reshaping tensor $\hat T$ to matrix $\hat M$ corresponds to the contents of the shaded area in \fref{fig:ch3c:treeDeco3}.

\subsection{Mapping Between Tree Decompositions\label{sec:ch3c:gammamap}}

Suppose now that we have a tensor $\hat T$ in matrix form $\hat M_1$, which is associated with
a particular choice of tree decomposition $\mathcal{T}_1$, and we wish to transform it into another matrix form $\hat M_2$, corresponding to another tree decomposition $\mathcal{T}_2$. As indicated, this process may frequently arise during the application of many common tensor network algorithms.
The new matrix $\hat{M}_2$ can be obtained from $\hat{M}_1$ by means of a series of reshaping (splitting/fusing) and permuting operations, as indicated in \fref{fig:ch3c:MtoM}, and this series of operations may be understood as defining a map $\Gamma$: 
\begin{figure}
\begin{center}
  \includegraphics[width=206.25pt]{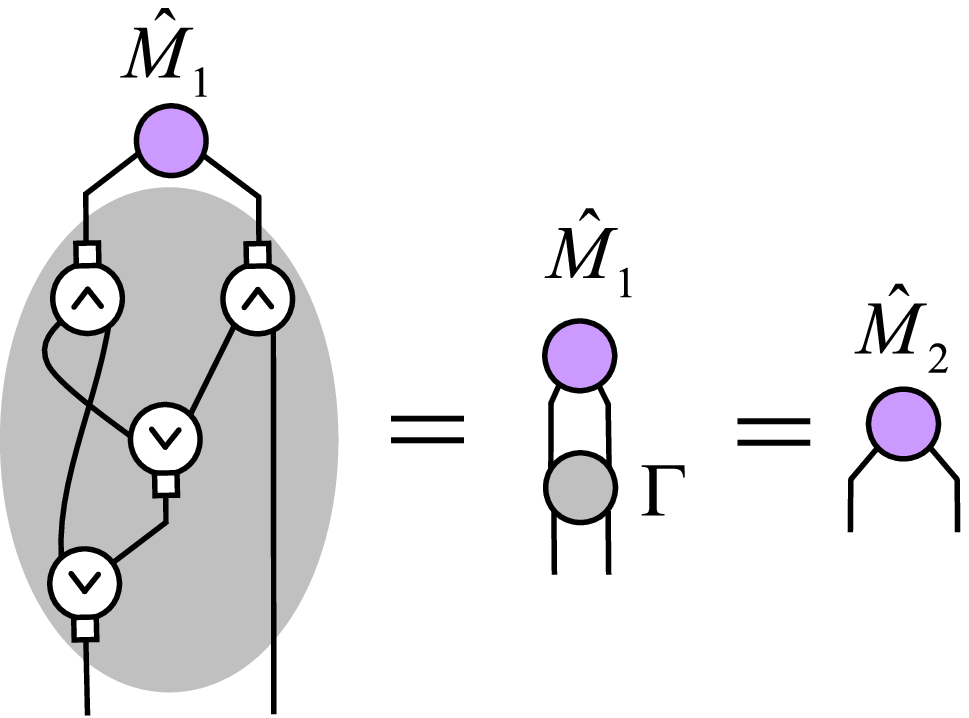}
\caption[A matrix $\hat{M}_1$ can be reorganized into another matrix $\hat{M}_2$ by means of fusion tensors, splitting tensors, and the permutation of indices. 
These operations define a one to one linear map $\Gamma$ that acts to reorganize the coefficients of $\hat{M}_1$, and depends only on the sequence of operations performed.]{%
A matrix $\hat{M}_1$ can be reorganized into another matrix $\hat{M}_2$ by means of fusion tensors, splitting tensors, and the permutation of indices. 
These operations define a one to one linear map $\Gamma$ that acts to reorganize the coefficients of $\hat{M}_1$. $\Gamma$ does not depend on the coefficients of $\hat{M}_1$, but solely on the sequence of operations performed. %
\label{fig:ch3c:MtoM}}
\end{center}
\end{figure}
\begin{equation}
	\hat{M}_2 = \Gamma(\hat{M}_1).
	\label{eq:ch3c:Gamma}
\end{equation}
The map $\Gamma$ is a linear map which depends only on the tree structure of $\mathcal{T}_1$ and $\mathcal{T}_2$, and is independent of 
the coefficients of $\hat{M}_1$. Moreover, $\Gamma$ is unitary, and 
it follows from the construction of fusing and splitting tensors and the behaviour of permutation of indices (which serves to relocate the coefficients of a tensor) that $\Gamma$ simply reorganizes the coefficients of $\hat{M}_1$ into the coefficients of $\hat{M}_2$ in a one-to-one fashion.

Therefore, one way to compute the matrix $\hat{M}_2$ from matrix $\hat{M}_1$ is by first computing the linear map $\Gamma$, which is independent of the specific coefficients in tensor $\hat{T}$, and by then applying it to $\hat{M}_1$.

\subsection{Precomputation Schemes For Iterative Tensor Network Algorithms}

\begin{figure}
\begin{center}
  \includegraphics[width=300pt]{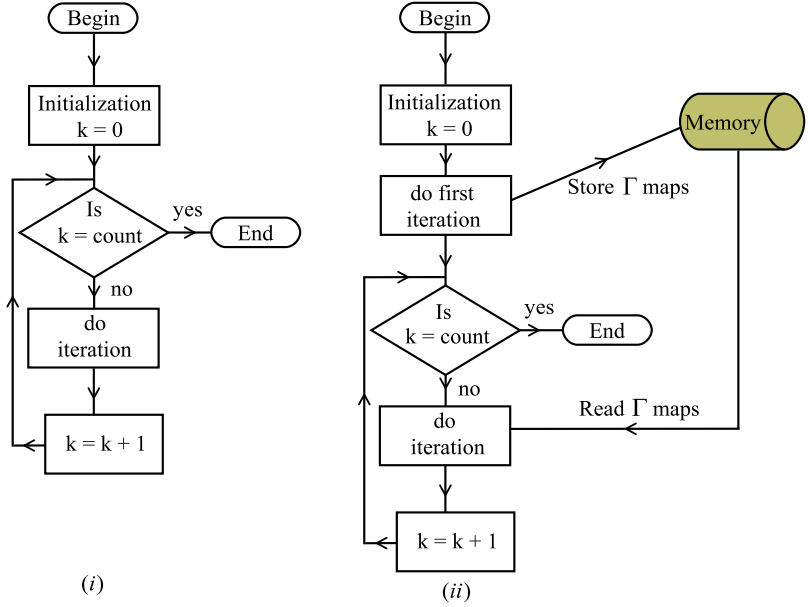}
\caption{Flow diagram for the execution of a predetermined number of iterations of a generic iterative tensor network algorithm, (i) without any precomputation and (ii) with precomputation of the operations $\Gamma$.\label{fig:ch3c:preCompute}}
\end{center}
\end{figure}

The observation that the map $\Gamma$ is independent of the specific coefficients in $\hat{M}_1$ is particularly useful in the context of iterative tensor network algorithms. It implies that, although the coefficients in $\hat M_1$ will change from iteration to iteration, the linear map $\Gamma$ in \Eref{eq:ch3c:Gamma} remains unchanged. It is therefore possible to calculate the map $\Gamma$ once, during the first iteration of the simulation, and then to store it in memory and re-use it during subsequent iterations.
We refer to such a strategy as a \textit{precomputation scheme}. Figure~\ref{fig:ch3c:preCompute} contrasts the program flow of a generic iterative tensor network algorithm with and without precomputation of the transformations $\Gamma$.

Using such a precomputation scheme, a significant speed-up of %
simulations can be obtained, at the price of storing potentially large amounts of precomputed data (as a single iteration of the algorithm may require the application of many different transformations $\Gamma$). Therefore a trade-off necessarily exists between the amount of speed-up that can be obtained and the memory requirement which this entails.
In this Section we describe two different precomputation schemes. The first one fully precomputes and stores all maps $\Gamma$, and is relatively straightforward to implement. This results in the maximal increase in simulation speed, but implementation requires a large amount of memory. The second scheme only partially precomputes the maps $\Gamma$, resulting in a moderate speed-up of simulations, but with memory requirements which are also similarly more modest.

\subsubsection{Maximal Precomputation Scheme}

As noted in \sref{sec:ch3c:gammamap}, applying the map $\Gamma$ to a matrix $\hat{M}_1$ simply reorganizes its coefficients to produce the matrix $\hat{M}_2$. Moreover, if the indices of matrices $\hat M_1$ and $\hat M_2$ are fused to yield vectors $\hat V_1$ and $\hat V_2$ then the map $\Gamma$ may be understood as a permutation matrix, and this in turn may be concisely represented as a string of integers $\Gamma = \gamma_1,\ldots, \gamma_{|\hat M_1|}$ such that entry $i$ of $\hat V_2=\Gamma\hat V_1$ is given by entry $\gamma_i$ of vector $\hat V_1$. Because all of the elements from which $\Gamma$ is composed are sparse, unitary, and composed entirely of 0's and 1's, the permutation 
to which $\Gamma$ corresponds may be calculated at a total cost of only $O(|\hat M_1|)$, where $|\hat M_1|$ counts only the 
elements of $\hat M_1$ which are not fixed to be zero by the 
symmetry constraints of \Eref{eq:ch3c:Tcanon}. In essence, for each element of the vector $\hat V_1$ one identifies the corresponding number and degeneracy indices $(n^{\hat M_1}_i,t^{\hat M_1}_i)$ on each leg $i\in\{1,2\}$ of matrix $\hat M_1$. One can then read down the figure, applying each table $\fuser$ or $\splitter$ in turn to identify the corresponding labels $(n',t')$ on the intermediate legs, until finally
the corresponding labels on the indices of $\hat M_2$ 
are obtained. 
There is then a further 1:1 mapping from each set of labels $(n^{\hat M_2}_1,t^{\hat M_2}_1)$, $(n^{\hat M_2}_2,t^{\hat M_2}_2)$ on $\hat M_2$ to the corresponding entry in $\hat V_2$, 
completing the definition of $\Gamma$ as a map
from $\hat V_1$ to $\hat V_2$.

Storing the map $\Gamma$ for a transformation such as the one shown in \fref{fig:ch3c:MtoM} 
imposes a memory cost 
of $O(|\hat M_1|)$. The application of this map also incurs a computational cost of $O(|\hat M_1|)$, but computational overhead is saved in not having to reconstruct the map $\Gamma$ on every iteration of the algorithm.

\subsubsection{Partial Precomputation Scheme}

The $O(|\hat{M}_1|)$ memory cost incurred in the previous scheme can be significant for large matrices. However, we may reduce this cost by replacing the single permutation $\Gamma$ employed in that scheme with multiple smaller operations which may also be precomputed. In this approach $\hat{M}_1$ is retained in matrix form rather than being reshaped into a vector, and we precompute permutations to be performed on its rows and columns. 

\begin{figure}
\begin{center}
  \includegraphics[width=262.5pt]{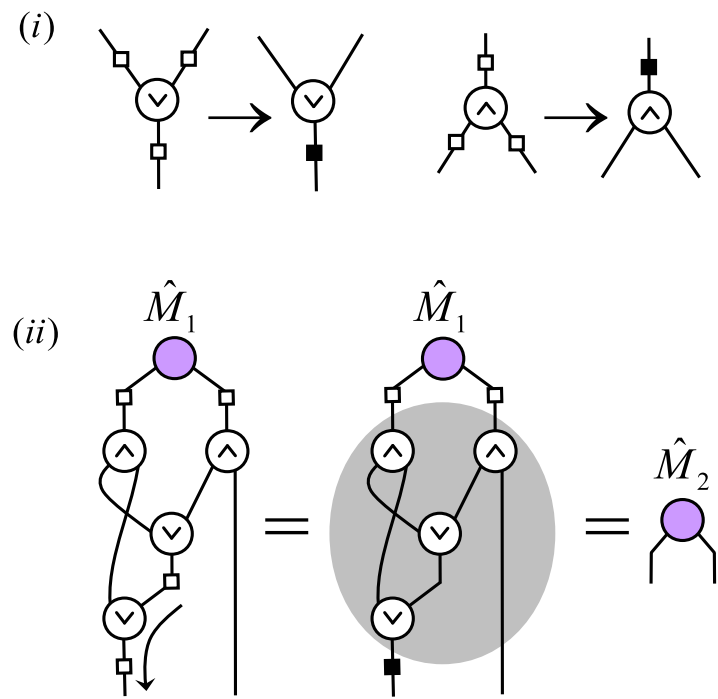}
\caption[Permutations applied to one or more legs of a fusion or splitting tensor can be replaced by an appropriate permutation on the coupled index. This process can be used to replace all permutations applied on internal indices of a diagram such as \protect{\fref{fig:ch3c:MtoM}} with net permutations on the indices of $\hat{M}_1$ and on the open indices of the network, as in shown in~(ii).]{(i)~Permutations applied to one or more legs of a fusion or splitting tensor can be replaced by an appropriate permutation on the coupled index. This process can be used to replace all permutations applied on internal indices of a diagram such as \protect{\fref{fig:ch3c:MtoM}} with net permutations on the indices of $\hat{M}_1$ and on the open indices of the network, as in shown in~(ii). The residual fusion and splitting operations, depicted as an arrow in a circle, simply perform the basic tensor product operation and its inverse, \eref{eq:ch3c:fuse}--\eref{eq:ch3c:split}, as described in \protect{\fref{fig:ch3c:u1fuse}}(iii) and \protect{\sref{apdx:ch3c:diag:fusesplit}}.\label{fig:ch3c:partial}}
\end{center}
\end{figure}

First, we decompose all the fusion and splitting tensors into two pieces in accordance with Fig. \ref{fig:ch3c:u1fuse}(iii). Next, we recognise that any permutations applied to one or more legs of a fusion or splitting tensor may always be written as a single permutation applied to the coupled index [\fref{fig:ch3c:partial}(i)]. 
We use this to replace all permutations on the intermediate indices of the diagram with equivalent permutations acting only on the indices of $\hat{M}_1$ and the open indices, as shown for a simple example in Fig. \ref{fig:ch3c:partial}(ii). The residual fusion and splitting operations, depicted by just a circle enclosing an arrow, then simply carry out fusion and splitting of indices as would be performed in the absence of symmetry \eref{eq:ch3c:fuse}-\eref{eq:ch3c:split}. These operations are typically far faster than their symmetric counterparts as they do not need to sort the entries of their output indices 
according to particle number.

In subsequent iterations, the matrix $\hat{M}_2$ is obtained from $\hat{M}_1$ by consecutively
\begin{enumerate}
\item permuting the rows and columns of $\hat{M}_1$ using the precomputed net permutations which act on the legs of $\hat M_1$; 
\item performing any elementary (non-symmetric) splitting, permuting of indices, and fusing operations, as described by the grey-shaded region in \fref{fig:ch3c:partial}(ii); 
\item permuting the rows and columns of the resulting matrix, using the precomputed net permutations which act on the open legs of \fref{fig:ch3c:partial}(ii).
\end{enumerate}
When matrix $\hat{M}_1$ is defined compactly, as in \eref{eq:ch3c:Tcanon}, so that elements which are identically zero by symmetry are not explicitly stored, 
a tensor $\hat T$ is constructed from multiple blocks identified by U(1) charge labels on their indices [$\hat T_{n_1n_2\ldots n_k}$ in \Eref{eq:ch3c:Tcanon}]. Under these conditions the elementary splitting, fusing, and permutation operations of step 2 above are applied to each individual block, but some additional computational overhead is incurred in determining the necessary rearrangements of these blocks arising out of the actions performed. This rearrangement may be computed on the fly, or may also be precomputed as a mapping between the arrangement of blocks in $\hat M_1$ and that in $\hat M_2$.

The memory required to store the precomputation data in this scheme is dominated by the size of the net permutations collected on the matrix indices, and is therefore of $O(\sqrt{|\hat{M}_1|})$. The overall cost of obtaining $\hat{M}_2$ from $\hat{M}_1$ is once again of $O(|\hat{M}_1|)$, but is in general higher than the previous scheme as this cost now involves two complete permutations of the matrix coefficients, as well as 
a reorganisation of the block structure of $\hat M_1$ which may possibly be computed at runtime.
Nevertheless, in situations where memory constraints are significant, partial precomputation schemes of this sort may be preferred.

\section{Supplement: Notes on the Implementation of Abelian Symmetries\label{sec:ch3:comp}}

As noted in \sref{sec:ch3a:symdc}, a symmetric tensor may be decomposed into a spin network and a collection of degeneracy tensors. This was seen again in Secs.~\ref{sec:ch3c:symmetry}--\ref{sec:ch3c:symTN} for the symmetry group U(1), with the spin network in this instance being trivial due to the Abelian nature of the group. For an Abelian symmetry, we may consequently understand this decomposition [\eref{eq:ch3c:Schur}, \eref{eq:ch3c:Tcanon}] as dividing a tensor into a number of blocks, labelled by the charges on each leg of the tensor, the majority of which are systematically zero. An example of this is seen in \Eref{eq:ch3c:ex2rev2} for a two-legged tensor, and is reproduced here with the different charge blocks highlighted and labelled:
\begin{equation}
\hat{T} = ~~
\begin{array}{cc|cccccc}
&&&\multicolumn{4}{c}{\mrm{charge}}\\
&&&0&\multicolumn{2}{c}{1}&2\\
\hline
\multirow{4}{*}{\begin{sideways}charge\end{sideways}}&0&\multirow{4}{*}{$\left(\begin{array}{c}\!\!\!\!\!\!\\ \!\!\!\!\!\!\\ \!\!\!\!\!\!\\ \!\!\!\!\!\!\end{array}\right.$} & \cellcolor[gray]{0.9}\alpha_0 & \cellcolor[gray]{0.7}0 & \cellcolor[gray]{0.7}0 & \cellcolor[gray]{0.9}0 & \multirow{4}{*}{$\left.\begin{array}{c}\!\!\!\!\!\!\\ \!\!\!\!\!\!\\ \!\!\!\!\!\!\\ \!\!\!\!\!\!\end{array}\right)$} \\
&\multirow{2}{*}{1}&& \cellcolor[gray]{0.7}0 & \cellcolor[gray]{0.9}\alpha_1 & \cellcolor[gray]{0.9}\beta_1 & \cellcolor[gray]{0.7}0 & \\
&&& \cellcolor[gray]{0.7}0 & \cellcolor[gray]{0.9}\gamma_1 & \cellcolor[gray]{0.9}\delta_1 & \cellcolor[gray]{0.7}0 & \\
&2&& \cellcolor[gray]{0.9}0 & \cellcolor[gray]{0.7}0 & \cellcolor[gray]{0.7}0 & \cellcolor[gray]{0.9}\alpha_2 &
\end{array}.
\end{equation}
For an Abelian tensor, this block decomposition may be trivially extended over an arbitrary number of legs, with each block being indexed by the associated charges on all legs \eref{eq:ch3c:tdec}, and this fact may be exploited in implementation of the symmetry.

\begin{figure}
\begin{center}
\includegraphics[width=425pt]{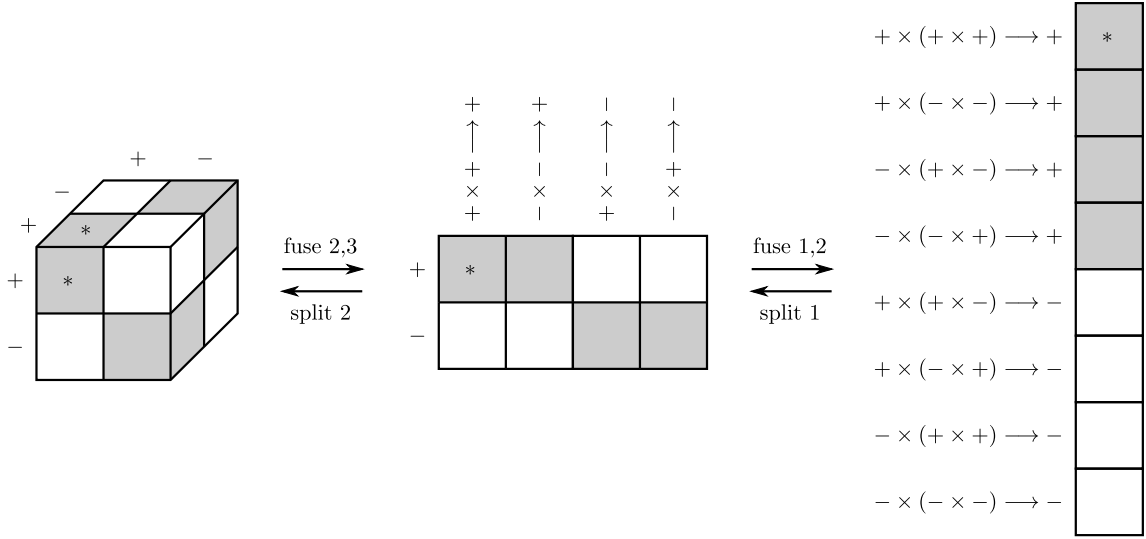}
\caption{A $Z_2$-symmetric tensor with three indices undergoes fusion, first of indices 2 and 3, then of indices 1 and 2 (as indicated by the upper arrows). This is achieved via a re-organisation of its block structure, and re-shaping of individual blocks. Blocks with non-trivial content in each diagram are shaded. We track the location of one individual block, marked $*$, and note that if in diagram~(i) it has dimension $x_1\times x_2\times x_3$, then in~(ii) it will have dimension $x_1\times(x_2x_3)$, and in~(iii), dimension $(x_1x_2x_3)\times 1$. The lower arrows indicate the reverse process.\label{fig:ch3c:blocks}}
\end{center}
\end{figure}

To illustrate how this is done, consider the fusing of two indices under the action of the simplest Abelian symmetry group, $Z_2$, and recall that fusing and splitting of indices may be thought of as consisting of two stages (see Secs.~\ref{sec:ch3c:symmetry:tp} and \ref{apdx:ch3c:diag:fusesplit}). In the first stage, the charges on the legs are combined according to a typical tensor product process, as per \Eref{eq:ch3c:fuse}, iterating rapidly over one charge and slowly over the other:
\begin{equation}
\begin{split}
+\times +&\longrightarrow +\\
+\times -&\longrightarrow -\\
-\times +&\longrightarrow -\\
-\times -&\longrightarrow +.
\end{split}
\end{equation}
In the second stage, the states of the tensor product space are re-ordered to collect like charges:
\begin{equation}
\begin{split}
+\times +&\longrightarrow +\\
-\times -&\longrightarrow +\\
+\times -&\longrightarrow -\\
-\times +&\longrightarrow -.
\end{split}\label{eq:ch3c:sortedfuse}
\end{equation}
Significantly, this entire process may be conducted at the level of blocks, and no mixing or re-ordering of the elements within individual blocks is required. If the entire tensor is maintained in the form of a sparsely-populated array of blocks in this manner, and blocks are \emph{not} concatenated on fusion, then both fusion and splitting may be performed without requiring the addressing of the contents of any individual blocks. This is illustrated in \fref{fig:ch3c:blocks}. [It is assumed either that any reshaping of individual blocks is deferred, in keeping with the philosophy of \sref{sec:ch3c:supplement}, or that an efficient representation of $n$-dimensional tensors employed---such as that used by \textsc{matlab}---for which reshape operations may be performed on the individual blocks at trivial cost independent of the size of the block.]
A similar treatment may be applied to all the primitives of set $\mc{P}$ (\sref{sec:ch3c:tensor:TN}).

Using this approach, the author was able to implement the $Z_2$-symmetric ternary 1D MERA with a demonstrable increase in performance over the standard MERA for bond dimensions $\chi>12$. Performance for Abelian symmetries with greater numbers of charge sectors, such as U(1), is anticipated to be even higher, effectively removing the need for the precomputation techniques discussed in \sref{sec:ch3c:supplement}. However, these techniques remain important for the exploitation of non-Abelian symmetries (\cref{sec:nonabelian} and \citealt{singhvidalinprep}) and anyons (\cref{sec:anyons} and \citealt{pfeifer2010}).

One note of caution: When combining indices using a symmetry-preserving ordering such as that given in \Eref{eq:ch3c:sortedfuse}, care must be taken to ensure that fusion is implemented in an associative manner, i.e. $d\equiv((a\times b)\times c)$ results in the same ordering of entries on index $d$ as the fusion $d\equiv(a\times(b\times c))$, if consistent results are to be obtained.

\subsection{Fermions\label{sec:ch3c:fermions}}

My initial development of efficient $Z_2$ symmetry algorithms was performed in parallel with Philippe Corboz, who adopted a similar computational strategy. To avoid inappropriate duplication of research efforts, it was decided that Philippe would then proceed to incorporate fermionic exchange behaviour, while I would instead study the extension of this technique to other symmetry groups, resulting in the present work on Abelian symmetries, non-Abelian symmetries, and anyons. The principle by which fermionic exchange statistics may be incorporated into this scheme is, however, easily understood as follows:

Recall that we have chosen to enumerate the legs of a tensor counterclockwise from the 9~o'clock position (\sref{sec:ch3c:tensor:tensor}). When a tensor network diagram involves crossings of pairs of legs, these crossings may be understood as a permutation operation which exchanges the ordering of the indices on the tensor. However, for fermionic statistics, we have a $Z_2$ symmetry and thus for any given index, each block in the tensor will have a charge label, $+$ or $-$, associated with that index. The permutation operation corresponding to index exchange is performed independently on each block in the normal fashion for a $Z_2$-symmetric tensor, subject to one simple modification: When the charge labels for a given block are $-$ on both indices, in addition to the permutation operation, the block is multiplied by $-1$. This is the essence of the ``swap gate'' formalism introduced by \citet{corboz2010a}---see also %
\citet{corboz2009a,pineda2010,barthel2009,pizorn2010}.

I note that even greater efficiency may be attained not only by writing each tensor as a grid of blocks, but also by associating with each block a numeric multiplier, initially 1. To multiply a block by $-1$, it now suffices to multiply the corresponding numeric factor by $-1$. In this manner the cost of the swap gate is reduced, requiring only one operation to change the sign of an entire block. Of course, this minus sign must eventually be applied in order to obtain numerical results, but if many swap gates are applied to a tensor before these numerical results are computed, then deferring (and combining) their evaluation in this manner may result in a significant saving in calculation time. A generalised version of this technique is also used to reduce computational cost in the study of anyonic systems, as described in \sref{sec:ch4:implementation}

\begin{figure}
\begin{center}
\includegraphics[width=300.0pt]{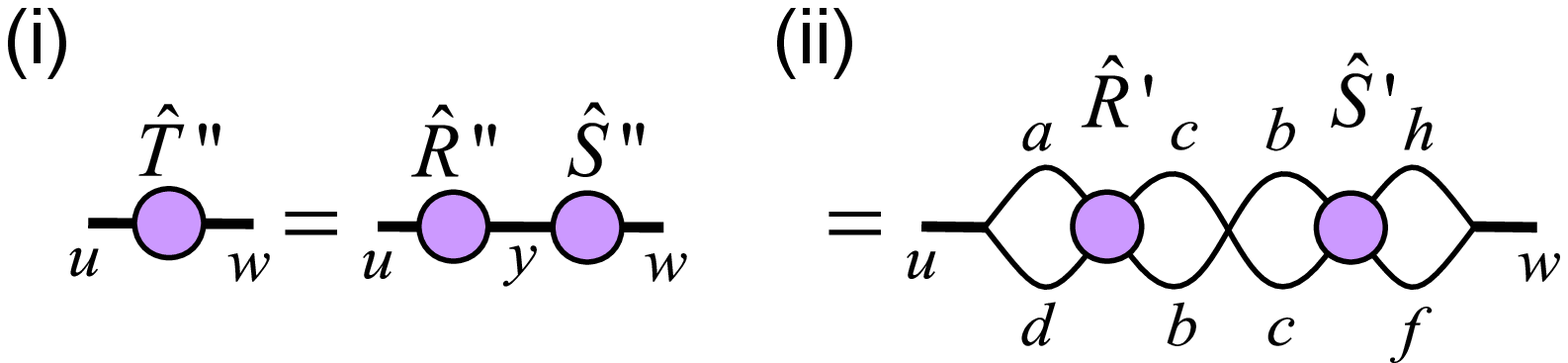}
\caption[For an Abelian group symmetry, the contraction of tensors $\hat R$ and $\hat S$ illustrated in \protect{\fref{fig:ch3c:multiply2}} is reduced to the matrix multiplication $\hat T''=\hat R''\times \hat S''$. However, there is an implicit index exchange present in this multiplication, and for fermions this will introduce additional factors of $-1$.]{For an Abelian group symmetry, the contraction of tensors $\hat R$ and $\hat S$ illustrated in \protect{\fref{fig:ch3c:multiply2}} is reduced to the matrix multiplication $\hat T''=\hat R''\times \hat S''$, which may be represented as shown in diagram~(i), above [or diagram~(3) of \protect{\fref{fig:ch3c:multiply2}}]. However, there is an implicit index exchange present in this multiplication, shown explicitly in~(ii), and for fermions this will introduce additional factors of $-1$.\label{fig:ch3c:modifiedmultiply2}}
\end{center}
\end{figure}

One caution is required when using the methods of \sref{sec:ch3:U1MERA} for the simulation of fermions. Note that when contracting a pair of tensors as per \sref{sec:ch3c:tensor:multiply}, the counterclockwise ordering of the indices to be contracted must be the same on each tensor. There will frequently be index permutations required to set up this ordering, which will introduce some exchange factors of $-1$. However, there are also exchange factors associated with the contraction itself, as seen in e.g. \fref{fig:ch3c:modifiedmultiply2},
and these too must be taken into account. For an alternative approach where exchange factors are associated only with the re-ordering of indices, and there are no such hidden factors associated with the contraction of two tensors, see the discussion in \sref{sec:ch5:fermions2}.

\chapter{Anyonic Tensor Networks\label{sec:anyons}}

Having developed a formalism for the exploitation of internal symmetries in tensor networks and exploited it for Abelian symmetry groups, it was natural to seek to apply this formalism to more general classes of physical systems. As a colleague (Sukhwinder Singh) was already working on the implementation of SU(2) symmetries for the MPS and MERA, 
with obvious extension to spin systems having
any non-Abelian internal symmetry group, it was decided that I would instead study whether our approach could be extended to permit the study of systems of anyons. 

Anyonic systems do not in general exhibit an internal symmetry group; instead their behaviours and statistics are typically described by a more general structure known as a Unitary Braided Tensor Category (UBTC). \nomenclature{\tbf{UBTC}}{Unitary Braided Tensor Category: The mathematical structure used to describe a system of anyons (see \protect{\cref{sec:anyons}}). There is also a UBTC associated with any symmetry group (see \protect{\cref{sec:nonabelian}}).}
To specify an anyon model using a UBTC, one must declare a set of charge labels (including a vacuum charge), fusion rules describing how they combine (which must be associative), a set of basis transformations known as $F$ moves, and a tensor $R^{ab}_c$ which describes the exchange statistics of the charges. These properties will all be described in detail in \sref{sec:ch4:anyonstates}. For any group $\mc{G}$, it is possible to construct an associated UBTC where the $F$ moves are derived from the 6-$j$ symbols, and the $R^{ab}_c$ tensor reflects the universal braid matrix for the irreps of the group. In fact, this UBTC provides a complete description of the group at the level of representations, up to (but not including) explicit construction of representations of the irreps themselves (see \cref{sec:nonabelian}). However, although every group may be associated with a UBTC, the converse does not hold, and UBTCs may also be used to describe more general structures such as quantum doubles [e.g. D(D$_3$)], and quantum groups [e.g. $q$-deformed SU(2), or SU(2)$_k$]. It is the UBTCs associated with these more general structures which may be used to describe systems with anyonic statistics.

Because of the close association between groups and UBTCs, it was tempting to ask whether our internal symmetry formalism, which could be applied to quantum systems with a group-based mathematical structure, could be extended to address systems based on any UBTC. Of particular interest are the UBTCs associated with quantum groups, as these are associated with many interesting anyon models such as SU(2)$_3$, for which the integer subalgebra describes a class of anyons known as ``Fibonacci anyons'', capable of supporting universal quantum computation simply through particle exchange. However, many of the charges in these models may not be associated with explicit matrix representations, as in many cases the fusion rules imply that the dimensions of these irreps must be non-integer, or even irrational. Consequently the problem could not be viewed as one of decomposing a known system into charge sectors as per \cref{sec:abelian}, but instead had to be formulated directly in the graphical language of UBTCs which constitutes a natural description of an anyonic system.

In this Chapter, I describe a formalism permitting exactly this, whereby any tensor network Ansatz or algorithm may be constructed for a system of anyons, or indeed for any other model describable in terms of a UBTC. This Chapter will only concern itself explicitly with systems of anyons on the disc, though a subsequent treatment of anyons on surfaces of higher genus is planned.

(The same formalism may even be applied to 1D systems described by a Unitary Tensor Category without including a notion of particle exchange, though the author is as yet unaware of any physical systems of interest which take such a form.)

\begin{center}
\rule{0.75\linewidth}{0.3mm}
\end{center}

Sections~\ref{sec:ch4:intro}--\ref{sec:ch4:summary} of this Chapter have previously been published as \citeauthor*{pfeifer2010}, \emph{Physical Review B}, \textbf{82}, 115126, 2010, \copyright~(2010) by the American Physical Society.

\clearpage

\section{Introduction\label{sec:ch4:intro}}

The study of anyons offers one of the most exciting challenges in contemporary physics. Anyons are exotic quasiparticles with non-trivial exchange statistics, which makes them difficult to simulate. However, they are of great interest as some species offer the prospect of a highly fault-tolerant form of universal quantum computation\citecomma{kitaev2003,nayak2008} and it has been suggested\prbtext{\citep{xia2004}} that the simplest such species may appear in the fractional quantum Hall state with filling fraction $\nu$ = 12/5\pratext{ \citep{xia2004}}. Despite the current strong interest in the development of practical quantum computing, our ability to study the collective behaviour of systems of anyons remains limited.

The study of interacting systems of anyons using numerical techniques was pioneered by \textcitecomma{feiguin2007} using exact diagonalisation for {1D} systems of up to 37 anyons, and the Density Matrix Renormalisation Group algorithm (DMRG)\pratext{ }\citep{white1992} for longer chains. Also related is work by \textcitecomma{sierra1997} later extended by \textcitecomma{tatsuaki2000} which applies a variant of DMRG to spin chain models having $SU(2)_k$ symmetry. Some of these models are now known to correspond to $SU(2)_k$ anyon chains\citecomma{trebst2008} and using this mapping these systems may also be studied using the Bethe Ansatz\pratext{ }\citep{alcaraz1987} and quantum Monte Carlo\citestop{todo2001}

However, all of these methods have their limitations. Exact diagonalisation has a computational cost which is exponential in the number of sites, strongly limiting the size of the systems which may be studied. DMRG is capable of studying larger system sizes, but is typically limited to {1D} or quasi-{1D} systems (e.g. ladders). Mapping to a spin chain is useful in one dimension but is substantially less practical in two.
There are therefore good reasons to desire
a formalism which will allow the application of other tensor network algorithms %
to systems of anyons. 
Many of these tensor networks, such as 
Projected Entangled Pair States (PEPS)\citecomma{verstraete2004,nishino1998,gu2008,xie2009,jordan2008}
and the 2D versions of Tree Tensor Networks (TTN)\pratext{ }\citep{tagliacozzo2009} and of the Multi-scale Entanglement Renormalisation Ansatz (MERA)\pratext{ }\citep{cincio2008,evenbly2009b,evenbly2010}
have been designed specifically to accurately describe two-dimensional systems. 

In one dimension, 
many previously studied systems of interacting anyons display extended critical phases\pratext{ }\prbtext{,}\cite[e.g.][]{feiguin2007,trebst2008}\pratext{,} which are characterised by correlators exhibiting polynomial decay\citestop{difrancesco1997} 
Whereas DMRG favours accurate representation of short range correlators at the expense of long-range accuracy, 
the {1D} MERA\pratext{ }\citep{vidal2007,vidal2008a} is ideally suited to this situation as its hierarchical structure naturally encodes the renormalisation group flow at the level of operators and wavefunctions\citecomma{vidal2007,vidal2008a,vidal2010,chen2010} and hence %
accurately reproduces correlators across a wide range of length scales\citestop{vidal2007,vidal2008a,giovannetti2008,pfeifer2009,evenbly2009} The development of a general formalism for anyonic tensor networks is therefore also advantageous for the study of {1D} anyonic systems.

This Chapter describes how any tensor network algorithm may be adapted to systems of anyons in one or two dimensions using structures which explicitly implement the quantum group symmetry of the anyon model. As a specific example I demonstrate the construction of the anyonic {1D} MERA, which I then apply to an infinite chain of interacting Fibonacci anyons at criticality.
The approach which I present is %
completely general, and can be applied to any species of anyons and any tensor network Ansatz.

\section{Anyonic States\label{sec:ch4:anyonstates}}

\begin{figure}
\begin{center}
\includegraphics[width=300.0pt]{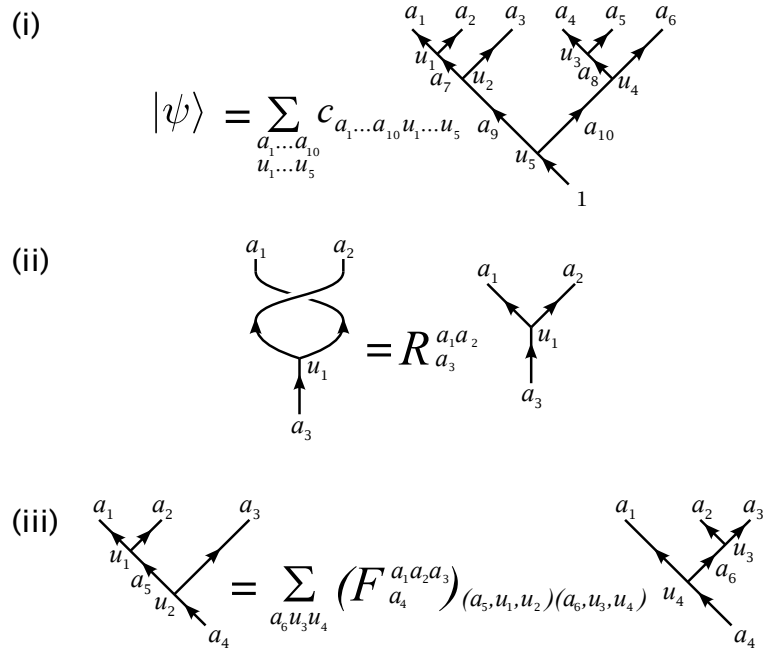}
\caption[(i)~Example representation of a state $|\psi\ra$ in a fusion tree basis for a system of $6$ anyons. (ii)~Braiding may be used to change the ordering of the leaves of a fusion tree basis, or to represent anyon exchange. (iii)~$F$-moves convert between the bases associated with different fusion trees.]{(i)~Example representation of a state $|\psi\ra$ in a fusion tree basis for a system of $6$ anyons. Labels $a_i$ indicate charges associated with edges of the fusion tree graph, and labels $u_i$ are degeneracies associated with vertices. The structure of the tree corresponds to a choice of basis, and does not affect the physical content of the theory.
(ii)~Braiding may be used to change the ordering of the leaves of a fusion tree basis, or to represent anyon exchange.
(iii)~$F$-moves convert between the bases associated with different fusion trees.
\label{fig:ch4:anyons}}
\end{center}
\end{figure}

Consider a lattice $\mc{L}_0$ of $n$ sites populated by anyons. In contrast to bosonic and fermionic systems, for many anyon models the total Hilbert space $\mbb{V}_{\mc{L}_0}$ can not be divided into a tensor product of local Hilbert spaces. Instead, a basis is defined by introducing a specific fusion tree [e.g. \fref{fig:ch4:anyons}(i)]. The fusion tree is always constructed on a linear ordering of %
anyons, and while the {1D} lattice naturally exhibits such an ordering, for 2D lattices some linear ordering %
must be imposed.
Each line is then labelled with a charge index $a_i$ such that the labels are consistent with the fusion rules of the anyon model,
\begin{equation}
a\times b\rightarrow \sum_{c} N_{ab}^{c} \,c.\label{eq:ch4:fusion}
\end{equation}
For anyon types where some entries of the multiplicity tensor $N_{ab}^c$ take values greater than 1, a label $u_i$ is also affixed to the vertex which represents the fusion process to distinguish between the different copies of charge $c$.
The edges of the graph which are connected to a vertex only at their lower end are termed ``leaves'' of the fusion tree, and we will associate these leaves with the charge labels $a_1\ldots a_n$. Different orderings of the leaves on a fusion tree may be interconverted by means of braiding [\fref{fig:ch4:anyons}(ii)], and different fusion trees, corresponding to different bases of states, may be interconverted by means of $F$ moves [\fref{fig:ch4:anyons}(iii)]\citestop{kitaev2006,bonderson2007}
In some situations it may also be useful to associate a further index $b_i$ %
with each of the leaves of the fusion tree. For example, if the leaves are equated with the sites of a physical lattice, then this additional index may be used to enumerate additional non-anyonic degrees of freedom associated with that lattice. %
For simplicity we will usually leave %
these extra indices $b_1\ldots b_n$ implicit, as we have done in \fref{fig:ch4:anyons}, 
as they do not directly participate in anyonic manipulations such as $F$ moves and braiding.

Let the total number of charge labels on the fusion tree be given by $m$, where $m\geq n$. 
For Abelian anyons the fusion rules uniquely constrain all $a_i$ for $i>n$, and provided there are no constraints on the total charge, the total Hilbert space reduces to a product of local Hilbert spaces $\mbb{V}$, such that $\mbb{V}_{\mc{L}_0} = \mbb{V}^{\otimes n}$. For non-Abelian anyons, additional degrees of freedom arise because some fusion rules admit multiple outcomes, permitting certain $a_i\ (i>n)$ to take on multiple values while remaining consistent with the fusion rules, and the resulting Hilbert space does not necessarily admit a tensor product structure.

We will now associate a parameter $\nu_{i,a_i}$ with each charge on the fusion tree, which we will term the degeneracy. This parameter corresponds to the number of possible fusion processes by which charge $a_i$ may be obtained at location $i$. Where charge $a_k$ arises from the fusion of charges $a_i$ and $a_j$, then $\nu_{k,a_k}$ will satisfy
\begin{equation}
\nu_{k,a_k} = \sum_{a_i,a_j} \nu_{i,a_i}\nu_{j,a_j} N^{a_k}_{a_i a_j}.\label{eq:ch4:compounddegens}
\end{equation}
For systems where the only degrees of freedom are anyonic, degeneracies on the physical lattice $\mc{L}_0$ (i.e. $\nu_{i,a_i}$, $1\leq i\leq n$) will take values of 0 or 1 depending on whether a charge $a_i$ is permitted on lattice site $i$. Higher values of $\nu_{i,a_i}$ may be used on the physical lattice 
if there is also a need to represent additional 
non-anyonic degrees of freedom, %
enumerated by indices $b_1\ldots b_n$.

\begin{figure}
\begin{center}
\includegraphics[width=300.0pt]{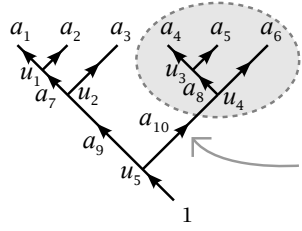}
\caption[A fusion tree whose leaves carry the charge labels
$a_1$ to $a_6$. An edge which is not a leaf, labelled with charge $a_{10}$, is indicated by a large grey arrow. The portion of the fusion tree extending from edge $a_{10}$ out to the leaves
is indicated by the grey ellipse. If a degeneracy index $\mu_{10}$ is associated with charge $a_{10}$, then for a given value of $a_{10}$, index $\mu_{10}$ will enumerate all compatible labellings of the highlighted portion of the fusion tree.]{%
The leaves of this fusion tree carry the charge labels
$a_1$ to $a_6$. %
An edge which is not a leaf, labelled with charge $a_{10}$, is indicated by the large grey arrow. 
The portion of the fusion tree extending from edge $a_{10}$ out to the leaves
is indicated by the grey ellipse. If a degeneracy index $\mu_{10}$ is associated with charge $a_{10}$, then for a given value of $a_{10}$, index $\mu_{10}$ will 
enumerate all compatible labellings of the highlighted portion of the fusion tree.\label{fig:ch4:exampleedges}}
\end{center}
\end{figure}

\begin{figure}
\begin{center}
\includegraphics[width=300.0pt]{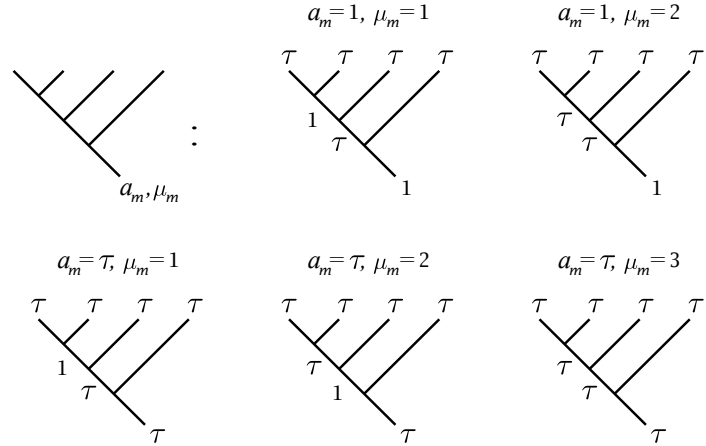}
\caption[Example enumeration of states according to $a_m$ and $\mu_m$ for a fusion tree describing four Fibonacci anyons.]{Example enumeration of states according to $a_m$ and $\mu_m$ for a fusion tree describing %
four Fibonacci anyons. The Fibonacci anyon model has one non-vacuum charge label ($\tau$) and one non-trivial fusion rule, $\tau\times\tau\rightarrow 1+\tau$. Because the charges $1$ and $\tau$ are both self-dual, no arrows are required on diagrammatic representations of Fibonacci anyon fusion trees.\label{fig:ch4:exampleenum}}
\end{center}
\end{figure}

Up to this point we have parameterised our Hilbert space in terms of explicit labellings of the fusion tree. We now adopt a different approach: Consider an edge $i$ of the fusion tree which is not a ``leaf''. %
As well as labelling this edge with a charge $a_i$ we may introduce a second index $\mu_i$, running from 1 to $\nu_{i,a_i}$. Each pair of values $\{a_i,\mu_i\}$ may be associated with a unique charge labelling for the portion of the fusion tree 
from edge $i$ out to the leaves,
with these labellings being compatible with the fusion rules in the presence of a charge of $a_i$ on site $i$ (for an illustration of this, see \fref{fig:ch4:exampleedges}). 
Provided we know the structure of the fusion tree above $i$ and have a systematic means of associating labellings of that portion of the tree with values of $\mu_i$, then in lieu of stating the values of all $a_j$ for edges $j$ involved in that portion of the tree, we may simply specify the value of the degeneracy index $\mu_i$. 
In this way we may specify an entire state in the form
\begin{equation}
|\psi\ra = \sum_{\mu_{m}} c_{a_{m}\mu_{m}} |a_{m},\mu_{m}\ra\label{eq:ch4:statepsi_pre}
\end{equation}
where $a_{m}$ is the total charge obtained on fusing all the anyons. %
The index $\mu_{m}$, which is the degeneracy index associated with the total charge of the fusion tree, may be understood as systematically enumerating all possible labellings of the entire fusion tree including charge labels, vertex labels, and any labels associated with additional non-anyonic degrees of freedom. %
For an example, see \fref{fig:ch4:exampleenum}.
Note that for a given edge $i$, the value of the degeneracy $\nu_{i,a_i}$ may vary with the charge $a_i$ and consequently the range of the degeneracy index $\mu_{m}$ in Eq.~\eref{eq:ch4:statepsi_pre} is dependent on the value of the charge $a_{m}$.

The notation of Eq.~\eref{eq:ch4:statepsi_pre} should be contrasted with that of
\fref{fig:ch4:anyons}(i). In the latter, the number of indices on $c$ depends upon the number of charge labels on the fusion tree, whereas in the former, the tensor describing the state is always indexed by just one pair of labels---charge and degeneracy---which will prove advantageous in constructing a tensor network formalism for systems of anyons.

We now choose to restrict our attention to systems having the identity charge. We may do this without loss of generality as a state on $n$ lattice sites with a total charge $a_{m}$ may always be equivalently represented by a state on $n+1$ lattice sites whose total charge is the identity, with a charge $\overline{a_{m}}$ on lattice site $n+1$. This additional charge annihilates the total charge $a_{m}$ of sites $1\ldots n$ to give the vacuum. The expression for $|\psi\ra$ then becomes
\begin{equation}
|\psi\ra = \sum_{\mu_{m'}} c_{1\mu_{m'}} |1,\mu_{m'}\ra\label{eq:ch4:statepsi}
\end{equation}
where $\mu_{m'}$ ranges from 1 to the dimension of the Hilbert space of the system of $n$ sites with total charge $a_m$.
Consequently we may represent the state $|\psi\ra$ of a system of anyons by means of the vector $c_{1\mu_{m'}}$. For simplicity of notation, we will take greek indices from the beginning of the alphabet to correspond to pairs of indices $\{a_i,\mu_i\}$ consisting of a charge index and the associated degeneracy index. The vector $c_{1\mu_{m'}}$ will therefore be denoted simply $c^\alpha$, with the understanding that in this case 
the charge component $a_{m'}$ of multi-index $\alpha$ takes 
only
the value 1. (Multi-index $\alpha$ is raised as we will shortly introduce a diagrammatic formalism in which vector $c$ is represented by an object with a single upward-going leg. In this formalism, upward- and downward-going legs may be associated with upper and lower multi-indices respectively.)

\section{Anyonic Operators}

We will divide our consideration of anyonic operators into two parts. First we shall consider operators which map a state on some Hilbert space $\mc{H}$ into another state on the same Hilbert space. When applied to a state represented by $c^\alpha$, such an operator leaves the degeneracies of the charges in multi-index $\alpha$ unchanged. We will therefore call these
\emph{degeneracy-preserving} anyonic operators.
Then we will consider those operators which map a state on some Hilbert space $\mc{H}$ into a state on some other Hilbert space $\mc{H}'$. These operators may represent processes which modify the environment, for example by adding or removing lattice sites, and also play an important part in anyonic tensor networks, for instance taking the role of isometries in the TTN and MERA. 
As these operators can change the degeneracies of charges in a multi-index $\alpha$, we will call them \emph{degeneracy-changing} anyonic operators.
More generally, the degeneracy-preserving anyonic operators may be considered a subclass of the degeneracy-changing anyonic operators for which $\mc{H}=\mc{H}'$.

\subsection{Degeneracy-Preserving Anyonic Operators\label{sec:ch4:anyonops}}

We begin with those operators which map
states on some Hilbert space $\mc{H}$ into other states on the same Hilbert space $\mc{H}$. 
Examples of these operators include Hamiltonians, reduced density matrices, and unitary transformations such as the disentanglers of the MERA.

\begin{figure}
\begin{center}
\includegraphics[width=300.0pt]{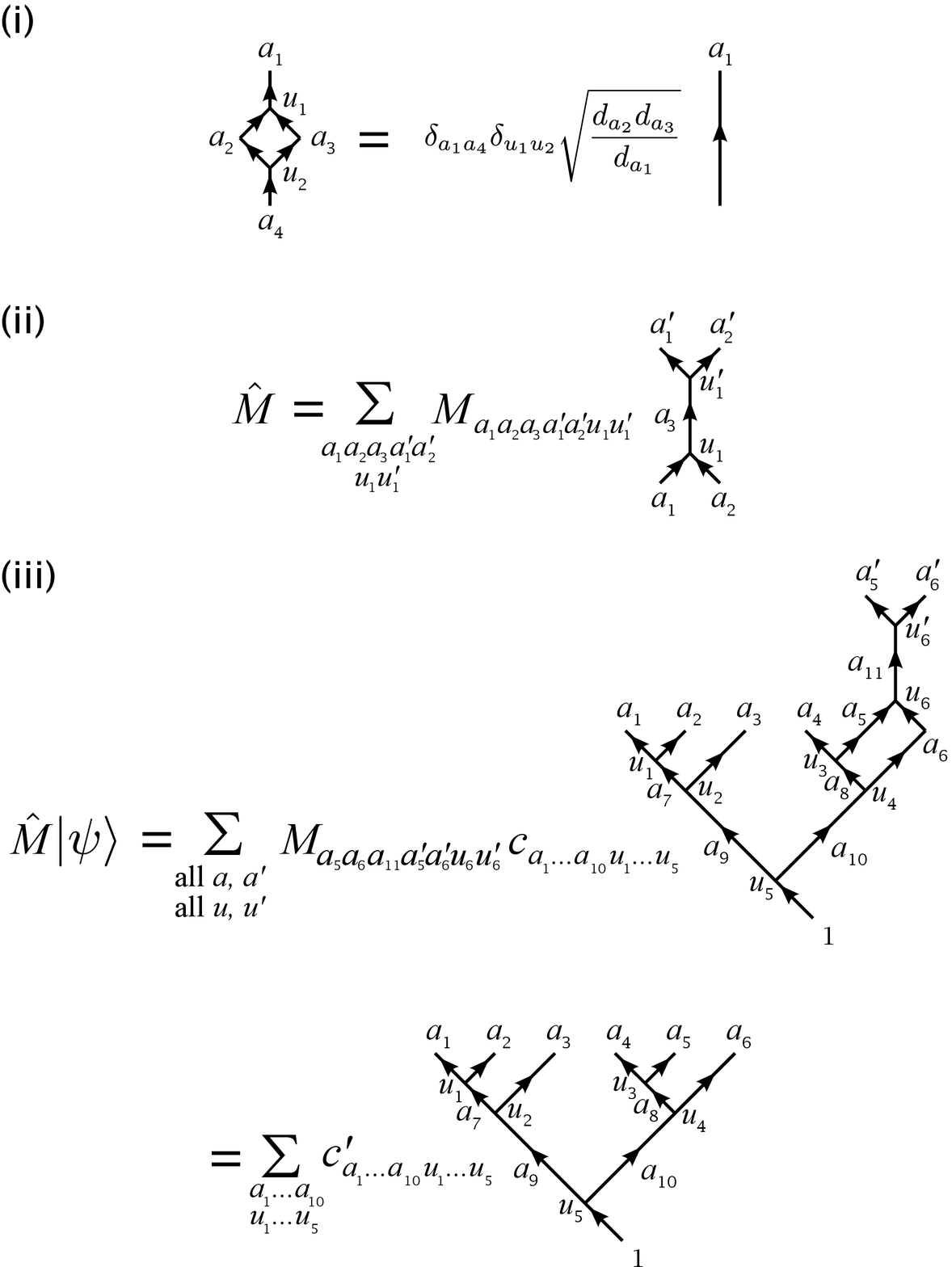}
\caption[(i)~Loops are eliminated from diagrams by replacing them with an equivalent numerical factor determined by the normalisation convention. (ii)~Definition of a simple two-site anyonic operator. (iii)~Application of an operator to a state is performed by connecting the diagrams' free legs, performing $F$ moves, and eliminating loops (and in more complex examples, also braiding).]{
(i)~Loops are eliminated by replacing them 
with an equivalent numerical factor determined by the normalisation convention. 
The factor given here corresponds
to the diagrammatic isotopy convention employed in \protect{\rcite{bonderson2007}}.
(ii)~Definition of a simple two-site anyonic operator. 
(iii)~Application of an operator to a state is performed by connecting the diagrams' free legs. By performing $F$ moves and eliminating loops (and in more complex examples, also braiding) it is possible to obtain an expression for the resulting state in the original basis. 
\label{fig:ch4:operators}}
\end{center}
\end{figure}

First, we introduce splitting trees. The space of splitting trees is dual to the space of fusion trees. While the space of fusion trees consists of labelled directed graphs whose number of branches increases monotonically when read from bottom to top, the space of splitting trees consists of labelled directed graphs whose number of branches increases monotonically when read from top to bottom. 
An inner product is defined by connecting the leaves of fusion and splitting trees which have equivalent linear orderings of the leaves (%
braiding first %
if necessary), 
then eliminating all loops as per \fref{fig:ch4:operators}(i), with $F$ moves performed as required. %

Anyonic operators
may always be written as a sum over fusion and splitting trees, such as the two-site operator $\hat M$
shown in \fref{fig:ch4:operators}(ii), and for degeneracy-preserving anyonic operators it is %
always possible to choose the splitting tree to be the adjoint of the fusion tree.
To apply an operator %
to a state the two corresponding diagrammatic representations are connected as shown in \fref{fig:ch4:operators}(iii), and closed loops may be eliminated as shown in \fref{fig:ch4:operators}(i). 
Sequences of $F$ moves, braiding, and loop eliminations may be performed until the diagram has been reduced once more to a fusion tree without loops on a lattice of $n$ sites.

Much as the state of an anyonic system may be represented by a vector $c^\alpha$, anyonic operators may be represented by a matrix $\Mop$. Each value of $\alpha$ corresponds to a pair $\{a_i,\mu_i\}$ where $a_i$ is a possible charge of the central edge of the operator diagram [e.g. $a_3$ in \fref{fig:ch4:operators}(ii)], and $\mu_i$ is a value of the degeneracy index associated with charge $a_i$. We will denote the degeneracy of $a_i$ by $\nu_{a_i}$. Similarly, values of $\beta$ correspond to pairs $\{a_j,\mu_j\}$ where $a_j$ has degeneracy $\nu_{a_j}$.
For degeneracy-preserving anyonic operators the charge indices
$a_i$ and $a_j$ necessarily take on the same range of values, and $\nu_{a_i}=\nu_{a_j}$ when $a_i=a_j$. The values of $\nu_{a_i}$ may equivalently be calculated from either the fusion tree making up the top half or the splitting tree making up the bottom half of the operator diagram.

\begin{figure}
\begin{center}
\includegraphics[width=300.0pt]{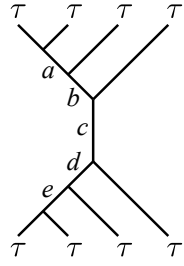}
\caption{An operator acting on four Fibonacci anyons. The values of the coefficients $M_{abcde}$ may be specified as a block-diagonal matrix $M_\alpha^{~\beta}$, for example as in \protect{\tref{tab:ch4:examplematrix}}.\label{fig:ch4:exampleoperator}}
\end{center}
\end{figure}

A well-defined anyonic operator $\hat M$ must respect the (quantum) symmetry group of the anyon model, and consequently all entries in $\Mop$ for which $a_i\not=a_j$ will be zero. However, in contrast with $c^\alpha$ we do not require that $a_i=a_j=1$. When $\hat M$ is a degeneracy-preserving operator, matrix $\Mop$ is therefore a square matrix of side length
\begin{equation}
\ell_M=\sum_{a_i} \nu_{a_i},
\end{equation}
which may be organised to exhibit a structure which is block diagonal in the charge indices $a_i$ and $a_j$, and for which the blocks are also square. 
As an example consider \fref{fig:ch4:exampleoperator}, which shows an operator acting on four Fibonacci anyons. An example matrix $\Mop$ for an operator of this form is given in \tref{tab:ch4:examplematrix}, from which the entries of $M_{abcde}$ can be reconstructed, e.g. $M_{\tau 1 \tau 1 \tau}=3$.

\begin{table}
\begin{center}
\begin{equation*}
\Mop=\quad\begin{array}{cc|ccccccc}
&&&\multicolumn{5}{c}{a_j,\mu_j}&\\
&&&1,1&1,2&\tau,1&\tau,2&\tau,3&\\
\hline
\rule{0pt}{3.1ex}\multirow{5}{*}{$a_i$,$\mu_i$}&1,1&
\multirow{5}{*}{$\left(\begin{array}{c}\!\\ \!\\ \!\\ \!\\ \!\end{array}\right.$}& 1&0.5&0&0&0 &
\multirow{5}{*}{$\left.\begin{array}{c}\!\\ \!\\ \!\\ \!\\ \!\end{array}\right)$}\\
&1,2&& 0.5&1&0&0&0 &\\
&\tau,1&& 0&0&1&2&-1 &\\
&\tau,2&& 0&0&2&3&-1 &\\
&\tau,3&& 0&0&1&1&1 &
\end{array}
\end{equation*}
~

\rule{0pt}{13ex}
\begin{tabular}{|cc|ccccc|}
\bottomrule
$a_i$&$\mu_i$&~&$a$&$b$&$c$&\\
\hline
1&1&&1&$\tau$&1&\\
1&2&&$\tau$&$\tau$&1&\\
$\tau$&1&&1&$\tau$&$\tau$&\\
$\tau$&2&&$\tau$&1&$\tau$&\\
$\tau$&3&&$\tau$&$\tau$&$\tau$&\\
\toprule
\end{tabular}
~~~~
\begin{tabular}{|cc|ccccc|}
\bottomrule
$a_j$&$\mu_j$&~&$e$&$d$&$c$&\\
\hline
1&1&&1&$\tau$&1&\\
1&2&&$\tau$&$\tau$&1&\\
$\tau$&1&&1&$\tau$&$\tau$&\\
$\tau$&2&&$\tau$&1&$\tau$&\\
$\tau$&3&&$\tau$&$\tau$&$\tau$&\\
\toprule
\end{tabular}
\caption[Matrix representation $M_\alpha^{~\beta}$ for an example operator of the form shown in \protect{\fref{fig:ch4:exampleoperator}}.]{Matrix representation $M_\alpha^{~\beta}$ for an example operator of the form shown in \protect{\fref{fig:ch4:exampleoperator}}. Multi-index $\alpha$ corresponds to index pair $\{a_i,\mu_i\}$ and multi-index $\beta$ corresponds to pair $\{a_j,\mu_j\}$. Subject to an appropriate ordering convention for $\mu_i$ and $\mu_j$, these indices may be related to the fusion tree labels $a,b,c,d,e$ of \protect{\fref{fig:ch4:exampleoperator}} as shown. Note that as $c$ is the charge on the central leg of \protect{\fref{fig:ch4:exampleoperator}}, all nonzero entries of $M_\alpha^{~\beta}$ satisfy $a_i=a_j=c$.\label{tab:ch4:examplematrix}}
\end{center}
\end{table}

\subsection{Degeneracy-Changing Anyonic Operators\label{sec:ch4:degenexp}}

We now introduce the second class of anyonic operators, which map states in some Hilbert space $\mc{H}$ into some other Hilbert space $\mc{H}'$. These operators may reduce or increase the degeneracy of any charge present in the spaces on which they act, and may even project out entire charge sectors by setting their degeneracy to zero.
When these operators are written in the conventional notation of \fref{fig:ch4:operators}, the fusion and splitting trees will not be identical. Further, we may choose to allow combinations of degeneracies which do not naturally admit complete decomposition into individual anyons. For example, a degeneracy-changing operator may map a state on five Fibonacci anyons (having total degeneracies $\nu_1=3$, $\nu_\tau=5$) into a state having degeneracies $\nu_1=2$, $\nu_\tau=2$. As these degeneracies do not admit decomposition into an integer number of nondegenerate anyons, it is necessary %
to associate an index $u_i$ with the single open leg of the fusion tree. This index behaves identically to the vertex indices $u_i$ of \fref{fig:ch4:anyons}, serving to enumerate the different copies of each individual charge, and as with the vertex indices of \fref{fig:ch4:anyons}, it is absorbed into the degeneracy index $\mu_i$.

As a further example, a state having degeneracies $\nu_1=4$, $\nu_\tau=4$ could be associated with a fusion tree having either one leg, or two legs each with degeneracies $\nu_1=0$, $\nu_\tau=2$. Again, indices $u_i$ would have to be associated with each open leg.

Matrix representations of degeneracy-changing anyonic operators may also %
be constructed, and when they %
are written in block diagonal form, the matrices and their blocks may be rectangular rather than square. Degeneracy-changing anyonic operators therefore 
represent a generalisation of the degeneracy-preserving anyonic operators discussed in \sref{sec:ch4:anyonops}. It is worth noting that the presence of indices $u_i$ on the open legs of the fusion or splitting trees of an operator do not automatically imply that it is a degeneracy-changing anyonic operator: The defining characteristic of a degeneracy-preserving anyonic operator is that it maps a state in a Hilbert space $\mc{H}$ into a state in the same Hilbert space $\mc{H}$, and consequently both the matrix as a whole and all of its blocks are square. Thus a degeneracy-preserving anyonic operator may act on states having additional indices $u_i$ on their open legs, and the resulting state may be expressed in the form of the same fusion tree, with the same additional indices on the open legs.

Operators which change degeneracies may %
represent physical processes which change the accessible Hilbert space of a system. %
As we will see in \sref{sec:ch4:MERAconstr}, 
they may also be used in tensor network algorithms as part of an efficient representation of particular states or subspaces of a Hilbert space, for example the ground state or the low energy sector of a local Hamiltonian.

\begin{figure}
\begin{center}
\includegraphics[width=300.0pt]{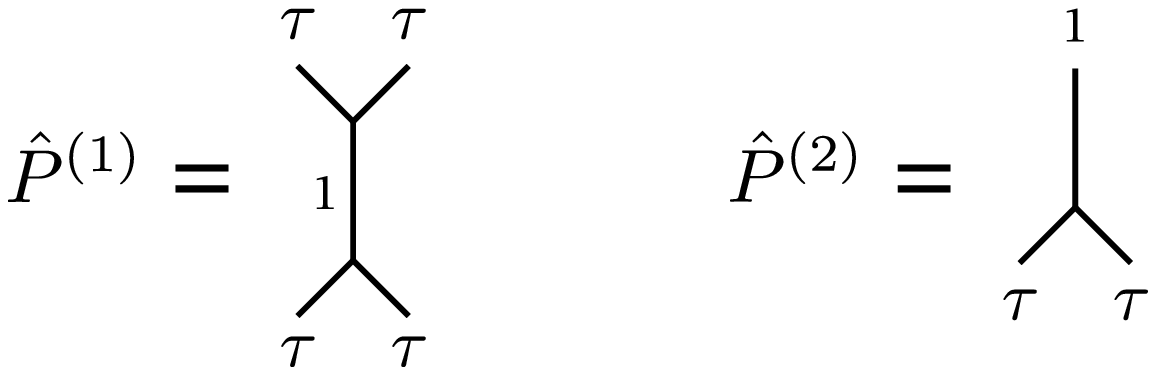}
\caption{Diagrammatic representation of operators $\hat P^{(1)}$ \eref{eq:ch4:P1} and $\hat P^{(2)}$ \eref{eq:ch4:P2}. The charges on all leaves are non-degenerate.\label{fig:ch4:projectionops}}
\end{center}
\end{figure}

This distinction between degeneracy-changing and degeneracy-preserving anyonic operators is %
clearly seen with a simple example. Let $|\psi\ra$ be a state on six Fibonacci anyons. This state can be parameterised by a vector $c^\alpha$, which has five components. We now define two projection operators, $\hat P^{(1)}$ and $\hat P^{(2)}$ (\fref{fig:ch4:projectionops}), each of which acts on the fusion space of anyons $\tau_1$ and $\tau_2$. 
Operator $\hat P^{(1)}$ is degeneracy-preserving, and projects $c^\alpha$ into the subspace in which anyons $\tau_1$ and $\tau_2$ fuse to the identity. Its matrix representation is
\begin{equation}
P_{\p{(1)}\,\alpha}^{(1)\,\p{\alpha}\beta} =
\left(
\begin{array}{cc}
1&0\\0&0
\end{array}
\right)\label{eq:ch4:P1}
\end{equation}
where the first value of each multi-index corresponds to a charge of 1, and the second to a charge of $\tau$.
Operator $\hat P^{(2)}$ performs the same projection, but is degeneracy-changing. Its matrix representation is written
\begin{equation}
P_{\p{(2)}\,\alpha}^{(2)\p{\,\alpha}\beta}=(~1~0~).%
\label{eq:ch4:P2}
\end{equation}
Both operators perform equivalent projections, in the sense that %
\begin{equation}
\la\psi|\hat P^{(1)\dagger}\hat P^{(1)}|\psi\ra = \la\psi|\hat P^{(2)\dagger}\hat P^{(2)}|\psi\ra.
\end{equation} 
When $\hat P^{(1)}$ acts on $|\psi\ra$ it leaves the Hilbert space unchanged, and hence the vector $c'^\alpha$ describing state $|\psi'\ra=\hat P^{(1)}|\psi\ra$ is once again a five-component vector, although in an appropriate basis some components will now necessarily be zero. In contrast $\hat P^{(2)}$ explicitly reduces the dimension of the Hilbert space, and the vector $c''^\alpha$ describing state $|\psi''\ra=\hat P^{(2)}|\psi\ra$ is of length two, describing a fusion tree on only four Fibonacci anyons (as both $\tau_1$ and $\tau_2$ have been eliminated). One consequence of this distinction is that while $(\hat P^{(1)})^2=\hat P^{(1)}$, the value of $(\hat P^{(2)})^2$ is undefined.

\section{Anyonic Tensor Networks}

\subsection{Diagrammatic Notation\label{sec:ch4:tensordiagrammatic}}

The diagrammatic notation conventionally employed in the study of anyonic systems, and used here in Figs.~\ref{fig:ch4:anyons} and \ref{fig:ch4:operators}, is well suited to the complete description of anyonic systems, as it provides a physically meaningful depiction of the entire Hilbert space. 
However, the number of parameters required for such a description grows exponentially in the system size, and because it is necessary to explicitly assign every index to a specific charge or degeneracy, specification of a tensor network rapidly becomes inconveniently verbose [for example see \fref{fig:ch4:operators}(iii)]. %

\begin{figure}
\begin{center}
\includegraphics[width=300.0pt]{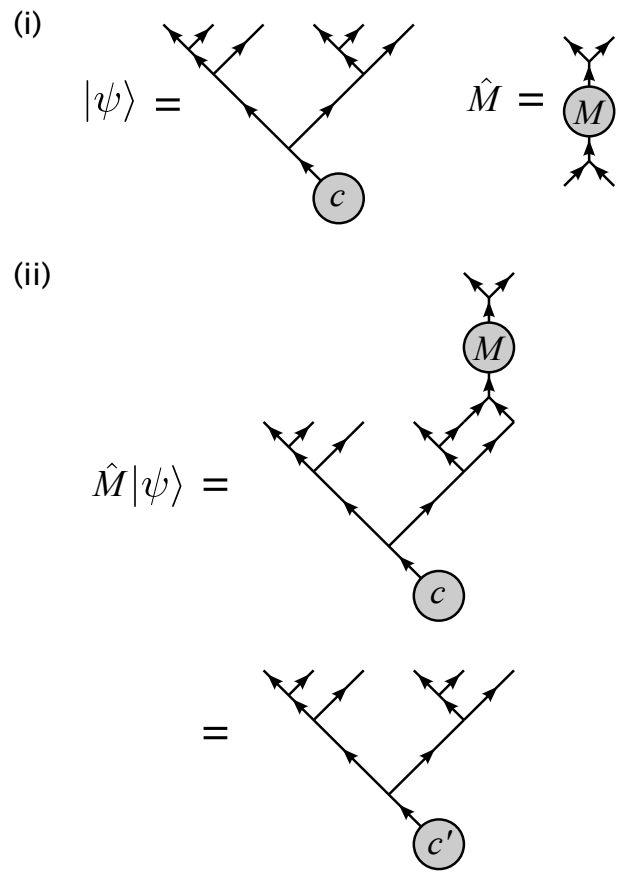}
\caption[(i)~Diagrammatic representation of a state $|\psi\ra$ and two-site operator $\hat M$ expressed in terms of degeneracy indices. (ii)~Application of $\hat M$ to state $|\psi\ra$. These diagrams represent the same state, operator, and process as \protect{\fref{fig:ch4:anyons}}(i) and \protect{\fref{fig:ch4:operators}}(ii)-(iii).]{(i)~Diagrammatic representation of a state $|\psi\ra$ and two-site operator $\hat M$ expressed in terms of degeneracy indices. (ii)~Application of $\hat M$ to state $|\psi\ra$. Grey shapes represent tensors with charge and degeneracy multi-indices, with each leg of the shape corresponding to one charge and degeneracy index pair. These diagrams represent the same state, operator, and process as \protect{\fref{fig:ch4:anyons}}(i) and \protect{\fref{fig:ch4:operators}}(ii)-(iii).\label{fig:ch4:blobtensors}}
\end{center}
\end{figure}

In the preceding Sections, we developed techniques whereby anyonic states and operators could be %
represented as vectors and matrices, bearing only one or two multi-indices apiece. We now introduce the graphical notation which complements this description, and in which we will formulate anyonic tensor networks. Figure~\ref{fig:ch4:blobtensors}(i) gives the graphical representations of a state $|\psi\ra$ associated with a vector $c^\alpha$, and of an operator $\hat M$ associated with a matrix $\Mop$.
The circle marked $c$ corresponds to the vector $c^\alpha$, and the circle marked $M$ corresponds to the matrix $\Mop$. In general, grey circles correspond to tensors, and the number of legs on the circle corresponds to the number of multi-indices on the associated tensor. Each multi-index is also associated with a fusion or splitting tree structure, which is specified graphically. For reasons to be discussed shortly, we will require that no tensor ever have more than three multi-indices.
As the legs of the grey shapes are each associated with a multi-index, they carry both degeneracy and charge indices. Consequently it is not necessary to explicitly assign labels to the fusion/splitting trees, as these labellings are contained implicitly in the degeneracy index (for example see \tref{tab:ch4:examplematrix}, where specifying the values of $\{a_i,\mu_i\}$ and $\{a_j,\mu_j\}$ is equivalent to fully labelling the fusion and splitting trees of \fref{fig:ch4:exampleoperator}). 

The fusion or splitting tree associated with a particular multi-index may be manipulated in the usual way by means of braids and $F$ moves, recalling that each component of the tensor is associated with a particular labelling of the fusion and splitting trees via the corresponding values of the multi-indices. %
Manipulations performed upon a particular tree thus generate unitary matrices which act upon the multi-index that corresponds to the labellings of that particular tree.

The application of an operator to a state is, unsurprisingly, performed by connecting the appropriate diagrams, as shown in \fref{fig:ch4:blobtensors}(ii). For operators of the type discussed in \sref{sec:ch4:anyonops}, the outcome is necessarily a new state in the same Hilbert space, which consequently can be described by a new state vector $c'^\alpha$, as shown. However, in general an operator $\hat M$ will not act on the entire Hilbert space of the system, and so will be described by a tensor constructed on the fusion space of some subset of lattice sites, and not on the system as a whole. Operator $\hat M$ acting on state $|\psi\ra$ in \fref{fig:ch4:blobtensors}(ii) is an example of this. Because $c^\alpha$ describes a six-site system but $\Mop$ is constructed on the fusion space of two sites, 
the multi-indices of $\Mop$ span a significantly smaller Hilbert space than that of $c^\alpha$ and we cannot simply write 
\begin{equation}
c'^\beta = c^{\alpha}M_\alpha^{\phantom{\alpha}\beta}
\end{equation}
(using Einstein notation, where repeated multi-indices are assumed to be summed).
Instead, we must understand how to expand the matrix representation of an operator on some number of sites $x$, to obtain its matrix representation as an operator on $x'$ sites, where $x'>x$.

\subsection{Site Expansion of Anyonic Operators\label{sec:ch4:siteexpanyop}}

The multiplicity tensor $N^{c}_{ab}$ describes the fusion of two charges without degeneracies. It is easily extended to incorporate degeneracies of the charges, 
and we will denote this expanded multiplicity tensor $\tilde N^\gamma_{\alpha\beta u}$ 
where %
multi-indices $\alpha$, $\beta$, and $\gamma$ are associated with the pairs 
$\{a,\mu_a\}$, $\{b,\mu_b\}$, and $\{c,\mu_c\}$ respectively, and for given values of $\alpha$, $\beta$, and $\gamma$, $u$ runs from 1 to $N^c_{ab}$. 
The degeneracies associated with charges $a$, $b$, and $c$ are denoted $\nu_{a}$, $\nu_{b}$, and $\nu_c$ respectively. As with $\mu_a$, $\mu_b$, and $\mu_c$, there is an implicit additional index on each degeneracy $\nu_x$ representing the edge of the tree on which %
charge $x$ resides. 
The values of $\nu_a$ and $\nu_b$ may be chosen arbitrarily (for example, $\nu_{a}|_{a=1}$ may differ from $\nu_{b}|_{b=1}$), but the degeneracies associated with the values of $c$ must satisfy
\begin{equation}
\nu_{c}=\sum_{a,b} \nu_{a}\nu_{b}N^{c}_{ab}
\end{equation}
in accordance with Eq.~\eref{eq:ch4:compounddegens}. When this constraint is satisfied, 
every quadruplet of indices $\{a,\mu_a,b,\mu_b\}$ corresponding to a unique pair of choices for $\alpha$ and $\beta$ may be associated with $N^c_{ab}$ distinct
pairs of indices $\{c,\mu_c\}$ for each $c\in a\times b$. These pairs $\{c,\mu_c\}$ are enumerated by the additional index $u$. %
This defines a 1:1 mapping between sets of values on $\{a,\mu_a,b,\mu_b,u\}$ and pairs $\{c,\mu_c\}$,
and we set the corresponding entries in $\tilde N^\gamma_{\alpha\beta u}$ to 1, with all other entries being zero.
A simple example is given in \tref{tab:ch4:exampleN}.
\begin{table}
\begin{center}
\begin{tabular}{|c|c|}
\bottomrule
Pair & Assigned pentuplet \\
\{$c,\mu_c$\} & \{$a,\mu_a,b,\mu_b,u$\} \\
\hline
$1,~1$ & $1,~1,~1,~1,~1$ \\
$1,~2$ & $\tau,~1,~\tau,~1,~1$ \\
$1,~3$ & $\tau,~1,~\tau,~2,~1$ \\
$\tau,~1$ & $1,~1,~\tau,~1,~1$ \\
$\tau,~2$ & $1,~1,~\tau,~2,~1$ \\
$\tau,~3$ & $\tau,~1,~1,~1,~1$ \\
$\tau,~4$ & $\tau,~1,~\tau,~1,~1$ \\
$\tau,~5$ & $\tau,~1,~\tau,~2,~1$\\
\toprule
\end{tabular}
\caption[Construction of $\tilde N^\gamma_{\alpha\beta u}$ for a fusion vertex for Fibonacci anyons.]{Construction of $\tilde N^\gamma_{\alpha\beta u}$ for a fusion vertex for Fibonacci anyons.
In this example $a$ may take charges $1$ and $\tau$ each with degeneracy 1, and $b$ may take charges $1$ and $\tau$ with degeneracies 1 and 2 respectively. By Eq.~\protect{\eref{eq:ch4:compounddegens}}, charge $c$ may therefore take values $1$ and $\tau$ with degeneracies 3 and 5 respectively.
A correspondence between the values of multi-index $\gamma$ and of multi-indices $\alpha$ and $\beta$ is established in some systematic manner, with each assignation satisfying $c\in a\times b$, and for Fibonacci anyons the index $u$ is trivial as all multiplicities $N^c_{ab}$ are zero or one.
An example %
assignation is shown in the table. The corresponding entries of $\tilde N$ are then set to %
1, with all other entries zero. For example, the fourth row indicates that $\tilde N^{(\tau,1)}_{(1,1)(\tau,1)1}=1$.\label{tab:ch4:exampleN}}
\end{center}
\end{table}

By virtue of their derivation from $N^{c}_{ab}$, the object $\tilde N^\gamma_{\alpha\beta u}$ and its conjugate $\tilde N^{\dagger\alpha\beta u}_{\phantom{\dagger}\gamma}$ 
represent application of the anyonic fusion rules, and may be associated with vertices of the splitting and fusion trees. Under the isotopy invariance convention there is an additional factor of $[d_{c}/(d_{a}d_{b})]^\frac{1}{4}$ associated with the fusion of %
charges $a$ and $b$ into $c$, where $d_{x}$ is the quantum dimension of charge $x$, and similarly for splitting, but we will account for these factors separately (see \sref{sec:ch4:blockstructure}). Thus constructed, the tensors $\tilde N$ satisfy $\tilde N^\gamma_{\alpha\beta u}\tilde N^{\dagger\alpha\beta u}_{\p\dagger\epsilon}=\delta_\epsilon^{\p\epsilon\gamma}$.

When used as a representation of the fusion rules, the generalised multiplicity tensor $\tilde N^\gamma_{\alpha\beta u}$ and its conjugate $\tilde N^{\dagger\alpha\beta u}_{\phantom{\dagger}\gamma}$ permit us to increase or decrease the number of multi-indices on a tensor in a manner which is consistent with the fusion rules of the quantum symmetry group. This process is reversible provided the symmetry group is Abelian or, for a non-Abelian symmetry group, provided the total number of multi-indices on the tensor does not at any time exceed three.
In constructing and manipulating a tensor network for a system of anyons, we will require only objects which respect the fusion rules of the anyon model. It is a defining property of such objects that when the number of multi-indices they possess is reduced to 1 by repeated application of $\tilde N$ and $\tilde N^\dagger$, non-zero entries may be found only in the vacuum sector. We imposed this requirement for states in \sref{sec:ch4:anyonstates}, and it is equivalent to the restriction we imposed on anyonic operators in \sref{sec:ch4:anyonops}.
In \rcite{singh2010a} (\sref{sec:ch3:globalsym} of this Thesis), an equivalent condition was observed for tensors remaining unchanged under the action of a Lie group, and these tensors were termed \emph{invariant}. When working with invariant tensors, we may separately evaluate the components of the tensors acting on the degeneracy spaces (e.g. the nonzero blocks of $M_\alpha^{\phantom{\alpha}\beta}$), and the factors arising from loops and vertices of the associated spin network (see \sref{sec:ch4:blockstructure} for details). This property greatly simplifies the contraction of pairs of tensors. 

\begin{figure}
\begin{center}
\includegraphics[width=300.0pt]{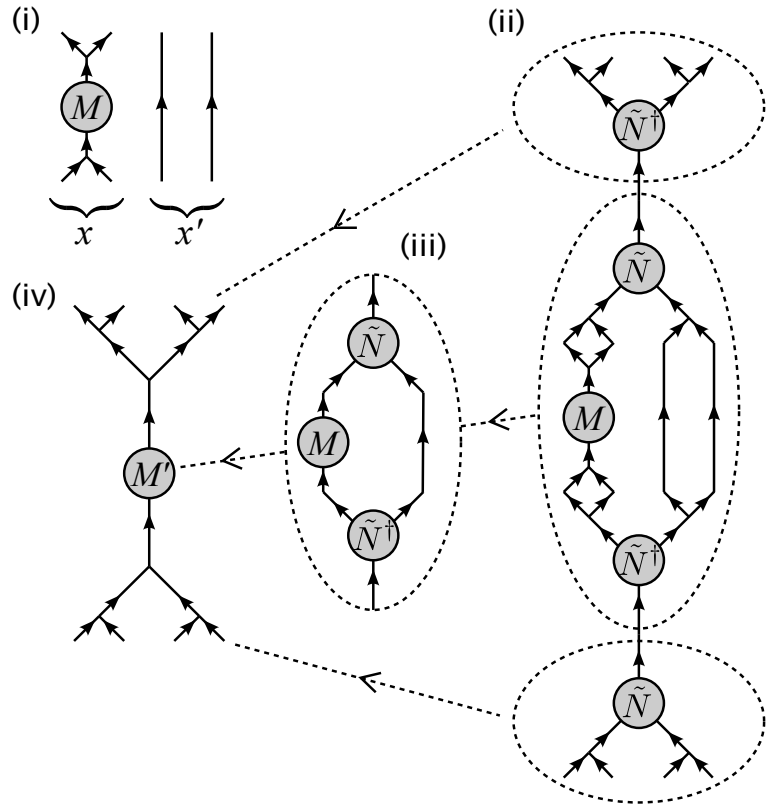}
\caption[``Raising'' of an operator $\hat M$ from sites $x$ to sites $x+x'$.]{``Raising'' of an operator $\hat M$ from sites $x$ to sites $x+x'$: (i)~Operator $\hat M$ defined only on sites denoted $x$. (ii)~Resolutions of the identity are inserted above and below $\hat M$, being constructed from tensors $\tilde N$ and $\tilde N^\dagger$. The central portion of this diagram is identified as corresponding to the new matrix $M'^{~\beta}_\alpha$ which describes $\hat M$ on $x+x'$. (iii)~Loop and vertex factors in the central region are evaluated separately and eliminated. (iv)~The tensor network corresponding to the new central portion is contracted. The $\tilde N$ and $\tilde N^\dagger$ tensors outside the central region become vertices of the fusion and splitting trees associated with $M'^{~\beta}_\alpha$. Together the trees and the matrix $M'^{~\beta}_\alpha$ constitute the raised version of $\hat M$. \label{fig:ch4:raiseoperator}}
\end{center}
\end{figure}

In addition to increasing or decreasing the number of legs of a tensor, we may also use $\tilde N$ to ``raise'' the matrix representation of an operator from the space of $x$ sites to the space of $(x+x')$ sites. This is shown in \fref{fig:ch4:raiseoperator}, and the matrix representation of the raised operator is given by
\begin{equation}
M_{\p{\prime}\alpha}^{\prime\p{\alpha}\beta} = M_\gamma^{\p{\alpha}\delta}\tilde N^{\dagger \gamma\epsilon u}_{\p{\dagger}\alpha} \tilde N^\beta_{\delta\epsilon u}\label{eq:ch4:Mprime}
\end{equation}
where multi-index $\epsilon$ describes the fusion space of all sites in $(x+x')$ but not in $x$. 
Because the numeric factors associated with loops and vertices (and braiding where applicable) are handled separately, 
no factors of quantum dimensions appear in Eq.~\eref{eq:ch4:Mprime}.

To act an operator $\hat M$ on a state $|\psi\ra$ in the matrix representation, we therefore connect the diagrams for $\hat M$ and $|\psi\ra$, eliminate all loops, and then raise the  matrix representation of the operator $\hat M$ using Eq.~\eref{eq:ch4:Mprime}, repeatedly if necessary, until the resulting matrix $M_{\p{\prime}\alpha}^{\prime\p{\alpha}\beta}$ may be applied directly to the state vector $c^\alpha$. 
Similarly it is possible to combine the matrix representations of operators, by connecting their diagrams appropriately, eliminating loops, and performing any required raising so that both operators act on the same fusion space. Their matrix representations can then be combined to yield the matrix representation of the new operator:
\begin{equation}
M^{(1\times2)\p{\alpha}\beta}_{\p{(1\times2)}\alpha} = M^{(1)\p{\alpha}\gamma}_{\p{(1)}\alpha} M^{(2)\p{\gamma}\beta}_{\p{(2)}\gamma},\label{eq:ch4:contractmatrices}
\end{equation}
and the fusion/splitting tree associated with this new operator is obtained as shown in \fref{fig:ch4:raiseoperator}.

Note that as yet, we have not described how two objects may be combined if their multi-indices are both up or both down, and are connected by a curved line. To contract such objects together, it is necessary to understand how bends act on the central matrix of an operator. Once this is understood, the bend can be absorbed into one of the central matrices, so that the connection is once again between an upper and a lower multi-index as in Eq.~\eref{eq:ch4:contractmatrices}. This process is described in \sref{sec:ch4:anyonmanip}.

\subsection{Manipulation of Anyonic Operators\label{sec:ch4:anyonmanip}}

\begin{figure}
\begin{center}
\includegraphics[width=300.0pt]{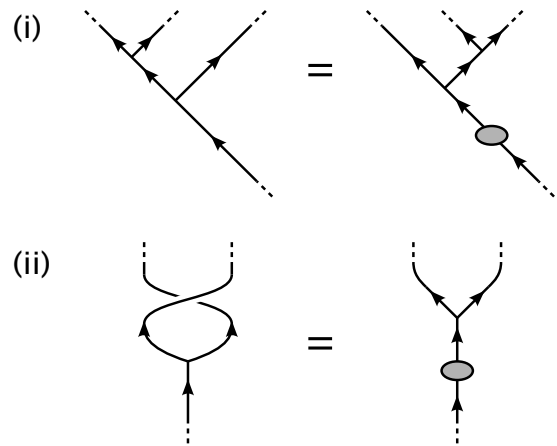}
\caption{(i)~$F$ move, and (ii)~braiding, performed on a section of fusion tree in the diagrammatic notation of \protect{\sref{sec:ch4:tensordiagrammatic}}.\label{fig:ch4:anyonopmanip1}}
\end{center}
\end{figure}

As observed in \sref{sec:ch4:siteexpanyop}, when we describe a system entirely in terms of objects invariant under the action of the symmetry group, we may account separately for the numerical normalisation factors associated with the spin network. %
However, as well as affecting these numerical factors, transformations of the fusion or splitting tree of an anyonic operator will typically also generate unitary matrices which act on the matrix representation of the operator. These matrices respect the symmetry of the anyon model, and thus can be written as block-diagonal matrices where each block is %
a unitary matrix acting on %
a particular charge sector. %
In terms of the diagrammatic notation of \sref{sec:ch4:tensordiagrammatic}, $F$ moves and braids therefore result in the insertion of a unitary matrix, as shown in \fref{fig:ch4:anyonopmanip1}. These matrices, whose entries are derived from the tensors $(F^{abc}_{d})_{(euv)(fu'v')}$ and $R^{ab}_c$ respectively, are raised if required, as described in \sref{sec:ch4:siteexpanyop}, and then contracted with $\Mop$, the matrix representation of the operator. 
To compute the unitary matrices involved, it suffices to recognise that $F$ moves and braids are unitary transformations in the space of labelled tree diagrams.
Identifying the leg on which the unitary matrix is to be inserted, 
the relevant region of the space of labelled diagrams is then enumerated by the multi-index which can be associated with this leg (compare \fref{fig:ch4:exampleedges}).

\begin{figure}
\begin{center}
\includegraphics[width=300.0pt]{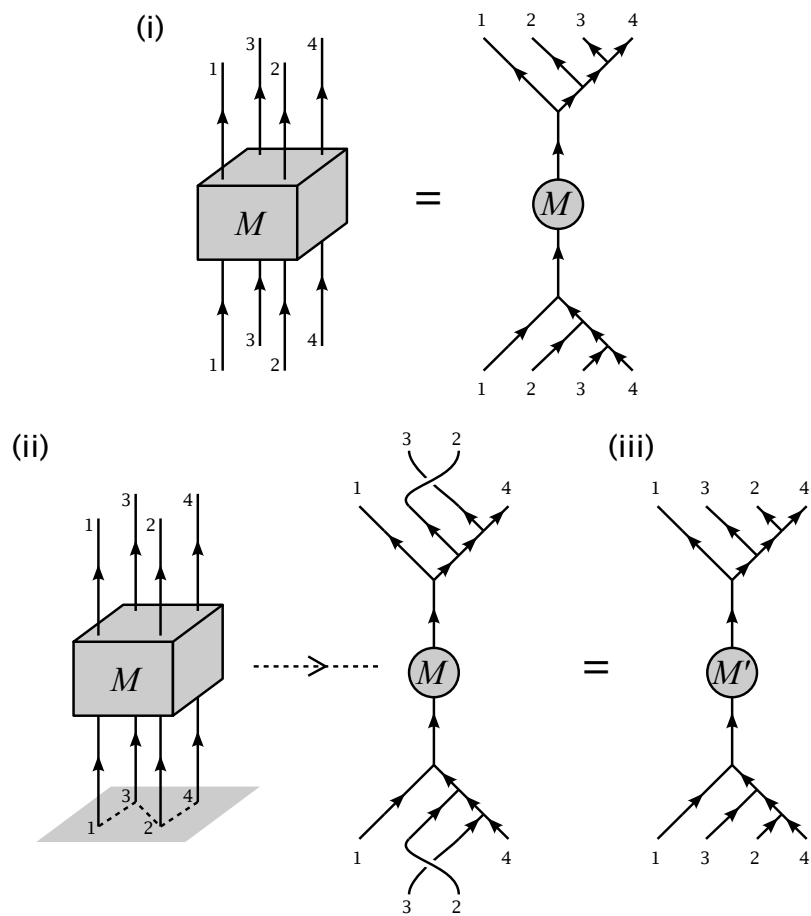}
\caption[(i)~An operator $\hat M$ acting on sites on a 2D lattice is defined with respect to some arbitrary linear ordering of these sites. (ii)~When manipulating the tensor network, it may be convenient for the lattice to be linearised according to some alternative scheme. To apply $\hat M$ to a different linearisation of the lattice may require braiding. (iii)~The unitary matrices corresponding to the required $F$ moves and braiding operations may be absorbed into $\hat M$, defining a new operator $\hat M'$ on the linearised lattice.]{(i)~An operator $\hat M$ acting on sites on a 2D lattice is defined with respect to some arbitrary linear ordering of these sites. (ii)~When manipulating the tensor network, it may on occasion be computationally convenient for the lattice to be linearised according to some alternative linearisation scheme. In this example, the imposed linearisation scheme is indicated by the dotted line. To apply $\hat M$ to a different linearisation of the lattice may require braiding. The orientation of the braids can be determined by putting the fusion tree of~(i) onto the 2D lattice, then smoothly deforming the lattice into a chain in accordance with the linearisation prescription. (iii)~The unitary matrices corresponding to the required $F$ moves and braiding operations may be absorbed into $\hat M$, defining a new operator $\hat M'$ on the linearised lattice.\label{fig:ch4:absorbbraid}}
\end{center}
\end{figure}

Braiding is of particular importance when working in two dimensions, as an operator will necessarily be defined with respect to some arbitrary linear ordering of its legs, and when manipulating a tensor network it may be necessary to map between this original definition and other equivalent definitions, corresponding to different leg orderings. %
For example, let $\hat M$ be a four-site anyonic operator as shown in \fref{fig:ch4:absorbbraid}(i), which we wish to apply to a 2D lattice. For the indicated linearisation of this lattice, application of $\hat M$ will require braiding as shown in \fref{fig:ch4:absorbbraid}(ii).
By evaluating the unitary transformations corresponding to these braids and absorbing them into $\Mop$, we may define a new operator $\hat M'$ which acts directly on the linearised lattice without any intervening manipulations of the fusion/splitting trees.

We will also frequently wish to deal with tensor legs which bend vertically through 180$^\circ$. 
If working with an anyon model that has non-trivial Frobenius--Schur indicators, then 
indicator flags must be applied to all bends. %
Like $F$ moves and braiding, the reversal of a Frobenius--Schur indicator flag is a unitary transformation%
, and once again this leads to the introduction of a unitary matrix which can be absorbed into a nearby existing tensor. %
However, we may wish to perform other operations on bends, such as absorbing them into fusion vertices or the central matrices of anyonic operators. We may also need to move a matrix $\Mop$ across a bend. We must therefore develop the description of bends in the new diagrammatic formalism.

\begin{figure}
\begin{center}
\includegraphics[width=300.0pt]{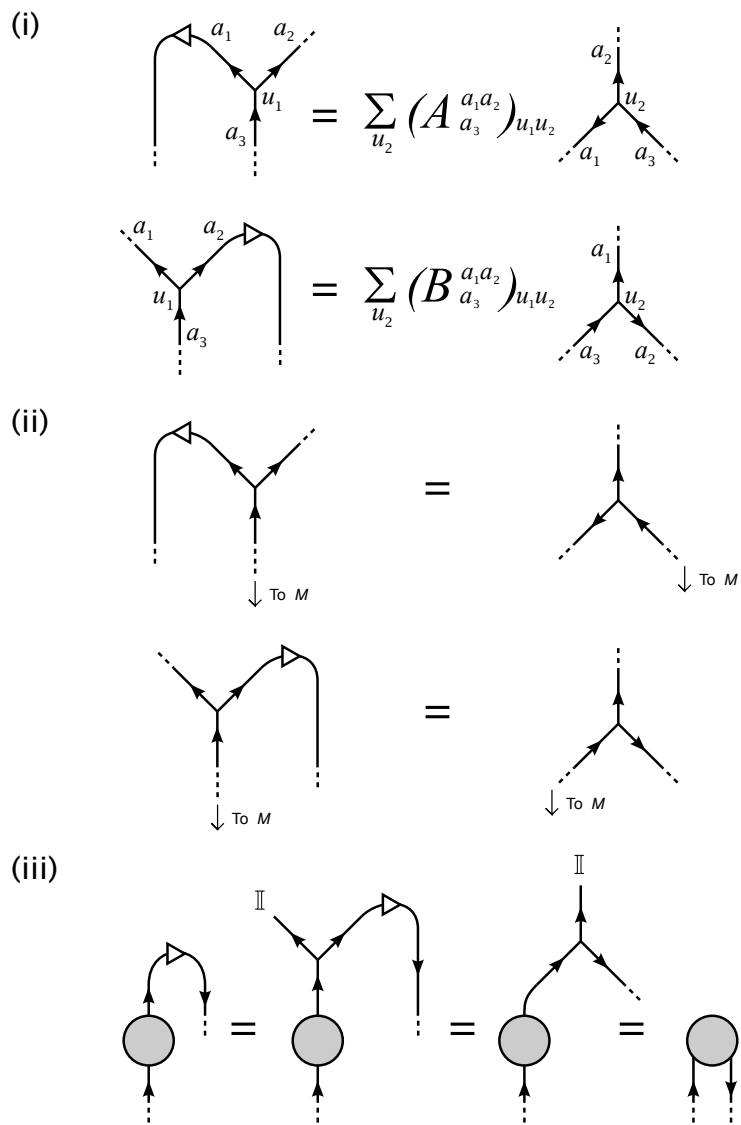}
\caption[Vertical bending of legs (i) in the standard diagrammatic notation, and (ii) in the diagrammatic notation of \protect{\sref{sec:ch4:tensordiagrammatic}}. (iii)~Legs on the matrix representations of states and operators may also absorb bends.]{Vertical bending of legs (i) in the standard diagrammatic notation, and (ii) in the diagrammatic notation of \protect{\sref{sec:ch4:tensordiagrammatic}}. White triangles represent Frobenius--Schur indicator flags. (iii)~Legs on the matrix representations of states and operators may also absorb bends. %
\label{fig:ch4:anyonopmanip2}}
\end{center}
\end{figure}

In \rcite{bonderson2007} a prescription for absorbing bends into fusion vertices is given in terms of tensors $(A^{ab}_c)_{uv}$ and $(B^{ab}_c)_{uv}$, derived from the $F$ moves, and corresponding to clockwise and counter-clockwise bends respectively. 
The absorption of a clockwise or counterclockwise bend into a fusion vertex is reproduced in 
\fref{fig:ch4:anyonopmanip2}(i), and results in a vertex fusing upward- and downward-going legs.  
We now assign new tensors $(\tilde N^\mrm{CW})^{\dagger\alpha u}_{\p\dagger\gamma\beta}$ and $(\tilde N^\mrm{CCW})^{\dagger\beta u}_{\p\dagger\alpha\gamma}$ to such vertices, such that writing these transformations in the notation of \sref{sec:ch4:tensordiagrammatic} is trivial. This is shown in \fref{fig:ch4:anyonopmanip2}(ii).

Explicit expressions for the new vertex tensors $(\tilde N^\mrm{CW})^\dagger$ and $(\tilde N^\mrm{CCW})^\dagger$ may be obtained by recognising that \fref{fig:ch4:anyonopmanip2}(i) describes the action of unitary transformations %
on $\tilde N^{\dagger\alpha\beta u}_{\p\dagger\gamma}$. When the bend is counterclockwise, the corresponding unitary matrix is derived from $(A^{ab}_c)_{uv}$, and when the bend is clockwise, the unitary matrix is derived from $(B^{ab}_c)_{uv}$. We will denote these unitary matrices $A_\gamma^{\p\gamma\delta}$ and $B_\gamma^{\p\gamma\delta}$ respectively.
We then have
\begin{eqnarray}
(\tilde N^\mrm{CW})^{\dagger\alpha u}_{\p\dagger\gamma\beta}&=&A_\gamma^{\p\gamma\delta}\tilde N^{\dagger\alpha\epsilon u}_{\p\dagger\delta}\delta_{\epsilon\beta}\label{eq:ch4:Nbend}\\
(\tilde N^\mrm{CCW})^{\dagger\beta u}_{\p\dagger\alpha\gamma}&=&B_\gamma^{\p\gamma\delta}\tilde N^{\dagger\epsilon\beta u}_{\p\dagger\delta}\delta_{\epsilon\alpha}.\label{eq:ch4:Nbend2}
\end{eqnarray}
and conjugation describes equivalent vertices $\tilde N^{CW}$ and $\tilde N^{CCW}$ when a bend is absorbed into a splitting tree. 

\begin{figure}
\begin{center}
\includegraphics[width=300.0pt]{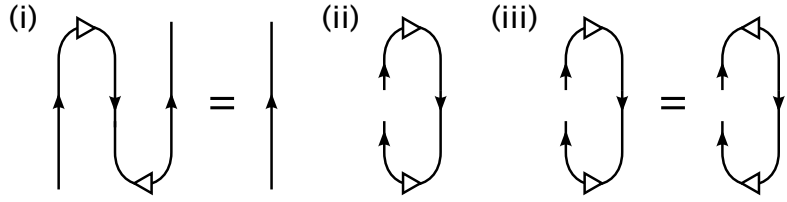}
\caption{Opposing pairs of Frobenius--Schur indicators (i) on a pair of bends equivalent to the identity, and (ii) on a pair of bends such as might be used when computing a quantum trace. (iii)~As a%
n anyon model can always be specified such that the Frobenius--Schur indicators are $\pm 1$, reversing a pair of contiguous opposed Frobenius--Schur indicator flags is always free%
.
\label{fig:ch4:frobeniusschur}}
\end{center}
\end{figure}

\begin{figure}
\begin{center}
\includegraphics[width=300.0pt]{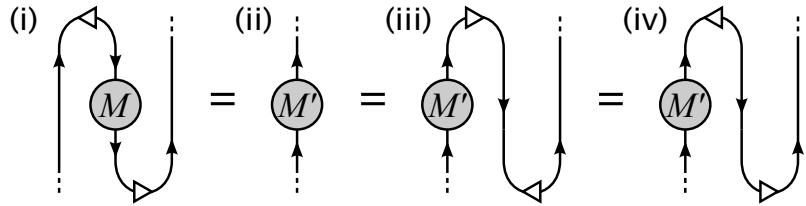}
\caption[Moving a matrix across a bend in a tensor network diagram.]{Moving a matrix across a bend in a tensor network diagram. (i)~Initial diagram. (ii)~Bends are absorbed into the matrix. (iii)~New bends are introduced, in accordance with \protect{\fref{fig:ch4:frobeniusschur}(i)}. (iv)~A pair of contiguous, opposed Frobenius--Schur indicators are reversed, as per \protect{\fref{fig:ch4:frobeniusschur}(iii)}. The initial and final matrices $M$ and $M'$ are related as specified in Eq.~\protect{\eref{eq:ch4:movepastbend}}.\label{fig:ch4:movepastbend}}
\end{center}
\end{figure}

\begin{figure}
\begin{center}
\includegraphics[width=300.0pt]{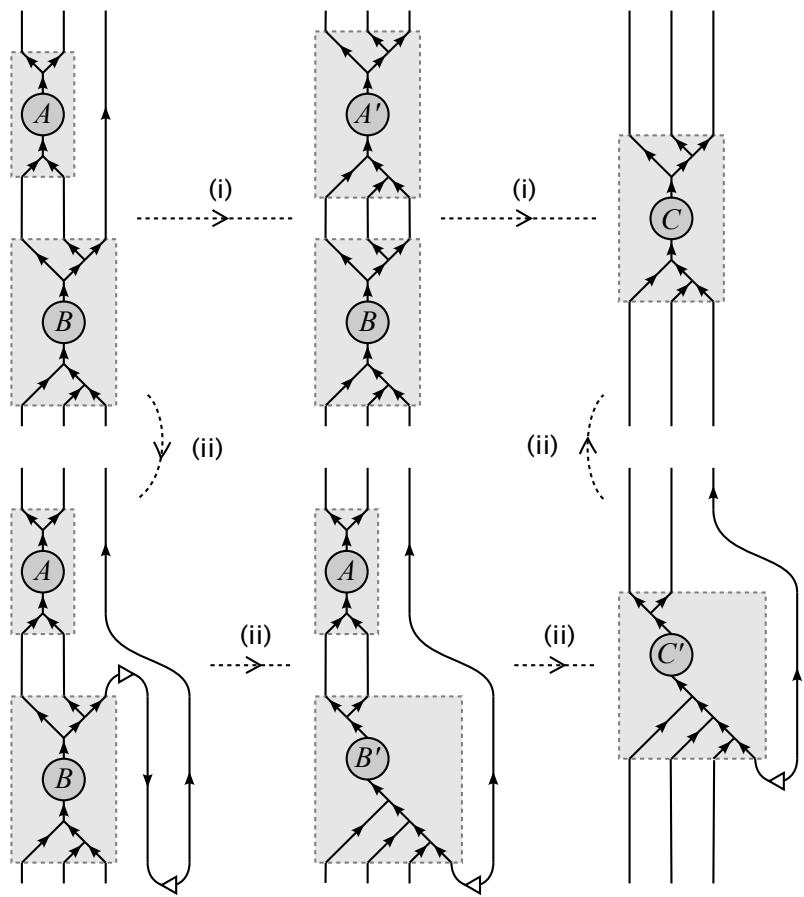}
\caption{The use of bends may permit the more efficient contraction of pairs of anyonic operators. In the sequence of events marked (i), operator $\hat A$ is first raised to the space of three sites then contracted with $\hat B$. In sequence (ii) the operators are instead contracted using bends. For many anyon models the latter approach offers a significant computational advantage.\label{fig:ch4:controps_bends}}
\end{center}
\end{figure}

Knowing how the absorption of bends acts on a vertex tensor, we may readily infer how the same process acts on the matrix representation of an operator. In \fref{fig:ch4:anyonopmanip2}(iii) we see a bend absorbed into the matrix $\Mop$, resulting in a new object with two lower multi-indices, $M'_{\alpha\beta}$. First we exploit the freedom to introduce fusion with the trivial charge (denoted $\mbb{I}$), with degeneracy 1. The corresponding $\tilde N^\dagger$ object takes only one value on its upper left multi-index, and is fully defined by $\tilde N^{\dagger\mbb{I}\beta 1}_{\p\dagger\gamma} = \delta^{\p\gamma\beta}_\gamma$. Absorbing the bend into this fusion vertex as per Eq.~\eref{eq:ch4:Nbend} yields $A_\gamma^{\p\gamma\delta}\delta_\delta^{\p\delta\epsilon}\delta_{\epsilon\beta} = A_{\gamma\beta}$, which may be then combined with $\Mop$ to give
\begin{equation}
M'_{\alpha\beta} = M_\alpha^{\p\alpha\gamma}A_{\gamma\beta}.
\end{equation}
In conjunction with the relationships given in \fref{fig:ch4:frobeniusschur}, this gives us the ability to move a matrix past a bend. 
An example of this is given in \fref{fig:ch4:movepastbend}, for which $M$ and $M'$ are related according to
\begin{equation}
M'^{\p\alpha\beta}_\alpha = A_{\alpha\gamma}M_\delta^{\p\delta\gamma}\varkappa_\epsilon^{\p\epsilon\delta}B^{\dagger\epsilon\beta}\label{eq:ch4:movepastbend}
\end{equation}
where $\varkappa_\epsilon^{\p\epsilon\delta}$ represents reversal of the Frobenius--Schur indicator flag on the lower bend.
Finally, bending may also allow more efficient contraction of pairs of anyonic operators, as shown in \fref{fig:ch4:controps_bends}.

\begin{figure}
\begin{center}
\includegraphics[width=300.0pt]{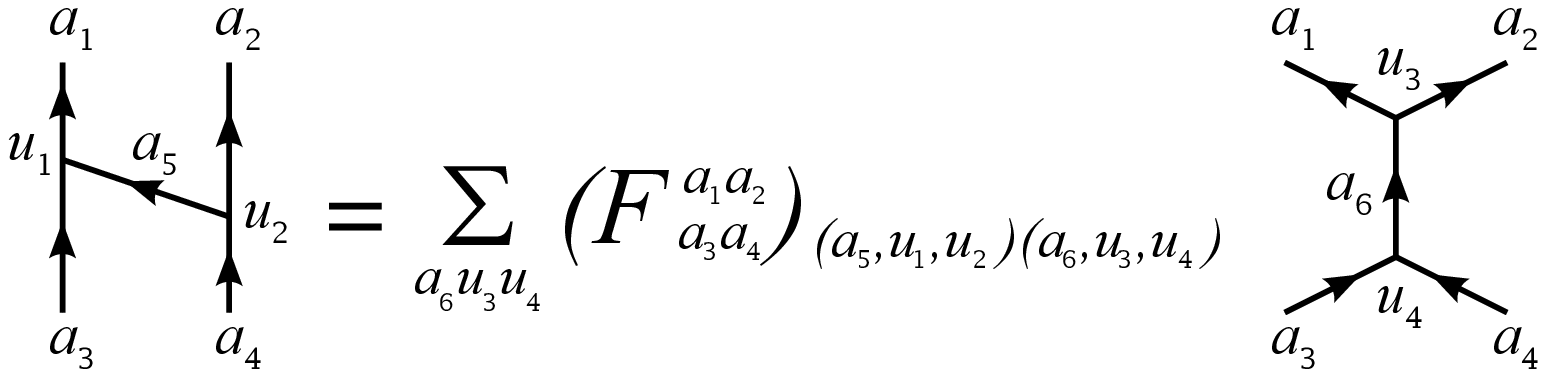}
\caption{Now that we may bend legs up and down it is customary to introduce a further type of $F$ move, derived by applying bends to the one presented in \protect{\fref{fig:ch4:anyons}}(iii). %
\label{fig:ch4:newfmove}}
\end{center}
\end{figure}

Having described the action of bends, it is customary also to introduce a second type of $F$ move which is described by the tensor $(F^{a_1a_2}_{a_3a_4})_{(a_5u_1u_2)(a_6u_3u_4)}$ (\fref{fig:ch4:newfmove}). 
This tensor may be derived from $(F^{a_1a_2a_3}_{a_4})_{(a_5u_1u_2)(a_6u_3u_4)}$ by bending, and as with $(F^{a_1a_2a_3}_{a_4})_{(a_5u_1u_2)(a_6u_3u_4)}$ these $F$ moves perform a transformation of the fusion tree, accompanied by the introduction of a unitary matrix which can be absorbed into the matrix representation of the operator. These unitary matrices %
correspond to the consecutive application of a bend, an $F$ move of the original type%
, and a second bend whose action is the inverse of the first.

\subsection{Constructing a Tensor Network\label{sec:ch4:constensnet}}

\begin{figure}
\begin{center}
\includegraphics[width=300.0pt]{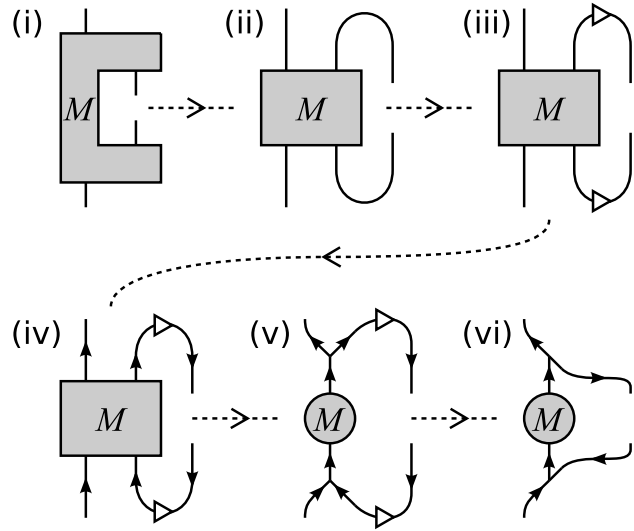}
\caption[Construction of an anyonic tensor corresponding to a normal tensor with more than three legs.]{Construction of an anyonic tensor corresponding to a normal tensor with more than three legs. (i)~The original tensor. (ii)~If required, any concavities are eliminated by introducing bends. (iii)~Frobenius--Schur indicators are assigned to the bends. (iv)~Directions are assigned to all legs, consistent with the rest of the network. %
(v)~Legs are collected together into fusion and splitting trees. The central object, representing degrees of freedom of the tensor, now has less than four legs. (vi)~If desired, bends can be re-absorbed into the fusion and splitting trees.
\label{fig:ch4:generalTN}}
\end{center}
\end{figure}

Now that we have developed a formalism for anyonic tensors,
we may convert an existing tensor network algorithm for use with anyons. 
First, the tensor network must be drawn in such a manner that every leg has a discernible vertical orientation. Although these orientations may be changed during manipulation of the tensor network, an initial assignment of upward or downward direction is required. Second, all tensors must be represented by entirely convex shapes, such as circles or regular polygons. For existing tensor network algorithms such as MERA and PEPS, this requirement is trivial. However, it is conceivable that future algorithms might involve superoperator-type objects whose graphical representations interleave upward- and downward-pointing legs. Concavities on these objects may be eliminated by replacing some of their upward-pointing legs with downward-pointing legs (or vice versa), followed by a bend [\fref{fig:ch4:generalTN}(i)-(ii)]. 
A similar treatment may be applied to any superoperators which arise during manipulations of the tensor network, introducing a pair of bends as in \fref{fig:ch4:frobeniusschur}(i) and then absorbing one into the matrix representation of the object.

If working with an anyon model that has non-trivial Frobenius--Schur indicators, then indicator flags must %
be applied to all bends. 
Initial choices are a matter of convenience, and it is frequently possible to assign these indicators in opposed pairs, as shown in \fref{fig:ch4:frobeniusschur}. If these paired indicators are not flipped or are only flipped in adjacent opposed pairs %
during subsequent manipulations of the tensor network, then they may %
frequently be left implicit.

Next, if there exist charges in the anyon model which are not self-dual, a direction (represented by a solid arrow) must be assigned to every multi-index. Any tensor with more than three legs (e.g. $M$ in \fref{fig:ch4:generalTN}) is then replaced by %
a trivalent tensor network
consisting of a core object, e.g. $M\alphabeta$, which contains the free parameters of the tensor, and as many copies of $\tilde N$ or $\tilde N^\dagger$ as are required to provide the correct output legs. 
These tensors $\tilde N$, $\tilde N^\dagger$ %
correspond to vertices in the fusion and splitting trees associated with $M\alphabeta$, yielding the corresponding anyonic tensor. Objects with three legs or less can be directly identified with an anyonic tensor object carrying the appropriate number of indices (i.e. three multi-indices and a vertex index $u$), though for consistency with the methods described in Sections~\ref{sec:ch4:siteexpanyop} and \ref{sec:ch4:anyonmanip} we point out that it is possible to similarly replace three-legged objects with anyonic operators consisting of a central matrix $\Mop$ and a fusion or splitting vertex, if desired.

Any bends introduced earlier may now be reabsorbed, so that some vertices now correspond to $(\tilde N^\mrm{CW})$, $(\tilde N^\mrm{CCW})$, $(\tilde N^\mrm{CW})^\dagger$, and $(\tilde N^\mrm{CCW})^\dagger$. This step, however, is optional as it may be more convenient for subsequent manipulations of the tensor network if the bends are left explicit. 
The anyonic tensors are 
then connected precisely as in the original Ansatz. %

Manipulations of the anyonic tensor network are equivalent to those performed on the spin version of the Ansatz, differing only
in that the degrees of freedom of the tensor network are now expressed entirely by the at-most-trivalent central objects, and certain topological elements such as braids and vertical bends must be accounted for in accordance with the prescriptions of \sref{sec:ch4:anyonmanip}. 
These changes may naturally imply minor changes to the manipulation algorithms, and we will see examples of this in the {1D} MERA. Similar considerations will %
apply %
to other tensor network algorithms. %

Our construction of an anyonic tensor network draws upon two important elements which have previously been observed in other, simpler, physical systems:
\begin{enumerate}
\item Tensors in the Ansatz exhibit a global symmetry, which may be non-Abelian. Exploiting a non-Abelian symmetry requires that the Ansatz be written in the form of a trivalent tensor network. This has previously been observed and implemented for non-Abelian Lie group symmetries such as SU(2)
(\citealp{singh2010a}, \sref{sec:ch3:globalsym} of this Thesis; \citealp{singhvidalinprep})%
.
\item Tensors in the Ansatz must be able to account for non-trivial exchange statistics. This has previously been observed in the simulation of systems of fermions\citecomma{corboz2010,kraus2010,pineda2010,corboz2010a,barthel2009,shi2009,pizorn2010,gu2010} where efficient implementation of particle statistics can be achieved through the use of ``swap gates''\citestop{pineda2010,corboz2010a,barthel2009,pizorn2010} 
\end{enumerate}
In both cases, anyonic tensor networks extend the concepts introduced in previous work. The symmetry structure of an anyon model may be a quantum group, for example a member of the series $SU(2)_k$, $k\in\mbb{Z}^+$, rather than having to be a Lie group, %
and this permits representation of non-Abelian anyonic systems whose Hilbert space does not admit decomposition into a tensor product of local Hilbert spaces. Similarly, anyonic braiding may be implemented using a generalisation of the fermionic ``swap gate'' formalism. When braiding, particle exchange may introduce transformation by a unitary matrix rather than by a sign, and efficient implementation of the resulting swap gates is particularly important for the simulation of 2D systems. 

Although anyonic systems pose a number of unique challenges, we see that these are addressed by developments based on existing techniques, and we therefore anticipate that the resulting generalisations of existing tensor network Ans\"atze %
should still be capable of accurately representing the states of an anyonic system.

\subsection{Contraction of Anyonic Tensor Networks}

The techniques described in Secs.~\ref{sec:ch4:siteexpanyop} and \ref{sec:ch4:anyonmanip} ($F$ moves, braids, bending of legs, elimination of loops, diagrammatic isotopy, flipping of Frobenius--Schur indicator flags, and the use of $\tilde N^{(\dagger)}$ tensors) suffice to contract any network of anyonic tensors written in the form of matrices with degeneracy indices, and unlabelled trees. Through careful application of these techniques, and avoiding at all times processes which would yield a tensor with more than three legs, the matrix representations of any pair of contiguous tensors in a network may always be brought into conjunction such that their multi-indices can be contracted in the manner of Eq.~\eref{eq:ch4:contractmatrices}, and any tensor network may be contracted by means of a sequence of such pairwise contractions.

That a tensor network may represent a system of anyons in this way
is possible because throughout the anyonic tensor network, each value of a degeneracy index is associated with a specific labelling of the corresponding unlabelled tree. Consequently it is always possible %
to fully reconstruct any operation in terms of the more verbose representation of \fref{fig:ch4:operators}.

An anyonic tensor network is therefore fully specified merely by the unlabelled tree (with Frobenius--Schur indicator flags if required), and the values and locations of the matrix representations of its tensors, written in the degeneracy index form. %

\section{Example: The {1D} MERA}

\subsection{Construction\label{sec:ch4:MERAconstr}}

\begin{figure}
\begin{center}
\includegraphics[width=300.0pt]{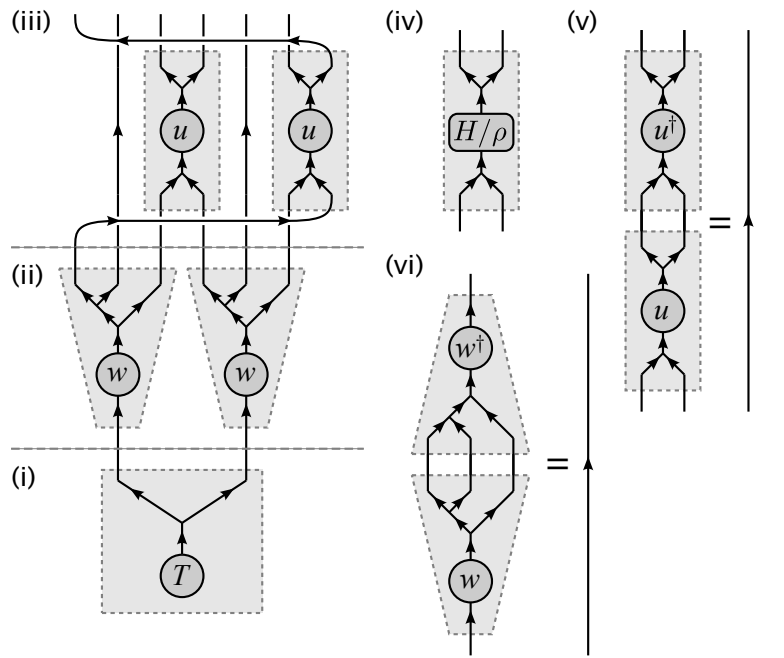}
\caption[Construction of a {1D} ternary MERA on a periodic lattice from anyonic operators.]{Construction of a {1D} ternary MERA on a periodic lattice from anyonic operators. (i)~The ``top'' tensor, $\hat{T}$. (ii)~Isometries, $\hat{w}$. (iii)~Disentanglers, $\hat{u}$. The fusion tree representing an anyonic state (or ket) is usually drawn with the lattice sites at the top, so this MERA has been constructed ``upside down'' when compared with the diagrams in \protect{\citet{evenbly2009}} and \protect{\citet{vidal2010}}. This is unimportant, and we could equally well have decided to follow the convention usually adopted in tensor network algorithms, labelled the tensors in~(i)-(iii) by $T^\dagger$, $w^\dagger$, and $u^\dagger$, and identified diagram~(i)-(iii) as a bra.
(iv)~Structure of a 2-site term in the Hamiltonian, $\hat{h}$, or a 2-site reduced density matrix, $\hat{\rho}$. 
(v)~Disentanglers and (vi) isometries satisfy the relationships $\hat u^\dagger\hat u=\mbb{I}$, $\hat w^\dagger\hat w=\mbb{I}$.
\label{fig:ch4:MERAoperators}}
\end{center}
\end{figure}

To construct an anyonic MERA for a {1D} lattice with $n$ sites, where $n$ satisfies $n=2\times3^k,~k\in\mbb{Z^+}$, we begin with a ``top'' tensor on a two-site lattice $\mc{L}_\tau$ whose matrix representation is of a computationally convenient size. [The top tensor is named for its position in the usual diagrammatic representation of the MERA, where diagrams with open legs at the bottom correspond to a ket. For anyons the converse convention applies, and consequently in \fref{fig:ch4:MERAoperators}(i) the ``top'' tensor is ironically located at the bottom.]

To each leg of the top tensor, we now append an isometry [\fref{fig:ch4:MERAoperators}(ii)]. The matrix representations of the isometries consist of rectangular blocks, as described in \sref{sec:ch4:degenexp}, and we choose isometries whose fusion trees have three legs, so as to construct a ternary MERA\citestop{evenbly2009} Next, disentanglers are applied above the isometries.
For periodic boundary conditions this must be performed in a manner which respects the anyonic braiding rules, as shown in \fref{fig:ch4:MERAoperators}(iii). 
We identify the open legs of the resulting network as the sites of a lattice $\mc{L}_{\tau-1}$, and the rows of disentanglers and isometries may be understood as a coarse-graining transformation taking a finer-grained lattice $\mc{L}_{\tau-1}$ into a coarser-grained lattice $\mc{L}_\tau$, similar to the standard MERA. 
Note that the geometry of the periodic lattice is reflected by the connections of the disentanglers.
Specifically, whether the outside legs are braided over or under the other lattice sites reflects whether the lattice closes towards or away from the observer. 
 
The application of anyonic isometries and disentanglers is now repeated $k$ times [\fref{fig:ch4:MERAoperators}(i)-(iii) corresponds to $k=1$], until the Ansatz has $n$ legs. The final row of isometries should be chosen such that each of their upper legs have the same charges and degeneracies as the sites of the physical lattice $\mc{L}_0$, and the open legs above the last row of disentanglers are identified with the physical lattice. For coarse-grained lattices $\mc{L}_1$ to $\mc{L}_\tau$, the dimensions of the lattice sites correspond to the lower legs of the isometries and are chosen for computational convenience, subject to the requirement that %
each charge sector is sufficiently large to adequately reproduce the physics of the low-energy portion of the Hilbert space. For all other legs, their charges and degeneracies are determined by requiring consistency with Eq.~\eref{eq:ch4:compounddegens}. Initial choices of which charges to represent on the ``top'' tensor and on the lower legs of the isometries, and with what degeneracies, must be guided either by prior knowledge about the physical system, or by balancing computational convenience against the inclusion of a broad and representative range of possible charges. When used in a numerical optimisation algorithm, the choice of relative weightings for the different charge sectors may often be refined by examination of the spectra of the reduced density matrices on the coarse-grained lattices, after initial optimisation of the tensor network is complete.

This concludes construction of the MERA for a state on a finite, periodic {1D} anyonic lattice. That this tensor network does represent an anyonic state is easily seen by sequentially raising tensors, performing $F$ moves, and combining tensors, until the entire network is reduced to a single vector whose length is equal to the dimension of the physical Hilbert space, and an associated fusion tree. These then represent the state of the system as per Eq.~\eref{eq:ch4:statepsi}. The structure of this tensor network closely resembles that of the normal MERA, according to the identifications given in \fref{fig:ch4:MERAoperators}, and consequently we anticipate that it will share many of the same properties, including the ability to reproduce polynomially decaying correlators in strongly correlated physical systems.
Open lattices may also be easily represented by omitting the braided disentanglers at the edge of the diagram.

We also note that in common with the MERA for spins, the anyonic MERA may be understood as a quantum circuit, although one which carries anyonic charges in its wires. Any junction in the fusion/splitting trees may be associated with a $\tilde N$ or $\tilde N^\dagger$ tensor, and the entire network may be considered as the application of a series of gates to a Hilbert space of fixed dimension beginning mostly (or entirely, if the top tensor is considered to be the first gate) in the vacuum state, with individual gates introducing entanglement across some limited number of wires.

\subsection{Energy Minimisation\label{sec:ch4:MERAoptimisation}}

The anyonic MERA can be used as a variational Ansatz to compute the ground state of a local Hamiltonian. The Hamiltonian is introduced as a sum over nearest neighbour interactions, each term having the form of \fref{fig:ch4:MERAoperators}(iv), and 
optimisation of the tensor network is carried out in the usual manner\citestop{evenbly2009} Also as per usual, Hamiltonians involving larger interactions, such as next-to-nearest neighbour, can be accommodated by means of an initial exact $n$-into-one coarse-graining of the physical lattice.

\begin{figure}
\begin{center}
\includegraphics[width=300.0pt]{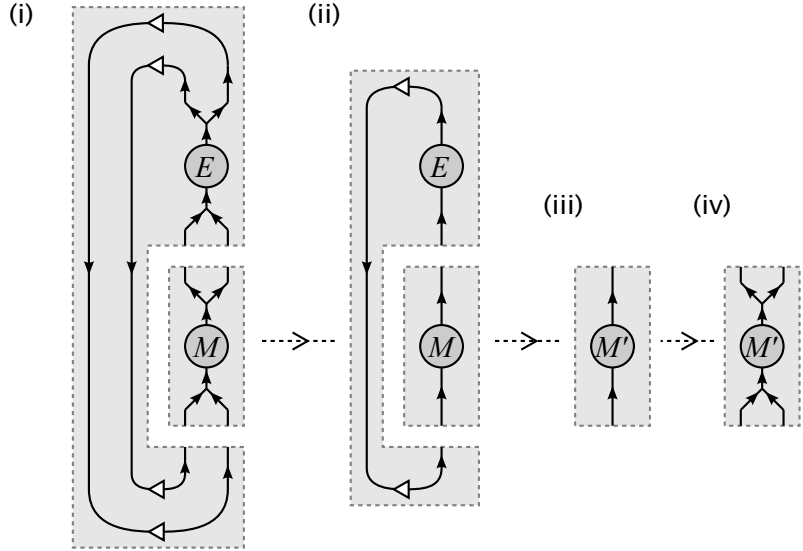}
\caption[Optimisation of an anyonic operator $\hat{M}$ with respect to an environment represented by an operator $\hat{E}$.]{(i)~Anyonic operator $\hat E$ constitutes the environment of operator $\hat M$. %
Factors arising from the fusion and splitting trees should be evaluated and absorbed into matrices $E$ and $M$, following which (ii) matrix $E$ constitutes the environment of matrix $M$. After (iii) updating the matrix $M$ %
to $M'$, (iv) the fusion and splitting trees of $\hat M$ should be reinstated, the numerical factors associated with this process being the inverse of the fusion tree factors previously absorbed into matrix $M$. Frobenius--Schur flags in (i)-(ii) are represented by white triangles, and are not to be confused with the black arrows which indicate the orientation of lines in the fusion/splitting trees.\label{fig:ch4:environment}}
\end{center}
\end{figure}

As in \rcite{evenbly2009}, optimisation of the MERA then consists of repeatedly lifting the Hamiltonian from $\mc{L}_0$ to the coarse-grained lattices, updating their isometries and disentanglers, and lowering the reduced density matrix, or the top tensor and its conjugate. When lifting the Hamiltonian or lowering the reduced density matrix, then the diagrams in \rcite{evenbly2009} taken in conjunction with the key given in \fref{fig:ch4:MERAoperators} serve to describe networks of anyonic operators which, when contracted to a single operator, yield the lifted form of the Hamiltonian or lowered form of the reduced density matrix respectively. Similarly, when optimising disentanglers or isometries, the diagrams of \rcite{evenbly2009} and the identifications in \fref{fig:ch4:MERAoperators} indicate how to construct an anyonic operator which constitutes the environment of the anyonic operator being optimised. However, once the admissible ranges of charges and degeneracies on each leg have been fixed, the only optimisable content of an anyonic operator is its matrix representation. 
Consequently, the fusion and splitting tree contributions should be evaluated and absorbed into the operator and its environment,
reducing them both 
to their matrix representations, denoted $M$ and $E$ respectively (see \fref{fig:ch4:environment}).
If the singular value decomposition of $E$ is written
$E=USW^\dagger$, then the updated matrix content $M$ of the anyonic operator being optimised is given by $-WU^\dagger$, minimising the value of $\mrm{Tr}(EM)$ subject to the usual constraint for disentanglers and isometries that $\hat M\hat M^\dagger=\mbb{I}$ %
[\fref{fig:ch4:MERAoperators}(v)-(vi)]. 
The fusion/splitting tree content of the operator can then be restored, along with any appropriate numerical factors that may be required.

\begin{figure}
\begin{center}
\includegraphics[width=300.0pt]{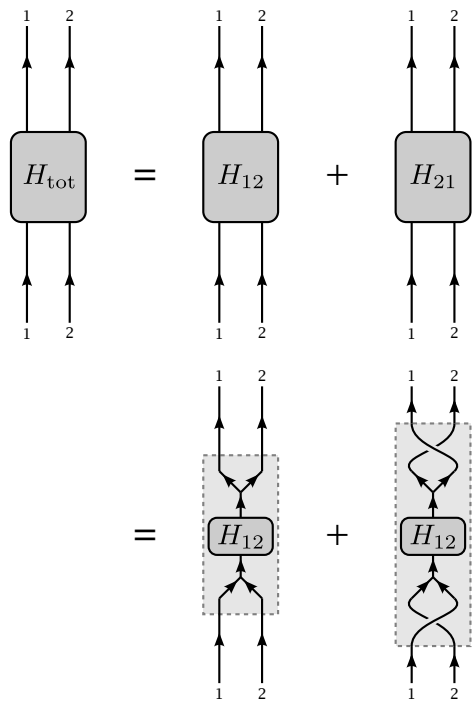}
\caption{Definition of $\hat H_{21}$ in terms of $\hat H_{12}$, on the most coarse-grained lattice ($\mc{L}_\tau$) of the translation-invariant periodic ternary MERA. Lattice $\mc{L}_\tau$ is a two-site periodic lattice.\label{fig:ch4:H21}}
\end{center}
\end{figure}

As with the standard MERA, the ``top'' tensor is constructed by diagonalising the total Hamiltonian on the most coarse-grained lattice, $\hat H_{\mrm{tot}}$ on $\mc{L}_\tau$. As $\mc{L}_\tau$ is a two-site lattice, the total Hamiltonian $\hat H_\mrm{tot}$ is a sum of two terms, $\hat H_{12}$ and $\hat H_{21}$. For the translation-invariant anyonic MERA, we may formally define $\hat H_{21}$ in terms of $\hat H_{12}$ as shown in \fref{fig:ch4:H21}, and the top tensor $\hat T$ (together with any factors arising from the chosen normalisation scheme) then corresponds to the lowest-energy eigenstate of $\hat H_\mrm{tot}$.

\subsection{Scale-Invariant MERA}

Having identified the anyonic counterparts of the tensors of the standard MERA, and described how these tensors may be lifted, lowered, and optimised, the algorithm for the scale-invariant MERA described in \rcite{pfeifer2009} may also be implemented for anyonic systems, simply by applying the dictionary of \fref{fig:ch4:MERAoperators} and the techniques described in \sref{sec:ch4:MERAoptimisation}. As with optimisation of $\hat{u}$ and $\hat{w}$, the computation of the top reduced density matrix (which is a descending eigenoperator of the scaling superoperator with eigenvalue 1) may be understood as a calculation of the matrix component $\rho_\alpha^{\p{\alpha}\beta}$ of the reduced density matrix $\hat{\rho}$. The ascending eigenoperators of the scaling superoperator, or local scaling operators of the theory, may also be computed in this manner.

\subsection{Results\label{sec:ch4:results}}

To demonstrate the effectiveness of the anyonic generalisation of the MERA, we applied it to a {1D} critical system of anyons whose physical properties are already well known: The golden chain\citestop{feiguin2007} This model consists of a string of Fibonacci anyons subject to a local interaction. Fibonacci anyons have only two charges, $1$ (the vacuum) and $\tau$, and one non-trivial fusion rule ($\tau\times\tau\rightarrow 1+\tau$). %
The simplest local interactions for a chain of Fibonacci $\tau$ anyons are nearest neighbour interactions favouring fusion of pairs into either the $1$ channel (termed antiferromagnetic, or AFM), or the $\tau$ channel (termed ferromagnetic, or FM). Both choices correspond to critical Hamiltonians, associated with the conformal field theories $\mc{M}(4,3)$ and $\mc{M}(5,4)$ for AFM and FM couplings respectively. 
Individual lattice sites are each associated with a charge of $\tau$.
\nomenclature{\tbf{AFM}}{Antiferromagnetic.}
\nomenclature{\tbf{FM}}{Ferromagnetic.}

\begin{figure}
\begin{center}
\includegraphics[width=300.0pt]{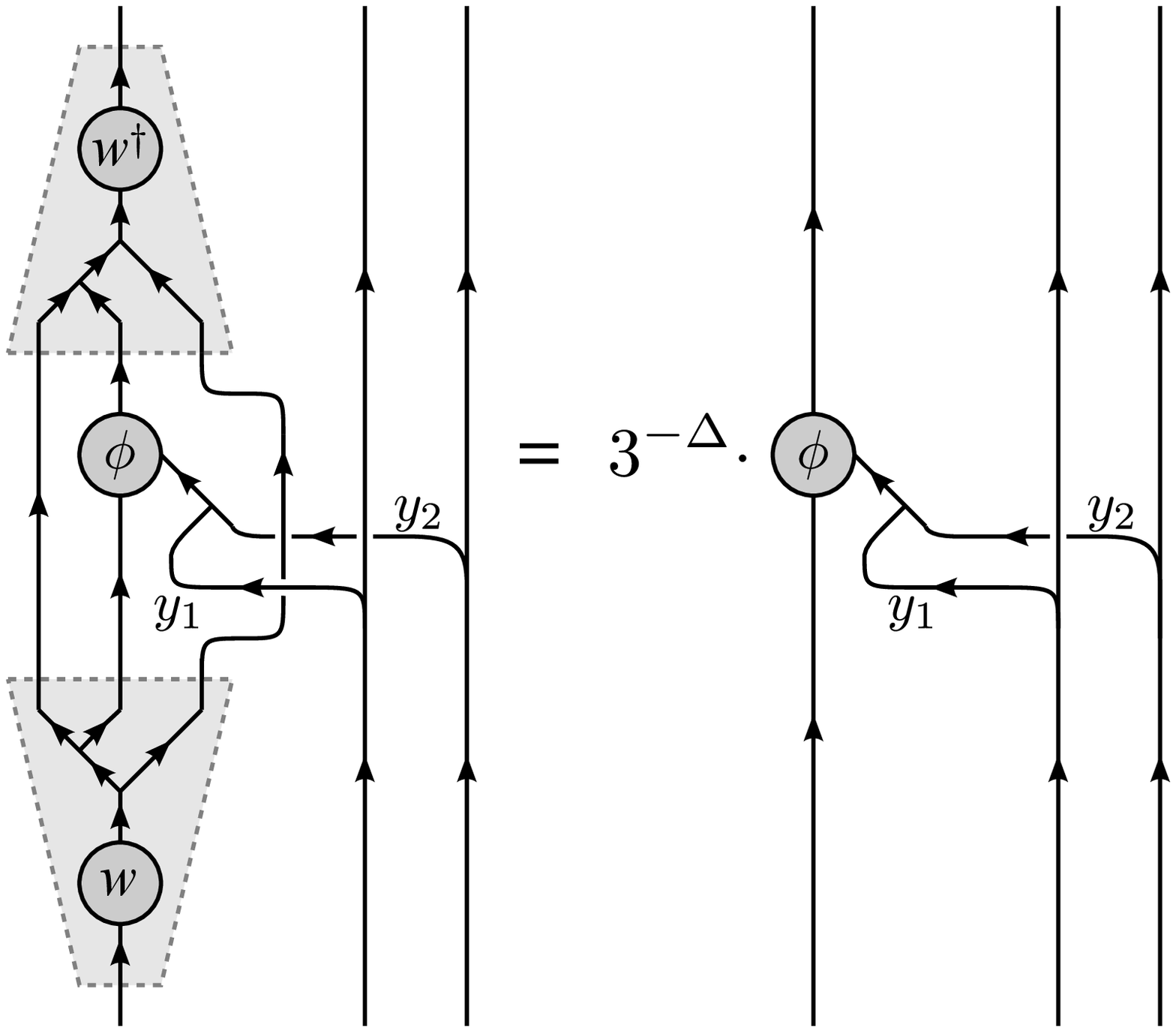}
\caption[Determination and classification of eigenoperators ($\phi$) and associated scaling dimensions ($\Delta$) for the one-site scaling superoperator of the anyonic {1D} MERA.]{Determination of eigenoperators ($\phi$) and associated scaling dimensions ($\Delta$) for the one-site scaling superoperator of the anyonic {1D} MERA. Eigenoperators may be classified according to the charges on edges $y_1$ and $y_2$. One interpretation of these labels is %
that, in
addition to the sites of the {1D} lattice, there may exist free charges lying in front of and behind the anyon chain. The labels $y_1$ and $y_2$ then represent the transfer of charge between these regions and the {1D} lattice. 
\label{fig:ch4:scalingsuperop}}
\end{center}
\end{figure}

\begin{figure}
\begin{center}
\includegraphics[width=300.0pt]{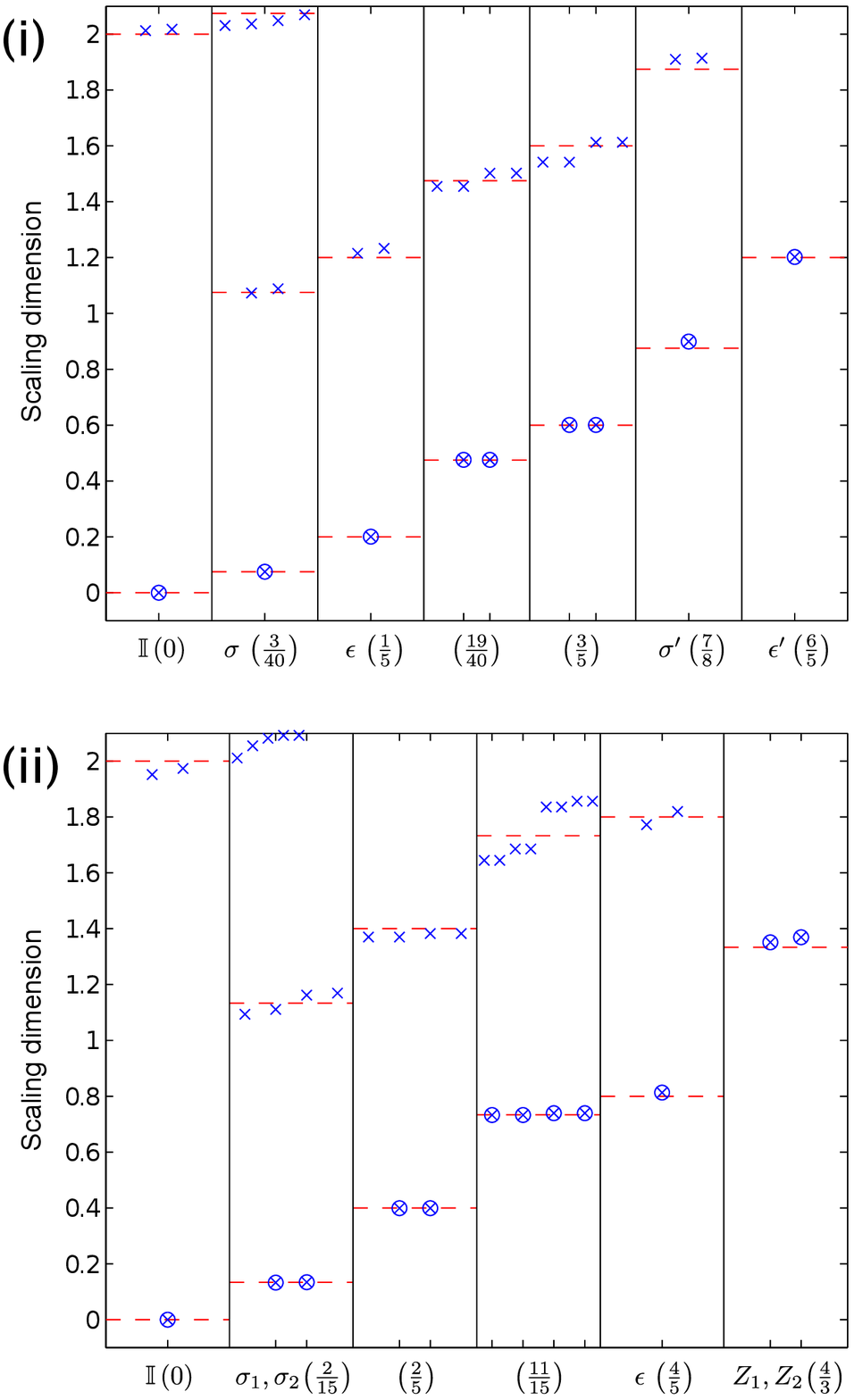}
\caption[Scaling dimensions of leading primary operators and their descendants, computed for (i)~antiferromagnetic and (ii)~ferromagnetic local Hamiltonians \protect{\eref{eq:ch4:Hamiltonians}} on the golden chain.]{Scaling dimensions of leading primary operators and their descendants, computed for %
(i)~antiferromagnetic and (ii)~ferromagnetic local Hamiltonians \protect{\eref{eq:ch4:Hamiltonians}} on the golden chain. %
Results are grouped into conformal towers, with a slight horizontal spread 
introduced to show the degeneracies of the descendant fields.
A circled cross indicates a primary field, and a plain cross indicates a descendant. Dashed lines indicate values predicted from CFT. 
\label{fig:ch4:results}}
\end{center}
\end{figure}

The AFM and FM Hamiltonians act on pairs of adjacent Fibonacci anyons. On a pair of lattice sites each carrying a charge of $\tau$, the matrix representations of the AFM and FM Hamiltonians are written
\begin{equation}
(H\alphabeta)_\mrm{AFM}=\left(\begin{array}{cc}\!\!-1\,\,&0\\\!\!\p{-}0\,\,&0\end{array}\right)
\quad
(H\alphabeta)_\mrm{FM}=\left(\begin{array}{cc}0&\p{-}0\\0&-1\end{array}\right)\label{eq:ch4:Hamiltonians}
\end{equation}
where a multi-index value of 1 corresponds to the vacuum charge, 2 corresponds to $\tau$, and the charges are non-degenerate. %
We optimised a scale-invariant MERA on the golden chain for each of these Hamiltonians%
, %
and computed local scaling operators using the tensor network given in \fref{fig:ch4:scalingsuperop}. 
The operators calculated using this diagram may be classified according to the values of the charge labels $y_1$ and $y_2$, and the scaling dimensions and conformal spins which we obtained are given in Tables~\ref{tab:ch4:results} and \ref{tab:ch4:results2}, and \fref{fig:ch4:results}.
\begin{table}
\begin{center}
\begin{tabular}{|c|c|c|}
\bottomrule
\multicolumn{3}{|c|}{$y_1=y_2=1$}\\
\hline
Exact & Numerics & Error \\
\hline
$0$ & $0$ & $0\%$\\ %
$7/8$ & $0.8995$ & $+2.80\%$\\ %
$7/8+1$ & $1.9096$ & $+1.85\%$\\
$7/8+1$ & $1.9141$ & $+2.09\%$\\
$0+2$ & $2.0124$ & $+0.62\%$\\
$0+2$ & $2.0181$ & $+0.90\%$\\
\toprule
\end{tabular}
~
\begin{tabular}{|c|c|c|}
\bottomrule
\multicolumn{3}{|c|}{$y_1=y_2=\tau$}\\
\hline
Exact & Numerics & Error \\
\hline
$3/40$ & $0.0751$ & $+0.19\%$\\ %
$1/5$  & $0.2006$ & $+0.28\%$\\ %
$3/40+1$ & $1.0730$ & $-0.19\%$\\
$3/40+1$ & $1.0884$ & $+1.25\%$\\
$6/5$   & $1.2026$ & $+0.21\%$\\ %
$1/5+1$ & $1.2156$ & $+1.30\%$\\
\toprule
\end{tabular}

~

~

\begin{tabular}{|c|c|c|}
\bottomrule
\multicolumn{3}{|c|}{$y_1=1$, $y_2=\tau$}\\
\hline
Exact & Numerics & Error \\
\hline
$19/40$ & $0.4757$ & $+0.14\%$\\ %
$3/5$   & $0.6009$ & $+0.15\%$\\ %
19/40+1 & $1.4549$ & $-1.37\%$\\
19/40+1 & $1.5022$ & $+1.85\%$\\
$3/5+1$ & $1.5414$ & $-3.66\%$\\
$3/5+1$ & $1.6129$ & $+0.80\%$\\
\toprule
\end{tabular}
~
\begin{tabular}{|c|c|c|}
\bottomrule
\multicolumn{3}{|c|}{$y_1=\tau$, $y_2=1$}\\
\hline
Exact & Numerics & Error \\
\hline
$19/40$ & $0.4757$ & $+0.14\%$\\
$3/5$   & $0.6009$ & $+0.15\%$\\
19/40+1 & $1.4549$ & $-1.37\%$\\
19/40+1 & $1.5022$ & $+1.85\%$\\
$3/5+1$ & $1.5414$ & $-3.66\%$\\
$3/5+1$ & $1.6129$ & $+0.80\%$\\
\toprule
\end{tabular}
\caption[Computed scaling dimensions for Fibonacci anyons with antiferromagnetic nearest neigbour interactions on an infinite chain.]{Scaling dimensions for Fibonacci anyons with antiferromagnetic nearest neigbour interactions on an infinite chain. Numerical values were computed using an anyonic MERA with maximum degeneracies for charges $1$ and $\tau$ of 3 and 5 respectively (denoted $\chi=[3,5]$), and are grouped according to their classification by the values of $y_1$ and $y_2$ in \protect{\fref{fig:ch4:scalingsuperop}}.
\label{tab:ch4:results}}
\end{center}
\end{table}
\begin{table}
\begin{center}
\begin{tabular}{|c|c|c|}
\bottomrule
\multicolumn{3}{|c|}{$y_1=y_2=1$}\\
\hline
Exact & Numerics & Error \\
\hline
$0$ & $0$ & $0\%$\\ %
$4/3$ & $1.3514$ & $+1.36\%$\\ %
$4/3$ & $1.3695$ & $+2.71\%$\\ %
$0+2$ & $1.9519$ & $-2.41\%$\\
$0+2$ & $1.9742$ & $-1.29\%$\\
$1+4/3$ & $2.2570$ & $-3.27\%$\\
\toprule
\end{tabular}
~
\begin{tabular}{|c|c|c|}
\bottomrule
\multicolumn{3}{|c|}{$y_1=y_2=\tau$}\\
\hline
Exact & Numerics & Error \\
\hline
$2/15$ & $0.1329$ & $-0.35\%$\\ %
$2/15$  & $0.1339$ & $+0.44\%$\\ %
$4/5$ & $0.8134$ & $+1.67\%$\\ %
$2/15+1$ & $1.0937$ & $-3.49\%$\\
$2/15+1$   & $1.1108$ & $-1.99\%$\\
$2/15+1$ & $1.1622$ & $+2.55\%$\\
\toprule
\end{tabular}

~

~

\begin{tabular}{|c|c|c|}
\bottomrule
\multicolumn{3}{|c|}{$y_1=1$, $y_2=\tau$}\\
\hline
Exact & Numerics & Error \\
\hline
$2/5$ & $0.3993$ & $-0.18\%$\\     %
$11/15$   & $0.7327$ & $-0.09\%$\\ %
$11/15$ & $0.7392$ & $+0.80\%$\\
$2/5+1$ & $1.3699$ & $-2.15\%$\\
$2/5+1$ & $1.3823$ & $-1.26\%$\\
11/15+1 & $1.6450$ & $-5.10\%$\\
\toprule
\end{tabular}
~
\begin{tabular}{|c|c|c|}
\bottomrule
\multicolumn{3}{|c|}{$y_1=\tau$, $y_2=1$}\\
\hline
Exact & Numerics & Error \\
\hline
$2/5$ & $0.3993$ & $-0.18\%$\\
$11/15$   & $0.7327$ & $-0.09\%$\\
$11/15$ & $0.7392$ & $+0.80\%$\\
$2/5+1$ & $1.3699$ & $-2.15\%$\\
$2/5+1$ & $1.3823$ & $-1.26\%$\\
11/15+1 & $1.6450$ & $-5.10\%$\\
\toprule
\end{tabular}
\caption[Computed scaling dimensions for Fibonacci anyons with ferromagnetic nearest neigbour interactions on an infinite chain.]{Scaling dimensions for Fibonacci anyons with ferromagnetic nearest neigbour interactions on an infinite chain. Numerical values were computed using an anyonic MERA with maximum degeneracies for charges $1$ and $\tau$ of 3 and 5 respectively (denoted $\chi=[3,5]$), and are grouped according to their classification by the values of $y_1$ and $y_2$ in \protect{\fref{fig:ch4:scalingsuperop}}. %
\label{tab:ch4:results2}}
\end{center}
\end{table}

Comparison of the AFM case with existing results in the literature %
show that the scaling dimensions obtained when $y_1=y_2$ 
correspond to those obtained when studying a system of anyons with a toroidal fusion diagram\citestop{feiguin2007} For a system of anyons on the torus it is possible to define an additional topological symmetry\pratext{ }\citep{feiguin2007} and classify local scaling operators according to whether or not they respect this symmetry. Operators satisfying $y_1=y_2=1$ correspond to those which respect the topological symmetry, and those satisfying $y_1=y_2=\tau$ do not. We will discuss the interpretation of the different sectors and their relationship to anyons on the torus in a forthcoming paper\citestop{pfeifer2010a}

When $y_1\not=y_2$ the scaling operators obtained are chiral, with those obtained from $y_1=1,~y_2=\tau$ and $y_1=\tau,~y_2=1$ believed to form conjugate pairs.

\section{Summary\label{sec:ch4:summary}}

Numerical study of systems of interacting anyons is difficult due to their non-trivial exchange statistics. To date, study of these systems has been restricted to exact diagonalisation, Matrix Product States (MPS) for {1D} systems, or special-case mappings to equivalent spin chains. This paper shows how any tensor network Ansatz may be translated into a form applicable to systems of anyons, opening the door for the study of large systems of interacting anyons in both one and two dimensions. As an example, this paper demonstrates how the MERA may be implemented for a {1D} anyonic system. This Ansatz is particularly important as many
{1D} systems of anyons are known which exhibit extended critical phases\pratext{ }\prbtext{.}\cite[see e.g.][]{feiguin2007,trebst2008,trebst2008a}\pratext{.} The structure of the MERA is known to be particularly well suited to reproducing long range correlations, and the scale-invariant MERA has the additional 
advantage of providing simple and direct means of computing the scaling dimensions and matrix representations of local scaling operators. %

We applied the scale invariant MERA to infinite chains of Fibonacci anyons under antiferromagnetic and ferromagnetic nearest neighbour couplings, and identified a large number of local scaling operators. Our results for the scaling dimensions are in agreement with those previously obtained by exact diagonalisation of closely related systems, and for the relevant primary fields they are within $2.8\%$ of the theoretical values obtained from conformal field theory. 
We thus demonstrate that an anyonic MERA with $\chi=[3,5]$ permits conclusive identification of the relevant conformal field theory, and gives a level of accuracy comparable to that of the scale invariant MERA on a spin chain (\tref{tab:ch2:results} of this Thesis; \citealp{pfeifer2009}).

The anyonic generalisation of the {1D} MERA presented here is useful in its own right, but the greatest significance of the approach described is that it is equally applicable to 2D tensor network Ans\"atze, and hence opens the door to studying the collective behaviour of large systems of anyons in two dimensions by numerical means, in situations where analytical solutions may not be possible.

\emph{Note---}Simultaneous with the work presented in this Chapter, the 1D MERA for systems of anyons was also independently constructed by \citet{konig2010} on the torus. These authors provide proof of principle by computing ground state energies and two-point correlators for finite systems of Fibonacci anyons with $\chi=[1,1]$ ($s=2$ in their notation), with errors in the energy on the order of a few percent. Once again we see that even for small values of $\chi$, the anyonic MERA is capable of providing an accurate description of the low-energy behaviour of a system of interacting anyons.

\section{Some Notes on Implementation\label{sec:ch4:implementation}}

\subsection{Block Structure\label{sec:ch4:blockstructure}}

Recall that in \sref{sec:ch3:comp} we saw that an efficient way of storing Abelian symmetric tensors was as a number of blocks, and that operations on these tensors such as fusing and splitting legs amounted to nothing more than a rearrangement of these blocks. It is possible to implement a similar scheme for anyonic tensor networks, although with some important differences.

\begin{figure}
\begin{center}
\raisebox{-70pt}{\includegraphics[width=200.0pt]{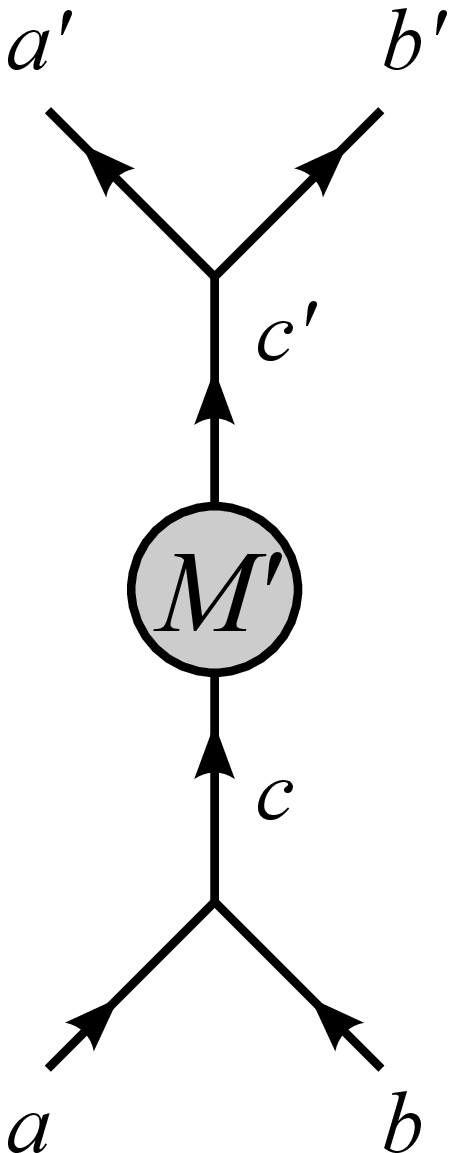}}
~
\begin{tabular}{|c|c|}
\hline
Block number & Dimensions \\
\hline\hline
$\mc{B}_1$ & $1\times 1$ \\
$\mc{B}_2$ & $1\times 4$ \\
$\mc{B}_3$ & $4\times 1$ \\
$\mc{B}_4$ & $4\times 4$ \\
$\mc{B}_5$ & $2\times 2$ \\
$\mc{B}_6$ & $2\times 2$ \\
$\mc{B}_7$ & $2\times 2$ \\
$\mc{B}_8$ & $4\times 2$ \\
$\mc{B}_9$ & $4\times 2$ \\
$\mc{B}_{10}$ & $2\times 4$ \\
$\mc{B}_{11}$ & $2\times 4$ \\
$\mc{B}_{12}$ & $4\times 4$ \\
\hline
\end{tabular}
\\\rule{0pt}{1.8ex}\\
\begin{tabular}{|cr|c|c|c|c|c|}
\cline{3-7}
\multicolumn{2}{c|}{} & \multicolumn{5}{c|}{(a,b,c)~~} \\
\multicolumn{2}{c|}{} & \multicolumn{1}{c}{1,1,1} & \multicolumn{1}{c}{$\tau$,$\tau$,1} &
\multicolumn{1}{c}{1,$\tau$,$\tau$} & \multicolumn{1}{c}{$\tau$,1,$\tau$} & 
\multicolumn{1}{c|}{$\tau$,$\tau$,$\tau$}\\
\cline{1-7}
\multirow{10}{*}{\begin{sideways}(a',b',c')~~~\end{sideways}} & \multirow{2}{*}{1,1,1} & $\mc{B}_1\quad 1$ & $\mc{B}_2\quad 1$ & & & \\
& & & & & &\\
\cline{3-7}
& \multirow{2}{*}{$\tau$,$\tau$,1} & $\mc{B}_3\quad 1$ & $\mc{B}_4\quad 1$ & & & \\
& & & & & &\\
\cline{3-7}
& \multirow{2}{*}{1,$\tau$,$\tau$} & & & $\mc{B}_5\quad \frac{1}{\sqrt{2}}$ & $\mc{B}_7\quad\rmi$ & $\mc{B}_{10}\quad 1$ \\
& & & & $\mc{B}_6\quad \frac{1}{\sqrt{2}}$ & & \\
\cline{3-7}
& \multirow{2}{*}{$\tau$,1,$\tau$} & & & $\mc{B}_7\quad -\rmi$ & $\mc{B}_5\quad\frac{1}{\sqrt{2}}$ & $\mc{B}_{11}\quad 1$ \\
& & & & & $\mc{B}_6\quad \frac{-1}{\sqrt{2}}$ & \\
\cline{3-7}
& \multirow{2}{*}{$\tau$,$\tau$,$\tau$} & & & $\mc{B}_8\quad 1$ & $\mc{B}_9\quad 1$ & $\mc{B}_{12}\quad 1$ \\
& & & & & & \\
\cline{1-7}
\end{tabular}
\caption[Map-based implementation of the central object $M_\alpha^{~\beta}$ of an anyonic tensor.]{Map-based implementation of the central object $M_\alpha^{~\beta}$ of an anyonic tensor: 
The diagram shows an operator $\hat M$ which acts on two adjacent sites of a 1D lattice, where each site may carry charge $\mbb{I}$ with degeneracy 1, or charge $\tau$ with degeneracy 2. The associated tables show how the central object $M_\alpha^{~\beta}$ of this particular operator is assembled in terms of its constituent blocks.
If appropriate, a block may appear in more than one location in the map which describes $M_\alpha^{~\beta}$, and a location on the map may contain more than one block.\label{fig:ch4:mapstructure}}
\end{center}
\end{figure}

As with the Abelian symmetric tensors of \cref{sec:abelian}, we construct the central objects (e.g. $M\alphabeta$) of our tensors from a number of blocks. However, in contrast to the Abelian tensors, we no longer simply assemble these blocks into the final object. Instead, we give each block a unique identifying index, and introduce a \emph{map}. This map then records the locations of the blocks, with each entry in the map corresponding to a unique labelling of all fusion trees (\fref{fig:ch4:mapstructure}). Note that a given entry in the map may contain a list of more than one block, and each block listed is associated with a numeric multiplier. To reconstruct the tensor from the map and blocks, each entry in the map is assembled by summing over the relevant blocks, each multiplied by their associated numeric multiplier.

This system may at first seem unwieldly. However, it has a number of advantages. First, $F$ moves may be performed quickly and efficiently, with linear recombinations of the blocks of the tensor being performed simply by modifying the map. Second, braiding will permute the entries within individual blocks; efficiency gains may be made %
by deferring these permutations for as long as possible, letting them accumulate, and then determining and performing a single, cumulative operation on each block. By being able to perform $F$ moves without accessing the contents of individual blocks, it becomes unnecessary to evaluate these deferred permutations when performing $F$ moves (or indeed any other unitary operations depending only on the charges, such as the reversal of Frobenius--Schur indicators), and this can lead to greater computational efficiency. 

Further, the numerical factors associated with this braiding may also be introduced at the level of the map. This is typically an advantage when a particular charge labelling of the fusion trees is multiply degenerate, as is common at higher levels of the MERA. Each charge labelling is still only associated with a single block, and so applying these numerical factors at the level of the map requires less operations than applying them directly to the numerical content of the tensor.

Finally, in \sref{sec:ch4:siteexpanyop} it was mentioned that it could be convenient to separate out the handling of normalisation factors associated with the diagrammatic isotopy convention. This takes place at two levels. First, when performing an operation such as ``raising'' an anyonic tensor to act on the fusion space of a larger number of sites, it is frequently possible to observe that factors coming from the introduction of vertices exactly cancel those arising due to the presence of loops; this is the reason why no factors of quantum dimensions $d_a$ appear in \eref{eq:ch4:Mprime}. Second, when drawing an anyonic tensor with fusion trees, the vertex normalisation factors may be kept associated \emph{with the vertex}. They therefore do not enter into $M\alphabeta$ except when $M\alphabeta$ absorbs a vertex during the process of splitting or fusion of multi-indices. During splitting, when a vertex is absorbed, the associated factors of $[d_c/(d_ad_b)]^\frac{1}{4}$ are applied to the coefficients in the map, and not the entries of the tensor itself. During fusion, again the coefficients are applied only to the map, only this time they account for both the vertex itself and the loop which it makes with the central object of the tensor.

With appropriate care, any effects of operations on the contents of individual blocks may be deferred until an operation is performed which by its nature must access the contents of the blocks, such as a singular value decomposition or a matrix multiplication.

\subsection{Precomputation}

If an algorithm employs repeated application of the same series of tensor manipulations, for example the repeated iterations of optimisation for the MERA, then many calculations involved in these operations may be stored, and recycled on subsequent iterations. Examples are the matrices generated by $F$ moves and braids (\fref{fig:ch4:anyonopmanip1}), and the permutations of the numerical elements of each block which are generated by braiding.
It is now that the ability to defer permutations of the elements of a tensor really comes into its own, as it is only necessary to store the cumulative operation for subsequent iterations, and not each individual step.

It should be recognised that this Section describes only one particular scheme for the efficient implementation of anyonic tensors, and that this approach is by no means necessarily the only means of achieving this. However, for those interested in pursuing this approach further, additional discussion of the philosophy of precomputation [in this instance, as applied to U(1)-symmetric tensors] may be found in the Appendix of \citet{singh2011a}, and \sref{sec:ch3c:supplement} of this Thesis.

\subsection{Reminder: The Important Difference Between Fusion and Splitting Trees}

Given the implementation-oriented nature of this final Section of the Chapter, it seems appropriate to include a timely reminder of the difference between fusion and splitting trees. Recall that a tree assembled from fusion vertices describes the state of a system, $|\psi\ra$, and one assembled from splitting vertices describes a state in the dual space, $\la\psi|$, where Hermitian conjugation is performed by vertical reflection of a tree diagram and complex conjugation of its coefficients.

Note well that this same rule for Hermitian conjugation applies also to the vertex tensors $\tilde N_{\alpha\beta u}^\gamma$ and $\tilde N_\gamma^{\dagger\alpha\beta u}$, and thus fusing and splitting of legs takes place differently depending on whether it is acting on a fusing or a splitting tree. This important distinction will affect not only the arrangement of the charge blocks, but also of the entries within each block itself, and the correct implementation of these processes is one of the cornerstones for implementation of an anyonic tensor network.

\chapter{Non-Abelian Symmetries of Spin Systems\label{sec:nonabelian}}

\section{Unitary Braided Tensor Categories and Group Symmetries}

The formalism developed in \cref{sec:anyons} constitutes a methodology for performing tensor network simulations of any physical system which admits a description in terms of a UBTC. As mentioned in the introduction to that Chapter, it is also possible to associate a UBTC with a group $\mc{G}$, where the $F$ moves are related to the 6-$j$ symbols of the group, and the tensor $R^{ab}_c$ is related to the choice of universal braid matrix (which describes the exchange properties of the irreps). Note that the choice of universal braid matrix is not in general unique; in fact, we have already seen two systems with $Z_2$ symmetry but different braiding. The first was the spin-0 formulation of the Ising model of \Eref{eq:ch3c:HIsing}, 
\begin{equation}
\hat H_\mrm{Ising}=-\sum_s \sigma_x^{(s)}\sigma_x^{(s+1)}-h\sigma_z^{(s)},\tag{\ref{eq:ch3c:HIsing}}
\end{equation}
where the $Z_2$ symmetry is associated with a $\pi$-radian rotation. If we write the charge labels of $Z_2$ as 0 $(\equiv +)$ and 1 ($\equiv -$), then the fusion rules may be written 
\begin{equation}
a\times b\rightarrow (a+b)|_2\label{eq:ch5:abfuse}
\end{equation}
(where $|_2$ denotes that the addition is performed modulo 2), and the nonzero entries in $R^{ab}_c$ are given by
\begin{equation}
R^{ab}_{(a+b)|_2}=1.\label{eq:ch5:Z2Rab1}
\end{equation}
The second example of a system with $Z_2$ symmetry is any system of fermions, as discussed in \sref{sec:ch3c:fermions}, for example the fermionic formulation of the Ising model:
\begin{equation}
\hat H'_\mrm{Ising}=-J\sum_s\left(c^{\dagger(s)}c^{(s+1)}+c^{\dagger(s+1)}c^{(s)}+c^{\dagger(s)}c^{\dagger(s+1)}+c^{(s+1)}c^{(s)}-2gc^{\dagger(s)}c^{(s)}+g\right).
\end{equation}
Here, the charges of $Z_2$ correspond to parity, indicating the presence or absence of a fermion at a site $s$. The operators $c^{\dagger(s)}$ and $c^{(s)}$ in the Hamiltonian are fermionic creation and annihilation operators, and the non-zero entries in $R^{ab}_c$ for the associated UBTC 
are given by
\begin{equation}
R^{ab}_{(a+b)|_2}=(-1)^{ab}.\label{eq:ch5:Z2R-}
\end{equation}
As $Z_2$ is Abelian, the non-zero $F$ moves in both examples are simply
\begin{equation}
\left[F^{abc}_{(a+b+c)|_2}\right]_{(a+b)|_2\,(b+c)|_2} = 1.\label{eq:ch5:Z2F}
\end{equation}
[There are no vertex indices $u_1$, $u_2$, $u_3$, $u_4$ (see \fref{fig:ch4:anyons}) in this expression, as all fusion products are non-degenerate.]

For a non-Abelian group such as SU(2), both the $F$ tensor and $R^{ab}_c$ may be more complicated. The charges in SU(2) are the non-negative half-integers, with fusion rules
\begin{equation}
a\times b\longrightarrow \sum_{c=|a-b|}^{a+b} c,\label{eq:ch5:SU2fusionrules}
\end{equation}
and we may write the $F$ moves for SU(2) as
\begin{equation}
\left(F^{abc}_d\right)_{ef} = (-1)^{(a+b+c+d)}\sqrt{(2e+1)(2f+1)}\left\{\begin{array}{ccc}a&b&e\\c&d&f\end{array}\right\}\label{eq:ch5:SU2F}
\end{equation}
where $\left\{\begin{array}{ccc}a&b&e\\c&d&f\end{array}\right\}$ denotes the 6-$j$ symbol
\begin{align}
\begin{split}
\left\{\begin{array}{ccc}a&b&e\\c&d&f\end{array}\right\} &= \Delta(a,b,e)\ \Delta(e,c,d)\ \Delta(b,c,f)\ \Delta(a,f,d)\\
&\times\sum_z\left[\frac{(-1)^z(z+1)!}{(z-a-b-e)!(z-e-c-d)!(z-b-c-f)!(z-a-f-d)!}\right.\\
&~~~~~\times\left.\frac{1}{(a+b+c+d-z)!(a+e+c+f-z)!(b+e+d+f-z)!}\right],
\end{split}\\
\Delta(a,b,c)&=\sqrt{\frac{(-a+b+c)!(a-b+c)!(a+b-c)!}{(a+b+c+1)!}},
\end{align}
with the sum running over all integer values of $z$ such that the factorials are of non-negative numbers. %
When representing quantum mechanical spin, the half-integer charges are fermionic, and thus the $R^{ab}_c$ tensor is given by
\begin{equation}
R^{ab}_c = (-1)^{(c-a-b)}\label{eq:ch5:SU2Rabc}
\end{equation}
for any combination of $a$, $b$, and $c$ permitted by the fusion rules \eref{eq:ch5:SU2fusionrules}. 

Given the $F$ tensor and the particle exchange tensor $R^{ab}_c$, we may apply the UBTC formalism of \cref{sec:anyons} to any quantum mechanical system which exhibits a group symmetry. For Abelian symmetry groups, %
much of this machinery is redundant and we may prefer the simpler approach outlined in \cref{sec:abelian}. For non-Abelian symmetries, however, this provides a %
useful means of %
exploiting those symmetries.

Expressions \eref{eq:ch5:abfuse}, \eref{eq:ch5:Z2Rab1}, and \eref{eq:ch5:Z2R-}--\eref{eq:ch5:SU2Rabc} in this Section are adapted from expressions found in Chapter~5 of \citet{bonderson2007}.

\section{Fermions revisited\label{sec:ch5:fermions2}}

Before presenting the application of the formalism of \cref{sec:anyons} to an example of a system exhibiting a non-Abelian symmetry, we will first consider its application to a %
system of fermions. In \sref{sec:ch3c:fermions} we saw that adding fermionic statistics to a symmetric tensor network algorithm involved introducing extra factors of $-1$ into permutation operations, and also into the multiplications used to perform tensor contraction. This process is counter-intuitive and an approach would be preferable in which factors associated with particle exchange arise only during index permutation. This may achieved by representing fermionic systems using the $F$ and $R^{ab}_c$ tensors of Eqs.~\eref{eq:ch5:Z2F} and \eref{eq:ch5:Z2R-} in the UBTC formalism of \cref{sec:anyons}. However, in truth, the full machinery of the UBTC tensor network formalism is not necessary. Instead, it suffices to simply assign a vertical orientation to each leg and to contract pairs of upgoing legs using the expanded multiplicity tensor $\tilde N^\gamma_{\alpha\beta}$, and pairs of downgoing legs using its Hermitian conjugate $\tilde N^{\dagger\alpha\beta}_{\gamma}$, as per \sref{sec:ch4:siteexpanyop}. When contracting two tensors together, the counterclockwise ordering of indices on one of these tensors is now the \emph{opposite} of that on the other, e.g.
\begin{equation}
\rbimg{-55pt}{130pt}{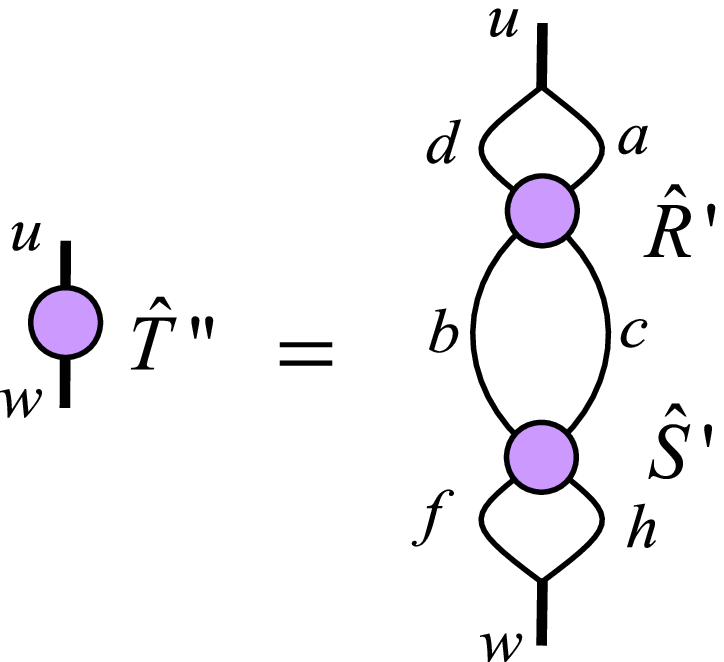}\quad,
\end{equation}
and there are no longer any concealed particle exchanges within the equivalent of \fref{fig:ch3c:multiply2}(3),
\begin{equation}
\rbimg{-30pt}{105pt}{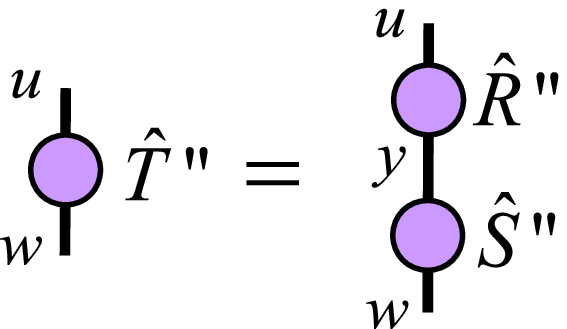}\quad,
\end{equation}
which is now just a simple matrix multiplication.

\section{The 1D Spin-$\frac{1}{2}$ Heisenberg (Anti)Ferromagnet}

The procedure for studying systems with non-Abelian symmetries using a UBTC-based tensor network is largely the same as that described in \cref{sec:anyons}, with only a couple of minor differences.
First, until now I have implicitly employed a convention where every fusion tree has a total charge of $\mbb{I}$, and represents a single state. It is also possible to use a fusion tree with a total charge $a\not=\mbb{I}$ to represent a single state, where a total charge of $a$ is taken to imply the existence of an ancillary system not explicitly considered, and having charge $\bar a$ (for example, for a system of anyons on a finite disc this ancillary charge may live on the boundary). However, in this Section it is instead preferable to use a fusion tree with a total charge of $a$ to represent a subspace of the Hilbert space having dimension $d_a$.\footnote{There is a subtlety here, in that for a system with SU(2) symmetry, we might choose to define the Hilbert space such that an orthonormal basis is given by eigenstates of a complete set of SU(2)-symmetric commuting operators. Under this choice, a fusion tree always corresponds to a single state, regardless of total charge. More commonly, however, for systems exhibiting a non-Abelian symmetry we define the Hilbert space with respect to a complete set of commuting operators on the \emph{microscopic} degrees of freedom of the system. These measurements are not necessarily SU(2)-symmetric, e.g. measurement of spin in the $z$ basis for a Heisenberg spin chain. In the resulting basis of this example, a pair of spin-$1/2$ fermions can have total spin~0 in precisely one way, or total spin~1 in three orthogonal ways, corresponding to the SU(2)-symmetry-breaking measurement of $z$-axis spin permitting resolution of a three-dimensional subspace for the spin-1 triplet [which has a total SU(2) ``charge'' of 1, with $d_1=3$].}
Second, if we are considering systems with only fermionic and/or bosonic statistics, it is not necessary to specify the orientation of a braid. Thus we may denote particle exchange simply by line crossings,
\begin{equation}
\rbimg{-30pt}{224pt}{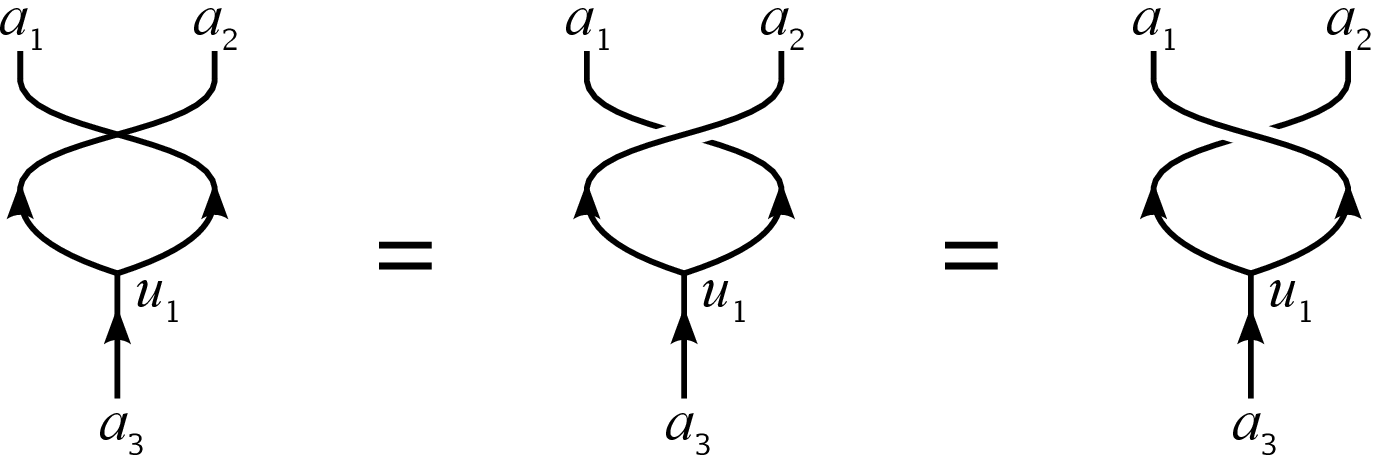}\quad.
\end{equation}

As examples of systems with SU(2) invariance, we shall consider the 1D spin-$\frac{1}{2}$ Heisenberg  antiferromagnet and ferromagnet, with periodic boundary conditions. The former model exhibits a nearest-neighbour Hamiltonian which favours neighbouring pairs of particles occupying the singlet state, with total spin 0,
\begin{equation}
\rbimg{-42pt}{100pt}{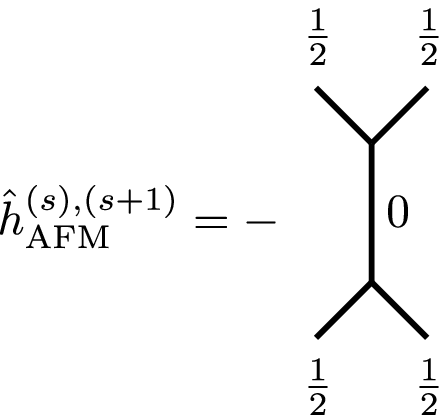}\quad,
\end{equation}
whereas the latter favours occupation of the triplet sector, with spin 1,
\begin{equation}
\rbimg{-42pt}{100pt}{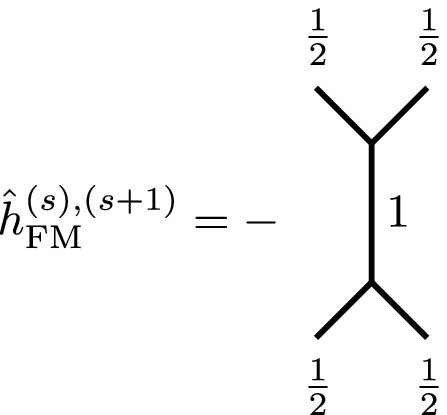}\quad.
\end{equation}
On two sites, the total Hamiltonians may be written as
\begin{align}
\rbimg{-66pt}{215pt}{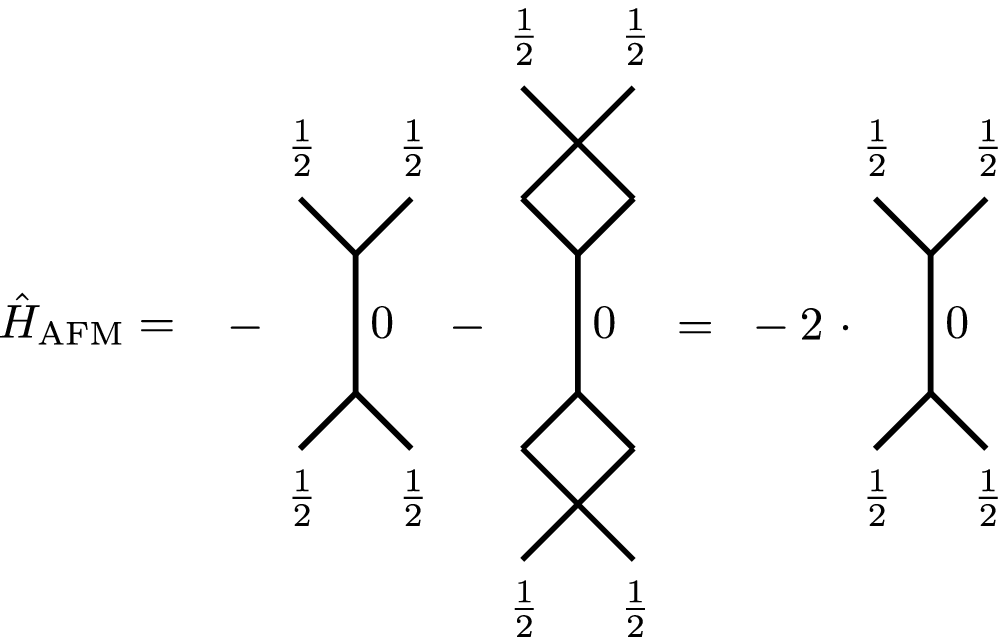}\quad,\\
\rbimg{-66pt}{215pt}{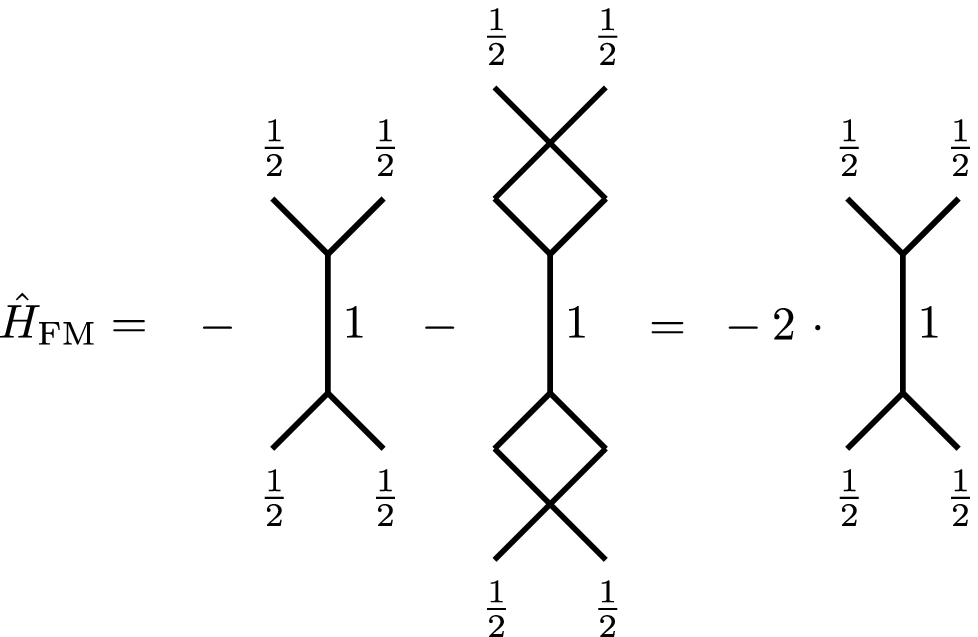}\quad,
\end{align}
and the diagrams
\begin{equation}
\rbimg{-20pt}{30pt}{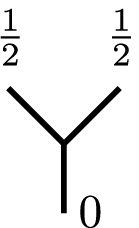}\qquad\textrm{and}\qquad\rbimg{-20pt}{30pt}{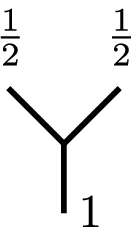}
\end{equation}
correspond to eigensubspaces of these Hamiltonians, with the former representing the spin-0 singlet state [total SU(2) ``charge'' 0; $d_0=1$, hence a state] and the latter representing the 3-dimensional spin-1 triplet sector [total SU(2) ``charge'' 1; $d_1=3$, hence a 3-dimensional subspace]. For the given antiferromagnetic Hamiltonian these diagrams have eigenvalues $-2$ and 0 respectively, whereas for the ferromagnetic Hamiltonian these eigenvalues are reversed.

This concludes a very simple demonstration of the application of UBTC tensor networks for the exploitation of non-Abelian symmetries. A fuller treatment will be provided in \citet{singhvidalinprep}.

%In general, it is recommended that for any system having non-trivial $F$ moves (including anyonic systems based on Abelian groups, such as $Z_2$ semions) 
\chapter{Summary and Outlook\label{sec:summary}}

In this Thesis, we have seen how symmetries---both spatial and internal---may be exploited in tensor network algorithms. We began in \cref{sec:SIMERA} with the exploitation of scale invariance, constructing an Ansatz which naturally reflects the entanglement structure present in quantum critical systems. Applying this Ansatz to the critical Ising and three-state Potts models in 1D, we 
were able to extract most of the conformal data of the CFTs which describe the continuum limit of these theories, namely the scaling dimensions, central charge, and the coefficients of the operator product expansion. 

In \cref{sec:abelian} we described the mathematical background behind internal symmetries of lattice models, and developed basic techniques to exploit Abelian symmetries in tensor network algorithms. These techniques enabled us to address specific symmetry sectors of models such as the $XX$ and Heisenberg models, and to simulate these systems at a substantially reduced computational cost. In their own rights, these techniques therefore substantially increase the power of tensor network algorithms as tools for the study of symmetric systems.

In \cref{sec:anyons}, we turned our attention to systems of anyons. The simulation of anyonic systems is acknowledged to be challenging: Like fermions, the study of anyons suffers from the sign problem,\footnote{The sign problem may be paraphrased as ``For the class of Hamiltonians obtained by taking a bosonic system whose ground state energy may be computed in polynomial time and introducing fermionic particle exchange statistics, does there exist an algorithm similarly capable of computing the ground state energy of these fermionic Hamiltonians in polynomial time?'' This question, which is stated more formally \citet{troyer2005}, remains unanswered, and as demonstrated by \citeauthor{troyer2005},
is %
in fact NP-hard. Specific solutions are known for many problems and problem groups, but there exist many other such fermionic systems whose ground state may at present only be computed for a cost exponential in the system size, even though their bosonised equivalents may be solved in polynomial time. There are of course bosonic systems (such as frustrated systems, and spin glasses) which are also exponentially hard, and their fermionic counterparts tend to be exponentially hard as well. It is the presence of the middle ground, where fermions are ``harder'' than bosons, to which the term ``the sign problem'' is usually %
applied.}
and the problem of developing a general algorithm to compute the ground state of \emph{arbitrary} fermionic systems in polynomial time is therefore known to be \emph{at least} NP-hard \citep{troyer2005}%
. Nevertheless, by means of a non-trivial generalisation of the techniques introduced in \cref{sec:abelian}, we were able to develop a formalism of tensor networks for anyons, allowing us to compute a close approximation to the ground state of an anyonic system in polynomial time, provided the entanglement structure of that ground state may be effectively represented by an appropriate tensor network algorithm (see the discussion on entanglement in \sref{sec:ch2:sc-inv}).

Finally, in \cref{sec:nonabelian} it was seen that the formalism developed for anyons in \cref{sec:anyons} may also be applied to exploit non-Abelian symmetries of spin systems, such as the SU(2) symmetry of the Heisenberg (anti)ferromagnet, or indeed to exploit Abelian symmetries in the presence of possibly non-trivial exchange statistics. In fact, the formalism of \cref{sec:anyons} may be used to study systems of bosons, fermions, Abelian and non-Abelian anyons, and to exploit the presence of Abelian and non-Abelian internal symmetries of the Hamiltonian in any of these systems.\footnote{With minimal modification, the formalism may even be applied to systems where particle exchange is not possible, and the system is a 1D open chain described by a unitary tensor category admitting no solutions to the hexagon equation; however, the author is as yet unaware of any interesting physical models of this form.}

The exploitation of symmetries is a powerful tool, vastly increasing the reach and power of 
tensor network algorithms as a condensed matter technique. 
Of all the developments described above, perhaps the most exciting is the 
extension of tensor network algorithms to
anyons, opening the door to the study of a vast array of condensed matter systems,
many of which have never been studied before. Many of the questions to %
be asked are of great significance---for example, consider the Fibonacci anyons studied in \sref{sec:ch4:results}. They can implement universal quantum computation through braiding alone, are believed to appear as quasiparticles in the $\nu=12/5$ fractional quantum Hall state, and yet we are only beginning to understand their phase diagrams %
under even the simplest of interactions \citep[e.g.][]{trebst2008}. Anyonic tensor networks are a powerful tool for asking fundamental questions about such systems, 
and could be of vital importance to the coming quantum revolution in information %
processing.
They also provide an unrivalled opportunity to gain insight into this fascinating and comparatively little-understood area of condensed matter physics.

This is an exciting time to be working on anyons!

%This is a great time to be working on anyons.

% --- End of main body of thesis
% --- Generate bibliography etc.
% Note: Bibliography style can be set up in header.tex, or here.

% Specify bibliographystyle. Necessary when using bibtex.
% Should be natbib-compliant if using natbib.
%\bibliographystyle{alpha}
%\bibliographystyle{unsrt}
\bibliographystyle{plainnat_dotfill} % Natbib default styles are plainnat, abbrvnat, unsrtnat
  % The _dotfill versions look nice when using the pagebackref option with the hyperref
  % package: They put a line of dots between the citation and the backreferences.
  % They also include support for the eprint field in BibTeX entries of type "misc".

\bibliography{thesis,Extra} % Generate bibliography (using BibTeX) - list all .bib files here

\end{document}